\documentclass[11pt]{article}
\usepackage[utf8]{inputenc}
\usepackage{geometry}
\usepackage{amsmath}
\usepackage{amssymb}
\usepackage{graphicx}
\usepackage{bm}
\usepackage{mathtools}
\usepackage{hyperref}
\usepackage{tikz}
\usepackage{setspace}
\usepackage[round]{natbib}
\usepackage{subcaption}
\usepackage{bbm}
\usepackage{mathbbol}
\usepackage{cleveref}
\usepackage{enumitem}
\usepackage{tabularx}

\usetikzlibrary{arrows, automata}
\usepackage{fullpage}
\newtheorem{proposition}{\small\sc Proposition}
\newtheorem{assumption}{\small\sc Assumption}

\newcommand{\indep}{\rotatebox[origin=c]{90}{$\models$}}

\newcommand{\E}{\mathbbm{E}}

\newcommand{\1}{\mathbbm{1}}

\DeclareMathOperator{\var}{var}
\DeclareMathOperator{\cov}{cov}

\title{Targeted Function Balancing: Flexible balancing weights that leverage strategic imbalance to estimate causal effects\thanks{This research project is affiliated with the Los Angeles Education Research Institute and has benefited from its research and partnership infrastructure. We thank Chad Hazlett for advising us throughout the course of this project. We also thank Kenai Burton-Heckman for leading the development of the \texttt{R} Package \href{https://github.com/bleth/tfb}{\texttt{tfb}}, which implements the method proposed here. 
We also thank Erin Hartman, Onyebuchi Arah, Mark Handcock, Avi Feller, Yixin Wang, the UCLA Practical Causal Inference Lab, Meredith Phillips, Mollie Rudnick, Emilio Chavez Herrerias, the Los Angeles Unified School District's Office of Research and Program Evaluation, and anonymous reviewers for their valuable comments and suggestions. } 
}
\date{\today}

\author{
    \hspace{-0.7in}
    Leonard Wainstein\thanks{Assistant Professor, Mathematics and Statistics Department, Reed College; Email: \href{mailto:lwainstein@reed.edu}{lwainstein@reed.edu}} 
    \hspace{1in}
    He Bai\thanks{University of Massachusetts Amherst; Email: \href{mailto:hbai@umass.edu}{hbai@umass.edu}}
} 

\begin{document}

\setcounter{page}{0}
\maketitle

\thispagestyle{empty}

\begin{abstract}
When estimating the average treatment effect of a binary intervention with weights, investigators are often burdened with how best to leave imbalance in the distributions of observed covariates. This requires that investigators proceed cautiously, as imbalance in key (nonlinear transformations of) covariates potentially introduces bias into the ultimate estimated effect. This paper introduces Targeted Function Balancing (TFB), a flexible covariate balancing weights approach that simplifies this problem for the researcher. TFB first regresses an outcome on covariates, and then selects weights that balance functions (of the covariates) that are probabilistically near the resulting regression function. This yields balance in the regression function's predicted values and the covariates, with the regression function's estimated variance determining how much balance in the covariates is sufficient. Notably, TFB demonstrates that intentionally leaving imbalance in some covariates can increase efficiency without introducing bias, challenging traditions that warn against imbalance in any variable. Though, TFB still does prioritize balance in the covariates that are most influential on the outcome. Additionally, TFB is entirely defined by a regression function and its estimated variance, turning the problem of how best to balance the covariates into how best to model the outcome. Kernel regularized least squares, the LASSO, and Bayesian Additive Regression Trees are considered as regression estimators. 
We apply TFB to estimate the effects of taking mathematics courses in twelfth grade in the Los Angeles Unified School District. The \texttt{R} Package \href{https://github.com/bleth/tfb}{\texttt{tfb}} implements TFB. 
\end{abstract}

\vspace{0.1in}
\vspace{0.2in}

\pagebreak
\clearpage

\onehalfspacing

\newpage
\setcounter{page}{1}

\section{Introduction}

When estimating the causal effect of an intervention, $D$, on an outcome of interest, $Y$, in observational settings, weighting is commonly employed. Given covariates, $X$, investigators seek weights that render the control (i.e., untreated) group more similar to the treated group on $X$. Traditionally, low ``imbalance", or the difference in the means between the treated and the control groups, of $X$ after weighting has been used as a benchmark for success for the chosen weights (e.g., \citealp{stuart2010matching}). For example, using a setting investigated by \cite{wainstein2023access, wainstein2023success} and that we return to below, to estimate the effects of taking a mathematics course in twelfth grade (the treatment), it would seem ideal to choose weights that allow the (weighted) averages of key student-level characteristics (the covariates), such as GPAs and standardized test scores, to be roughly equal among those who took math and those who did not. While we push back on this convention here, it has provided motivation for a leading family of weighting methods that we focus on: ``balancing" weights, which directly seek to reduce the imbalance in $X$, or seek ``balance" in $X$ (e.g., \citealp{hainmueller2012entropy, chan2016globally, zubizarreta2015stable, wang2020minimal}).

Often, the investigator must decide how to leave imbalance in $X$ when weighting.\footnote{These decisions can be made explicitly, for example by manually choosing imbalance thresholds for each covariate in \cite{zubizarreta2015stable}. However, they are often implicit from the choice of weighting method. For example, both \cite{hazlett2018kernel} and  \cite{kallus2020generalized} can balance the same high-dimensional basis set, but leave imbalance in it in different ways.} While some balancing methods do equate the means of $X$ in both groups, or balance $X$ ``exactly" (e.g., \citealp{hainmueller2012entropy}; \citealp{chan2016globally}), this is not feasible when the dimension of $X$ is large. This may happen because the dimension of $X$ is large to start, which is increasingly more likely in the current age of data-driven decision-making, as increasingly more information is being recorded. Or, the investigator may be following modern advice (e.g., \citealp{hazlett2018kernel, kallus2020generalized}) to instead, or additionally, balance nonlinear transformations of $X$ (e.g., polynomial terms, or basis functions) to make the ultimate estimate more robust to nonlinearities in the relationships between $Y$, $D$, and $X$. Because imbalance in certain elements of $X$ can be more damaging than in others, depending on their influence on $Y$, one must proceed cautiously when prioritizing balance in select dimensions of $X$, if at all. This also requires the investigator to exercise caution when adding to $X$, or exchanging it with, its high-dimensional transformations, as the average imbalance across $X$ typically increases as does its dimension. For example, when estimating the effects of taking twelfth grade math on college enrollment, it would seem essential to prioritize balance in academics-related variables, such as GPAs and standardized test scores. But how do we prioritize balance in these variables versus, for example, expected educational attainment? Is it worth seeking balance in the squared or interaction terms of these variables if it means the weights must leave more imbalance in their untransformed terms?

Our proposed method, Targeted Function Balancing (TFB), simplifies the problem of how to best leave imbalance in $X$ by reframing it as one that is more straightforward, particularly with modern advances in machine learning: how best to model $Y$ as a function of $X$ (and $D$). 
We begin by noting that while balance in $X$, or a transformation of it, may be sufficient for an unbiased estimator, it is not required. In fact, when estimating the average treatment effect on the treated with a weighted difference in means, all that is required is exact balance in the conditional expectation function of $Y$ given $X$ and without treatment, $f_0 (X) = \E(Y \ | \ D = 0, X)$, i.e., equality of the means of $f_0 (X)$ in the treated and control groups after weighting. From this perspective, balance in $X$ is excessive. To illustrate, were $f_0$ linear in $X$, balance in one linear function of $X$ (i.e., $f_0$) would be sufficient, but weights that balance $X$ also balance \textit{all} linear functions of it --- an inefficient strategy at best, and can hinder proper balance in $f_0 (X)$. This perspective motivates TFB, which more directly targets balance in $f_0 (X)$ by estimating it with $\hat{f}_0 (X)$, and choosing weights that balance functions near $\hat{f}_0$, in hopes that $f_0$ is among them. This yields balance in $X$ \textit{and} $\hat{f}_0 (X) $, with the estimated variance of $\hat{f}_0$ determining how much balance in $X$  is sufficient. 

TFB has several notable characteristics. First, and foremost, TFB may \textit{intentionally} leave imbalance in dimensions of $X$ because they have little to no influence on $Y$, or to offset imbalances in other dimensions. While this strategic imbalance is in contrast to traditions that advocate for good balance on every covariate (e.g., \citealp{stuart2010matching}), we demonstrate that it can yield efficiency gains without introducing bias. Additionally, TFB still prioritizes good balance on the dimensions of $X$ that are most influential on $Y$. Second, TFB's weights are entirely defined by the choice of regression estimator (i.e., $\hat{f}_0$) and an estimator for its variance, thus turning the problem of how best to balance the covariates into how best to model the outcome. Third, and finally, TFB specifies a choice of potentially nonlinear transformations of $X$ to balance through the choice of regression estimator. This includes basis functions from a reproducing kernel Hilbert space, building on the developing literature of weights that draw on them (e.g., \citealp{kallus2020generalized}; \citealp{hazlett2018kernel}; \citealp{wong2017kernel}; \citealp{hirshberg2019minimax}; \citealp{huang2007correcting}; \citealp{zhao2019covariate}; \citealp{tarr2021estimating}), by estimating $f_0$ with kernel regularized least squares (\citealp{hainmueller2014kernel}). Further, TFB is not limited to regression estimators with linear representations --- for example, we apply generalized linear models and Bayesian Additive Regression Trees (\citealp{hill2011bayesian}) in our demonstrations here. However, TFB offers a solution to feasibility concerns from balancing a high-dimensional $X$ by narrowing the function space that the weights seek balance in, that is, from all (linear) functions of $X$ (and its tranformations), to those probabilistically near $\hat{f}_0$ per its estimated variance. 

Combining a $\hat{f}_0$ and weights to estimate treatment effects is a well-studied research area. It is essential to the augmented estimator, the typical form of doubly robust estimators (e.g., \citealp{robins1994estimation}; \citealp{robins1995semiparametric}; \citealp{van2006targeted}; \citealp{tan2010bounded}; \citealp{rotnitzky2012improved}; \citealp{chernozhukov2018double}), which are consistent when the investigator has correctly specified $f_0$ or the propensity score (\citealp{rosenbaum1983central}), $\pi(X) = P(D=1 | X)$. Augmented estimators have also been applied with balancing weights (e.g. \citealp{athey2018approximate}, \citealp{hirshberg2017augmented}). In fact, TFB performs similarly to an augmented estimator in practice, 
and we find that TFB exhibits double robustness, due to the connection between propensity score and balancing weights (\citealp{zhao2017entropy, wang2020minimal}). Other methods that use a $\hat{f}_0$ in conjunction with weights include those that allow a $\hat{f}_0$ to inform which dimensions of $X$ to prioritize balance in, or to use to estimate $\pi$ for (inverse) propensity score weights (e.g., \citealp{shortreed2017outcome}; \citealp{kuang2017estimating}; \citealp{ning2020robust}), and methods that expand on prognostic score theory from \cite{hansen2008prognostic} (e.g., \citealp{leacy2014joint}; \citealp{antonelli2018doubly}), which supports seeking balance in $\hat{f}_0 (X)$. More recent work (e.g., \citealp{shen2025forest, jin2025cross}) has investigated allowing outcome information to inform balancing weights, specifically. TFB contributes to this literature by letting the estimated variance of $\hat{f}_0$ determine how to prioritize balance in $\hat{f}_0 (X)$ over $X$, and in certain dimensions of $X$ over others.

To outline, Section~\ref{sec:background} details notation and assumptions, and reviews reproducing kernel Hilbert spaces, which appear throughout, and Section~\ref{sec:motivation} expands on the motivation for TFB. Section~\ref{sec:tfb} introduces TFB, demonstrates its merits through simulation and in an application to the 2016 American Causal Inference Conference challenge data (\citealp{dorie2019automated}). 
Section~\ref{sec:lausd} then applies TFB to estimate the effects of taking mathematics courses in twelfth grade in the Los Angeles Unified School District. Section~\ref{sec:conclusion} concludes. We have also developed the \href{https://github.com/bleth/tfb}{\texttt{tfb}} package to implement TFB in the \texttt{R} programming language.

\section{Background}\label{sec:background}

\subsection{Notation and assumptions}
\paragraph{Notation.}
Let $i \in \{1, \dots, n\}$ index the units of observation and let $p(\cdot)$ be the density function of an arbitrary random variable. Define $D \in \{0, 1\}$ to be an indicator variable for receiving the treatment, with $n_1$ the number of treated units and $n_0 = n - n_1$ the number of control units. Additionally, let $\mathbf{D} = [D_1 \ \dots \ D_n]^{\top}$ be the vector of treatment statuses for the sample, and let the first $n_0$ units make up the control group (i.e., $1 \leq i \leq n_0$ are the indices for the control units and $n_0 + 1 \leq i \leq n$ are the indices for the treated units). Then, let $X$ be a $P$-dimensional vector of covariates and let $\mathbf{X}$ be the matrix of $X$ for the sample. 
	\begin{align}
    		X &= \begin{bmatrix} X^{(1)} \\ \vdots \\ X^{(P)} \end{bmatrix} \in \mathbbm{R}^{P}  \ , \  \mathbf{X} = \begin{bmatrix} X^{\top}_{1} \\ \vdots \\ X_{n}^{\top}  \end{bmatrix} \in \mathbbm{R}^{n \times P} 
	\end{align}

In accordance with the potential outcomes framework \citep{splawa1990application, rubin1974estimating}, $Y (0)$ is the outcome if control and $Y (1)$ is the outcome if treated, so $Y = Y (D)$ is observed. Implicit here is the stable unit treatment value assumption (SUTVA), that the potential outcomes for unit $i$ are not functions of the treatment statuses of other units, and that each treatment status is administered the same across the units. Further, the tuples $( X_i, D_i,  Y_i (0), Y_i (1) )$ are assumed independent and identically distributed (iid). 

Of particular importance to TFB is the conditional expectation function of $Y (d)$ given $X$, defined here as $f_d (X) = \E[ Y (d) \ | \ X]$, where $\E (\cdot)$ is the expectation over $p(\cdot)$ (i.e., the super-population). This allows the model:
	\begin{align}\label{eq:f_d}
		& Y (d) = f_d (X) + \epsilon (d) \ \ \text{where} \ \ \E[ \epsilon (d) \ | \ X ] = 0  \ \ \text{and} \ \ \var[ \epsilon (d) \ | \ X ] = \sigma^2 (d, X)
	\end{align}	
where $\epsilon (d) $ is an error term associated with potential outcome $Y (d)$. Although TFB is not limited to models with linear representations, we will often consider the setting where $f_d$ is linear in $X$, 
	\begin{align}\label{eq:f_d_linear}
	f_d (X) = X^{\top} \beta_d \ \ \text{where} \ \ \beta_d = [ \beta_{d}^{(1)} \ \cdots \ \beta_{d}^{(P)} ]^{\top} \in \mathbbm{R}^{P} 
	\end{align}
(\ref{eq:f_d_linear}) appears restrictive because $X$ has been defined thus far as a vector of untransformed covariates. However, exchanging $X$ with, or allowing it to include, its nonlinear transformations (e.g., polynomial terms, or basis functions) makes this representation of $f_d$ quite general, and is the setting in which the linear form of TFB largely operates.

\paragraph{Assumptions.}
This paper assumes throughout the absence of unobserved confounding in the relationship between $D$ and $Y$, otherwise known as conditional ignorability:
   \begin{assumption}[Conditional Ignorability]\label{asm:ci}
        $( Y (0), Y (1) ) \ \indep \ D \ | \ X$
    \end{assumption}
Note that the conditional independence in Assumption~\ref{asm:ci} extends to the $\epsilon (d) $ and $D$ (i.e., 
\newline 
$( \epsilon (0), \epsilon (1) ) \ \indep \ D \ | \ X$), as (\ref{eq:f_d}) implies that $\epsilon (d) = Y (d) - f_d (X)$. 

Unobserved confounders, which would violate Assumption~\ref{asm:ci}, are often a concern in real-data settings, but are beyond the scope of this paper. However, sensitivity analysis to unobserved confounders is a growing field of research, particularly for estimators that use weights --- see \cite{hong2021did}, \cite{soriano2021interpretable}, or \cite{wainstein2026sensitivity} for examples.

\paragraph{Estimands and estimators.} 
This paper primarily considers the average treatment effect on the treated (ATT),
	\begin{align}
		\mathrm{ATT} = \E[Y (1) - Y (0) \ | \ D = 1] 
	\end{align}
However, Appendix~\ref{app:ate_atc} adapts TFB to estimate the average treatment effect, $\mathrm{ATE} = \E[Y (1) - Y (0)]$, and the average treatment effect on the controls, $\mathrm{ATC} = \E[Y (1) - Y (0) \ | \ D = 0]$.

Were it observed, the sample average treatment effect on the treated (SATT),
	\begin{align}
		\mathrm{SATT} &= \frac{1}{n_1} \sum_{i:D_i=1} Y_{i} (1) - \frac{1}{n_1} \sum_{i:D_i=1} Y_{i} (0) 
	\end{align}
would be 
the ideal estimator for the ATT, but $Y(0)$ is unobserved for treated units without strong assumptions. The class of estimators studied here thus replaces $\frac{1}{n_1} \sum_{i:D_i=1} Y_{i} (0)$ in SATT with a weighted average of the $Y_i (0)$ for the control units, $\frac{1}{n_0} \sum_{i:D_i=0} w_i Y_i (0)$ for non-negative weights $w$. 
The resulting weighted difference in means estimator has the form
	\begin{align}
		\hat{\tau}_{\mathrm{wdim}} (\mathbf{w}) = \frac{1}{n_1} \sum_{i:D_i=1} Y_{i} (1) -  \frac{1}{n_0} \sum_{i:D_i=0} w_i Y_i (0)
	\end{align}
where $\mathbf{w}$ is an arbitrary vector of non-negative weights for control units,
	\begin{align}\label{eq:wvector}
		\mathbf{w} = [ w_1 \ \cdots \ w_{n_0} ]^{\top} \in \mathbb{R}^{n_0}_{\geq 0}
	\end{align}
These weights will have mean 1 (i.e., $\frac{1}{n_0} \sum_{i : D_i = 0} w_i = 1$) in most settings considered here. However, this condition will not be required, unless otherwise noted. Section~\ref{sec:motivation} discusses the conditions that $\mathbf{w}$ must satisfy for $\hat{\tau}_{\mathrm{wdim}} (\mathbf{w})$ to be unbiased.

\subsection{Reproducing kernel Hilbert spaces}\label{subsec:rkhs}

We often use reproducing kernel Hilbert spaces (RKHS) to characterize a large space of functions from which $f_0$, or an arbitrarily close approximation, could come. This section therefore provides a brief review of the relevant points in RKHS. For a more extensive review, see \cite{wainwright2019high}.

Background on ``positive semi-definite kernel functions" is first required. A positive semi-definite kernel function, $\mathcal{K}: \mathbb{R}^P \times \mathbb{R}^P \rightarrow \mathbb{R}$, is a symmetric function such that for all possible choices of $\{ x_1, \dots, x_m \} \subseteq \mathbb{R}^{P}$ for arbitrary $m$, the $m \times m$ matrix with $ij$th element equal to $\mathcal{K} (x_i, x_j)$ is positive semi-definite. Per \cite{hazlett2018kernel}, it is helpful to interpret $\mathcal{K}$ as a similarity score. For example, the $\mathcal{K}$ primarily employed here is the Gaussian kernel, $\mathcal{K} (x_i, x_j) = e^{-\frac{||x_i - x_j||_2^2}{b}}$ for some bandwidth parameter $b$. This kernel's range is 0 to 1, with $\mathcal{K} (x_i, x_j) = 1$ only when $x_i = x_j$ (i.e., when $x_i$ and $x_j$ are as similar as can be), and  $\mathcal{K} (x_i, x_j)$ approaches 0 as $||x_i - x_j||_2 \rightarrow \infty$ (i.e., as $x_i$ and $x_j$ diverge).

A RKHS, $H(\mathcal{K})$, is then the unique Hilbert space of functions $h : \mathbb{R}^{P} \rightarrow \mathbb{R}$ that contains $\mathcal{K}(\cdot, x)$ for $\forall x \in \mathbb{R}^P$, and has the inner product $\langle \cdot , \cdot \rangle_{H(\mathcal{K})} : H(\mathcal{K}) \times H(\mathcal{K}) \rightarrow \mathbb{R}$ satisfying the \ref{eq:reproducing_prop}:
	\begin{align} \label{eq:reproducing_prop}
		\langle \mathcal{K}(\cdot, x) , h \rangle_{H(\mathcal{K})} = h(x) \ \ \  \text{for} \ \ \  \forall{h} \in H(\mathcal{K}), \ \forall x \in \mathbb{R}^P \tag{Reproducing Property}
	\end{align}
and norm $|| h ||_{H(\mathcal{K})} = \sqrt{\langle h , h \rangle_{H(\mathcal{K})}}$. If $f_0 \in H(\mathcal{K})$, then the \ref{eq:reproducing_prop} allows the representation $f_0 (X_i) = K_i \alpha_0$ where $K_i = [\mathcal{K} (X_i, X_1) \ \cdots \ \mathcal{K} (X_i, X_n) ]$ for some $\alpha_0 \in \mathbb{R}^n$ and all $i$ (see \citealp{wainwright2019high} for proof). Thus, within the sample, any function in $H(\mathcal{K})$ can be expressed as a linear function of $K_i$, which holds unit $i$'s similarity scores to each unit in the data. This representation is powerful with certain choices of $\mathcal{K}$. For example, for some $\mathcal{K}$ (e.g., the Gaussian kernel), 
a function in $H(\mathcal{K})$ can approximate any smooth function to an arbitrarily close degree.
\section{Motivation for TFB}\label{sec:motivation}

We now motivate TFB by reframing existing approaches in the weighting literature as decisions on a set of potential $f_0$ to seek balance in. TFB builds on this idea by letting a $\hat{f}_0$, an estimate of $f_0$, and its estimated variance inform this decision. Section~\ref{subsec:wprops} first reviews the role that $f_0$ plays in the conditions the weights must satisfy for $\hat{\tau}_{\mathrm{wdim}}$ to be unbiased. 

\subsection{Desirable conditions for the weights}\label{subsec:wprops}

We first define what we refer to as the ``weighting condition" (WC),
\begin{align}\label{eq:wc5}
		\E [ w f_0 (X) \ | \ D = 0 ] = \E [ f_0 (X)\ | \ D = 1 ]	\tag{WC}
\end{align}
and its empirical analog, which we refer to as the ``empirical weighting condition" (EWC),	\begin{align*}\label{eq:ewc}
		\frac{1}{n_0} \sum_{i:D_i=0} w_i f_0 (X_i) = 	\frac{1}{n_1} \sum_{i:D_i=1} f_0 (X_i) \tag{EWC}
	\end{align*}
In words, WC and EWC require that the means of $f_0 (X)$ be equated in expectation or empirically. 

EWC is given particular focus in TFB, so we discuss it first. 
First define the imbalance, or difference in the means (after weighting), in a function of $X$ as
	\begin{align}\label{eq:imbal}
		\mathrm{imbal}( \mathbf{w}, g(\mathbf{X}), \mathbf{D} ) = \frac{1}{n_1} \sum_{i:D_i = 1} g(X_i) - \frac{1}{n_0} \sum_{i:D_i = 0} w_i g(X_i)
	\end{align} 
where $g$ is an arbitrary vector-valued function. How far the weights are from satisfying EWC, to be referred to as the ``EWC Bias", then has an intuitive form as the imbalance in $f_0 (X)$:
	\begin{align}\label{eq:ewc_bias}
		\mathrm{EWC \ Bias} = \mathrm{imbal}( \mathbf{w}, f_0 (\mathbf{X}), \mathbf{D} )
	\end{align}
Proposition~\ref{prop:ewc} below states that, under certain conditions, the expectation of the EWC Bias is exactly the bias of $\hat{\tau}_{\mathrm{wdim}} (\mathbf{w})$, and thus $\hat{\tau}_{\mathrm{wdim}} (\mathbf{w})$ is unbiased for the ATT under EWC:
\begin{proposition}[Unbiasedness under EWC]\label{prop:ewc} Given (i) Assumption~\ref{asm:ci} and (ii) that each pair of $w_i$ and $Y_i (0)$ are independent given $\mathbf{X}$ and $\mathbf{D}$ (i.e., $w_i \indep Y_i(0) \ | \ \mathbf{X}, \mathbf{D}$ for $\forall i$),
$$
    \E[ \hat{\tau}_{\mathrm{wdim}} (\mathbf{w}) - \mathrm{ATT}] = \E ( \mathrm{EWC \ Bias} )
$$
and, consequently, $\hat{\tau}_{\mathrm{wdim}} (\mathbf{w})$ is unbiased for the ATT (i.e., $\E[ \hat{\tau}_{\mathrm{wdim}} (\mathbf{w})  ] = \mathrm{ATT}$) under EWC.
\end{proposition}
Proof is given in Appendix~\ref{app:ewcprop_pf}. Of note is that Proposition~\ref{prop:ewc} requires each \textit{pair} of $w_i$ and $Y_i (0)$ to be independent given $\mathbf{X}$ and $\mathbf{D}$. As will be discussed later (in Section~\ref{subsec:tfb.weights}), TFB's weights depend on $Y$, but can be implemented in a way such that this condition holds. Many methods in the literature (e.g., \citealp{hainmueller2012entropy}; \citealp{zubizarreta2015stable}; \citealp{hazlett2018kernel}) choose weights that are ``honest", or entirely defined by $\mathbf{X}$ and $\mathbf{D}$, which leaves \textit{all} of the $w_i$ independent of \textit{all} of the $Y_i$ given $\mathbf{X}$ and $\mathbf{D}$, a stronger condition than is required for Proposition~\ref{prop:ewc}.

EWC (and WC) do not require \textit{complete} overlap in the conditional distributions of $X$, or
    \begin{align}\label{eq:psoverlap} 
    	0 <\pi(X = x) < 1 \ \ \text{for} \ \ \forall x
    \end{align}
where $\pi(X) = p(D = 1 \ | \ X)$ is the propensity score. This is traditionally assumed in settings with (inverse) propensity score weights (\citealp{rosenbaum1983central}). Instead, satisfying EWC with positive weights with mean 1 requires a far weaker overlap condition: that the average $f_0 (X)$ in the treated group (i.e., $\frac{1}{n_1} \sum_{i:D_i=1} f_0 (X_i)$) is within the range of $f_0 (X)$ among the control group. We thus forgo the traditional overlap assumption in (\ref{eq:psoverlap}). See also Appendix~\ref{app:overlap} for a simulated example where TFB is effectively unbiased, but traditional overlap is violated.\footnote{Note that this simulated setting is unrealistic in that TFB applies a model for $Y$ that correctly extrapolates in regions of the $X$ distribution where there is no overlap. As will be discussed later, we largely recommend applying nonparametric modeling techniques with TFB, which often do not extrapolate by design. For example, kernel regularized least squares with a Gaussian kernel (\citealp{hainmueller2014kernel})  tends to predict the average $Y$ for  values of $X$ that are beyond the data, so this modeling technique would struggle with TFB in settings where traditional overlap is violated. However, there are nonparametric techniques that can extrapolate, such as Gaussian processes with non-stationary kernels (e.g., see \citealp{cho2026inference}). We also recommend LASSO regression with TFB, after adding to $X$ numerous non-linear transformations, which can extrapolate.} 

While not as central to TFB, WC is a similarly desirable condition. Proposition~\ref{prop:wc} states that weights that satisfy WC yield an unbiased $\hat{\tau}_{\mathrm{wdim}} (\mathbf{w})$ under certain conditions. 
\begin{proposition}[Unbiasedness under WC]\label{prop:wc} Given (i) Assumption~\ref{asm:ci} and (ii) $w_i = w (X_i, D_i)$ for a weight function $w(\cdot)$, the resulting $\hat{\tau}_{\mathrm{wdim}} (\mathbf{w})$ is unbiased for the ATT if the weights satisfy WC.
\end{proposition}
Proof is given in Appendix~\ref{app:wcprop_pf}. Although less of a focus for TFB, Proposition~\ref{prop:wc} again emphasizes the importance of mean balance in $f_0$, this time in expectation.

\subsection{Building blocks}\label{subsec:litreview}

This section reviews two families of weighting procedures in the literature: (inverse) propensity score and balancing weights, the latter of which includes TFB. Within each family, focus is given to the tension between the assumptions on $f_0$ required for unbiasedness and the weights' feasibility.
 
WC is satisfied by $w_i = \frac{p(X_i \ | \ D_i = 1)}{p(X_i \ | \ D_i = 0)} = \frac{p(D_i = 0)}{p(D_i=1)} \frac{\pi(X_i)}{1 - \pi (X_i)}$. This offers motivation for (inverse) propensity score weights \citep{rosenbaum1983central} --- were $\hat{\pi} (X)$ to consistently estimate $\pi (X)$, then $\hat{w}_i = \frac{n_0}{n_1} \frac{\hat{\pi} (X_i)}{1 - \hat{\pi} (X_i)}$ yields a consistent $\hat{\tau}_{\mathrm{wdim}} (\hat{\mathbf{w}})$.\footnote{A $\hat{\tau}_{\mathrm{wdim}} (\hat{\mathbf{w}})$ with $\hat{w}_i = n_0 (\sum_{i:D_i=0} \frac{\hat{\pi}(X_i)}{1-\hat{\pi}(X_i)})^{-1} \frac{\hat{\pi}(X_i)}{1-\hat{\pi}(X_i)}$ is also consistent, allowing $\frac{1}{n_0}\sum_{i:D_i = 0} \hat{w}_i = 1$.} Of note is that propensity score weights satisfy WC (approximately) for \textit{any} $f_0$. However, this comes at the price of a specification of $\pi$. An incorrect specification may lead to biased $\hat{\tau}_{\mathrm{wdim}} (\hat{\mathbf{w}})$ (e.g., \citealp{kang2007demystifying}). Additionally, when there are $\hat{\pi} (X_i)$ that are near 1 (or 0, when estimating the ATC or ATE), the resulting weights can be extreme, yielding instability in $\hat{\tau}_{\mathrm{wdim}} (\hat{\mathbf{w}})$ (\citealp{robins2000inference, kang2007demystifying}).

But satisfying WC or EWC for any $f_0$ is unnecessary if one has prior knowledge about $f_0$. For example, if $f_0 (X) =X^\top \beta_0$ then weights that achieve exact balance in $X$,
	\begin{align}\label{eq:exactbal2}
		 \mathrm{imbal} (\mathbf{w}, \mathbf{X}, \mathbf{D}) = 0
	\end{align}
satisfy EWC. This motivates balancing weights, which attempt to make the imbalance in $X$ small. One subfamily, originating from calibration in the survey literature (e.g., \citealp{deming1940least, deville1992calibration}), is:
\begin{align}\label{eq:approxbal_w1}
	\hat{\mathbf{w}}_{\text{BAL1}} = \underset{\mathbf{w}}{\text{argmin}} \ B(\mathbf{w}) \ \ \text{where} \ \ | \mathrm{imbal} (\mathbf{w}, \mathbf{X}, \mathbf{D}) | \leq \delta \tag{BAL1}
	\end{align}
for some $\delta = [\delta^{(1)} \ \dots \ \delta^{(P)}] \geq 0$, and $B(\mathbf{w})$ is a function that controls the variance of $\mathbf{w}$ or distance from base weights. The constraints $\frac{1}{n_0} \sum_{i:D_i=0} w_i = 1$ and $\mathbf{w} \geq 0$ are typically added.

A common approach is to require exact balance on $X$ through \ref{eq:approxbal_w1} by setting $\delta = 0$ (e.g., \citealp{chan2016globally}). A notable example is Entropy Balancing (\citealp{hainmueller2012entropy}), which chooses $B(\mathbf{w})$ to be the (negative) entropy, $\frac{1}{n_0} \sum_{i:D_i=0} w_i \mathrm{log}(w_i)$. However, exact balance exhibits two shortcomings --- one of specification and one of feasibility. On specification, the resulting $\hat{\tau}_{\mathrm{wdim}} (\hat{\mathbf{w}}_{\text{BAL1}})$ is only assured to be unbiased when $f_0 (X) = X^\top \beta_0$. On feasibility, weights that achieve the exact balance in (\ref{eq:exactbal2}) may not exist (e.g., when $X$ is of large dimension) or may be extreme, resulting in a high variance $\hat{\tau}_{\mathrm{wdim}} (\hat{\mathbf{w}}_{\text{BAL1}})$. Further, these issues compound each other. One may expand $X$ to include its nonlinear transformations, finding exact balance has become infeasible.

These concerns have motivated \textit{approximate} balancing weights. Within the \ref{eq:approxbal_w1} framework, \cite{zubizarreta2015stable} and \cite{wang2020minimal} relax the exact balance conditions by letting $\delta >0$. If $f_0 (X) = X^{\top} \beta_0$, then  $\delta$ presents a bias-variance tradeoff: smaller $\delta^{(\ell)}$ reduce the bias of the resulting $\hat{\tau}_{\mathrm{wdim}} (\hat{\mathbf{w}}_{\text{BAL1}})$, but also shrink the space of solutions for the weights, which may increase the variance of $\hat{\tau}_{\mathrm{wdim}} (\hat{\mathbf{w}}_{\text{BAL1}})$. The choice of $\delta$ is thus paramount to the performance of \ref{eq:approxbal_w1}. Additionally, if $f_0 (X) = X^{\top} \beta_0$, weights found by \ref{eq:approxbal_w1} bound the EWC Bias for the ``worst-case" $\beta_0$, or the $\beta_0$ that yields the most bias, as
    \begin{align}\label{eq:approxbal_w.worstcase}
		| \text{EWC Bias} | &= | \mathrm{imbal} (\hat{\mathbf{w}}_{\text{BAL1}}, \mathbf{X}, \mathbf{D})^{\top} \beta_0  | \leq ||\beta_0||_2 | | \mathrm{imbal} (\hat{\mathbf{w}}_{\text{BAL1}}, \mathbf{X}, \mathbf{D}) | |_2 \leq ||\beta_0||_2 || \delta ||_2
	\end{align}
where the first inequality in (\ref{eq:approxbal_w.worstcase}) becomes an equality when $\beta_0$ is in the direction of the leftover imbalance. Thus, $X$ and $\delta$ can be thought of as defining a space of functions from which the worst-case $f_0$ may come (i.e., $\mathcal{F} = \{ x^{\top} \beta \ | \ |\beta^{(\ell)}| \leq \delta^{(\ell)} \frac{||\beta_0 ||_2}{||\delta||_2} \}$). Another subfamily of balancing procedures incorporates this idea directly, finding low variance weights that minimize the EWC Bias for the worst-case $f_0$ over some space of functions, $\mathcal{F}$,
	\begin{align}\label{eq:approxbal_w2}
	\hat{\mathbf{w}}_{\text{BAL2}} =  \underset{\mathbf{w}}{\text{argmin}} \biggr[ \underset{f_0 \in \mathcal{F}}{\text{max}}( \text{EWC Bias} )^2 + B(\mathbf{w}) \biggr] \tag{BAL2}
	\end{align}
where the constraints $\frac{1}{n_0} \sum_{i:D_i=0} w_i = 1$ and $\mathbf{w} \geq 0$ are usually added. Analogous to $X$ and $\delta$ in \ref{eq:approxbal_w1}, the parametrization of $\mathcal{F}$ determines the complexity of $f_0$ that \ref{eq:approxbal_w2} aims to balance, and the size of $\mathcal{F}$ determines how to prioritize minimizing the squared EWC Bias relative to $B(\mathbf{w})$.\footnote{For example, if $\mathcal{F} = \{ x^{\top} \beta \ | \ || \beta||_2 \leq C \}$, the objective function in \ref{eq:approxbal_w2} becomes $C^2 || \mathrm{imbal} (\mathbf{w}, \mathbf{X}, \mathbf{D}) ||_2^2 +  B(\mathbf{w})$. The resulting procedure attempts to balance the worst-case linear function of $X$, which is determined by overall balance in $X$. $C$ determines how much balance in $X$ is prioritized over $B(\mathbf{w})$.}

By removing the requirement of exact balance, approximate balancing weights can be made to appear robust to a wide range of $f_0$ by allowing a high-dimensional $X$ or, analogously in \ref{eq:approxbal_w2}, a flexible $\mathcal{F}$. For example, some (e.g., \citealp{kallus2020generalized}; \citealp{hazlett2018kernel}; \citealp{wong2017kernel}, \citealp{hirshberg2019minimax}; \citealp{zhao2019covariate}; \citealp{tarr2021estimating}; \citealp{shen2025forest}) assume $f_0$ to be of a RKHS, applying the representation  $f_0 (X_i)= K_i \alpha_0$, and approximately balance the $K_i$. However, the optimal ways to choose $\delta$ in \ref{eq:approxbal_w1} or $\mathcal{F}$ in \ref{eq:approxbal_w2} are unsolved problems, despite data-driven proposals available (e.g., \citealp{wang2020minimal} and \citealp{chattopadhyaybalancing2020} for \ref{eq:approxbal_w1}; \citealp{kallus2020generalized} and \citealp{wong2017kernel} for \ref{eq:approxbal_w2}). Furthermore, \ref{eq:approxbal_w2} estimators that use a data-driven $\mathcal{F}$ often center it at the zero function, $f (x) = 0$. \cite{kallus2020generalized}, for example, uses a ball in a RKHS: $\mathcal{F} = \{h \in H(\mathcal{K}) \ | \ ||h||_{H(\mathcal{K})} \leq C\}$ for a data-driven $C$. The worst-case function in $\mathcal{F}$ may then be in the opposite direction of, or orthogonal to, $f_0$ in $H(\mathcal{K})$ (i.e., $\langle \cdot, f_0 \rangle_{H(\mathcal{K})} = 0$).\footnote{An orthogonal function when $f_0 (X_i) = X_i^{\top} \beta_0$ is perhaps illustrative. If $X = [X^{(1)} \ X^{(2)}]^{\top}$ and $\beta_0 = [1 \ 0]^{\top}$, a function orthogonal to $f_0$ would be $f_{\perp} (X) = X^{\top} \beta_{\perp}$ where $\beta_{\perp} = [0 \ 1]^{\top}$. Therefore, while only $X^{(1)}$ has any marginal effect on $Y (0)$, balancing the worst-case linear function of $X_i$ would involve balancing functions that disregard any marginal effect of $X^{(1)}$ on $Y (0)$.} Balance on these orthogonal functions would not grant further bias reduction, were balance in $f_0 (X)$ to remain constant. However, requiring good balance on these functions would, at least, increase the variance of the resulting estimator, and could, at worst, distract the weights from obtaining good balance on $f_0 (X)$. Further, as $\mathcal{F}$ is allowed to be larger and more flexible, this inefficiency could be exacerbated. This concern provides the main motivation for TFB, which can be framed as \ref{eq:approxbal_w2} and seeks a $\mathcal{F}$ that contains only functions close to $f_0$ with a user-chosen probability (e.g., 0.95).

\section{Framework for TFB}\label{sec:tfb}

\subsection{Approximating the EWC Bias}\label{subsec:tfi}

This section introduces Targeted Function Imbalance (TFI), a balance diagnostic that TFB minimizes over the weights, and approximates the EWC Bias using a $\hat{f}_0$ and its estimated variance. 

To motivate TFI, we consider the setting where $f_0 (X) = X^{\top} \beta_0$. Recall that exchanging $X$ with, or allowing it to include, its nonlinear transformations (e.g., basis functions) makes this representation of $f_0$ quite general. The magnitude of the EWC Bias then simplifies to
	\begin{align}\label{eq:linTFImbal_1}
		| \text{EWC Bias} | = | \mathrm{imbal}(\mathbf{w}, \mathbf{X}, \mathbf{D})^{\top} \beta_0|
	\end{align}
Then, letting $\hat{\beta}_0$ be an arbitrary estimate of $\beta_0$, consider (\ref{eq:linTFImbal_1}) for the worst-case $\beta_0$ \textit{close} to $\hat{\beta}_0$ with some high level of probability (e.g., 0.95):	\begin{align}\label{eq:linTFImbal_3}
		\underset{ \beta_0 \in (\hat{\beta}_0 - S_q)}{\text{max}} | \mathrm{imbal}( \textbf{w}, \textbf{X}, \textbf{D})^{\top} \beta_0 | = \underset{ \hat{\beta}_0 - \beta_0 \in S_q}{\text{max}} | \mathrm{imbal}( \textbf{w}, \textbf{X}, \textbf{D})^{\top} (\beta_0 - \hat{\beta}_0 ) +  \mathrm{imbal}( \textbf{w}, \textbf{X}, \textbf{D})^{\top}\hat{\beta}_0 |	
	\end{align}
\noindent where $S_q \subseteq \mathbb{R}^{P}$ is a set that contains 0 and $p(\hat{\beta}_0 - \beta_0 \in S_q) \approx q$ for some $q$ ($q$ will be left arbitrary here, but will be discussed in Section~\ref{subsec:tfb.weights}). In other words, $S_q$ is set of small deviations such that, with probability approximately $q$, $\beta_0$ can be recovered by subtracting one of these small deviations from $\hat{\beta}_0$. Clearly, $S_q$ will be defined by the distribution of $\hat{\beta}_0$. 

Proceeding further thus requires a choice of $\hat{\beta}_0$. To motivate the final expression for TFI, consider the setting where $\hat{\beta}_0$ is the OLS estimate using only the control units. Letting $\hat{V}_{\beta_0}$ be the sandwich variance estimator~\citep{white1980heteroskedasticity},
then $\hat{V}_{\beta_0}^{-\frac{1}{2}} (\hat{\beta}_0 - \beta_0) \overset{d}{\rightarrow} \mathcal{N}(0, I_P)$ under iid data and mild conditions. Here, a logical $S_q$ is one that, after being transformed by $\hat{V}_{\beta_0}^{-\frac{1}{2}}$, becomes a ball centered at the origin in which a $\mathcal{N}(0, I_P)$ falls with probability $q$. This yields $S_q = \{ \hat{V}_{\beta_0}^{\frac{1}{2}} x \in \mathbb{R}^{P} \ | \ ||x||_2 \leq \sqrt{Q_q (\mathcal{X}^2_P)} \}$,  where $Q_q ( \mathcal{X}^2_P )$ is the $q$th quantile of a chi-squared random variable with $P$ degrees of freedom, simplifying (\ref{eq:linTFImbal_3}) to TFI: 
    \begin{align}\label{eq:linTFImbal}
		\mathrm{TFI}(\mathbf{w}, \mathbf{X}, \mathbf{D}, \hat{\beta}_0, \hat{V}_{\beta_0}, q) = \sqrt{Q_q (\mathcal{X}^2_P)} \times || \hat{V}_{\beta_0}^{\frac{1}{2}} \mathrm{imbal}(\mathbf{w}, \mathbf{X}, \mathbf{D}) ||_2 +  |\mathrm{imbal}(\mathbf{w}, \mathbf{X}, \mathbf{D})^{\top}\hat{\beta}_0|
	\end{align}

Although the above derivation uses OLS as motivation, we instead experiment below applying TFI with non-OLS, machine learning or nonparametric regressions, simply replacing $\hat{\beta}_0$ with a different estimate, and $\hat{V}_{\beta_0}$ with a suitable variance estimator. 
As will be discussed, applying these regressions allows TFB to exhibit desirable tendencies in finite samples even without clean asymptotic guarantees from OLS under a correctly specified linear model.

\subsection{Choosing weights in TFB}\label{subsec:tfb.weights}

This section introduces TFB, which finds low variance weights that minimize TFI. TFB, like \cite{kallus2020generalized}, considers minimizing the conditional mean squared error of $\hat{\tau}_{\mathrm{wdim}} (\textbf{w})$ as an estimate of the SATT,
	\begin{align}\label{eq:cmse_1}
		\E \biggr( (\hat{\tau}_{\mathrm{wdim}} (\textbf{w}) - \mathrm{SATT})^2  \ \biggr| \ \textbf{X}, \textbf{D} \biggr) 
		= (\text{EWC Bias})^2  + \frac{1}{n_0^2}\sum_{i:D_i = 0} w_i^2 \sigma^2 (0, X_i) +  \text{Constant}
	\end{align}
where the constant does not depend on the weights. Define TFB then as 
	\begin{align}\label{eq:tfb}
		\hat{\textbf{w}}_{\mathrm{TFB}} = \underset{\mathbf{w}}{\text{argmin}}\biggr[ [ \mathrm{TFI}(\textbf{w}, \textbf{X}, \textbf{D}, \hat{\beta}_0, \hat{V}_{\beta_0}, q) ]^2  + \frac{\hat{\sigma}^2 (0) }{n_0^2} || \textbf{w} ||_2^2 \biggr] \ \text{where} \ \textbf{w} \geq 0 \ \text{and} \ \frac{1}{n_0} \sum_{i:D_i=0}w_i = 1 \tag{TFB} 
	\end{align}
where the objective function in the above approximates the EWC Bias in (\ref{eq:cmse_1}) with TFI, and substitutes the $\sigma^2 (0, X_i) $  with $\hat{\sigma}^2 (0) = \frac{1}{n_0} \sum_{i:D_i = 0} (Y_i - X_i^{\top} \hat{\beta}_0)^2$. The objective function is convex, and, unless otherwise noted, $\texttt{MOSEK}$ (through the \texttt{Rmosek} package in \texttt{R}) was used to solve this problem in all demonstrations here. 

A key consequence of this is that TFB's weights are not honest, because TFI is a function of $Y$ through $\hat{\beta}_0$ (and $\hat{V}_{\beta_0}$). This is in contrast with many seminal methods in the weighting literature (e.g., \citealp{hainmueller2012entropy}; \citealp{imai2014covariate}; \citealp{zubizarreta2015stable}). However, dissecting TFB's objective function is illustrative for envisioning its strengths, which all stem from consulting the outcomes. The $|\mathrm{imbal}(\textbf{w}, \textbf{X}, \textbf{D})^{\top}\hat{\beta}_0|$ term penalizes imbalance in $\hat{f}_0 (X) = X^{\top} \hat{\beta}_0$, which equals the EWC Bias if $f_0  = \hat{f}_0$. However, even if $f_0 (X) = X^{\top} \beta_0$, it is unlikely that $\hat{\beta}_0 = \beta_0$, so the $\sqrt{Q_q (\mathcal{X}^2_P)} \times || \hat{V}_{\beta_0}^{\frac{1}{2}} \mathrm{imbal}(\textbf{w}, \textbf{X}, \textbf{D}) ||_2$ term penalizes imbalance in $X$, with the estimated variance of each $\hat{\beta}_0^{(\ell)}$ determining how to prioritize balance in the $X^{(\ell)}$.\footnote{Note that the coefficients' covariance also plays a role, adding penalties and rewards for the signs of the leftover imbalances. To illustrate, let $X$ be two-dimensional. Then,
	\begin{align*}
		|| \hat{V}_{\beta_0}^{\frac{1}{2}} \mathrm{imbal} (\textbf{w}, \textbf{X}, \textbf{D}) ||_2 &=  \biggr( \widehat{\var}(\hat{\beta}_0^{(1)}) [\mathrm{imbal} (\textbf{w}, \textbf{X}^{(1)}, \textbf{D})]^2 + \widehat{\var}(\hat{\beta}_0^{(2)}) [\mathrm{imbal} (\textbf{w}, \textbf{X}^{(2)}, \textbf{D})]^2 \\
		&+ 2 \widehat{\cov} (\hat{\beta}_0^{(1)}, \hat{\beta}_0^{(2)})  \mathrm{imbal} (\textbf{w}, \textbf{X}^{(1)}, \textbf{D}) \mathrm{imbal} (\textbf{w}, \textbf{X}^{(2)}, \textbf{D}) \biggr)^{1/2}
	\end{align*}
Therefore, if $\widehat{\cov} (\hat{\beta}_0^{(1)}, \hat{\beta}_0^{(2)}) > 0$, then the imbalances in the dimensions of $X$ are penalized more if they are of the same sign, and less if they are of the opposite sign. If $\widehat{\cov} (\hat{\beta}_0^{(1)}, \hat{\beta}_0^{(2)}) < 0$, then imbalances are penalized more if they are of the opposite sign, and less if they are of the same sign.} This yields two desirable properties. First, TFB balances $X$ to safeguard against uncertainty in $\hat{\beta}_0$ and only to a degree deemed necessary by its estimated variance. Section~\ref{sec:dgp1} demonstrates that this, while simultaneously seeking balance in $\hat{f}_0 (X)$, can yield intentional imbalance in certain dimensions of $X$ that offsets imbalance in others, granting efficiency gains without inviting bias. Second, the estimated variance of each $\hat{\beta}_0^{(\ell)}$ determines a hierarchy in balancing the $X^{(\ell)}$, meaning TFB incorporates variable selection. Appendix~\ref{sec:dgp2} shows how estimating $\beta_0$ with sparsity (e.g., the LASSO) can exploit this to great effect. 
Finally, $\frac{\hat{\sigma}^2 (0) }{n_0^2} || \mathbf{w} ||_2^2$ controls the resulting estimator's variance.

These qualities are not all unique to TFB. Prognostic score theory (\citealp{hansen2008prognostic}) also supports seeking balance in $\hat{f}_0 (X)$, inspiring some to incorporate $\hat{f}_0$ in propensity score matching schemes (e.g., \citealp{leacy2014joint}; \citealp{antonelli2018doubly}). Both \cite{kuang2017estimating} and \cite{ning2020robust} emphasize balance in $\hat{f}_0 (X)$, and, along with \cite{shortreed2017outcome}, develop methods that use $\hat{f}_0$ to inform variable selection, either in which dimensions of $X$ to prioritize balance in, or which dimensions to use to estimate $\pi$ for propensity score weights. TFB's contribution to this literature is its allowing $\hat{V}_{\beta_0}$ to guide how to prioritize balance in $\hat{f}_0 (X)$ over $X$, and in certain dimensions of $X$ over others. Additionally, \ref{eq:approxbal_w1} and \ref{eq:approxbal_w2} could integrate TFB's properties by carefully choosing $\delta$ and $\mathcal{F}$, respectively --- in fact, TFB is a special case of \ref{eq:approxbal_w2} where $B( \mathbf{w} ) = \frac{\hat{\sigma}^2 (0) }{n_0^2} || \mathbf{w} ||_2^2$ and $\mathcal{F} = \{ x^{\top} \beta \ | \ || \hat{V}_{\beta_0}^{-\frac{1}{2}} (\hat{\beta}_0 - \beta) ||_2 \leq \sqrt{Q_q (\mathcal{X}^2_P)} \}$. In real-data scenarios, investigators could apply content knowledge to tune $\delta$ and $\mathcal{F}$, potentially closing any performance gap between choosing arbitrary values and applying TFB. Nevertheless, TFB's advantage is that it is defined by $\hat{f}_0$ and $\hat{V}_{\beta_0}$, turning the problem of how best to balance  $X$ into how best to model $Y (0)$.

\paragraph{Connection to augmented and doubly robust estimation.}

Using estimates of $f_d$ in conjunction with weights is not novel --- it is essential for ``augmented estimators", the typical form of estimators that do so (e.g., \citealp{abadie2011bias};  \citealp{athey2018approximate}; \citealp{hirshberg2017augmented}). This includes ``doubly robust" estimators (\citealp{robins1994estimation}; \citealp{robins1995semiparametric}; \citealp{van2006targeted}; \citealp{chernozhukov2018double}) which employ propensity score weights, and are consistent when $f_0$ or $\pi$ has been correctly specified.\footnote{\cite{tan2010bounded} provides an illuminating review of doubly robust estimators that can be expressed as $\hat{\tau}_{\mathrm{aug}}$.} Recall from Proposition~\ref{prop:ewc} that the expectation of the EWC Bias is exactly the bias of $\hat{\tau}_{\mathrm{wdim}}$. Augmented estimators, $\hat{\tau}_{\mathrm{aug}}$,  bias-correct $\hat{\tau}_{\mathrm{wdim}}$ by estimating the EWC Bias with a $\hat{f}_0$, and subtracting the resulting estimate from $\hat{\tau}_{\mathrm{wdim}}$:
	\begin{align}\label{eq:augment}
		\hat{\tau}_{\mathrm{aug}} ( \textbf{w}, \hat{f}_0) &= \hat{\tau}_{\mathrm{wdim}} ( \textbf{w} ) - \underbrace{ \mathrm{imbal}( \textbf{w} , \hat{f}_0 ( \textbf{X} ), \textbf{D} ) }_{\widehat{\text{EWC Bias}}}
	\end{align}	
Noting that $\hat{f}_0 (X) = X^{\top} \hat{\beta}_0$ in TFB reveals its connection to augmented estimators, as the estimated EWC Bias subtracted from $\hat{\tau}_{\mathrm{wdim}} ( \textbf{w} )$ to form $\hat{\tau}_{\mathrm{aug}} ( \textbf{w}, \hat{f}_0) $ is $\widehat{\text{EWC Bias}} = \mathrm{imbal}( \textbf{w},  \textbf{X}, \textbf{D})^{\top} \hat{\beta}_0$, the magnitude of which TFB penalizes.  Therefore, often, $\widehat{\text{EWC Bias}} \approx 0$ in TFB, implying $\hat{\tau}_{\mathrm{wdim}} (\hat{ \textbf{w} }_{\mathrm{TFB}}) \approx \hat{\tau}_{\mathrm{aug}} (\hat{ \textbf{w} }_{\mathrm{TFB}}, \hat{f}_0)$, as we find in our demonstrations below. This also suggests that TFB is doubly robust due to the equivalence between balancing and propensity score weights (\citealp{zhao2017entropy, wang2020minimal}). In Appendix~\ref{app:tfb.dr}, we show that TFB's weights are in fact propensity score weights, where $\pi$ is fit with shrinkage, and specified as a function of the terms TFB attempts to balance: $\hat{f}_0 (X)$ and $X$. Appendix~\ref{app:tfb.dr} also leverages this fact to demonstrate TFB's double robustness in a simulated scenario. In Section~\ref{sec:acic}, we see that this double robustness can allow TFB to be more efficient than estimation purely through regression. However, TFB can avoid pitfalls of 
propensity score weights (e.g., extreme weights) that hinder certain doubly robust estimators in finite samples, which we demonstrate in Appendix~\ref{app:tfb_versus_dr}.

\paragraph{Sample splitting.} 

Recall that the central motivation for TFB is to minimize (an approximation of) the EWC Bias. Further, Proposition~\ref{prop:ewc} states that the bias of $\hat{\tau}_{\mathrm{wdim}}$ is exactly the expectation of the EWC Bias. However, this result requires that each pair of $w_i$ and $Y_i$ be independent of each other, given $\textbf{X}$ and $\textbf{D}$. This is violated by TFB as defined, but is remedied by ``sample splitting", as in \cite{chernozhukov2018double}. In Appendix~\ref{subsec:samplesplitting}, we explore the possibility of sample splitting with TFB, and our \href{https://github.com/bleth/tfb}{\texttt{tfb}} package can implement TFB with and without sample splitting. However, we find in Appendix~\ref{subsec:samplesplitting} that while TFB's performance with and without sample splitting converges as the sample size increases, sample splitting detracts from TFB's performance in smaller samples. Thus, despite potential theoretical guarantees (e.g., \citealp{jin2025cross}), unless otherwise noted, we apply TFB without sample splitting throughout. Nevertheless, note that none of our results below meaningfully change after sample splitting (these results are available upon request). 

\paragraph{Models without linear representations.}

Although we have defined TFB in the context of regression functions that have linear representations, the extension of TFB to regression functions that do \textit{not} have a linear representation (e.g., certain generalized linear models, regression trees, or neural networks) is simple. Letting $\hat{f}_0 ( \textbf{X} ) = [\hat{f}_0 (X_1) \ \cdots \ \hat{f}_0 (X_n)]^{\top}$ be the vector of predicted values from a general regression model, simply apply TFB with $\textbf{X} = \textbf{I}_n$ (i.e., the $n\times n$ identity matrix), $\hat{\beta}_0 = \hat{f}_0 (\textbf{X})$, and $\hat{V}_{\beta_0}$ set to $\hat{V}_{\hat{f}_0 (\textbf{X})}$, an estimated variance of the prediction vector $\hat{f}_0 (\textbf{X})$. Thus, while the linear form of TFB balances linear functions of $X$ in the neighborhood of an estimated function that is parameterized by the coefficient vector $\beta_0$, this extended form of TFB instead balances \textit{prediction vectors} in the neighborhood of a chosen $\hat{f}_0 (\textbf{X})$.

A consequence of this extension is that TFB's objective function is less intuitive at first glance  because the scaled imbalance term, $|| \hat{V}_{\beta_0}^{\frac{1}{2}} \mathrm{imbal}( \textbf{w}, \textbf{X}, \textbf{D} ) ||_2$, becomes $|| \hat{V}_{\hat{f}_0 (\textbf{X})}^{\frac{1}{2}} \mathrm{imbal}( \textbf{w}, \textbf{I}_n, \textbf{D}) ||_2$, and imbalance in $\textbf{I}_n$ is less intuitive than imbalance in $X$. However, the scaling by $\hat{V}_{\hat{f}_0 (\textbf{X})}^{\frac{1}{2}}$ gives this term an intuitive form in certain settings. For example, Appendix~\ref{app:glm} demonstrates that applying a generalized linear model results in TFB penalizing imbalance in \textit{functions} of $X$ in which $f_0$ is assumed to be approximately linear, according to a first-order Taylor approximation.

\paragraph{Implementation.}

As with any balancing method, the performance of \textit{linear} TFB is highly dependent on whether or not $f_0 (X) = X^{\top} \beta_0$. Thus, the key risk with linear TFB is not choosing a $X$ that contains the non-linear transformations of the covariates required for $f_0 (X) = X^{\top} \beta_0$ (see Appendix~\ref{app:miss}). Therefore, although it provided motivation for TFI, TFB is not recommended when $X$ is low-dimensional and OLS estimates $f_0$.
Instead, we recommend the linear form of TFB when $X$ is high-dimensional, either to start or because it is a high-dimensional transformation of a smaller set of variables, and a regularized regression estimates $f_0$. 
Similarly, for the \textit{non-linear} extension of TFB, we recommend applying machine learning regression methods that are flexible enough to capture a wide class of potential $f_0$.

These recommendations lead to the three forms of TFB that we primarily consider, and are listed in  Table~\ref{tab:abbreviations}.
The first form is TFB-K, which applies kernel regularized least squares (KRLS), setting each $X_i = K_{i}^{\top}$ and choosing $\hat{\beta}_0$ as the ridge regression-like coefficient estimate detailed by \cite{hainmueller2014kernel}. 
\cite{hainmueller2014kernel} also propose a closed-form $\hat{V}_{\beta_0}$, which we employ.\footnote{The \texttt{krls} function in \texttt{R} (from the \texttt{KRLS} package) is used to implement KRLS, and employs leave-one-out cross validation for its regularization tuning parameter $\lambda$. \cite{hainmueller2014kernel} prove that their proposed closed-form variance is valid under the additional assumption that $\epsilon (0)$ has constant variance given $X$. While this is unlikely, note that TFB-K does not require $\hat{V}_{\beta_0}$ for standard errors, but rather uses it to determine how to prioritize balance in $\hat{f}_0 (X)$ and the $K_i$ versus the variance of the weights. Further, Appendix~\ref{app:hetero} demonstrates that TFB-K can still perform well under heteroscedastic error. Therefore, we suspect that a violation to the constant variance assumption for $\hat{V}_{\beta_0}$ is not of major consequence.} 
The second form is TFB-L, which estimates $f_0$ with the \textit{modified} LASSO regression detailed by \cite{chatterjee2011bootstrapping}, allowing $\hat{V}_{\beta_0}$ to be found by residual bootstrapping.\footnote{The \texttt{glmnet} package in \texttt{R} is used to implement the standard LASSO. The $\lambda$ needed for $L_1$ regularization is found through cross-validation with the \texttt{cv.glmnet} function. The modified LASSO from \cite{chatterjee2011bootstrapping} is then found by forcing to 0 all estimated coefficients whose magnitude is below a thresholding parameter $a_n$. \cite{chatterjee2011bootstrapping} note that any $a_n \propto n^{-\zeta}$ for $\zeta \in (0, \frac{1}{2})$ allows the residual bootstrap to be consistent. We use $a_n = 5n^{-2/5}$ in our simulations below. While the residual bootstrap is inconsistent under mild conditions for the \textit{standard} LASSO's asymptotic variance when one of more components of $\beta_0$ are zero (\citealp{knight2000asymptotics, chatterjee2010asymptotic}), the modification detailed by \cite{chatterjee2011bootstrapping} corrects this inconsistency. Note, however, that any asymptotic inconsistency of the residual bootstrap does not keep TFB-L with a standard LASSO from performing well in later demonstrations --- see Appendix~\ref{app:lasso_dgp2}.} 
The third form is TFB-BART, which we implement as an example of the non-linear extension of TFB, and estimates $f_0$ with Bayesian Additive Regression Trees (BART; \citealp{hill2011bayesian}). TFB-BART sets the prediction vector $\hat{f}_0 ( \textbf{X} )$ to be the mean of the BART model's posterior distribution for $\hat{f}_0 ( \textbf{X} )$, and sets $\hat{V}_{\hat{f}_0 (\textbf{X})}$ to be the variance matrix of the posterior distribution for $\hat{f}_0 (\textbf{X})$.\footnote{In our experiments below, we regress $Y$ on both $X$ and $D$, and then obtain $\hat{f}_0 (X)$ by making predictions with the resulting BART model at $X$ and $D = 0$. The \texttt{BART} package in \texttt{R} is used to implement BART. We apply the \texttt{gbart} function with default parameters. Note that BART, in fact, implies a RKHS with a specific choice of kernel (e.g., \citealp{shen2025forest}). Thus, one could in theory apply BART through TFB's  linear framework. 
Nevertheless, we apply TFB's non-linear extension with BART to demonstrate this extension, and due to the convenience of using the posterior distribution of $\hat{f}_0 (\mathbf{X})$ to form $\hat{V}_{\hat{f} (\mathbf{X})}$.}
\begin{table}[!h]
\vspace{0.15in}
\begin{center}
\caption{Descriptions of TFB-K, TFB-L, and TFB-BART}\label{tab:abbreviations}
    \small
     \begin{tabular}{ c | c | c }
     \textbf{Form of TFB} & \textbf{Regression Estimator ($\hat{f}_0$)} & \textbf{Variance Estimator ($\hat{V}_{\beta_0}$)} \\
     \hline
     TFB-K & Kernel regularized least squares & Given in \cite{hainmueller2014kernel} \\ 
     \hline
     TFB-L & Modified LASSO & Residual bootstrap \\
     \hline
     TFB-BART & Bayesian Additive Regression Trees & Variance of posterior distribution of $\hat{f}_0 ( \textbf{X} )$ \\
     \end{tabular}
     \normalsize
\end{center}
\vspace{-0.1in}
\end{table}

Finally, note that TFB weights require a choice for the probability threshold $q$. An extensive discussion of this tuning parameter is left to Appendix~\ref{subsec:tfb.q}. However, to summarize, we find the specific choice of $q$ becomes inconsequential when $P$ is large or grows with $n$, as is the case for the forms of TFB we propose in Table~\ref{tab:abbreviations}. Thus, we recommend $q=0.95$ generally, and this is used in all demonstrations here.

\subsection{Demonstrations}\label{subsec:tfb.demonstrations}

This section illustrates TFB's merits in two demonstrations. Section~\ref{sec:dgp1} demonstrates TFB's strategic imbalance in a simulated data generating process (DGP). Section~\ref{sec:acic} then applies TFB to the 2016 American Causal Inference Conference (ACIC) challenge data (\citealp{dorie2019automated}) to demonstrate TFB's improvement over estimating average treatment effects purely through regression. Note that Appendix~\ref{sec:dgp2} provides a third demonstration of TFB that illustrates the role of $\hat{V}_{\beta_0}$ in TFB's strategic imbalance.


\subsubsection{Demonstration 1: Strategic imbalance}\label{sec:dgp1}


This first demonstration shows how TFB's strategic imbalance, both inducing imbalance in certain covariates to offset imbalance in other covariates and prioritizing balance in key covariates, allows TFB to outperform honest, canonical balancing methods in the DGP considered here, even after combining them with an outcome model for augmented estimation. For clarity, key aspects of the DGP (e.g., sample size, and overlap), which we refer to as ``DGP 1", are not varied here, but Appendix~\ref{app:dgp1_extension} examines TFB's performance after varying these aspects. We summarize here the findings from this more expanded demonstration.

\begin{figure}[!h]
	\vspace{0.15in}
	\begin{center}
	\caption{Illustration of overlap in $X$ in clusters from DGP 1}\label{fig.dgp1.clusters}
    \vspace{-.25in}
    \begin{subfigure}{.45\textwidth}
    \begin{center}
    \includegraphics[scale=0.50]{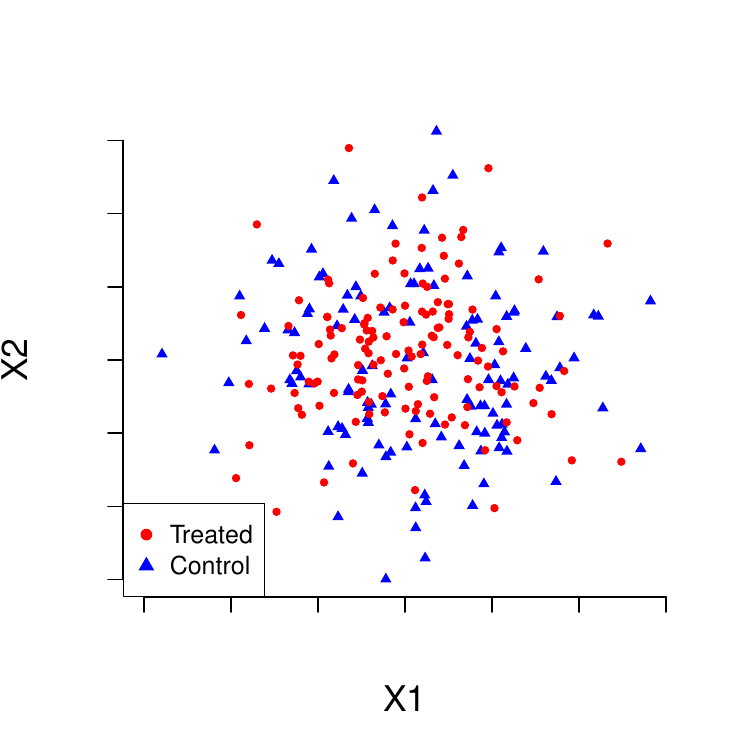} 
    \end{center}
    \vspace{-.25in}
    \subcaption{Easy-to-balance ($C = c_1$)}
    \end{subfigure}
    \begin{subfigure}{.45\textwidth}
    \begin{center}
    \includegraphics[scale=0.50]{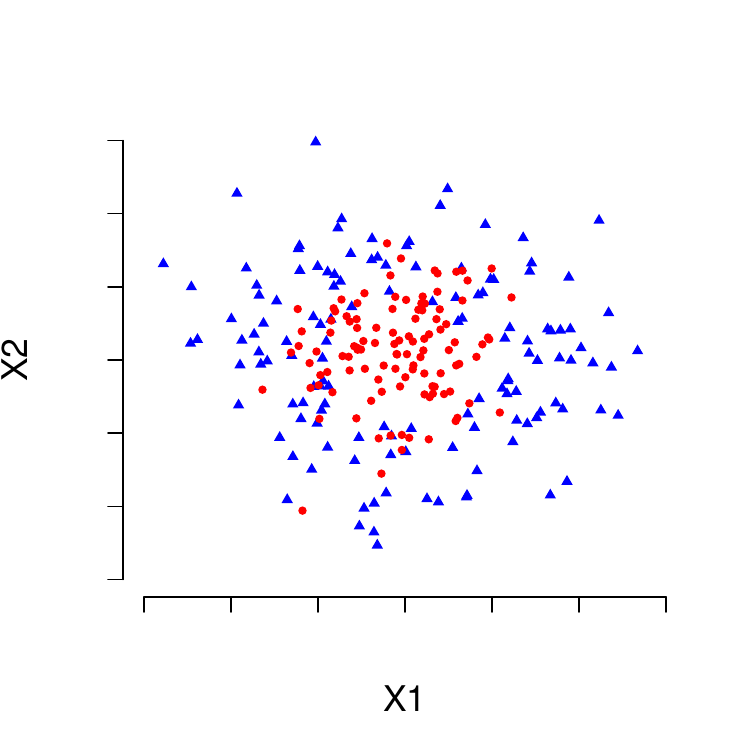} 
   \end{center}
    \vspace{-.25in}
    \subcaption{Hard-to-balance ($C \in \{c_2, c_3, c_4\}$)}
    \end{subfigure}
    \subcaption*{\textit{Note:} Plots of $X$ for \textit{(a)} the easy-to-balance cluster and \textit{(b)} one of the three hard-to-balance clusters from a single draw from DGP 1 with $n=1000$.}
	\end{center}
	\vspace{-0.25in}
	\end{figure}

Consider a DGP in which $X$ is drawn from a bivariate normal distribution around the random variable $C$, one of four equally likely (cluster) centers that are not known by the researcher,
	\begin{align}
		& X_i \overset{iid}{\sim} \mathcal{N} (C_i , \mathbf{I}_2) \ \ \text{where} \ \ C \in \{c_1= (0, 0), \ c_2= (0, 5), \ c_3 = (5, 0), \ c_4= (5, 5) \} \nonumber \\
  & \text{with} \ \ P(C = c_{\ell}) = \frac{1}{4}   \ \ \text{for} \ \ \ell = 1, \dots, 4
	\end{align}
Unknown to the researcher, however, $D$ and $Y$ depend on the proximity of $X$ to each center, given by $Z^{(\ell)}$:
	\begin{align}
		Z^{(\ell)} = \frac{1}{ ||X - c_{\ell} ||_2 + 1} 
	\end{align}
Each unit's proximity to its own center then determines the probability of treatment:
	\begin{align}\label{eq:dgp1_ps}
		\mathrm{log} \frac{\pi(X)}{1 - \pi(X)} &=  \1 \{C = c_1 \} (4)( Z^{(1)} - 0.47)  +  \sum_{\ell = 2}^{4}   \1 \{C = c_{\ell} \}  (20) (Z^{(\ell)} - 0.47)
	\end{align}
where centering $Z^{(\ell)}$ by 0.47 allows $P(D = 1) \approx 0.50$, and units closer to their center are more likely to be treated. Figure~\ref{fig.dgp1.clusters} shows that units with centers $c_2$, $c_3$, and $c_4$ show far more imbalance in the proximity to their center, and therefore their respective $Z^{(\ell)}$, than are units with center $c_1$. Units with center $c_1$ will thus be referred to as those in the ``easy-to-balance" cluster, and the other units will be referred to as those in the ``hard-to-balance" clusters. Finally, the outcome is linear in $Z^{(\ell)}$, 
	\begin{align}\label{eq:dgp1_y}
		Y = 10 Z^{(1)} +\sum_{\ell = 2}^4 Z^{(\ell)} + \epsilon, \ \ \ \epsilon \overset{iid}{\sim} N(0, 1.5) 
	\end{align}
where the proximity to $c_1$, captured by $Z^{(1)}$, is by far the most important contributor to $Y$, and there is no treatment effect (i.e., the ATT is 0). The choice of $\var(\epsilon) = 1.5$ results in a true $R^2 = \frac{\var(Y - \epsilon)}{\var(Y)}$ of approximately 0.60. 
We apply TFB-K and the honest kernel-based weights of \cite{hazlett2018kernel} and \cite{kallus2020generalized}, along with their augmented versions. 
    
\begin{figure}[!h]
	\vspace{0.15in}
	\begin{center}
	\caption{Performance of TFB in DGP 1}\label{fig.dgp1}
    \vspace{-.25in}
    \begin{subfigure}{.60\textwidth}
    \begin{center}
    \includegraphics[scale=0.40]{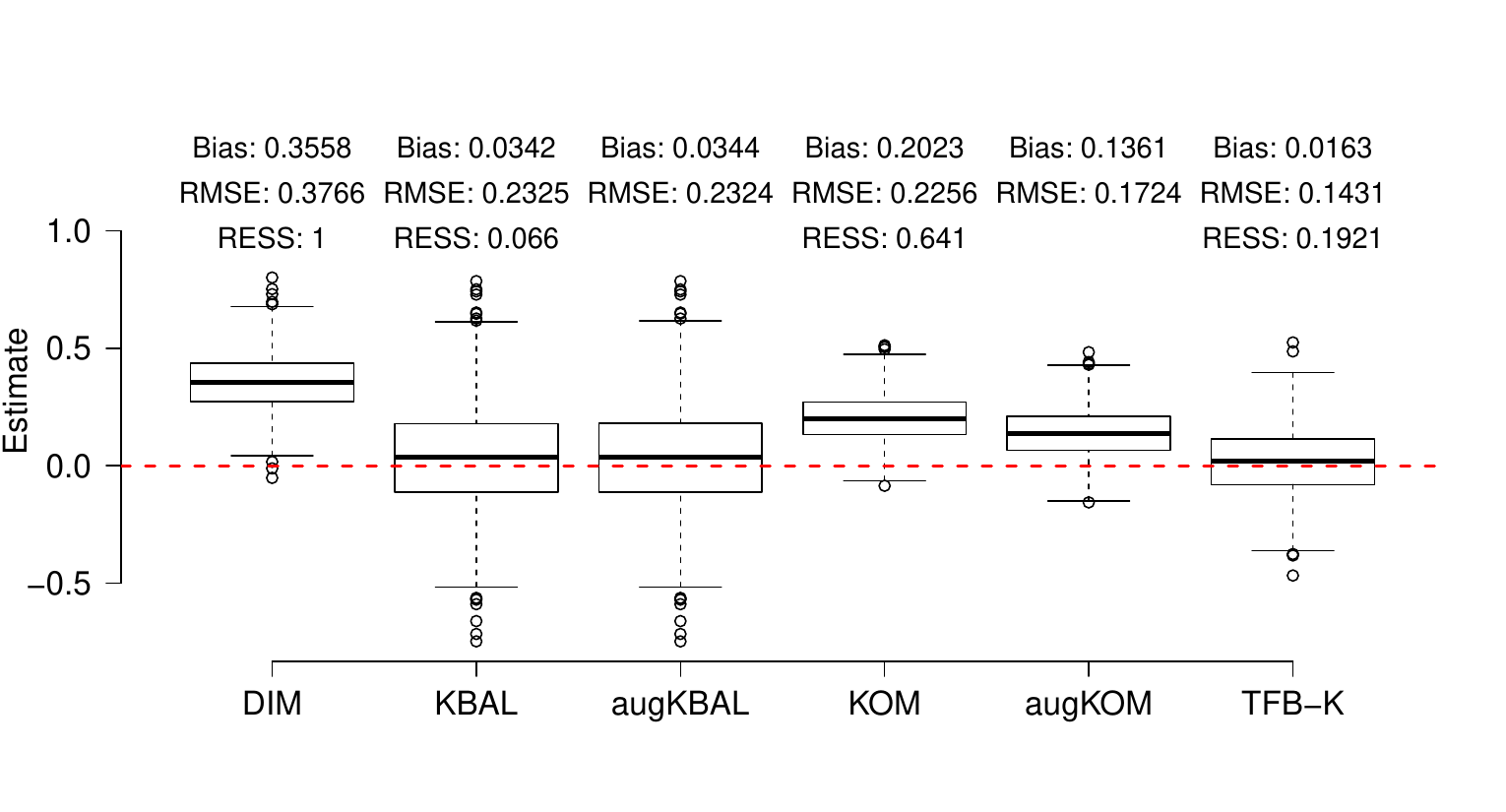} 
    \end{center}
    \vspace{-.25in}
    \subcaption{Bias, RMSE, and RESS}\label{fig.dgp1.bias}
    \end{subfigure}
    \begin{subfigure}{.38\textwidth}
    \begin{center}
    \includegraphics[scale=0.40]{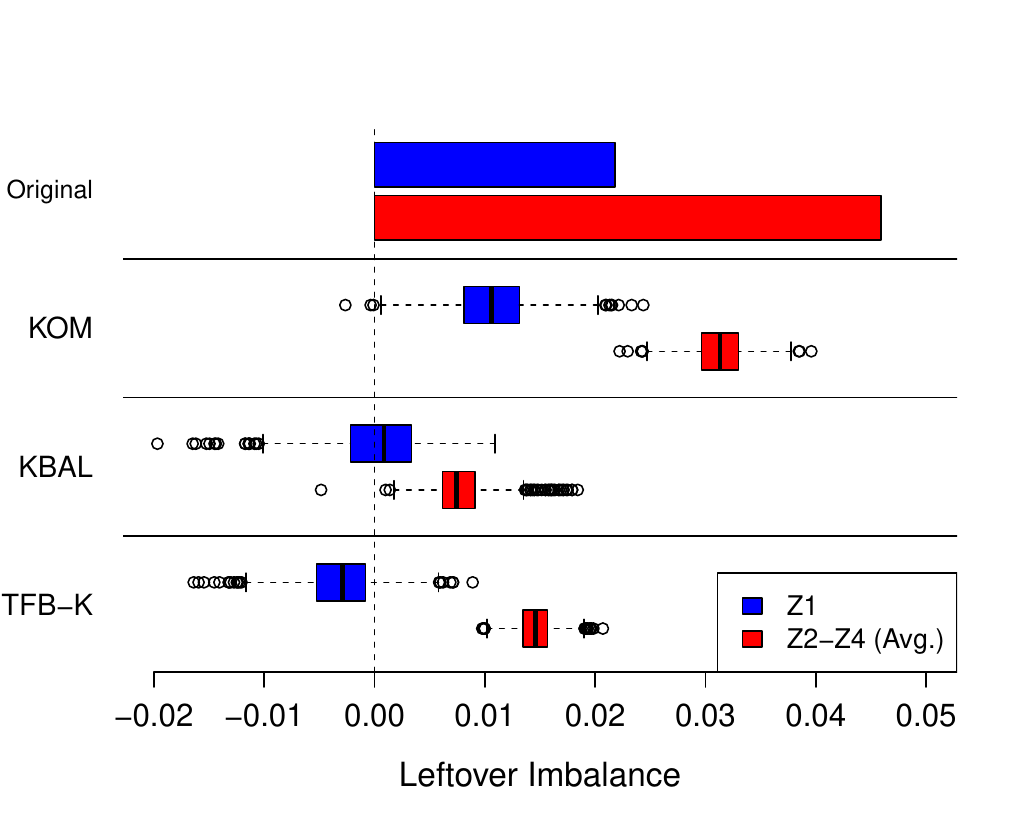} 
    \end{center}
    \vspace{-.25in}
    \subcaption{Leftover imbalance}\label{fig.dgp1.balance}
    \end{subfigure}
    \subcaption*{\textit{Note:} Comparison of TFB-K, Kernel Optimal Matching (KOM), Kernel Balancing (KBAL),  the augmented forms of KOM and KBAL (augKOM and augKBAL, respectively), and a difference in means (DIM) across 1000 draws from DGP 1 with $n=1000$. The augmented form of TFB-K is not included because it is numerically identical to  TFB-K here. \textit{(a)} Distributions of the estimates from each method, as well as their bias, RMSE, and RESS. \textit{(b)} Leftover imbalance from each method after weighting, where imbalance for $Z^{(2)}$, $Z^{(3)}$, and $Z^{(4)}$ is averaged at each iteration. }
	\end{center}
	\vspace{-0.25in}
	\end{figure}

Due to low overlap in the hard-to-balance clusters, balancing $Z^{(2)}$, $Z^{(3)}$, and $Z^{(4)}$ would be challenging in finite samples even were they known to the researcher.  However, balance in $f_0 (X)$ is obtainable even without balance in $Z^{(2)}$, $Z^{(3)}$, and $Z^{(4)}$ --- inducing imbalance in $Z^{(1)}$ \textit{of the opposite sign} could offset imbalance in $Z^{(2)}$, $Z^{(3)}$, and $Z^{(4)}$. For example, weights such that $\mathrm{imbal} (\textbf{w}, \textbf{Z}^{(1)}, \textbf{D}) = -\frac{1}{100}$ and $\mathrm{imbal} (\textbf{w}, \textbf{Z}^{(\ell)}, \textbf{D}) = \frac{1}{30}$ for $2 \leq \ell \leq 4$ achieve balance in $f_0 (X)$:
	\begin{align}
		\mathrm{imbal} (\textbf{w}, f_0 (\textbf{X}), \textbf{D}) = 10 (-\frac{1}{100}) + 1 (\frac{1}{30}) + 1 (\frac{1}{30}) + 1 (\frac{1}{30}) = 0
	\end{align}
This involves reversing the sign of the original imbalance in $Z^{(1)}$, which is positive (see Figure~\ref{fig.dgp1.balance}). However, due to high overlap in the easy-to-balance cluster, this may require lower variance weights than it would require to achieve exact balance in all the $Z^{(\ell)}$. Further, only \textit{slight} imbalance in $Z^{(1)}$ is needed due to the much larger influence of $Z^{(1)}$ on $Y$ than that of the other $Z^{(\ell)}$. Even though the $Z^{(\ell)}$ are unobserved  in reality and TFB-K must approximate them with the $K_{i}$, TFB-K should in theory be able to discern this by consulting a $\hat{f}_0$. Meanwhile, estimators that neglect $Y$ may increase their variance by balancing the hard-to-balance clusters and unimportant dimensions of the $K_i$, or be too distracted by them to properly balance $f_0 (X)$.

Figure~\ref{fig.dgp1} confirms these hypotheses, comparing TFB-K to Kernel Balancing (KBAL; \citealp{hazlett2018kernel}) and Kernel Optimal Matching (KOM; \citealp{kallus2020generalized}), along with their augmented versions (augKBAL and augKOM).\footnote{KBAL exactly balances (via Entropy Balancing) an $r$-dimensional approximation (i.e., using the first $r$ eigenvectors/eigenvalues) of the gram matrix, $\mathbf{K} = [K_1^{\top} \ \dots \ K_n^{\top}]^{\top}$, choosing the $r$ that minimizes the worst-case EWC Bias. The \texttt{kbal} function in \texttt{R} (from the \texttt{kbal} package) is used to implement KBAL. KOM fits into the \ref{eq:approxbal_w2} framework where $\mathcal{F}$ is a ball in a RKHS, and $B(\textbf{w}) = \lambda^2 || \textbf{w} ||_2^2$ with a data-driven choice of $\lambda$, described in \cite{kallus2020generalized}. We code KOM in \texttt{R} from scratch, using \texttt{MOSEK} (through the  \texttt{Rmosek} package in \texttt{R}) to solve the \ref{eq:approxbal_w2} problem, and the \texttt{optim} function in \texttt{R} to find $\lambda$. $\hat{f}_0$ is found by KRLS for augKBAL and augKOM. augKOM is a special case of the estimator studied by \cite{hirshberg2017augmented}. For each method, we use the Gaussian kernel 
with standardized $X$ and we forgo a kernel-selection procedure (e.g., \citealp{kallus2020generalized}) for simplicity. Additionally, KRLS with the Gaussian kernel and $b=2$ performs well here (the $\hat{f}_0 (X_i)$ and $f_0 (X_i)$ are correlated at 0.954 on average), and $b=P$ has been found to perform well empirically (e.g., \citealp{hainmueller2014kernel}; \citealp{scholkopf2002learning}).} Define bias, root mean squared error (RMSE), and relative effective sample size (RESS) as
	\begin{align*}
		& \text{Bias} = \frac{1}{M} \sum_{m=1}^{M} \biggr( \hat{\tau}_{\mathrm{wdim}} (\hat{\mathbf{w}}^{(m)}) - \mathrm{ATT} \biggr), \ \ \ \text{RMSE} = \sqrt{\frac{1}{M} \sum_{m=1}^{M} \biggr( \hat{\tau}_{\mathrm{wdim}} (\hat{\mathbf{w}}^{(m)}) - \mathrm{ATT} \biggr)^2  }, \\
		& \text{RESS} = \frac{1}{M}  \sum_{m=1}^{M} (n_0^{(m)})^{-1} \biggr( \frac{ (\sum_{i:D_i = 0} \hat{w}^{(m)}_i)^2}{\sum_{i:D_i = 0} (\hat{w}^{(m)}_i)^2}\biggr)
	\end{align*}
where $m$ indexes the iteration from $1$ to $M$, $\hat{\tau}_{\mathrm{wdim}} (\hat{\mathbf{w}}^{(m)})$ and $\hat{\mathbf{w}}^{(m)}$ are the estimate and weight vectors, respectively, from the $m$th iteration, and $n_0^{(m)}$ is the size of the control group in the $m$th iteration. In Figure~\ref{fig.dgp1.bias}, TFB-K shows minimal bias, unlike KOM, and though KBAL also shows little bias, TFB-K's RMSE is nearly two-thirds that of KBAL and KOM. Figure~\ref{fig.dgp1.balance} reveals why this is --- TFB-K leaves imbalance in $Z^{(1)}$ of the opposite sign of the original imbalance to offset imbalance in the other $Z^{(\ell)}$. Consequently, although KBAL obtains better balance in the $Z^{(\ell)}$, its weights are higher variance than are TFB-K's, as evidenced by TFB-K's higher RESS (see Figure~\ref{fig.dgp1.bias}). Additionally, TFB-K outperforms augKBAL and augKOM here, although augKOM reduces the gap between KOM and TFB-K.  
    
TFB-K's strategy here notably challenges traditional advice that one's weights should attain the best balance in $X$ possible given the setting (e.g., \citealp{stuart2010matching}). However, a more universal guiding principle is that, given equal bias, a more efficient estimator is preferable. Further, TFB should in general inhibit imbalances from becoming too extreme by using the variance of $\hat{f}_0$ to dictate how confidently it leaves imbalance. This is demonstrated in Figure~\ref{fig.dgp1.balance} --- balance is clearly a priority for TFB-K, as it successfully reduces the imbalance in each variable. Additionally, it obtains better balance in the $Z^{(\ell)}$ than does KOM because, tasked with minimal overlap in the hard-to-balance clusters, KOM instead prioritizes lower variance weights, as seen in its higher RESS.

Finally, recall that while key aspects of DGP 1 are not varied here for clarity, Appendix~\ref{app:dgp1_extension} examines how sensitive TFB's performance is to varying these aspects. Specifically, (i) sample size ($n$); (ii) level of overlap in $Z^{(1)}$ (through its coefficient in (\ref{eq:dgp1_ps})); and (iii) true $R^2$ for the model (through the variance of $\epsilon$ in (\ref{eq:dgp1_y})) are varied for a total of 18 different settings. To summarize the results of this more expansive simulation, TFB-K is no worse than the comparison estimators in these settings, and is often better when estimation is more challenging. Specifically, TFB-K has the lowest RMSE in the 12 settings where overlap is ``medium" or ``low", again demonstrating that TFB-K is more efficiently balancing the $Z^{(\ell)}$ when exact balance not feasible. TFB-K then either has the lowest RMSE, or is roughly tied for the lowest RMSE, in the 6 settings where overlap is ``high". Whether or not TFB-K leaves negative imbalance in $Z^{(1)}$ at a given sample size is sensitive to (i) the $R^2$ for the model and (ii) the level of overlap in $Z^{(1)}$. In particular, when the $R^2$ is higher, TFB-K more confidently leaves imbalance in $Z^{(1)}$ because $\hat{f}_0$ has lower variance. Further, with more overlap in $Z^{(1)}$, leaving negative imbalance is a viable strategy because it does not require outlandishly high variance weights. The settings in which TFB-K leaves negative imbalance in $Z^{(1)}$ coincide with the settings in which TFB-K shows little-to-no bias.

\subsubsection{Demonstration 2: Improvement over regression}\label{sec:acic}

This second demonstration shows that TFB is often superior to estimating treatment effects purely through regression, because seeking overall balance in the covariates protects TFB against an over-reliance on a poor model, and allows it to exhibit double robustness. Still, TFB's performance is connected to the accuracy of its outcome model. We apply TFB to the data from the 2016 American Causal Inference Conference (ACIC) Challenge, ``Is Your SATT Where It's At?" (\citealp{dorie2019automated}), additionally confirming TFB's validity in a wide variety of settings.  Note that we detail a more extensive analysis of TFB's performance on this data in Appendix~\ref{app:acic_extension}, and summarize the main findings here.

In this ACIC challenge, various methods for estimating average treatment effects were applied to common simulated datasets, and the performance of these methods was compared. The data come from 77 distinct DGPs, each simulated 100 times (for a total of 7,700 datasets).\footnote{These data were accessed through the \texttt{aciccomp} GitHub repository.} All DGPs use a common covariate matrix $\mathbf{X}$ that is based on real data from the Collaborative Perinatal Project (\citealp{niswander1972women}). But the DGPs vary in how $D$ and $Y(d)$ are generated (from $X$), varying settings such as the functional forms of the conditional expectation functions of the potential outcomes and the propensity score, covariate overlap, treatment effect heterogeneity, and others (see \citealp{dorie2019automated} for more detail).

We apply TFB-BART and TFB-K to these datasets and, following \cite{dorie2019automated} and \cite{cousineau2023estimating}, calculate their \textit{standardized} bias and RMSE in relation to SATT:
\begin{align*}
    \text{Std. Bias} &= \frac{1}{7700} \sum_{s=1}^{77} \sum_{m=1}^{100} \biggr( \frac{\hat{\tau}^{(\mathrm{TFB})}_{sm} - \mathrm{SATT}_{sm}}{\sigma^{(y)}_{sm}} \biggr) \\
    \text{Std. RMSE} &= \sqrt{\frac{1}{7700} \sum_{s=1}^{77} \sum_{m=1}^{100}\biggr( \frac{\hat{\tau}^{(\mathrm{TFB})}_{sm} - \mathrm{SATT}_{sm}}{\sigma^{(y)}_{sm}} \biggr)^2  }
\end{align*}
where $s$ indexes the DGP, $m$ indexes the iteration, and $\sigma^{(y)}$ is the standard deviation of the $Y_i$. Table~\ref{tab.acic} reports TFB-BART's and TFB-K's results, along with the results of the methods reported in \cite{cousineau2023estimating}, ordered by standardized RMSE. Table~\ref{tab.acic} also reports analogous results for an unweighted difference in means (DIM), and two g-computation (\citealp{robins1986new}) estimators that impute the unobserved $Y_i(0)$ in SATT with predictions from a regression model: BART and KRLS. The KRLS and BART models are the same models that TFB-K and TFB-BART use. 
\begin{table}[!h]
\vspace{0.05in}
\caption{Performance of Estimators in 2016 ACIC Challenge}\label{tab.acic}

\vspace{-0.15in}

	\begin{center}
	\scriptsize
     \begin{tabular}{ l | c | c | c  }
      & Std. Bias & Std. RMSE & Number of Datasets \\
     \hline
      \texttt{bart\_on\_pscore} &  0.001 & 0.014 &  7700 \\
     \hline
      \textcolor{red}{TFB-BART*} &  \textcolor{red}{0.001} & \textcolor{red}{0.015} &  \textcolor{red}{7700} \\
     \hline
      \texttt{bart\_tmle} &  0.000& 0.016&  7700 \\
     \hline
      \texttt{mbart\_symint} & 0.002 & 0.017&  7700 \\
     \hline
      \texttt{bart\_mchains} &  0.002 & 0.017&  7700 \\
     \hline
      \texttt{bart\_xval} & 0.002 & 0.017 &  7700 \\
     \hline
      \texttt{bart} &  0.002 & 0.018 &  7700 \\
     \hline
      \textcolor{red}{BART*} &  \textcolor{red}{0.002} & \textcolor{red}{0.018} &  \textcolor{red}{7700} \\
     \hline
      \texttt{sl\_bart\_tmle} & 0.003 & 0.029 & 7689  \\
     \hline
      \texttt{h2o\_ensemble} & 0.007 & 0.030 &  6683 \\
     \hline
      \texttt{bart\_iptw} &  0.002 & 0.032&  7700 \\
     \hline
      \texttt{sl\_tmle} & 0.007 & 0.032 &  7689 \\
     \hline
      \texttt{superlearner} & 0.006 & 0.039 &  7689 \\
     \hline
      \texttt{calcause} & 0.003 & 0.043 &  7694 \\
     \hline
      \texttt{tree\_strat} & 0.022 & 0.052 &  7700 \\
     \hline
      \texttt{balanceboost} & 0.020 & 0.054 &  7700 \\
     \hline
      \texttt{adj\_tree\_strat} & 0.027 & 0.074 &  7700 \\
     \hline
      \textcolor{red}{TFB-K*} &  \textcolor{red}{0.032} & \textcolor{red}{0.079} &  \textcolor{red}{7700} \\
     \hline
      \texttt{lasso\_cbps} &  0.027 & 0.082 &  7108 \\
     \hline
      \texttt{kbal} &  0.036 & 0.091 &  7700 \\
     \hline
      \texttt{sl\_tmle\_joint} & 0.010 & 0.102 &  7698 \\
     \hline
      \textcolor{red}{KRLS*} &  \textcolor{red}{0.042} & \textcolor{red}{0.103} &  \textcolor{red}{7700} \\
      \hline
      \texttt{cbps} & 0.041 & 0.107 &  7344 \\
     \hline
      \texttt{balancehd} & 0.041 & 0.107 & 7700  \\
     \hline
      \texttt{teffects\_psmatch} & 0.043 & 0.108 & 7506  \\
     \hline
      \texttt{sbw} & 0.041 & 0.110 & 4513  \\
     \hline
      \texttt{cbps\_exact} & 0.041 & 0.112 & 7700  \\
     \hline
      \texttt{ebal} & 0.041 & 0.117 & 4513  \\
     \hline
      \texttt{cbps\_over} & 0.044 & 0.125 & 7700  \\
     \hline
      \texttt{linear\_model} & 0.045 & 0.135 & 7700  \\
     \hline
      \texttt{mhe\_algorithm} & 0.045 & 0.135 &  7700 \\
     \hline
      \texttt{teffects\_ra} &  0.043 & 0.140 &  7685 \\
     \hline
      \texttt{genmatch} & 0.052 & 0.151 &  7700 \\
     \hline
      \texttt{teffects\_ipwra} & 0.044 & 0.166 & 7634  \\
     \hline
      \textcolor{red}{DIM*} & \textcolor{red}{0.085} & \textcolor{red}{0.215} & \textcolor{red}{7700}  \\
     \hline
      \texttt{teffects\_ipw} & 0.042 & 0.301 & 7665  \\
    \hline
     \end{tabular}
     \normalsize
	\end{center}
	\vspace{0.05in}
    \subcaption*{\textit{Note:} The ``Number of Datasets" column reports how many datasets (out of 7,700) were used for the results. The rows from TFB-BART, TFB-K, BART, KRLS, and DIM (starred (*), and in red) are new from our study. The other rows are copied from Tables 5 and 6 in \cite{cousineau2023estimating}.}
\vspace{-0.15in}
\end{table}

\cite{dorie2019automated} note that non-parametric machine learning methods excel in these simulated datasets, particularly those that incorporate tree-based methods such as BART. Thus, it is affirming that TFB-BART is one of the top performers in Table~\ref{tab.acic}, reporting the second lowest RMSE (0.015) that is just barely higher than that of the top estimator (0.014). Additionally, TFB-BART slightly improves on the BART g-computation estimator --- reducing the RMSE from 0.018 to 0.015. 

Because tree-based methods excel in these datasets, it is unsurprising that TFB-K is not among of the top performers  in Table~\ref{tab.acic} --- in fact, all of the estimators with lower RMSE than TFB-K incorporate tree-based methods when modeling the outcome or the propensity score. Additionally, we show Appendix~\ref{app:acic_extension} that TFB-K struggles, relative to TFB-BART, when the DGP involves step functions. More interesting is TFB-K's performance relative to KBAL (\texttt{kbal}) and KRLS given the performance of  \texttt{linear\_model}, the coefficient on $D$ from an OLS regression of $Y$ on $X$ and $D$. TFB-K has meaningfully lower RMSE \textit{and} bias than does KBAL, again showing the benefits of consulting a model for $Y$ to determine how to strategically balance the high dimensional $K_i$. Further, TFB-K's performance is impressive given the performance of KRLS, whose RMSE (0.103) resembles that of \texttt{linear\_model} (0.135) more than that of BART (0.018). In other words, TFB-K is only outperformed by methods that incorporate tree-based models, despite the outcome model it relies on not fitting these DGPs particularly well (as does BART). This shows the robustness that is gained from TFB-K's seeking overall balance on $K_i$ to prevent an over-reliance on its outcome model.

\begin{figure}[!h]
	\vspace{0.15in}
	\begin{center}
	\caption{Comparison of TFB and g-computation estimators in the 77 ACIC DGPs}\label{fig.acic}
    \vspace{-.25in}
    \begin{subfigure}{.45\textwidth}
    \begin{center}
    \includegraphics[scale=0.50]{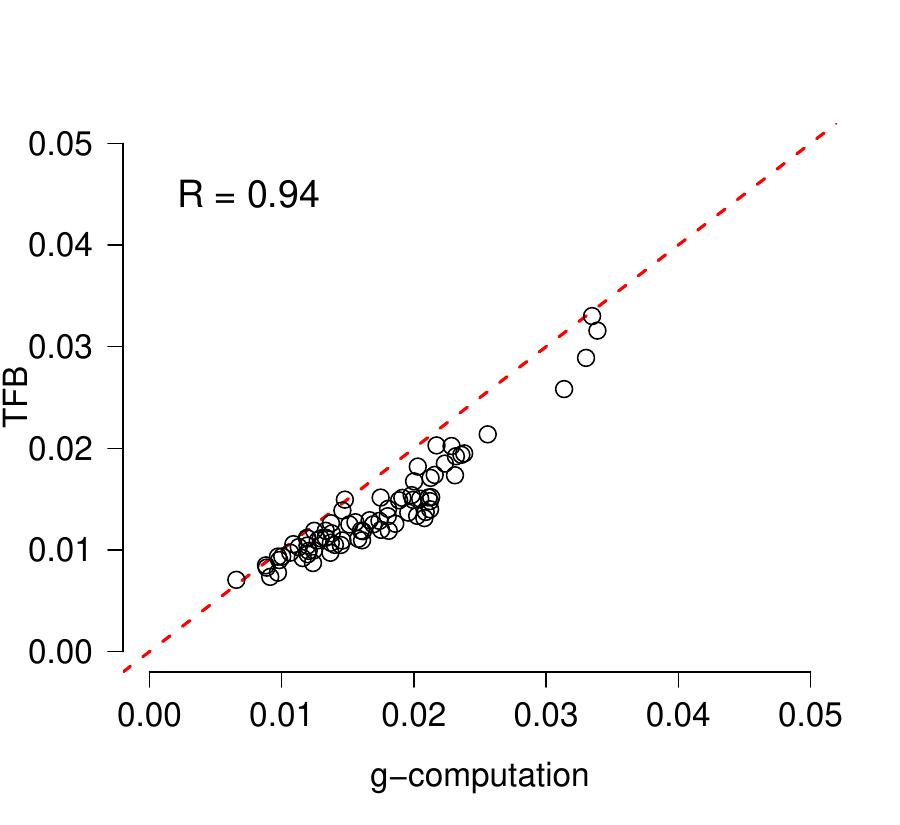} 
    \end{center}
    \vspace{-.25in}
    \subcaption{TFB-BART vs. BART}\label{fig.acic.bart}
    \end{subfigure}
    \begin{subfigure}{.45\textwidth}
    \begin{center}
    \includegraphics[scale=0.50]{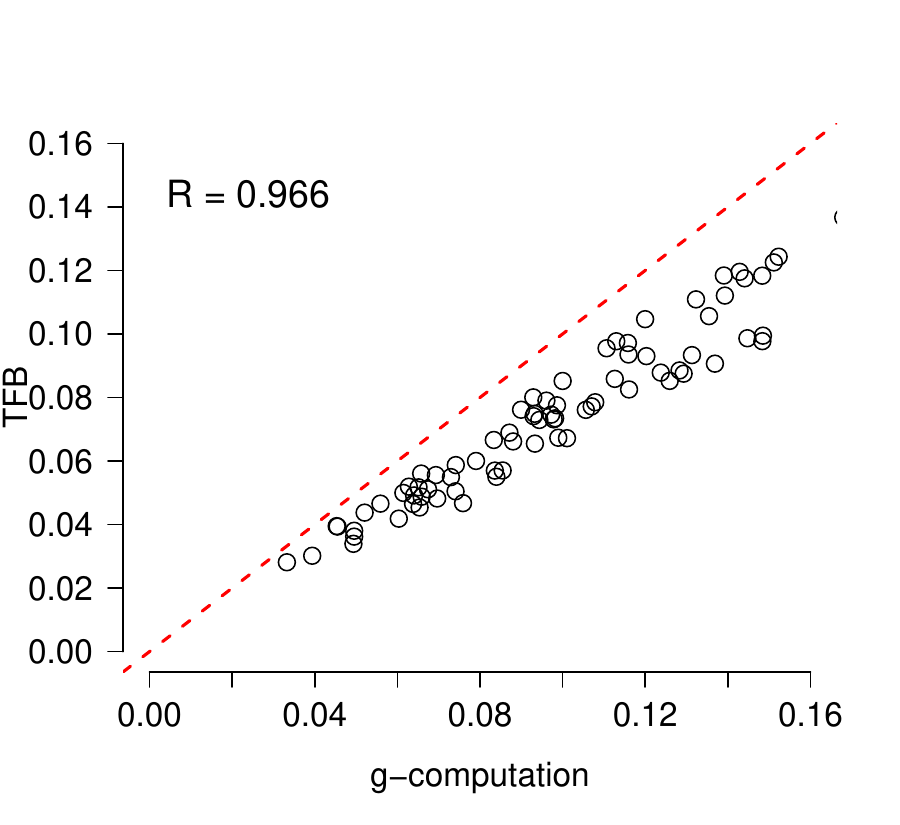} 
   \end{center}
    \vspace{-.25in}
    \subcaption{TFB-K vs. KRLS}\label{fig.acic.krls}
    \end{subfigure}
    \subcaption*{\textit{Note:} Plots standardized RMSE in each of the 77 DGPs, averaged over the corresponding 100 iterations, from the g-computation ($x$-axis) and TFB ($y$-axis) estimators. The red dashed line indicates equality. Note that the scale of the axes is larger when comparing TFB-K and KRLS (Figure~\ref{fig.acic.krls}).}
	\end{center}
	\vspace{-0.25in}
	\end{figure}

On this idea, a deeper comparison between the TFB estimators and the corresponding g-computation estimators is illuminating. Figure~\ref{fig.acic} compares TFB-BART with BART (Figure~\ref{fig.acic.bart}) and TFB-K with KRLS (Figure~\ref{fig.acic.krls}) on RMSE in each of the 77  DGPs, averaged over their 100 iterations. To start, each TFB estimator's performance is highly correlated with the performance of its corresponding g-computation estimator, indicating that TFB does, at least partially, rely on its $\hat{f}_0$ to estimate $f_0$ well. However, other than a few exceptions, the TFB estimators have nearly uniformly  lower RMSE than do the corresponding g-computation estimators across the DGPs. TFB's improvement also tends to be more drastic when the g-computation estimator struggles (see the points with higher values on the $x$-axis). This again shows that TFB's seeking overall balance on (transformations of) $X$ is protecting it against an over-reliance on $\hat{f}_0$.

Another factor for TFB's improvement over g-computation is the equivalence between TFB and propensity score weights noted in Section~\ref{subsec:tfb.weights}. Specifically, this equivalence allows TFB to exhibit double robustness. In Appendix~\ref{app:acic_extension}, we investigate which aspects of the 77 DGPs examined are predictive of the performance gap between TFB and its corresponding g-computation estimator. We leave the details to Appendix~\ref{app:acic_extension}, but we find that one of the most predictive aspects is the degree of ``alignment" (see \citealp{dorie2019automated}) in the DGP. In short, the alignment is how influential the true outcome model $f_0$ is in the true propensity score $\pi$. TFB tends to perform markedly better than the corresponding g-computation estimator when the alignment in the DGP is ``high", i.e., when $\pi$ is mostly a function of $f_0$. In these settings, because TFB's weights are propensity score weights where $\pi$ is modeled as a function of $\hat{f}_0$ (and $X)$, TFB is implying an (approximately) correctly specified $\pi$.


\section{Application 1: Twelfth grade math in L.A. Unified}\label{sec:lausd}

We now apply TFB to estimate the effects of taking mathematics courses in the Los Angeles Unified School District (L.A. Unified) of Los Angeles, California, United States of America (USA). This section uses data from the \href{https://laeri.luskin.ucla.edu/}{Los Angeles Education Research Institute} (LAERI) data archive—a longitudinal archive of L.A. Unified student, staff, and school-level administrative and survey data. For clarity, we summarize the main findings from our analyses here, and leave many of our results to Appendix~\ref{app:application1_additional}. Note also that we provide another application of TFB in Appendix~\ref{sec:applications}, assessing TFB's performance on the National Supported Work Demonstration  data (\citealp{lalonde1986evaluating}).

L.A. Unified is the second-largest school district in the USA. It serves over 500,000 students overall, and over 100,000 high school (grades 9 through 12)   students (\citealp{lausd2025finger}).  L.A. Unified students only need to complete three years of high school mathematics in order to graduate high school (\citealp{lausd2025pathways}). The three corresponding courses are, normatively, Algebra 1, Geometry, and Algebra 2, occurring in grades 9, 10, and 11, respectively.\footnote{L.A. Unified also allows other mathematics pathways. For example, students can follow accelerated pathways that could involve taking two or more of the courses in the same year (\citealp{lausd2025pathways}), or taking one or more of these courses before ninth grade (\citealp{lausd2017middle}). The district also allows alternative sequences, such as taking Financial Algebra 1 and/or 2 instead of traditional Algebra 1 and/or 2 (\citealp{lausd2025pathways}).} As a result, students can often choose whether or not to take a mathematics course (or a course in a related field, such as statistics) in twelfth grade. These choices are undoubtedly a function of student-level factors, such as students' ability in math (as measured by GPAs and standardized test scores), and school-level factors, such as course offerings and availability. 

\cite{wainstein2023access} originally estimated the effects of taking math in twelfth grade in L.A. Unified on end-of-high school and college enrollment and persistence outcomes in a research report that was published by LAERI, henceforth ``The LAERI Report". 
The authors estimated that taking math in twelfth grade increased the likelihood of students enrolling in college in the years they examined, despite a small decrease in their cumulative high school GPA. 
The authors also estimated the effects of taking different types of mathematics courses, or courses in related subjects. The authors focused on comparing Precalculus and Calculus courses with Statistics courses, as those were the most commonly taken courses in the district in the years they examined. But they also investigated the effects of taking two alternative mathematics-related courses: Transition to College Math and Statistics (TCMS) and Introduction to Data Science (IDS).\footnote{The goal of TCMS is to prepare students for college-level mathematics (\citealp{lausd2025specialized}), covering many topics in mathematics and statistics through applications to real-life problems (\citealp{wainstein2023access}). IDS develops students' computer programming/coding and data analysis skills (\citealp{lausd2025specialized}).} Ultimately, the authors found very few differences between taking these types of courses that were statistically significant at traditional thresholds (e.g., 0.05). Statistical power was an issue for investigations of TCMS and IDS, as relatively smaller numbers of students took these courses in the examined years. 

\subsection{Data and methods}\label{subsec:lausd_data}

We focus on a subset of the students studied in The LAERI Report: students who, by the end of eleventh grade, had completed Algebra 1, Geometry, and Algebra 2; earned at least a C in these math courses, and thus had completed the math requirements for eligibility for admission to a public four-year university in California (e.g., California State University or University of California campuses); and had not taken any math courses beyond Algebra 2 (e.g., Precalculus). The data we use describes 11,255 students who were in twelfth grade in academic years 2016-17 and 2017-18. About two-thirds (66.5\%) of these students took a mathematics course, or a course in a related field, in twelfth grade. Of those who took a course in mathematics (or a related field), the most common courses that these students took were Precalculus (63.2\%) and Statistics (21.4\%). Significantly fewer students took TCMS or IDS (8.4\% and 3.6\%, respectively).

We consider two outcomes: (1) cumulative weighted GPA at the end of twelfth grade and (2) enrolling in a two-year or four-year college within one year of high school graduation. We estimate the effects on these two outcomes of taking any mathematics course in twelfth grade versus not taking a math course. For this analysis,  we follow The LAERI Report in treating mathematics courses, such as Precalculus, and courses in related fields, such as Statistics, as being the same ``treatment". Thus, unless otherwise noted, when we use the terms ``math" or ``mathematics", we are including related fields (e.g., Statistics). However, we also compare the effects of taking Precalculus, Statistics, TCMS, and IDS. 

For covariates, the data includes various demographics variables (e.g., gender and race/ethnicity), academics-related variables (e.g., eleventh grade standardized test scores and GPAs), behavioral variables (e.g., behavioral grades and attendance rate), measures of self-perception (e.g., self-reported educational expectations), the academic year in which the student was in twelfth grade, and school attended in twelfth grade.\footnote{We utilize exactly the same covariates as does The LAERI Report, and point readers to that report for a complete list of these variables and how they are defined.} Table~\ref{tab.mthvnone.descriptives} reports the average values on a key \textit{subset} of the covariates for students in the analytic sample (see Appendix~\ref{app:application1_additional_samples}) when estimating the effects of taking any math course in twelfth grade. 
Table~\ref{tab.mthvnone.descriptives} offers evidence that these covariates (or rather, the constructs that underlie them) influenced students' math course-taking in twelfth grade: when compared to students who did not take math in twelfth grade, those who did take math, on average, had higher GPAs through eleventh grade, scored higher on state-wide standardized tests, were more likely to have taken the SAT/ACT, had higher behavioral grades, and were more likely to self-report to expect to attain a four-year degree. We see similar differences across students who took different types of mathematics courses in Appendix~\ref{app:application1_additional_descriptives}. 
\begin{table}[!h]
\vspace{0.05in}
\caption{Descriptive statistics for analytic sample when estimating the effects of taking any math course in twelfth grade}\label{tab.mthvnone.descriptives}

\vspace{-0.15in}

	\begin{center}
	\footnotesize
     \begin{tabular}{ |m{4.1cm} | m{1.2cm} m{1.2cm} | m{1.2cm} m{1.2cm} | m{1.2cm} m{1.2cm} |}
     \hline
     \multicolumn{1}{ |c| }{ \textbf{Covariate} } & \multicolumn{2}{ c| }{\textbf{All}} & \multicolumn{2}{  c | }{ \textbf{No Math} } & \multicolumn{2}{ c| }{ \textbf{Math} } \\
      & Mean & (SD) & Mean & (SD) & Mean & (SD)  \\
     \hline
     Cohort: 2017-18 & 0.531 & (0.499) & 0.530 & (0.499) & 0.532 & (0.499) \\
     Gender: Female & 0.540 & (0.498) & 0.548 & (0.498) & 0.536 & (0.499) \\
     Age (in Months) & 195.618 & (6.026) & 195.976 & (6.544) & 195.404 & (5.685) \\
     GPA through 11th & 2.870 & (0.566) & 2.687 & (0.543) & 2.979 & (0.552) \\
     ELA SBAC (Z-Score) & 0.042 & (0.778) & -0.166 & (0.792) & 0.166 & (0.742) \\
     Math SBAC (Z-Score) & -0.041 & (0.696) & -0.253 & (0.673) & 0.086 & (0.679) \\
     Took SAT/ACT & 0.322 & (0.467) & 0.225 & (0.418) & 0.380 & (0.485) \\
     WPA & 2.452 & (0.334) & 2.360 & (0.341) & 2.506 & (0.317) \\
     CPA & 2.605 & (0.269) & 2.546 & (0.279) & 2.641 & (0.256) \\
     Expect: Unsure & 0.091 & (0.288) & 0.104 & (0.305) & 0.084 & (0.277) \\
     Expect: HS or Less & 0.033 & (0.177) & 0.041 & (0.198) & 0.028 & (0.164) \\
     Expect: Certificate or 2-Year & 0.052 & (0.222) & 0.068 & (0.252) & 0.043 & (0.202) \\
     Expect: 4-Year or More & 0.556 & (0.497) & 0.517 & (0.500) & 0.579 & (0.494) \\
     Expect: Missing & 0.268 & (0.443) & 0.271 & (0.444) & 0.267 & (0.442) \\
     \hline
     $n$ & 9,993 &  & 3,736 &  & 6,257 &  \\
     \hline
     \end{tabular}

     \normalsize
	\end{center}
    \subcaption*{\textit{Note:} Average values (and standard deviations) for a subset of key covariates for all students in the analytic sample (``All"), students who did not take math in twelfth grade (``No Math"), and students who did take math in twelfth grade (``Math"). This analytic sample ($n$=9,993) differs from the full sample ($n$=11,255) because it drops schools in each year for which all students took math, or all students did not take math --- see Appendix~\ref{app:application1_additional_samples} for more detail. All covariates are recorded for the student at the end of their eleventh grade year.  ``Cohort: 2017-18" is an indicator variable for being in twelfth grade in the 2017-18 academic year. The SBAC is a math and English language arts (ELA) standardized test that was developed by the Smarter Balanced Assessment Consortium (SBAC), and, in the years examined, was taken by California students in eleventh grade. Work Effort GPA (WPA) and Cooperation  GPA (CPA) are two behavioral GPAs that are on a 1-3 scale. The ``Expect" variables indicate students' self-reported expected educational attainment.}
\vspace{-0.15in}
\end{table}

We apply TFB-K, TFB-L, and TFB-BART to account for this confounding.\footnote{Due to computational limitations, we use the \texttt{CLARABEL} and \texttt{ECOS} solvers through the \texttt{CVXR} package in \texttt{R}, instead of \texttt{MOSEK}, to solve TFB's optimization problem. Additionally, when applying TFB-K to estimate the effects of taking any math course, we apply the low-rank Nystrom approximation option in the \texttt{KRLS} package in \texttt{R}.} For TFB-L, we include as additional covariates all of the squared and pair-wise interactions terms of the academics-related covariates, the behavioral covariates, and the indicator variable for twelfth grade year. Following The LAERI Report, we apply the version of TFB that targets the ATE (see Appendix~\ref{app:ate_atc}) when comparing taking any math course in twelfth grade to not taking math, and when comparing taking Statistics to taking Precalculus. When comparing TCMS and IDS to Precalculus and Statistics, we estimate the ATT. 

We also compare TFB to some of the estimation strategies employed by The LAERI Report. The authors implemented linear, weighted least squares regression --- regressing the outcomes on the treatment and covariates --- with different sets of weights. Note that this estimator is in the augmented form, $\hat{\tau}_{\mathrm{aug}}$, discussed in Section~\ref{subsec:tfb.weights}. Their primary set of weights were propensity score matching weights through Preferential Within-Cluster Matching (CMatch; \citealp{arpino2016propensity}), which prioritizes matching students who attended the same school in twelfth grade. Some of the alternative weighting strategies that the authors employed were Kernel Balancing (KBAL; \citealp{hazlett2018kernel}); Kernel Matching (KMatch; \citealp{heckman1998matching}); and no weighting (i.e., OLS).

\subsection{Results: Any math course versus no math}\label{subsec:lausd_mathvsnone}

Table~\ref{tab.mthvnone.estimates} reports TFB-L's and TFB-BART's estimated effects of taking any math course in twelfth grade, versus not taking math, on the two outcomes examined. Overall, TFB-L's and TFB-BART's conclusions here are the same as those from The LAERI Report: taking math in twelfth grade increases the likelihood of enrolling in college, despite a small decrease in GPA. More specifically, TFB-L and TFB-BART estimate a \textit{decrease} in end-of-high school GPA by roughly $0.04$ GPA points, and an \textit{increase} in the probability of enrolling in college by about $4$ percentage points, which are very similar to the estimates from the other methods. Though, TFB-L's estimate for the college enrollment outcome (+0.035) is slight lower than the other estimates (+0.042-0.057). The estimated effects from all methods are statistically significant at the 0.01 level, and all except one (TFB-L for college enrollment) are statistically significant at the 0.001 level. 
 \begin{table}
\vspace{0.05in}
\caption{Estimated effects of taking any math course in twelfth grade}\label{tab.mthvnone.estimates}

\vspace{-0.15in}

	\begin{center}
	\footnotesize
     \begin{tabular}{ |m{2.1cm} | m{1.38cm} m{1.2cm} m{1.3cm} m{1.3cm} | m{1.38cm} m{1.2cm} m{1.3cm} m{1.3cm} |}
     \hline
     \multicolumn{1}{ |c| }{ \textbf{Method} } & \multicolumn{4}{ c| }{\textbf{End-of-High School GPA}} & \multicolumn{4}{ c| }{\textbf{College Enrollment}} \\
      & Estimate & (SE) & RESS: & RESS:  & Estimate & (SE) & RESS: & RESS: \\
      &  & & Control & Treated  & & & Control & Treated \\
     \hline
     Unadjusted & 0.232*** & (0.022) & 1.000 & 1.000 & 0.098*** & (0.010) & 1.000 & 1.000 \\
     \hline
     WLS: CMatch & -0.039*** & (0.004) & 0.415 & 0.569 & 0.049*** & (0.012) & 0.415 & 0.569 \\
     WLS: KBAL & -0.044*** & (0.004) & 0.478 & 0.645 & 0.057*** & (0.013) & 0.478 & 0.645 \\
     WLS: KMatch & -0.036*** & (0.004) & 0.447 & 0.662 & 0.044*** & (0.011) & 0.447 & 0.662 \\
     OLS & -0.042*** & (0.004) & 1.000 & 1.000 & 0.053*** & (0.010) & 1.000 & 1.000 \\
     \hline
     TFB-L & -0.038*** & (0.004) & 0.260 & 0.644 & 0.035** & (0.013) & 0.324 & 0.648 \\
     TFB-BART & -0.039*** & (0.003) & 0.681 & 0.879 & 0.042*** & (0.011) & 0.635 & 0.654 \\
     \hline
     \end{tabular}

     \normalsize
	\end{center}
    \subcaption*{\textit{Note:} The ``Unadjusted" estimates are unweighted differences in means. WLS = Weighted Least Squares. CMatch = Preferential Within-Cluster Matching, which was the primary estimation approach in The LAERI Report. KBAL = Kernel Balancing. KMatch = Kernel Matching. OLS = Unweighted Least Squares. SE = Standard Error. The non-TFB estimates and associated standard errors are pulled directly from The LAERI Report. Standard errors for TFB are estimated per Appendix~\ref{subsec:tfb.asymptotics}. * $p<0.05$, ** $p<0.01$, *** $p<0.001$.} 
\vspace{-0.15in}
\end{table}

\begin{figure}
\vspace{0.15in}
\begin{center}
\caption{Balance plot for estimating the effects of taking any math course in twelfth grade}\label{fig.mthvnone}
	\vspace{-0.50in}
    \begin{center}
    \begin{subfigure}{.47\textwidth}
    \includegraphics[scale=0.45]{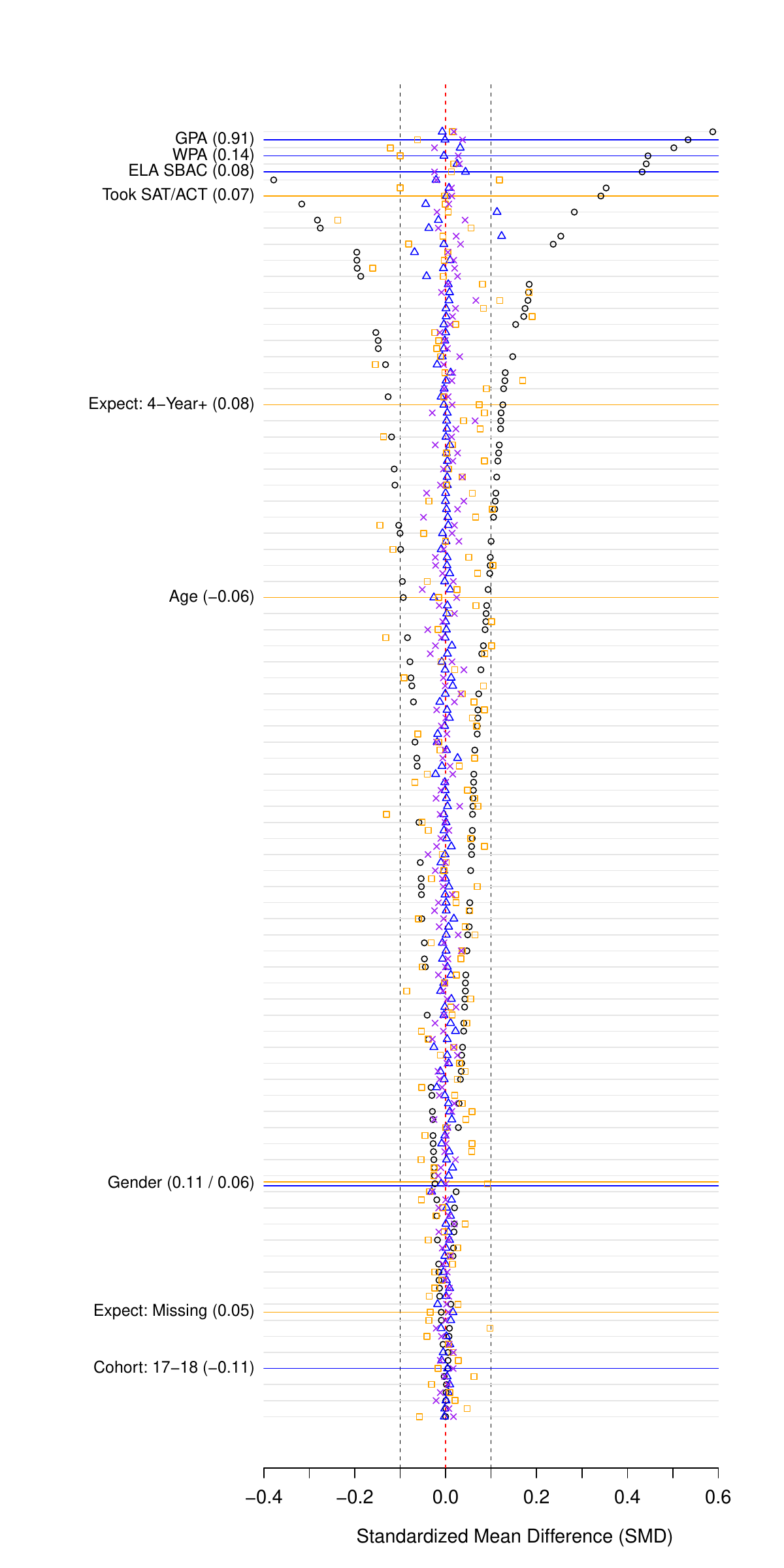} 
    \end{subfigure}
     \begin{subfigure}{.43\textwidth}
    \includegraphics[scale=0.45]{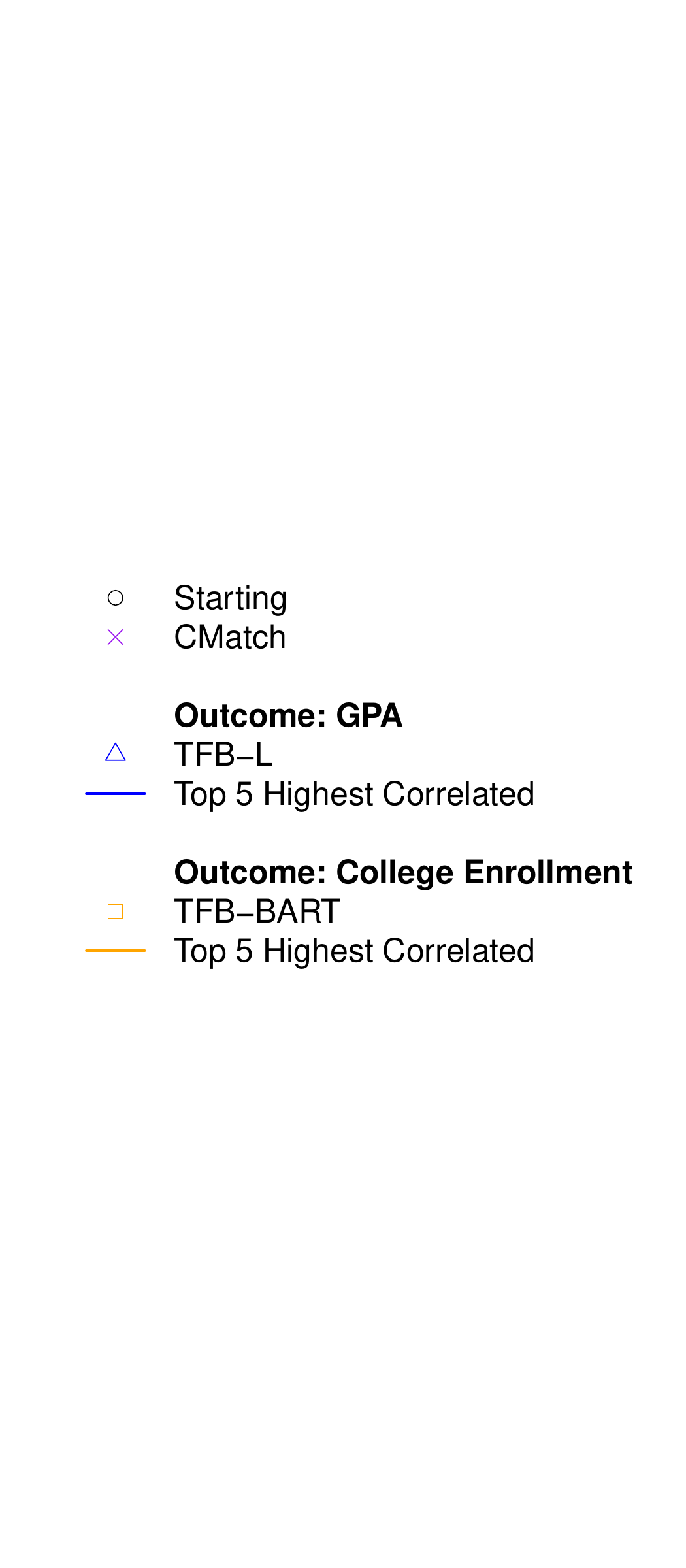} 
    \end{subfigure}
    \end{center}
    \subcaption*{\textit{Note:} SMDs before and after weighting for all covariates. For each outcome, only the five (non-school indicator) covariates with the highest absolute partial correlations (given the other covariates) with the outcome are labeled, along with their partial correlations in parentheses. The blue horizontal lines highlight the five most correlated covariates with the end-of-high school GPA outcome. The orange horizontal lines highlight the five most correlated covariates with the college enrollment outcome. ``Gender" is highlighted for both outcomes, and its partial correlations for end-of-high school GPA and college enrollment are both listed (in that order). Vertical black dotted lines indicate the common SMD benchmark of $\pm 0.10$  (\citealp{austin2009balance, normand2001validating}). The vertical red dotted line indicates a SMD of 0, or exact balance. Horizontal light gray lines are visual aids help the reader follow the SMD for a covariate across the weighting methods. }
\end{center}
\vspace{-0.25in}
\end{figure}

However, while TFB-L's and TFB-BART's conclusions are similar to the methods used in The LAERI Report, these applications of TFB reach this conclusion despite leaving more imbalance in the covariates after weighting than does The LAERI Report's primary weighting strategy, again showcasing TFB's strategic imbalance.  Figure~\ref{fig.mthvnone} reports the standardized mean difference (SMD) in each covariate for TFB-L, applied to the end-of-high school GPA outcome; for TFB-BART, applied to the college enrollment outcome; and for CMatch, The LAERI Report's primary estimation approach.\footnote{The SMD in a covariate is the difference in the (weighted) means of that covariate between the treatment and control groups, standardized by a combination of the (weighted) standard deviations in the treatment and control groups.} Note that only TFB-L is shown for the end-of-high school GPA outcome because the LASSO was the most predictive model for that outcome, and similarly, TFB-BART is only shown for the college enrollment outcome because BART was the most predictive model for that outcome (see Appendix~\ref{app:application1_additional_cvresults}). CMatch attains excellent balance on all of the covariates, with all SMDs within the commonly-used benchmark of $\pm 0.10$  (\citealp{austin2009balance, normand2001validating}). TFB-L here also obtains good balance on the covariates, but has SMDs over 0.10 for two covariates. TFB-BART is more aggressive, leaving several variables with SMDs over 0.10, and in several cases even increasing the imbalance in the covariate from where it started before weighting. Note also that imbalances for TFB are not addressed in post by augmented estimation, unlike CMatch which is ultimately combined with weighted least squares to produce its estimates in Table~\ref{tab.mthvnone.estimates}. 

Though while TFB-L and TFB-BART leave non-negligible imbalance in some covariates, none of these covariates are among the most predictive of the outcome in question. TFB-L, applied to the end-of-high school GPA outcome, attains excellent balance on the variables that have the highest partial correlation (given the other covariates) with that same outcome. In particular, TFB-L obtains near perfect balance on eleventh grade cumulative GPA, which has a partial correlation of 0.91 with the end-of-high school GPA outcome. TFB-BART, applied to the college enrollment outcome, also attains good balance on the variables that have the highest partial correlation with the college enrollment outcome. Additionally, because TFB-BART is not working as hard to attain good balance on all of the covariates, it appears to be prioritizing minimizing the weights' variance: TFB-BART has the highest overall RESS of all the weighting methods shown in Table~\ref{tab.mthvnone.estimates}. Note also that TFB-L and TFB-BART both seek balance in non-linear transformations of the covariates that are not pictured in Figure~\ref{fig.mthvnone}. In summary, the imbalance that TFB-L and TFB-BART leave here is not a mistake: these applications of TFB are clearly using their respective outcome models to determine what imbalance is acceptable. 

Finally, note that TFB-K was omitted from the results and discussion in this section. This is because KRLS does not model either outcome well --- in Appendix~\ref{app:application1_additional_cvresults}, we compare the five-fold cross-validation error of LASSO, KRLS, BART, and several other comparison models, and find that KRLS is by far the least predictive model. Accordingly, and in contrast to TFB-L and TFB-BART, the imbalance that TFB-K ultimately shows does appear to be a mistake, as it leaves high imbalance even on covariates that have high partial correlation with the outcome. Because of this, and because TFB's performance is, at least partially, tied to the effectiveness of its outcome model(s) (see Section~\ref{sec:acic}), we deem TFB-K unreliable here.\footnote{Because TFB-K fails here, it may be a surprise that KBAL produces roughly equivalent results to the other methods in Table~\ref{tab.mthvnone.estimates}. There are two main reasons for this. First, the specific implementation of KBAL in The LAERI Report first prioritizes exact balance on the untransformed covariates (using the \texttt{meanfirst} option in the \texttt{kbal} package in \texttt{R}) before attempting to balance the gram matrix. Second, these KBAL weights are ultimately combined with weighted least squares regression to produce the estimate in Table~\ref{tab.mthvnone.estimates}.} Nevertheless, we report TFB-K's leftover imbalance for this analysis in Appendix~\ref{app:application1_additional_balance}. Note that Appendix~\ref{app:application1_additional_balance} also reports the leftover imbalance in the covariates for TFB-BART when applied to the end-of-high school GPA outcome, and TFB-L when applied to the college enrollment outcome, which were omitted in Figure~\ref{fig.mthvnone}.


\subsection{Results: Comparing different math courses}\label{subsec:lausd_tcmsids}

Table~\ref{tab.othercourses.estimates} reports TFB's estimated effects of taking certain mathematics courses over others in twelfth grade. Note that only TFB-L's estimates are shown for the end-of-high school GPA outcome because the LASSO was the most predictive model for this outcome (see Appendix~\ref{app:application1_additional_cvresults}). Similarly, only TFB-BART's estimates are shown for the college enrollment outcome. See Appendix~\ref{app:application1_additional_estimates} for TFB-L's and TFB-BART's estimates for both outcomes, as well as the estimates from the methods from The LAERI Report.
 \begin{table}
\vspace{0.05in}
\caption{Estimated effects from TFB for taking different mathematics courses in twelfth grade}\label{tab.othercourses.estimates}

\vspace{-0.15in}

	\begin{center}
	\footnotesize
     \begin{tabular}{ |m{2cm} m{2cm}  m{1cm} | m{3.1cm} m{1.1cm} | m{3.1cm} m{1.1cm} | }
     \hline
     \multicolumn{3}{ |c| }{ \textbf{Course Comparison} } & \multicolumn{2}{ c| }{\textbf{End-of-High School GPA}} & \multicolumn{2}{ c| }{\textbf{College Enrollment}} \\
    $D=1$ & $D=0$  & $n$  & TFB-L Estimate & (SE) & TFB-BART Estimate & (SE) \\
     \hline
     Statistics & Precalculus & 3,807 & 0.013** & (0.005) & 0.001 & (0.019) \\
     TCMS & Precalculus & 1,357 & 0.015 & (0.012) & -0.021 & (0.048) \\
     IDS & Precalculus & 892 & 0.030** & (0.011) & -0.047 & (0.044) \\
     \hline
     TCMS & Statistics & 619 & 0.018 & (0.014) & -0.021 & (0.037) \\
     IDS & Statistics & 390 & 0.030* & (0.015) & -0.038 & (0.055) \\
     \hline
     \end{tabular}

     \normalsize
	\end{center}
    \subcaption*{\textit{Note:} 
    See Appendix~\ref{app:application1_additional_estimates} for complete results for TFB-L and TFB-BART, and for the comparison methods from The LAERI Report. See also Appendix~\ref{app:application1_additional_samples} for descriptions of the analytic samples that underlie these estimates. When comparing Statistics and Precalculus, the ATE is the target estimand. When comparing TCMS and IDS to Precalculus and Statistics, the ATT is the target estimand.  Standard errors are formed per Appendix~\ref{subsec:tfb.asymptotics}. * $p<0.05$, ** $p<0.01$, *** $p<0.001$.} 
\vspace{-0.15in}
\end{table}

Like The LAERI Report, TFB finds few statistically significant differences between taking different types of math courses in twelfth grade. The only statistically significant estimates in Table~\ref{tab.othercourses.estimates} pertain to the end-of-high school GPA outcome. TFB-L reports small, positive estimated effects of taking Statistics (+0.013 GPA points) or IDS (+0.030 GPA points) over Precalculus that are statistically significant at the 0.01 level. And TFB-L reports a small, positive estimated effect of taking IDS over Statistics (+0.030 GPA points). TFB-L reports positive estimated effects on end-of-high school GPA for taking TCMS over Precalculus or Statistics, but these estimates are not close to statistical significance at the 0.05 level. 

However, while Table~\ref{tab.othercourses.estimates} offers some evidence that taking TCMS or IDS over Precalculus or Statistics is not harmful, and is potentially beneficial, to one's end-of-high school GPA, the estimated effects on college enrollment, while not near statistical significance (at the 0.05 level), are negative and fairly large (e.g., -4.7 percentage points for taking IDS over Precalculus). Statistical power, given the small sample sizes, is clearly still an issue for TFB in the comparisons involving these courses, as TFB's standard errors for these college enrollment estimates are also large. 

Finally, we leave plots of TFB's covariate balance for these course comparisons and accompanying discussion to Appendix~\ref{app:application1_additional_balance}. However, to summarize, TFB-L and TFB-BART exhibit similar behavior as is depicted in Figure~\ref{fig.mthvnone}. While TFB-L and TFB-BART leave non-negligible imbalance in several covariates, they largely obtain good balance on the covariates that have higher partial correlations with the outcome in question.

\section{Conclusion}\label{sec:conclusion}

This paper has introduced TFB, a balancing weights framework that first regresses the outcome on the observed covariates, and then uses the resulting regression function and its estimated variance to determine how to balance the covariates. As a result, TFB seeks balance in the predicted values of the regression function and in the covariates, allowing the regression function's variance to determine how much of the latter is necessary. Our \texttt{R} Package \href{https://github.com/bleth/tfb}{\texttt{tfb}} implements TFB.

In simulated examples, TFB showed promising results when compared to applications of existing balancing methods that disregarded the outcome (i.e., are ``honest") and any potential background knowledge (e.g., which covariates are most influential on the outcome). TFB outperforms these existing methods, in our tests, even after combining them with an outcome model for augmented estimation. In real-data scenarios, investigators could apply background knowledge to tune these methods, potentially closing any performance gap that might exist between these methods and TFB. Nevertheless, TFB's advantage is that the choice of regression estimator and an estimator for its variance entirely determines how it balances the covariates. This turns the problem of how best to balance the covariates into how best to model the outcome. TFB does, however, require different weights for different outcomes. This is a practical disadvantage in that it complicates analyses, and can be computationally intensive. But it is also a potential advantage, particularly in settings with minimal overlap in the covariates ---  while it may be impossible to find a single set of weights that is appropriate for all outcomes, TFB may be able to find a different sets of weights that are individually appropriate for some (or all) of the outcomes.

One key characteristics of TFB is that it may intentionally leave imbalance in the covariates even when exact balance is feasible, either because certain dimensions appear conditionally unrelated to the outcome, or even to use imbalances in some dimensions to offset imbalances in others. Intentional imbalance is not a novel idea --- several existing methods also leave imbalance in the name of variance reduction (e.g., \citealp{zubizarreta2015stable}) or to prioritize balance in more predictive dimensions (e.g., \citealp{ning2020robust}). However, the motivation behind this characteristic of TFB does challenge traditional advice (e.g., \citealp{stuart2010matching}) that the ultimate goal of weighting is to equate the distributions of the covariates in the treated and control groups, with a particular focus on mean balance in the covariates as a benchmark for success. TFB instead argues that any balance in the covariates should only be a byproduct of balance in $f_0 (X) = \E[Y (0) \ | \ X]$, which is all that is required for unbiased estimation. It is understandable to be suspicious of this perspective in practice, given TFB's reliance on a $\hat{f}_0$, not $f_0$. However, note that it is theoretically sound, and is shown here to potentially yield efficiency gains. Further, recall that TFB still does value overall balance in the covariates, using it to safeguard against an \textit{over-}reliance on $\hat{f}_0$. We see this in our tests, finding that TFB consistently outperforms estimating the average treatment effect purely through regression (i.e., g-computation). Further, the connection between balancing and propensity score weights (\citealp{zhao2017entropy, wang2020minimal}) allows TFB to exhibit double robustness.
Nevertheless, intentional imbalance presents a challenge from a diagnostics perspective, as in practice one can never know if imbalance that TFB leaves is cause for concern or if it can be safely ignored. However, note that in demonstrations here, TFB achieved good balance on the variables that were most influential on the outcome when TFB modeled the outcome well. Thus, in practice, we suggest first confirming that TFB has obtained good balance on the variables the researcher expects to be the most influential on the outcome, determined by content knowledge, or by calculating the partial correlations between each variable and the outcome. Then, one should confirm that the variables that TFB leaves more imbalance in are the variables that one expects to be less influential. Section~\ref{sec:lausd} and Appendix~\ref{sec:applications} provide examples of this on real data. 

Because the accuracy of TFB's outcome model, for predicting the conditional expectations of the unobserved potential outcomes, can greatly influence TFB's performance, it is often best to apply flexible, machine-learning modeling techniques with TFB. Additionally, using machine learning, which often implies a high-dimensional transformation of $X$, increases the chance that TFB's implied propensity score is (approximately) correctly specified. \textit{Thus, we generally do not recommend a form of TFB that 
regresses the outcome with OLS in a low-dimensional setting.} We recommend fitting several flexible models for $Y$, and comparing their accuracy in a way that penalizes overfitting, such as cross-validation. Then, one can choose the best (or top $k$) models to apply with TFB. This was done in Section~\ref{sec:lausd} (and the accompanying Appendix~\ref{app:application1_additional}). At time of writing, our \href{https://github.com/bleth/tfb}{\texttt{tfb}} package can implement KRLS, elastic net regression (including the LASSO), and BART for the user. However, the package also offers a ``custom" model option that allows users to provide their own, manually-fit models (and associated variance matrices). Additionally, when applying TFB, a probability threshold close to 1 (e.g., $q = 0.95$) is recommended.

Finally, note that while this paper has framed TFB as an estimator for the average treatment effect on the treated (ATT), Appendix~\ref{app:ate_atc} adapts it to estimate the average treatment effect (ATE) and the average treatment effect on the controls (ATC). Appendix~\ref{subsec:samplesplitting} also explores the possibility of sample splitting with TFB, which our \href{https://github.com/bleth/tfb}{\texttt{tfb}} package can implement, if desired. Appendix~\ref{subsec:tfb.asymptotics} proposes a closed-form variance estimator for TFB.

\bibliographystyle{apalike}
\nocite{*}
\bibliography{TFB}

\begin{thebibliography}{}

\bibitem[Abadie and Imbens, 2006]{abadie2006large}
Abadie, A. and Imbens, G.~W. (2006).
\newblock Large sample properties of matching estimators for average treatment
  effects.
\newblock {\em Econometrica}, 74(1):235--267.

\bibitem[Abadie and Imbens, 2011]{abadie2011bias}
Abadie, A. and Imbens, G.~W. (2011).
\newblock Bias-corrected matching estimators for average treatment effects.
\newblock {\em Journal of Business \& Economic Statistics}, 29(1):1--11.

\bibitem[Antonelli et~al., 2018]{antonelli2018doubly}
Antonelli, J., Cefalu, M., Palmer, N., and Agniel, D. (2018).
\newblock Doubly robust matching estimators for high dimensional confounding
  adjustment.
\newblock {\em Biometrics}, 74(4):1171--1179.

\bibitem[Arpino and Cannas, 2016]{arpino2016propensity}
Arpino, B. and Cannas, M. (2016).
\newblock Propensity score matching with clustered data. an application to the
  estimation of the impact of caesarean section on the apgar score.
\newblock {\em Statistics in medicine}, 35(12):2074--2091.

\bibitem[Athey et~al., 2018]{athey2018approximate}
Athey, S., Imbens, G.~W., and Wager, S. (2018).
\newblock Approximate residual balancing: Debiased inference of average
  treatment effects in high dimensions.
\newblock {\em Journal of the Royal Statistical Society: Series B (Statistical
  Methodology)}, 80(4):597--623.

\bibitem[Austin, 2009]{austin2009balance}
Austin, P.~C. (2009).
\newblock Balance diagnostics for comparing the distribution of baseline
  covariates between treatment groups in propensity-score matched samples.
\newblock {\em Statistics in medicine}, 28(25):3083--3107.

\bibitem[Ben-Michael et~al., 2021a]{ben2021balancing}
Ben-Michael, E., Feller, A., Hirshberg, D.~A., and Zubizarreta, J.~R. (2021a).
\newblock The balancing act in causal inference.
\newblock {\em arXiv preprint arXiv:2110.14831}.

\bibitem[Ben-Michael et~al., 2021b]{ben2018augmented}
Ben-Michael, E., Feller, A., and Rothstein, J. (2021b).
\newblock The augmented synthetic control method.
\newblock {\em Journal of the American Statistical Association}, 0(0):1--15.

\bibitem[Chan et~al., 2016]{chan2016globally}
Chan, K. C.~G., Yam, S. C.~P., and Zhang, Z. (2016).
\newblock Globally efficient non-parametric inference of average treatment
  effects by empirical balancing calibration weighting.
\newblock {\em Journal of the Royal Statistical Society: Series B (Statistical
  Methodology)}, 78(3):673--700.

\bibitem[Chatterjee and Lahiri, 2010]{chatterjee2010asymptotic}
Chatterjee, A. and Lahiri, S.~N. (2010).
\newblock Asymptotic properties of the residual bootstrap for lasso estimators.
\newblock {\em Proceedings of the American Mathematical Society},
  138(12):4497--4509.

\bibitem[Chatterjee and Lahiri, 2011]{chatterjee2011bootstrapping}
Chatterjee, A. and Lahiri, S.~N. (2011).
\newblock Bootstrapping lasso estimators.
\newblock {\em Journal of the American Statistical Association},
  106(494):608--625.

\bibitem[Chattopadhyay et~al., 2020]{chattopadhyaybalancing2020}
Chattopadhyay, A., Hase, C.~H., and Zubizarreta, J.~R. (2020).
\newblock Balancing versus modeling approaches to weighting in practice.
\newblock {\em Statistics in Medicine}, 39(24):3227--3254.

\bibitem[Chen et~al., 2022]{chen2022debiased}
Chen, Q., Syrgkanis, V., and Austern, M. (2022).
\newblock Debiased machine learning without sample-splitting for stable
  estimators.
\newblock {\em Advances in Neural Information Processing Systems},
  35:3096--3109.

\bibitem[Chernozhukov et~al., 2018]{chernozhukov2018double}
Chernozhukov, V., Chetverikov, D., Demirer, M., Duflo, E., Hansen, C., Newey,
  W., and Robins, J. (2018).
\newblock Double/debiased machine learning for treatment and structural
  parameters.
\newblock {\em Econometrics Journal}, 21(1):C1--C68.

\bibitem[Cho et~al., 2026]{cho2026inference}
Cho, S., Kim, D., and Hazlett, C. (2026).
\newblock Inference at the data’s edge: Gaussian processes for estimation and
  inference in the face of extrapolation uncertainty.
\newblock {\em Political Analysis}, pages 1--20.

\bibitem[Cousineau et~al., 2023]{cousineau2023estimating}
Cousineau, M., Verter, V., Murphy, S.~A., and Pineau, J. (2023).
\newblock Estimating causal effects with optimization-based methods: A review
  and empirical comparison.
\newblock {\em European Journal of Operational Research}, 304(2):367--380.

\bibitem[Dehejia, 2023]{nswweb}
Dehejia, R. (2023).
\newblock data download.
\newblock \url{https://users.nber.org/~rdehejia/data/.nswdata2.html}.
\newblock Accessed: 2023-06-26.

\bibitem[Dehejia and Wahba, 1999]{dehejia1999causal}
Dehejia, R.~H. and Wahba, S. (1999).
\newblock Causal effects in nonexperimental studies: Reevaluating the
  evaluation of training programs.
\newblock {\em Journal of the American Statistical Association},
  94(448):1053--1062.

\bibitem[Deming and Stephan, 1940]{deming1940least}
Deming, W.~E. and Stephan, F.~F. (1940).
\newblock On a least squares adjustment of a sampled frequency table when the
  expected marginal totals are known.
\newblock {\em The Annals of Mathematical Statistics}, 11(4):427--444.

\bibitem[Deville and S{\"a}rndal, 1992]{deville1992calibration}
Deville, J.-C. and S{\"a}rndal, C.-E. (1992).
\newblock Calibration estimators in survey sampling.
\newblock {\em Journal of the American Statistical Association},
  87(418):376--382.

\bibitem[Dorie et~al., 2019]{dorie2019automated}
Dorie, V., Hill, J., Shalit, U., Scott, M., and Cervone, D. (2019).
\newblock Automated versus do-it-yourself methods for causal inference: Lessons
  learned from a data analysis competition.
\newblock {\em Statistical Science}, 34(1):43--68.

\bibitem[Farrell, 2015]{farrell2015robust}
Farrell, M.~H. (2015).
\newblock Robust inference on average treatment effects with possibly more
  covariates than observations.
\newblock {\em Journal of Econometrics}, 189(1):1--23.

\bibitem[Hainmueller, 2012]{hainmueller2012entropy}
Hainmueller, J. (2012).
\newblock Entropy balancing for causal effects: A multivariate reweighting
  method to produce balanced samples in observational studies.
\newblock {\em Political Analysis}, 20(1):25--46.

\bibitem[Hainmueller and Hazlett, 2014]{hainmueller2014kernel}
Hainmueller, J. and Hazlett, C. (2014).
\newblock Kernel regularized least squares: Reducing misspecification bias with
  a flexible and interpretable machine learning approach.
\newblock {\em Political Analysis}, 22(2):143--168.

\bibitem[Hansen, 2008]{hansen2008prognostic}
Hansen, B.~B. (2008).
\newblock The prognostic analogue of the propensity score.
\newblock {\em Biometrika}, 95(2):481--488.

\bibitem[Hazlett, 2020]{hazlett2018kernel}
Hazlett, C. (2020).
\newblock Kernel balancing: A flexible non-parametric weighting procedure for
  estimating causal effects.
\newblock {\em Statistica Sinica}, 30(1):1155--1189.

\bibitem[Heckman et~al., 1998]{heckman1998matching}
Heckman, J.~J., Ichimura, H., and Todd, P. (1998).
\newblock Matching as an econometric evaluation estimator.
\newblock {\em The review of economic studies}, 65(2):261--294.

\bibitem[Hill, 2011]{hill2011bayesian}
Hill, J.~L. (2011).
\newblock Bayesian nonparametric modeling for causal inference.
\newblock {\em Journal of Computational and Graphical Statistics},
  20(1):217--240.

\bibitem[Hirshberg et~al., 2021]{hirshberg2019minimax}
Hirshberg, D.~A., Maleki, A., and Zubizarreta, J.~R. (2021).
\newblock Minimax linear estimation of the retargeted mean.
\newblock {\em arXiv preprint arXiv:1901.10296v2}.

\bibitem[Hirshberg and Wager, 2021]{hirshberg2017augmented}
Hirshberg, D.~A. and Wager, S. (2021).
\newblock Augmented minimax linear estimation.
\newblock {\em The Annals of Statistics}, 49(6):3206--3227.

\bibitem[Hong et~al., 2021]{hong2021did}
Hong, G., Yang, F., and Qin, X. (2021).
\newblock Did you conduct a sensitivity analysis? a new weighting-based
  approach for evaluations of the average treatment effect for the treated.
\newblock {\em Journal of the Royal Statistical Society Series A: Statistics in
  Society}, 184(1):227--254.

\bibitem[Huang et~al., 2006]{huang2007correcting}
Huang, J., Smola, A.~J., Gretton, A., Borgwardt, K.~M., and Sch{\"o}lkopf, B.
  (2006).
\newblock Correcting sample selection bias by unlabeled data.
\newblock {\em Advances in Neural Information Processing Systems}, 19:601--608.

\bibitem[Imai and Ratkovic, 2014]{imai2014covariate}
Imai, K. and Ratkovic, M. (2014).
\newblock Covariate balancing propensity score.
\newblock {\em Journal of the Royal Statistical Society: Series B (Statistical
  Methodology)}, 76(1):243--263.

\bibitem[Jin and Zubizarreta, 2025]{jin2025cross}
Jin, Y. and Zubizarreta, J. (2025).
\newblock Cross-balancing for data-informed design and efficient analysis of
  observational studies.
\newblock {\em arXiv preprint arXiv:2511.15896}.

\bibitem[Kallus, 2020]{kallus2020generalized}
Kallus, N. (2020).
\newblock Generalized optimal matching methods for causal inference.
\newblock {\em Journal of Machine Learning Research}, 21:1--54.

\bibitem[Kang and Schafer, 2007]{kang2007demystifying}
Kang, J. D.~Y. and Schafer, J.~L. (2007).
\newblock Demystifying double robustness: A comparison of alternative
  strategies for estimating a population mean from incomplete data.
\newblock {\em Statistical Science}, 22(4):523--539.

\bibitem[Knight and Fu, 2000]{knight2000asymptotics}
Knight, K. and Fu, W. (2000).
\newblock Asymptotics for lasso-type estimators.
\newblock {\em The Annals of Statistics}, 28(5):1356--1378.

\bibitem[Kuang et~al., 2017]{kuang2017estimating}
Kuang, K., Cui, P., Li, B., Jiang, M., and Yang, S. (2017).
\newblock Estimating treatment effect in the wild via differentiated confounder
  balancing.
\newblock In {\em Proceedings of the 23rd ACM SIGKDD International Conference
  on Knowledge Discovery and Data Mining}, pages 265--274.

\bibitem[{L}.{A}.{ }Unified, 2017]{lausd2017middle}
{L}.{A}.{ }Unified (2017).
\newblock {M}iddle {S}chool {M}ath {P}lacement and {P}athways.
\newblock Viewed 3/10/2026 from
  \href{https://lausd-p12.edlioschool.com/apps/pages/index.jsp?uREC_ID=4397387&type=d&pREC_ID=2620106}{here}.

\bibitem[{L}.{A}.{ }Unified, 2025a]{lausd2025finger}
{L}.{A}.{ }Unified (2025a).
\newblock {F}ingertip {F}acts 2025-26.
\newblock Accessed 3/10/2026 from
  \href{https://4.files.edl.io/aef4/08/14/25/170225-cc3e26de-f440-4470-9e85-486022db3fa6.pdf}{here}.

\bibitem[{L}.{A}.{ }Unified, 2025b]{lausd2025pathways}
{L}.{A}.{ }Unified (2025b).
\newblock {REG}-069702.3 {H}igh {S}chool {M}athematics {P}lacement and
  {P}athways.
\newblock Accessed 3/10/2026 from
  \href{https://my.lausd.net/eLibrary/DocView.aspx?id=144076&dbid=0&repo=eLibrary&cr=1}{here}.

\bibitem[{L}.{A}.{ }Unified, 2025c]{lausd2025specialized}
{L}.{A}.{ }Unified (2025c).
\newblock {REQUESTING SPECIALIZED MATHEMATICS COURSES FOR} 2025-2026.
\newblock Accessed 3/19/2026 from
  \href{https://drive.google.com/file/d/1wXvCmJQVEQ_XcT2MCBxMbe9NQdBduZZ6/view}{here}.

\bibitem[LaLonde, 1986]{lalonde1986evaluating}
LaLonde, R.~J. (1986).
\newblock Evaluating the econometric evaluations of training programs with
  experimental data.
\newblock {\em The American Economic Review}, 76(4):604--620.

\bibitem[Leacy and Stuart, 2014]{leacy2014joint}
Leacy, F.~P. and Stuart, E.~A. (2014).
\newblock On the joint use of propensity and prognostic scores in estimation of
  the average treatment effect on the treated: A simulation study.
\newblock {\em Statistics in Medicine}, 33(20):3488--3508.

\bibitem[Nikolaev et~al., 2013]{nikolaev2013balance}
Nikolaev, A.~G., Jacobson, S.~H., Cho, W. K.~T., Sauppe, J.~J., and Sewell,
  E.~C. (2013).
\newblock Balance optimization subset selection ({BOSS}): An alternative
  approach for causal inference with observational data.
\newblock {\em Operations Research}, 61(2):398--412.

\bibitem[Ning et~al., 2020]{ning2020robust}
Ning, Y., Peng, S., and Imai, K. (2020).
\newblock Robust estimation of causal effects via a high-dimensional covariate
  balancing propensity score.
\newblock {\em Biometrika}, 107(3):533--554.

\bibitem[Niswander et~al., 1972]{niswander1972women}
Niswander, K.~R., Gordon, M.~J., and Gordon, M. (1972).
\newblock {\em The women and their pregnancies: the Collaborative Perinatal
  Study of the National Institute of Neurological Diseases and Stroke},
  volume~73.
\newblock National Institute of Health.

\bibitem[Normand et~al., 2001]{normand2001validating}
Normand, S.-L.~T., Landrum, M.~B., Guadagnoli, E., Ayanian, J.~Z., Ryan, T.~J.,
  Cleary, P.~D., and McNeil, B.~J. (2001).
\newblock Validating recommendations for coronary angiography following acute
  myocardial infarction in the elderly: a matched analysis using propensity
  scores.
\newblock {\em Journal of clinical epidemiology}, 54(4):387--398.

\bibitem[Pirracchio and Carone, 2018]{pirracchio2018balance}
Pirracchio, R. and Carone, M. (2018).
\newblock The balance super learner: A robust adaptation of the super learner
  to improve estimation of the average treatment effect in the treated based on
  propensity score matching.
\newblock {\em Statistical Methods in Medical Research}, 27(8):2504--2518.

\bibitem[Robins, 1986]{robins1986new}
Robins, J. (1986).
\newblock A new approach to causal inference in mortality studies with a
  sustained exposure period—application to control of the healthy worker
  survivor effect.
\newblock {\em Mathematical modelling}, 7(9-12):1393--1512.

\bibitem[Robins and Rotnitzky, 1995]{robins1995semiparametric}
Robins, J.~M. and Rotnitzky, A. (1995).
\newblock Semiparametric efficiency in multivariate regression models with
  missing data.
\newblock {\em Journal of the American Statistical Association},
  90(429):122--129.

\bibitem[Robins et~al., 1994]{robins1994estimation}
Robins, J.~M., Rotnitzky, A., and Zhao, L.~P. (1994).
\newblock Estimation of regression coefficients when some regressors are not
  always observed.
\newblock {\em Journal of the American Statistical Association},
  89(427):846--866.

\bibitem[Robins and Wang, 2000]{robins2000inference}
Robins, J.~M. and Wang, N. (2000).
\newblock Inference for imputation estimators.
\newblock {\em Biometrika}, 87(1):113--124.

\bibitem[Rosenbaum and Rubin, 1983]{rosenbaum1983central}
Rosenbaum, P.~R. and Rubin, D.~B. (1983).
\newblock The central role of the propensity score in observational studies for
  causal effects.
\newblock {\em Biometrika}, 70(1):41--55.

\bibitem[Rotnitzky et~al., 2012]{rotnitzky2012improved}
Rotnitzky, A., Lei, Q., Sued, M., and Robins, J.~M. (2012).
\newblock Improved double-robust estimation in missing data and causal
  inference models.
\newblock {\em Biometrika}, 99(2):439--456.

\bibitem[Rubin, 1974]{rubin1974estimating}
Rubin, D.~B. (1974).
\newblock Estimating causal effects of treatments in randomized and
  nonrandomized studies.
\newblock {\em Journal of Educational Psychology}, 66(5):688--701.

\bibitem[Sch{\"o}lkopf and Smola, 2002]{scholkopf2002learning}
Sch{\"o}lkopf, B. and Smola, A.~J. (2002).
\newblock {\em Learning with kernels: Support vector machines, regularization,
  optimization, and beyond}.
\newblock MIT Press.

\bibitem[Shen et~al., 2025]{shen2025forest}
Shen, A.~A., Ben-Michael, E., Feller, A., Keele, L., and Murray, J. (2025).
\newblock Forest kernel balancing weights: Outcome-guided features for causal
  inference.
\newblock {\em arXiv preprint arXiv:2512.11751}.

\bibitem[Shortreed and Ertefaie, 2017]{shortreed2017outcome}
Shortreed, S.~M. and Ertefaie, A. (2017).
\newblock Outcome-adaptive lasso: Variable selection for causal inference.
\newblock {\em Biometrics}, 73(4):1111--1122.

\bibitem[Smucler et~al., 2019]{smucler2019unifying}
Smucler, E., Rotnitzky, A., and Robins, J.~M. (2019).
\newblock A unifying approach for doubly-robust {$\ell_1$} regularized
  estimation of causal contrasts.
\newblock {\em arXiv preprint arXiv:1904.03737}.

\bibitem[Soriano et~al., 2021]{soriano2021interpretable}
Soriano, D., Ben-Michael, E., Bickel, P.~J., Feller, A., and Pimentel, S.~D.
  (2021).
\newblock Interpretable sensitivity analysis for balancing weights.
\newblock {\em arXiv preprint arXiv:2102.13218}.

\bibitem[Splawa-Neyman et~al., 1990]{splawa1990application}
Splawa-Neyman, J., Dabrowska, D.~M., and Speed, T. (1990).
\newblock On the application of probability theory to agricultural experiments.
  {E}ssay on principles. {S}ection 9.
\newblock {\em Statistical Science}, 5(4):465--472.

\bibitem[Stuart, 2010]{stuart2010matching}
Stuart, E.~A. (2010).
\newblock Matching methods for causal inference: A review and a look forward.
\newblock {\em Statistical Science}, 25(1):1--21.

\bibitem[Tan, 2010]{tan2010bounded}
Tan, Z. (2010).
\newblock Bounded, efficient and doubly robust estimation with inverse
  weighting.
\newblock {\em Biometrika}, 97(3):661--682.

\bibitem[Tarr and Imai, 2021]{tarr2021estimating}
Tarr, A. and Imai, K. (2021).
\newblock Estimating average treatment effects with support vector machines.
\newblock {\em arXiv preprint arXiv:2102.11926v2}.

\bibitem[van~der Laan and Rubin, 2006]{van2006targeted}
van~der Laan, M.~J. and Rubin, D. (2006).
\newblock Targeted maximum likelihood learning.
\newblock {\em The International Journal of Biostatistics}, 2(1):1--38.

\bibitem[Wainstein and Hazlett, 2026]{wainstein2026sensitivity}
Wainstein, L. and Hazlett, C. (2026).
\newblock Sensitivity of weighted least squares estimators to omitted
  variables.
\newblock {\em arXiv preprint arXiv:2508.02954}.

\bibitem[Wainstein et~al., 2023a]{wainstein2023access}
Wainstein, L., Miller, C.~E., Phillips, M., Yamashiro, K., and Melguizo, T.
  (2023a).
\newblock Twelfth grade math and college access.
\newblock {\em Los Angeles Education Research Institute}.
\newblock Available at
  https://laeri.luskin.ucla.edu/12thgrademathandcollegeaccess/.

\bibitem[Wainstein et~al., 2023b]{wainstein2023success}
Wainstein, L., Miller, C.~E., Phillips, M., Yamashiro, K., and Melguizo, T.
  (2023b).
\newblock Twelfth grade math and college success.
\newblock {\em Los Angeles Education Research Institute}.
\newblock Available at
  https://laeri.luskin.ucla.edu/12thgrademathandcollegesuccess/.

\bibitem[Wainwright, 2019]{wainwright2019high}
Wainwright, M.~J. (2019).
\newblock {\em High-dimensional statistics: A non-asymptotic viewpoint}.
\newblock Cambridge University Press, first edition.

\bibitem[Wang and Zubizarreta, 2020]{wang2020minimal}
Wang, Y. and Zubizarreta, J.~R. (2020).
\newblock Minimal dispersion approximately balancing weights: Asymptotic
  properties and practical considerations.
\newblock {\em Biometrika}, 107(1):93--105.

\bibitem[White, 1980]{white1980heteroskedasticity}
White, H. (1980).
\newblock A heteroskedasticity-consistent covariance matrix estimator and a
  direct test for heteroskedasticity.
\newblock {\em Econometrica}, 48(4):817--838.

\bibitem[Wong and Chan, 2018]{wong2017kernel}
Wong, R. K.~W. and Chan, K. C.~G. (2018).
\newblock Kernel-based covariate functional balancing for observational
  studies.
\newblock {\em Biometrika}, 105(1):199--213.

\bibitem[Zhao, 2019]{zhao2019covariate}
Zhao, Q. (2019).
\newblock Covariate balancing propensity score by tailored loss functions.
\newblock {\em The Annals of Statistics}, 47(2):965--993.

\bibitem[Zhao and Percival, 2017]{zhao2017entropy}
Zhao, Q. and Percival, D. (2017).
\newblock Entropy balancing is doubly robust.
\newblock {\em Journal of causal inference}, 5(1).

\bibitem[Zubizarreta, 2015]{zubizarreta2015stable}
Zubizarreta, J.~R. (2015).
\newblock Stable weights that balance covariates for estimation with incomplete
  outcome data.
\newblock {\em Journal of the American Statistical Association},
  110(511):910--922.

\end{thebibliography}

\newpage
\clearpage
\appendix


\section{Application 1: Extended results}\label{app:application1_additional}

This appendix includes results and discussion that accompany Section~\ref{sec:lausd}, which summarizes our investigation of the effects of taking mathematics courses in twelfth grade in L.A. Unified.

\subsection{Descriptive analysis: Comparing students who took different types of mathematics courses in twelfth grade}\label{app:application1_additional_descriptives}

In Section~\ref{subsec:lausd_data}, we compare students, on the observed covariates, who took a math course in twelfth grade and students who did not. In this section, we compare students who took Precalculus, Statistics, TCMS, and IDS. Table~\ref{tab.othercourses.descriptives} reports the average values of a key subset of the covariates for the students in our data who took each of these four courses in twelfth grade.

Among students who took these four courses, the Precalculus students, on average, had the highest cumulative GPAs through eleventh grade, scored the highest on the math portions of state-wide standardized tests, were the most likely to have taken the SAT or ACT, and were the most likely to self-report to expect to attain a four-year degree. In other words, among the students who took these four types of courses, the students who took Precalculus were the highest-performing, at least in math, and through eleventh grade had shown the most interest in enrolling in a four-year college. Of these four courses, Statistics students were, in terms of the measures considered here, the second highest-performing, and had shown the second most interest in enrolling in a four-year college. Additionally, a lower percentage of Precalculus and Statistics students were female than were TCMS or IDS students. 

TCMS and IDS students scored noticeably lower on their state-wide standardized tests than did Precalculus and Statistics students. They also had slightly lower cumulative GPAs through eleventh grade and lower behavioral grades. TCMS students scored slightly higher on their state-wide standardized tests than did IDS students. However, TCMS students were much less likely to have taken the SAT or ACT.
\begin{table}[!h]
\caption{Descriptive statistics for students who took different types of mathematics courses in twelfth grade}\label{tab.othercourses.descriptives}

	\begin{center}
	\footnotesize
     \begin{tabular}{ |m{4.1cm} | m{1.05cm} m{1.05cm} | m{1.05cm} m{1.05cm} | m{1.05cm} m{1.05cm} | m{1.05cm} m{1.05cm} |}
     \hline
     \multicolumn{1}{ |c| }{ \textbf{Covariate} } & \multicolumn{2}{ c| }{\textbf{Precalculus}} & \multicolumn{2}{  c | }{ \textbf{Statistics} } & \multicolumn{2}{ c| }{ \textbf{TCMS} } & \multicolumn{2}{ c| }{ \textbf{IDS} } \\
      & Mean & (SD) & Mean & (SD) & Mean & (SD) & Mean & (SD) \\
     \hline
     Cohort: 2017-18 & 0.503 & (0.500) & 0.569 & (0.495) & 0.766 & (0.424) & 0.463 & (0.500) \\
     Gender: Female & 0.526 & (0.499) & 0.537 & (0.499) & 0.600 & (0.490) & 0.582 & (0.494) \\
     Age (in Months) & 195.522 & (5.863) & 195.152 & (5.512) & 195.024 & (5.008) & 195.661 & (5.380) \\
     GPA through 11th & 3.022 & (0.547) & 2.898 & (0.555) & 2.737 & (0.516) & 2.768 & (0.530) \\
     ELA SBAC (Z-Score) & 0.170 & (0.741) & 0.180 & (0.772) & 0.058 & (0.697) & -0.038 & (0.776) \\
     Math SBAC (Z-Score) & 0.098 & (0.682) & 0.023 & (0.703) & -0.082 & (0.618) & -0.100 & (0.620) \\
     Took SAT/ACT & 0.410 & (0.492) & 0.377 & (0.485) & 0.229 & (0.421) & 0.377 & (0.486) \\
     WPA & 2.519 & (0.315) & 2.477 & (0.327) & 2.407 & (0.308) & 2.425 & (0.330) \\
     CPA & 2.647 & (0.254) & 2.625 & (0.270) & 2.555 & (0.252) & 2.584 & (0.274) \\
     Expect: Unsure & 0.075 & (0.263) & 0.082 & (0.275) & 0.117 & (0.322) & 0.116 & (0.320) \\
     Expect: HS or Less & 0.026 & (0.158) & 0.027 & (0.162) & 0.044 & (0.206) & - & - \\
     Expect: Certificate or 2-Year & 0.045 & (0.207) & 0.032 & (0.177) & 0.066 & (0.249) & - & - \\
     Expect: 4-Year or More & 0.598 & (0.490) & 0.586 & (0.493) & 0.576 & (0.495) & 0.526 & (0.500) \\
     Expect: Missing & 0.256 & (0.437) & 0.272 & (0.445) & 0.196 & (0.397) & 0.272 & (0.446) \\
     \hline
     $n$ & 4,733 &  & 1,600 &  & 632 &  & 268 &  \\
     \hline
     \end{tabular}

     \normalsize
	\end{center}
    \subcaption*{\textit{Note:} Average values (and standard deviations) for a subset of key covariates for students in the data who took Precalculus, Statistics, TCMS, and IDS. All covariates are recorded for the student at the end of their eleventh grade year.  ``Cohort: 2017-18" is an indicator variable for being in twelfth grade in the 2017-18 academic year. The SBAC is a math and English language arts (ELA) standardized test that was developed by the Smarter Balanced Assessment Consortium (SBAC), and, in the years examined, was taken by California students in eleventh grade. Work Effort GPA (WPA) and Cooperation  GPA (CPA) are two behavioral GPAs that are on a 1-3 scale. The ``Expect" variables indicate students' self-reported expected educational attainment. ``Expect: HS or Less" and ``Expect: Certificate or 2-Year" are omitted for IDS students because their averages (and standard deviations) imply cells with fewer than 11 students. }
\end{table}

\subsection{Methods: Analytic samples}\label{app:application1_additional_samples}

While the full dataset we have access to describes 11,255 students, we restrict our sample depending on the course-taking pattern we are investigating. We make the same sample restrictions as are made in The LAERI Report. This section describes the final samples we use to investigate each course-taking pattern, which we refer to as our ``analytic samples".

When investigating the effects of taking any math course in twelfth grade, we drop schools in each year in which all students took math, or all students did not take math. This yields 9,993 students (6,257 who took math, and 3,736 who did not take math), who are described in Table~\ref{tab.mthvnone.descriptives}, and analyzed in Section~\ref{subsec:lausd_mathvsnone}.

When comparing any two courses among Precalculus, Statistics, TCMS, and IDS, we first only retain students in the analytic sample who took exactly one of the courses being compared. For example, when comparing Precalculus and Statistics, we only retain students who took Precalculus or took Statistics, but did not take both courses. In other words, we drop students who did not take Precalculus or Statistics, or took both Precalculus and Statistics. Next, when comparing two courses, we only retain schools in each year in which at least one student took each course. For example, when comparing Precalculus and Statistics, we only retain schools in a given year in which at least one student took Precalculus, and at least one student took Statistics. After these sample restrictions, when comparing Precalculus and Statistics, the analytic sample has 3,807 students (2,396 who took Precalculus, and 1,411 who took Statistics). When comparing TCMS and Precalculus, the analytic sample has 1,357 students (583 who took TCMS, and 774 who took Precalculus). When comparing IDS and Precalculus, the analytic sample has 892 students (248 who took IDS, and 644 who took Precalculus). When comparing TCMS and Statistics, the analytic sample has 619 students (285 who took TCMS, and 334 who took Statistics). When comparing IDS and Statistics, the analytic sample has 390 students (104 who took IDS, and 286 who took Statistics). These analytic samples are analyzed in Section~\ref{subsec:lausd_tcmsids}, and underlie the causal estimates in Table~\ref{tab.othercourses.estimates}.

\subsection{Preliminary analysis: Predictive accuracy of KRLS, the LASSO, and BART}\label{app:application1_additional_cvresults}

Before applying TFB-L, TFB-K, and TFB-BART to this setting, we first assess the predictive accuracy of their corresponding outcome models: the LASSO, KRLS, and BART, respectively. We are particularly interested in out-of-sample predictive accuracy, so we implement five-fold cross validation. For the end-of-high school GPA outcome, we calculate RMSE for predicting the outcome in the cross-validated test datasets. For the college enrollment outcome, we again calculate RMSE, and additionally calculate the area under the Receiver Operating Characteristic (ROC) curve (AUC). For each course comparison, we apply these models and perform the cross-validation separately within each analytic sample (see Appendix~\ref{app:application1_additional_samples}). 

When fitting the LASSO, we model each outcome as a linear function of the covariates; the squared terms and pairwise interaction terms of the academics-related covariates, the behavioral covariates, and the indicator variable for twelfth grade year; the treatment variable; and the interaction terms of the treatment variable with all of the other regressors. This LASSO model is fit to the full dataset. When fitting KRLS, we fit separate KRLS models to the treatment and control groups, each using the Gaussian kernel with bandwidth $b = P$. For BART, we fit a model to the full dataset that predicts the outcome with the covariates and the treatment variable. Note that these are exactly the same implementations of the LASSO, KRLS, and BART that TFB-L, TFB-K, and TFB-BART employ, respectively.

We also measure the predictive accuracy for some comparison outcome models to give a baseline for the TFB-related outcome models. First, we apply a standard OLS that regresses each outcome on the entire dataset as a linear function of the covariates and the treatment variable. For the college enrollment outcome, we fit a logistic regression model (Logistic) that is the logistic regression equivalent of the OLS model. We also fit a LASSO Logistic regression model (LASSO Logistic) that is the logistic regression equivalent of the linear LASSO model.

\paragraph{End-of-high school GPA.} 
Table~\ref{tab.cvresults.gpa} reports the cross-validated RMSE for the end-of-high school GPA outcome for each course comparison's analytic sample. In summary, of the TFB-related models, the LASSO model is the most predictive for this outcome, with BART being a close second. However, KRLS struggles greatly at predicting this outcome. It is partly for this reason why TFB-K's estimated effects on this outcome are almost completely omitted in Section~\ref{sec:lausd} and Appendix~\ref{app:application1_additional_estimates}.  
\begin{table}[!h]
\caption{Five-fold cross validated RMSE of outcome models for the end-of-high school GPA outcome}\label{tab.cvresults.gpa}
	\vspace{-0.15in}
	\begin{center}
	\scriptsize
     \begin{tabular}{ |m{3.4cm} | m{1.2cm} m{1.2cm} m{1.2cm} m{1.2cm} | }
     \hline
     \textbf{Comparison} & \textbf{OLS}  & \textbf{LASSO}& \textbf{KRLS} & \textbf{BART} \\
     \hline
     Math vs. No Math & 0.1147 & \textcolor{red}{0.1143} & 0.2360 & 0.1159 \\
     Statistics vs. Precalculus & \textcolor{red}{0.1122} & 0.1131 & 0.2076 & 0.1157 \\
     TCMS vs. Precalculus & 0.1142 & \textcolor{red}{0.1117} & 0.2227 & 0.1183 \\
     IDS vs. Precalculus & \textcolor{red}{0.1214} & 0.1229 & 0.2032 & 0.1295 \\
     TCMS vs. Statistics & 0.1035 & \textcolor{red}{0.1017} & 0.2483 & 0.1105 \\
     IDS vs. Statistics & \textcolor{red}{0.1213} & 0.1227 & 0.2656 & 0.1327 \\
     \hline
     \end{tabular}

          \normalsize
	\end{center}
    \subcaption*{\textit{Note:}  For each row, the lowest RMSE is displayed in red text. RMSE is calculated for the full dataset for the ``Math vs. No Math" and ``Statistics vs. Precalculus" comparisons because the ATE is the target estimand in these comparisons. RMSE is calculated only for the control groups in the comparisons involving TCMS and IDS because the ATT is the target estimand in these comparisons.}
\end{table}

More specifically, the LASSO and OLS models perform very similarly, and are the models that have the lowest RMSE  for each of the course comparisons. BART performs slightly worse than LASSO and OLS, but is still fairly predictive. KRLS, however, performs very poorly, showing RMSEs that are roughly twice the RMSEs of the LASSO, OLS, and BART. The reason for this is because one of the covariates is cumulative GPA through eleventh grade, which is the most predictive covariate for this outcome by far. KRLS has difficulty noticing this, as all the covariates are put through the gaussian kernel before exploring their relationship with the outcome.

\paragraph{College enrollment.} Table~\ref{tab.cvresults.enrollment} reports the cross-validated RMSE and AUC for the college enrollment outcome for each course comparison's analytic sample. In summary, of the TFB-related models, BART is the most predictive model for this outcome overall. However, the LASSO and KRLS are not far behind, and even perform the best for certain course comparisons and in certain metrics. Thus, in terms of predictive accuracy, all of the models that TFB-L, TFB-BART, and TFB-K employ appear to be viable for this outcome.
\begin{table}[!h]
\caption{Five-fold cross validated RMSE and AUC of outcome models for the college enrollment outcome}\label{tab.cvresults.enrollment}
	\vspace{-0.15in}
	\begin{center}
	\scriptsize
     \begin{tabular}{ |m{3.4cm} | m{1.2cm} | m{1.2cm} m{1.2cm} m{1.2cm} m{1.2cm} m{1.2cm} m{1.2cm} |}
     \hline
     \textbf{Comparison} & \textbf{Measure} & \textbf{OLS} & \textbf{Logistic} & \textbf{LASSO} & \textbf{LASSO Logistic} & \textbf{KRLS} & \textbf{BART} \\
     \hline
     Math vs. No Math & RMSE & 0.4025 & 0.4027 & 0.4029 & 0.4026 & 0.4112 & \textcolor{red}{0.4007} \\
      & AUC & 0.6752 & 0.6730 & 0.6704 & 0.6722 & 0.6074 & \textcolor{red}{0.6830} \\
     \hline
     Statistics vs. Precalculus & RMSE & 0.3758 & 0.3774 & 0.3755 & 0.3748 & 0.3744 & \textcolor{red}{0.3726} \\
      & AUC & 0.6641 & 0.6568 & 0.6503 & 0.6551 & 0.6475 & \textcolor{red}{0.6696} \\
     \hline
     TCMS vs. Precalculus & RMSE & 0.3856 & 0.3850 & 0.3681 & 0.3694 & \textcolor{red}{0.3660} & 0.3663 \\
      & AUC & 0.5482 & 0.5534 & 0.5946 & 0.5885 & 0.5556 & \textcolor{red}{0.6034} \\
     \hline
     IDS vs. Precalculus & RMSE & 0.4113 & 0.4116 & 0.3939 & \textcolor{red}{0.3931} & 0.3955 & 0.4018 \\
      & AUC & 0.6091 & 0.5965 & \textcolor{red}{0.6508} & 0.6502 & 0.6308 & 0.6157 \\
     \hline
     TCMS vs. Statistics & RMSE & 0.3542 & 0.3600 & 0.3542 & 0.3511 & 0.3486 & \textcolor{red}{0.3415} \\
      & AUC & 0.6723 & 0.6403 & 0.5955 & 0.6459 & 0.5940 & \textcolor{red}{0.6949} \\
     \hline
     IDS vs. Statistics & RMSE & 0.4314 & 0.4452 & 0.3896 & 0.3934 & \textcolor{red}{0.3848} & 0.3909 \\
      & AUC & 0.5771 & \textcolor{red}{0.5916} & 0.5241 & 0.5064 & 0.5166 & 0.5708 \\
     \hline
     \end{tabular}

          \normalsize
	\end{center}
    \subcaption*{\textit{Note:}  For each row, the lowest RMSE or the highest AUC is displayed in red text. RMSE and AUC are calculated for the full dataset for the ``Math vs. No Math" and ``Statistics vs. Precalculus" comparisons because the ATE is the target estimand in these comparisons. RMSE and AUC are calculated only for the control groups in the comparisons involving TCMS and IDS because the ATT is the target estimand in these comparisons.}
\end{table}

More specifically, BART has the lowest RMSE and highest AUC in the analytic samples for Math vs. No Math, Statistics vs. Precalculus, and TCMS vs. Statistics. For the other three course comparisons, BART is usually not the top model. Nevertheless, its RMSE and AUC are often very close to the lowest RMSE and highest AUC, respectively.

In terms of RMSE, LASSO and KRLS perform very similarly to BART. And the fact that linear LASSO and KRLS are more traditionally applied to continuous outcomes, rather than binary outcomes like the college enrollment outcome examined here, does not appear to be seriously hindering them --- the LASSO and KRLS tend to have very similar RMSEs to the Logistic and LASSO Logistic models, and in several cases the RMSEs from the LASSO and KRLS are lower than the RMSEs from the logistic regression models. These models do, however, appear to struggle in terms of AUC at times. For example, in the Math vs. No Math comparison, the AUC from KRLS is significantly lower than the AUCs of the other methods. As another example, both the LASSO and KRLS have significantly lower AUC than the highest AUC for the TCMS vs. Statistics and IDS vs. Statistics comparisons.

\clearpage
\newpage

\subsection{Diagnostics: Balance plots for TFB}\label{app:application1_additional_balance}

In this section we report and discuss all of the covariate balance results for TFB-L, TFB-BART, and TFB-K for each outcome, and for each course comparison. As in Figure~\ref{fig.mthvnone}, we also report the starting imbalance and the leftover imbalance with Preferential Within-Cluster Matching (CMatch), which was the primary weighting approach in The LAERI Report.

\begin{figure}[!h]
	\vspace{0.15in}
	\begin{center}
	\caption{Balance plots for TFB-L, TFB-BART, and TFB-K for each outcome when estimating the effects of taking any math course in twelfth grade}\label{fig.balance.appendix.mthvnone}
    \vspace{-.60in}
    \begin{subfigure}{.95\textwidth}
    \begin{center}
    \includegraphics[scale=0.60]{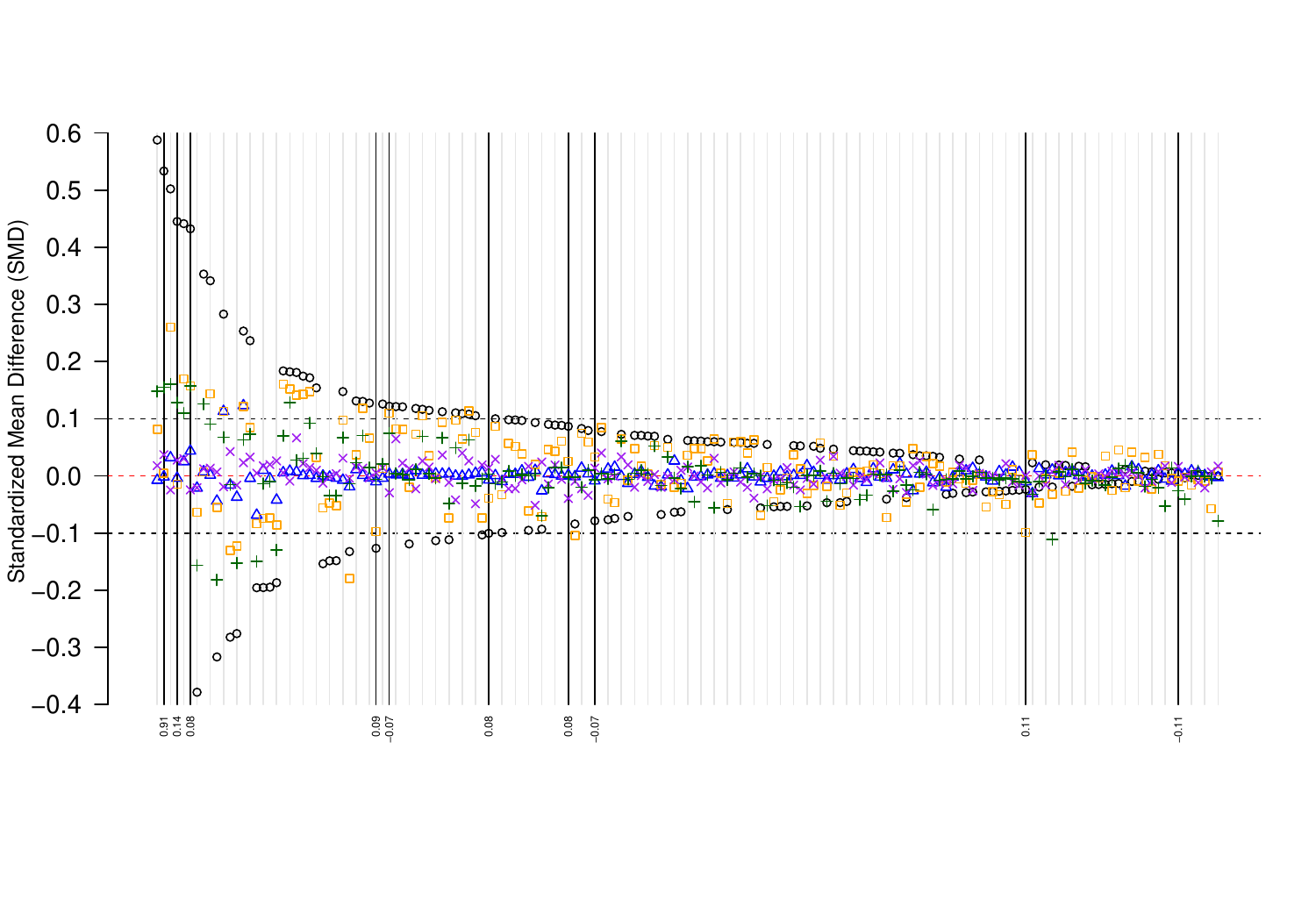} 
    \end{center}
    \vspace{-.90in}
    \subcaption{End-of-high school GPA}\label{fig.balance.appendix.mthvnone.gpa}
    \end{subfigure}
    \begin{subfigure}{.95\textwidth}
    \vspace{-1.1in}
    \begin{center}
    \includegraphics[scale=0.60]{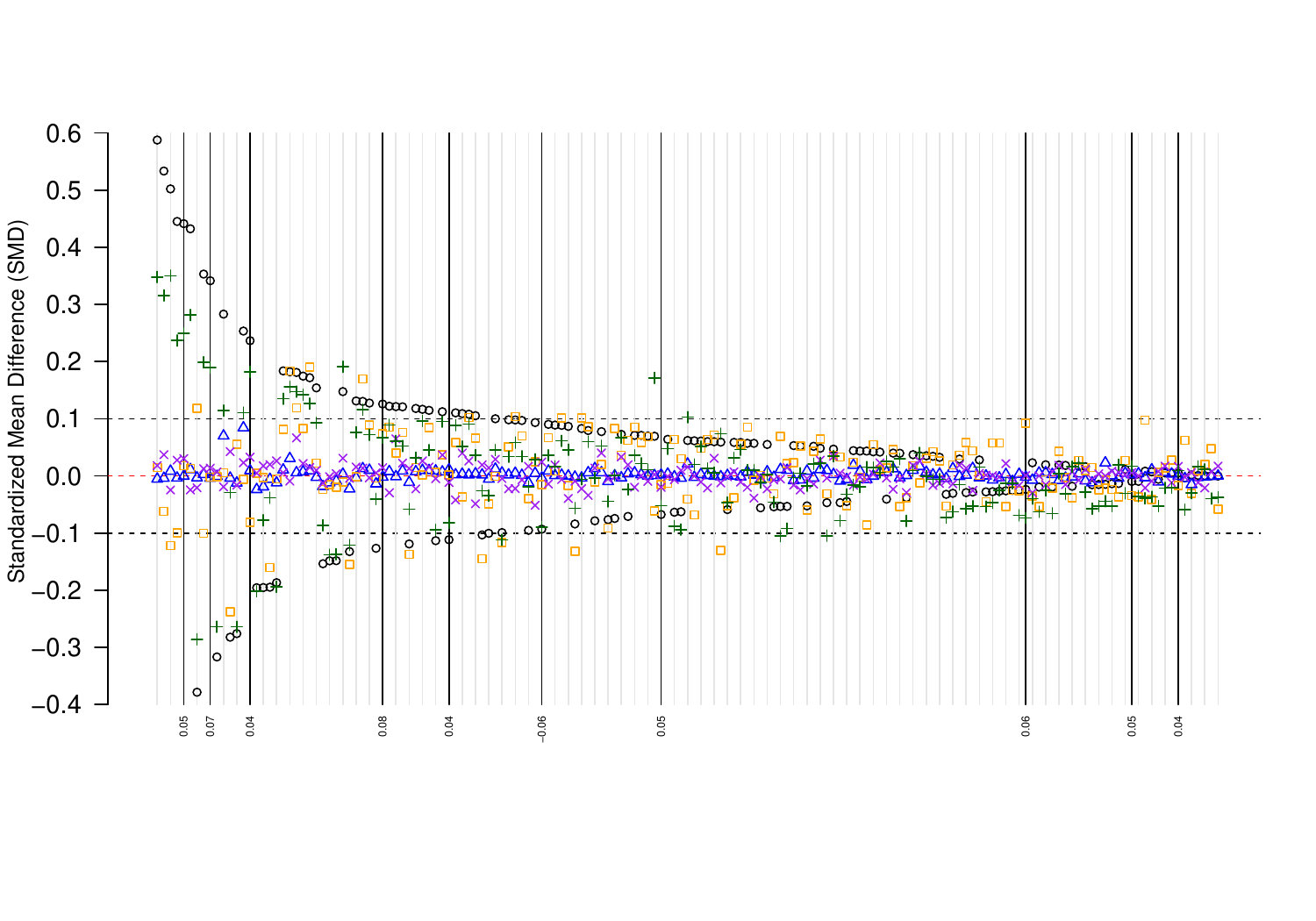} 
    \end{center}
    \vspace{-.90in}
    \subcaption{College enrollment}\label{fig.balance.appendix.mthvnone.enroll}
    \end{subfigure}
    \vspace{-0.35in}
    \begin{center}
    \includegraphics[scale=0.30]{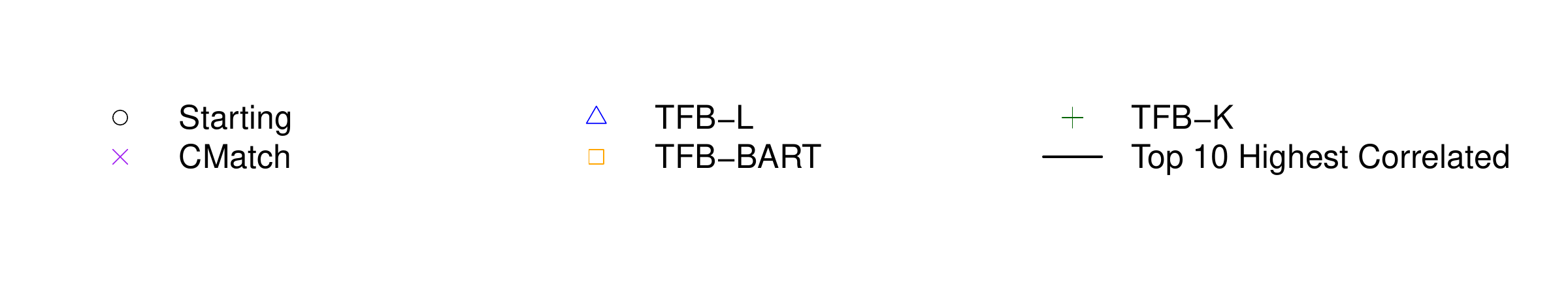} 
    \end{center}
    \vspace{-0.4in}
    \subcaption*{\textit{Note:} SMDs before and after applying TFB to estimate the effects of taking any math course in twelfth grade on the \textit{(a)} end-of-high school GPA outcome, and the \textit{(b)} college enrollment outcome. For each outcome, the ten (non-school indicator) covariates with the highest absolute partial correlations with the outcome are highlighted with black, vertical lines, and the corresponding partial correlation values are listed on the x-axis. Horizontal black dotted lines indicate the common SMD benchmark of $\pm 0.10$  (\citealp{austin2009balance, normand2001validating}). The horizontal red dotted line indicates a SMD of 0, or exact balance. Vertical light gray lines are visual aids help the reader follow the SMD for a covariate across the weighting methods. }
	\end{center}
	\vspace{-0.25in}
	\end{figure}

\paragraph{Math vs. No Math.} Figure~\ref{fig.balance.appendix.mthvnone} reports the SMDs before and after weighting when comparing taking math in twelfth grade to not taking math. Note that there is some overlap between what is shown in Figure~\ref{fig.balance.appendix.mthvnone} and what is shown in Figure~\ref{fig.mthvnone}. However, Figure~\ref{fig.balance.appendix.mthvnone} has more complete results on TFB's covariate balance. 

In summary, both TFB-L and TFB-BART attain acceptable balance on the covariates. While they tend to leave more imbalance in the covariates than does CMatch, the covariates they leave imbalance in tend to have lower partial correlations with the outcome. However, TFB-K's balance is unacceptable: for both outcomes it leaves non-negligible imbalance in covariates that do have higher partial correlations with the outcome. Thus, TFB-K's estimates are omitted in Table~\ref{tab.mthvnone.estimates}.

TFB-L for both outcomes obtains very good balance on the covariates. While it has SMDs outside the common SMD benchmark of $\pm 0.10$  (\citealp{austin2009balance, normand2001validating}) for two covariates in Figure~\ref{fig.balance.appendix.mthvnone.gpa}, it attains nearly exact balance on the covariates that have high correlations with the outcome in question. For example, TFB-L achieves essentially perfect balance on the top two covariates, which have partial correlations with the end-of-high school GPA outcome of 0.91 and 0.14. Thus, TFB-L appears to be viable for both these outcomes in this course-taking comparison.

TFB-BART is much more aggressive than is TFB-L in terms of the imbalance it leaves in the covariates, showing SMDs outside the  $\pm 0.10$ benchmark for several variables for both outcomes. However, for the most part, these variables are not among the covariates with the highest partial correlations with the outcome in question. In Figure~\ref{fig.balance.appendix.mthvnone.enroll}, TFB-BART attains SMDs within $\pm 0.10$ for all of the ten highlighted covariates with the highest partial correlations with the college enrollment outcome. In Figure~\ref{fig.balance.appendix.mthvnone.gpa}, TFB-BART does leave SMDs outside the $\pm 0.10$ benchmark for two of the ten highlighted covariates with the highest partial correlations with the end-of-high school GPA outcome. But it does drastically reduce the starting imbalance for one of those two variables, and the SMD is just barely above 0.10 for the other covariate. Additionally, TFB-BART does attain nearly perfect balance on the covariate with the highest partial correlation (0.91). Therefore, we deem TFB-BART to be viable for both of these outcomes in this course-taking comparison.

TFB-K, however, struggles for both of the outcomes. In Figure~\ref{fig.balance.appendix.mthvnone.gpa}, TFB-K reports a SMD over 0.10 for the variable with the highest partial correlation with the end-of-high school GPA outcome (0.91). In Figure~\ref{fig.balance.appendix.mthvnone.enroll}, TFB-K then reports high SMDs for two of the ten covariates with the highest partial correlations with the college enrollment outcome. Because of these imbalances, we believe that TFB-K is unreliable for both outcomes, and we omit TFB-K from Table~\ref{tab.mthvnone.estimates}.

\clearpage
\newpage

\begin{figure}[!h]
	\vspace{0.15in}
	\begin{center}
	\caption{Balance plots for TFB-L, TFB-BART, and TFB-K for each outcome when estimating the effects of taking Statistics over Precalculus in twelfth grade}\label{fig.balance.appendix.regstatvprecalc}
    \vspace{-.60in}
    \begin{subfigure}{.95\textwidth}
    \begin{center}
    \includegraphics[scale=0.60]{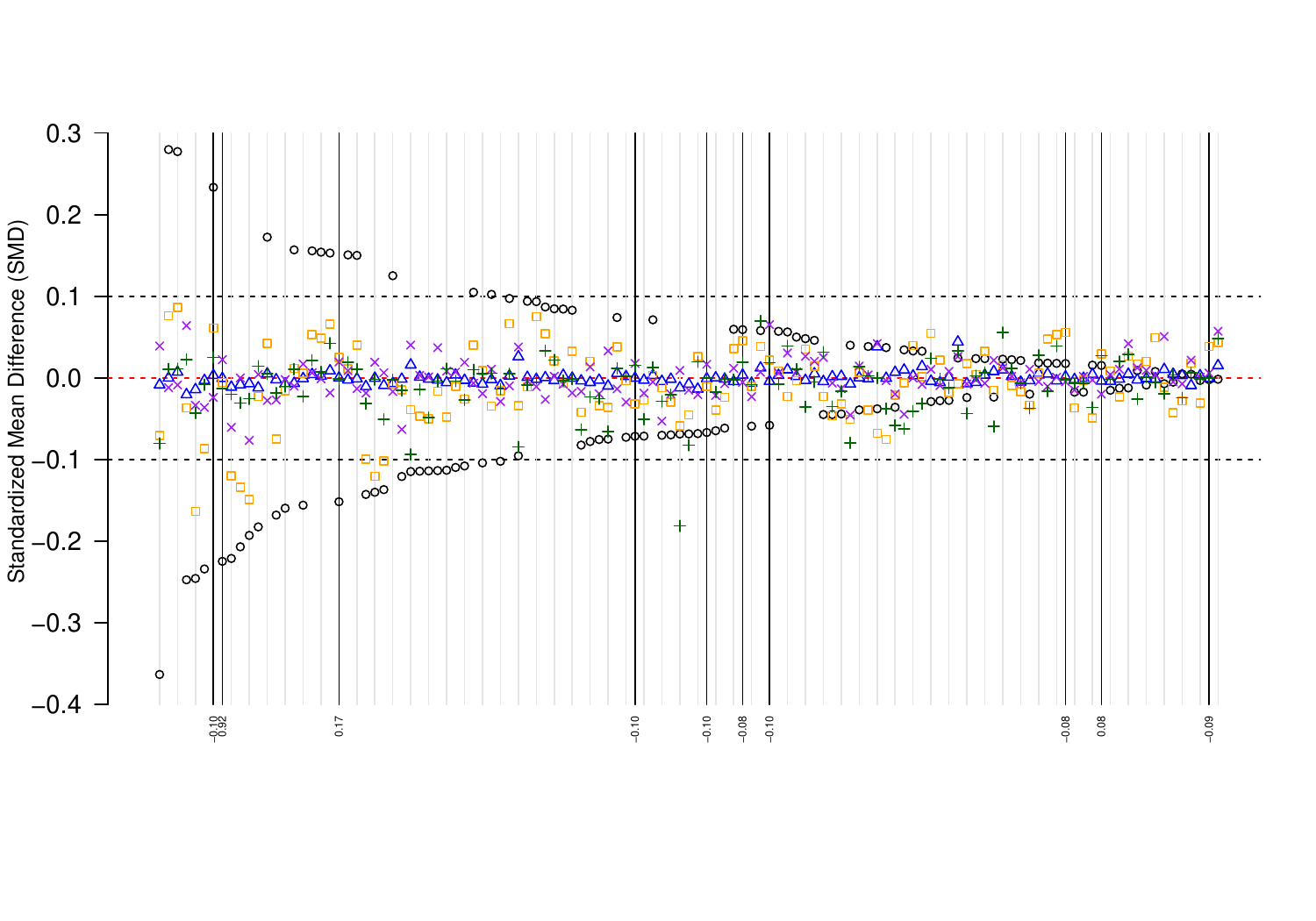} 
    \end{center}
    \vspace{-.90in}
    \subcaption{End-of-high school GPA}\label{fig.balance.appendix.regstatvprecalc.gpa}
    \end{subfigure}
    \begin{subfigure}{.95\textwidth}
    \vspace{-1.1in}
    \begin{center}
    \includegraphics[scale=0.60]{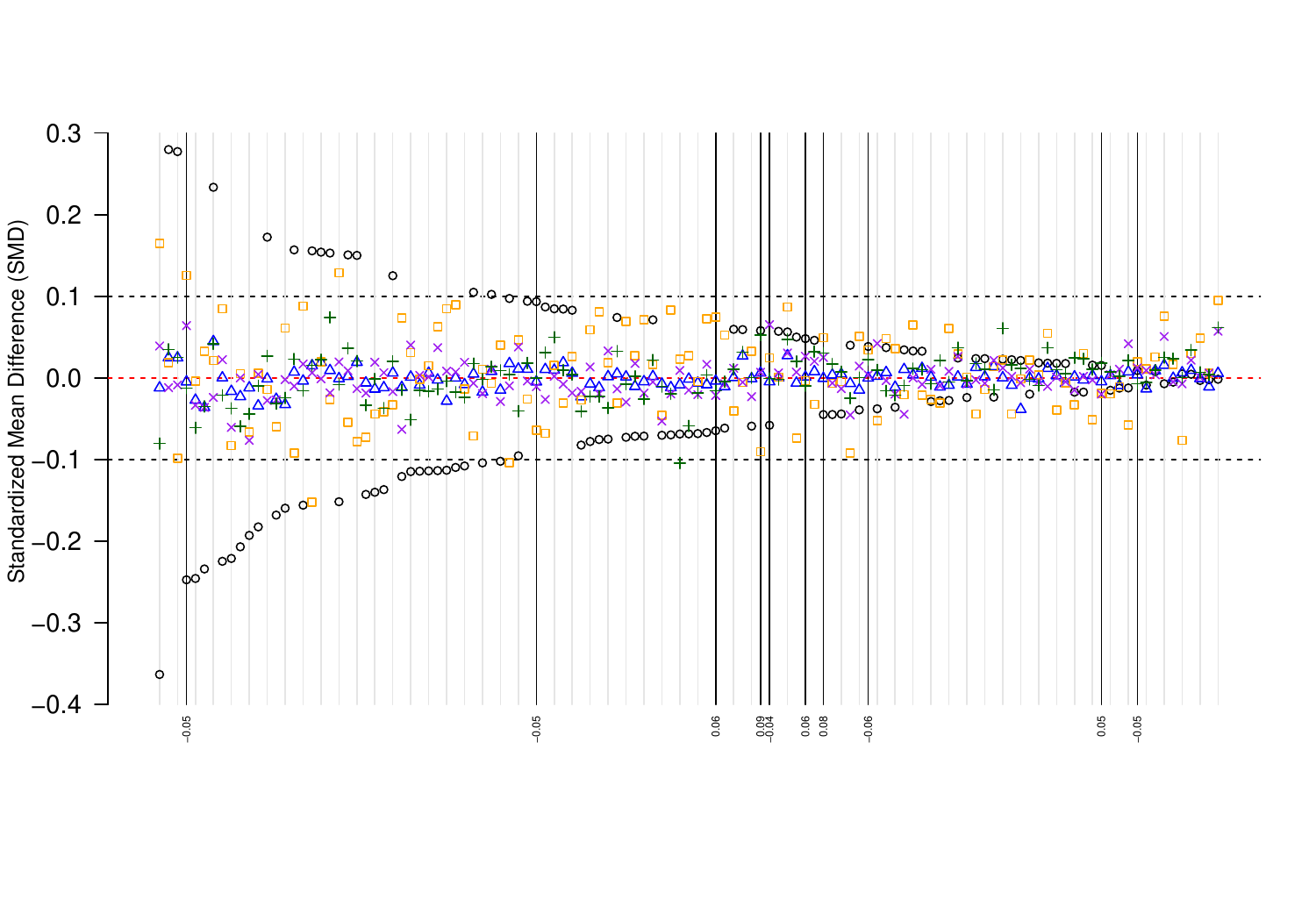} 
    \end{center}
    \vspace{-.90in}
    \subcaption{College enrollment}\label{fig.balance.appendix.regstatvprecalc.enroll}
    \end{subfigure}
    \vspace{-0.35in}
    \begin{center}
    \includegraphics[scale=0.30]{Figures_Tables/LAERI_balance_appendix_Legend.pdf} 
    \end{center}
    \vspace{-0.4in}
    \subcaption*{\textit{Note:} SMDs before and after applying TFB to estimate the effects of taking Statistics over Precalculus in twelfth grade on the \textit{(a)} end-of-high school GPA outcome, and the \textit{(b)} college enrollment outcome. For each outcome, the ten (non-school indicator) covariates with the highest absolute partial correlations with the outcome are highlighted with black, vertical lines, and the corresponding partial correlation values are listed on the x-axis. Horizontal black dotted lines indicate the common SMD benchmark of $\pm 0.10$  (\citealp{austin2009balance, normand2001validating}). The horizontal red dotted line indicates a SMD of 0, or exact balance. Vertical light gray lines are visual aids help the reader follow the SMD for a covariate across the weighting methods. }
	\end{center}
	\vspace{-0.25in}
	\end{figure}

\paragraph{Statistics vs. Precalculus.} Figure~\ref{fig.balance.appendix.regstatvprecalc} reports the SMDs before and after weighting when comparing taking Statistics to taking Precalculus in twelfth grade. In summary, all of TFB-L, TFB-BART, and TFB-K attain acceptable balance on the covariates. TFB-BART is again more aggressive in leaving noticeable imbalance in certain covariates, and leaves more imbalance overall than does CMatch.

TFB-L for both outcomes obtains very good balance on the covariates --- all of the covariates have SMDs within the $\pm 0.10$  benchmark. Despite additionally attempting to balance multiple squared and interaction terms, it also appears to obtain better balance overall than does CMatch. Thus, TFB-L is clearly viable for both outcomes in this course-taking comparison.

TFB-BART is again more aggressive than are TFB-L and CMatch in leaving imbalance in the covariates, showing SMDs outside the  $\pm 0.10$ benchmark for several variables for both outcomes. However, TFB-BART does report better balance than it did when comparing taking math and not taking math (see Figure~\ref{fig.balance.appendix.mthvnone}). In Figure~\ref{fig.balance.appendix.regstatvprecalc.gpa}, TFB-BART attains SMDs within $\pm 0.10$ for all of the ten highlighted covariates with the highest partial correlations with the end-of-high school GPA outcome. In Figure~\ref{fig.balance.appendix.regstatvprecalc.enroll}, TFB-BART leaves a SMD outside the $\pm 0.10$ benchmark for one of the ten highlighted covariates with the highest partial correlations with the college enrollment outcome. However, it greatly reduces the starting SMD for this variable, and appears to reverse the sign of the imbalance, which can be intentional and beneficial (e.g., TFB-K in Section~\ref{sec:dgp1}). Thus, TFB-BART appears to be viable for both outcomes in this course-taking comparison.

In contrast to when comparing taking any math course in twelfth grade to not taking math (see Figure~\ref{fig.balance.appendix.mthvnone}), TFB-K achieves acceptable balance for both outcomes here. For both outcomes, TFB-K attains SMDs within the $\pm 0.10$ benchmark for all of the covariates with the highest partial correlations with the outcome in question. Additionally, despite KRLS struggling to predict the end-of-high school GPA outcome in the analytic sample for this course comparison (see Appendix~\ref{app:application1_additional_cvresults}), in Figure~\ref{fig.balance.appendix.regstatvprecalc.gpa}, TFB-K achieves very good balance on the covariate with the highest partial correlation with this outcome (0.92). Thus, we deem TFB-K viable for both outcomes in this course-taking comparison. Note that this is the only course comparison for which we do not omit TFB-K's estimated effect on the end-of-high school GPA outcome (see Table~\ref{tab.regstatvprecalc.estimates}).

\clearpage
\newpage

\begin{figure}[!h]
	\vspace{0.15in}
	\begin{center}
	\caption{Balance plots for TFB-L, TFB-BART, and TFB-K for each outcome when estimating the effects of taking TCMS over Precalculus in twelfth grade}\label{fig.balance.appendix.tcmsvprecalc}
    \vspace{-.60in}
    \begin{subfigure}{.95\textwidth}
    \begin{center}
    \includegraphics[scale=0.60]{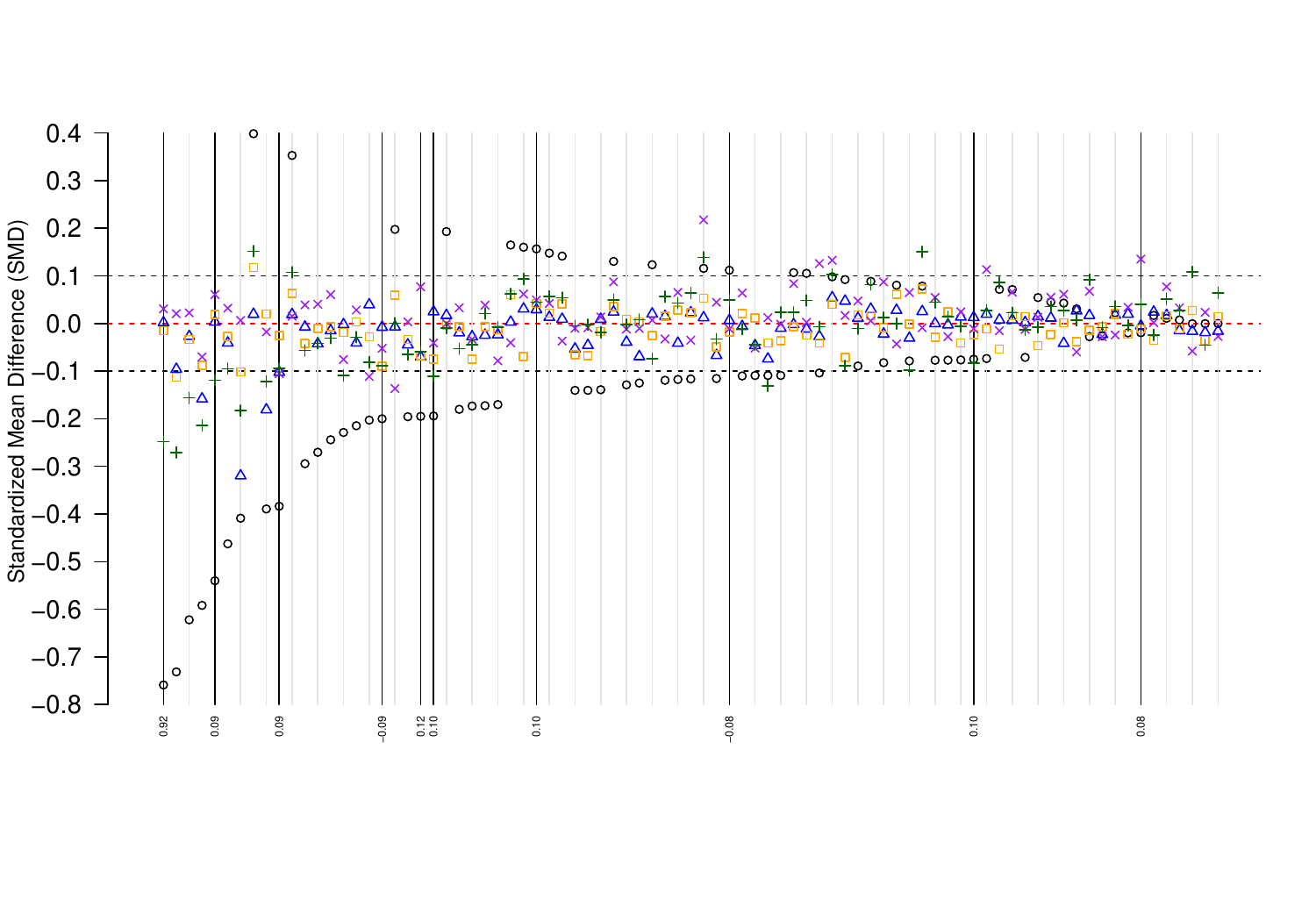} 
    \end{center}
    \vspace{-.90in}
    \subcaption{End-of-high school GPA}\label{fig.balance.appendix.tcmsvprecalc.gpa}
    \end{subfigure}
    \begin{subfigure}{.95\textwidth}
    \vspace{-1.1in}
    \begin{center}
    \includegraphics[scale=0.60]{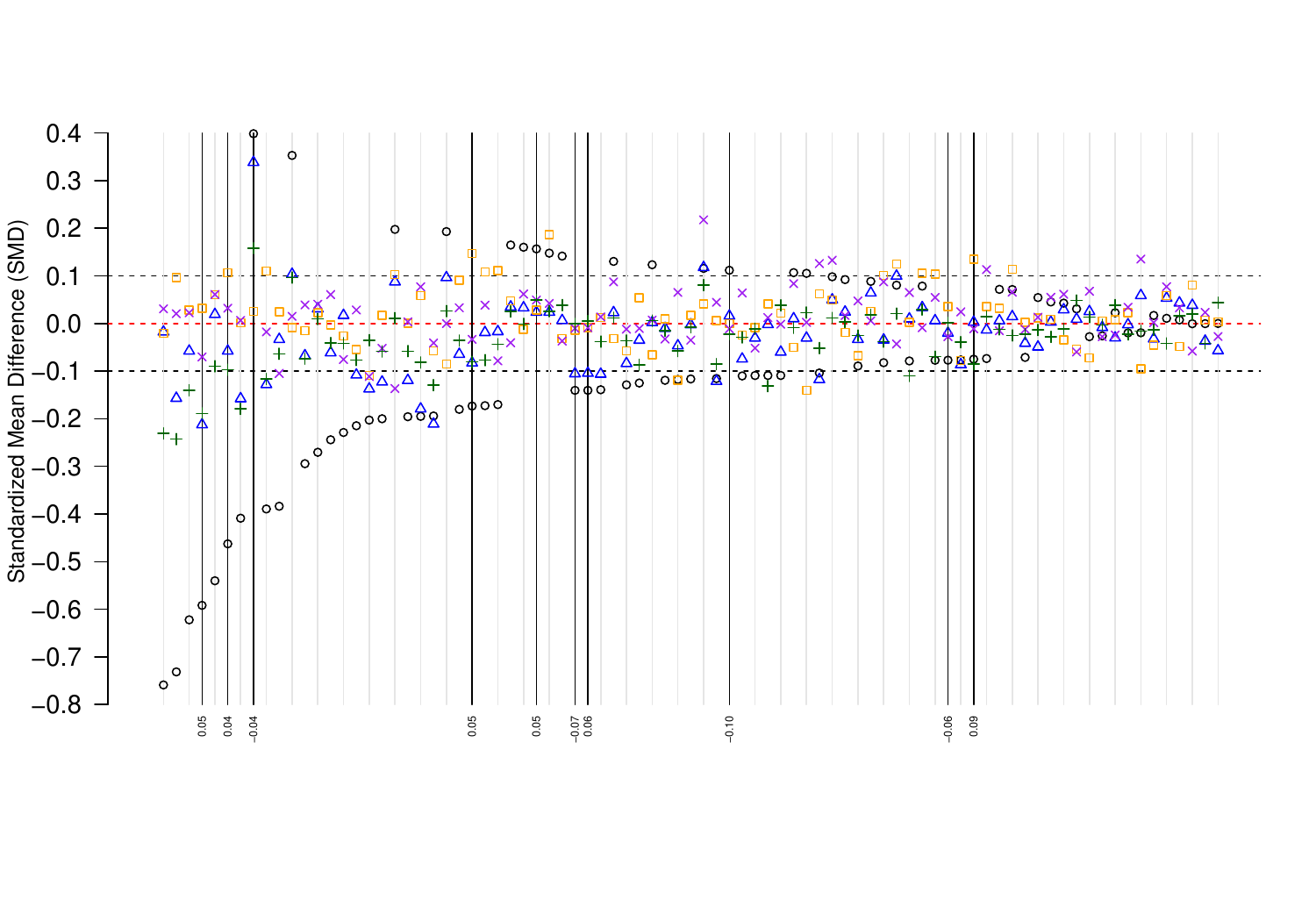} 
    \end{center}
    \vspace{-.90in}
    \subcaption{College enrollment}\label{fig.balance.appendix.tcmsvprecalc.enroll}
    \end{subfigure}
    \vspace{-0.35in}
    \begin{center}
    \includegraphics[scale=0.30]{Figures_Tables/LAERI_balance_appendix_Legend.pdf} 
    \end{center}
    \vspace{-0.4in}
    \subcaption*{\textit{Note:} SMDs before and after applying TFB to estimate the effects of taking TCMS over Precalculus in twelfth grade on the \textit{(a)} end-of-high school GPA outcome, and the \textit{(b)} college enrollment outcome. For each outcome, the ten (non-school indicator) covariates with the highest absolute partial correlations with the outcome are highlighted with black, vertical lines, and the corresponding partial correlation values are listed on the x-axis. Horizontal black dotted lines indicate the common SMD benchmark of $\pm 0.10$  (\citealp{austin2009balance, normand2001validating}). The horizontal red dotted line indicates a SMD of 0, or exact balance. Vertical light gray lines are visual aids help the reader follow the SMD for a covariate across the weighting methods. }
	\end{center}
	\vspace{-0.25in}
	\end{figure}

\paragraph{TCMS vs. Precalculus.} Figure~\ref{fig.balance.appendix.tcmsvprecalc} reports the SMDs before and after weighting when comparing taking TCMS to taking Precalculus in twelfth grade. In summary, all of TFB-L, TFB-BART, and TFB-K attain acceptable balance on the covariates when applied to the college enrollment outcome. TFB-L and TFB-BART are also viable when applied to the end-of-high school GPA outcome. However, TFB-K is deemed unreliable for the end-of-high school GPA outcome, and is thus omitted in Table~\ref{tab.tcmsvprecalc.estimates}. 

TFB-L for both outcomes obtains acceptable balance on the covariates.  In Figure~\ref{fig.balance.appendix.tcmsvprecalc.gpa}, TFB-L attains SMDs within $\pm 0.10$ for nine of the ten highlighted covariates with the highest partial correlations with the end-of-high school GPA outcome. However, the one covariate that is outside this range has a SMD that is just barely below -0.10. Further, TFB-L attains nearly perfect balance on the covariate with the highest partial correlation (0.92). In Figure~\ref{fig.balance.appendix.tcmsvprecalc.enroll}, TFB-L leaves SMDs outside the $\pm 0.10$ benchmark for four of the ten highlighted covariates with the highest partial correlations with the college enrollment outcome. But two of these SMDs are just barely outside of the benchmark range, and TFB-L is successful in reducing the starting imbalance in the other two covariates. Though the one SMD for TFB-L above 0.30 in Figure~\ref{fig.balance.appendix.tcmsvprecalc.enroll} is admittedly concerning. Nevertheless, we deem TFB-L to be viable for both of these outcomes in this course-taking comparison.

In contrast to when comparing math to no math, and Statistics to Precalculus, TFB-BART actually obtains better balance overall than does TFB-L in this course comparison. In Figure~\ref{fig.balance.appendix.tcmsvprecalc.gpa}, TFB-BART attains SMDs within $\pm 0.10$ for all ten of the highlighted covariates that have the highest partial correlations with the end-of-high school GPA outcome. In Figure~\ref{fig.balance.appendix.tcmsvprecalc.enroll}, TFB-BART leaves SMDs outside the $\pm 0.10$ benchmark for three of the ten highlighted covariates with the highest partial correlations with the college enrollment outcome. But one of these SMDs is just barely outside of the benchmark range, and the other two SMDs are not too far from the benchmark (and have switched the signs from the original SMDs). Thus, we believe TFB-BART is viable for both outcomes in this course-taking comparison.

TFB-K is definitely unreliable for the end-of-high school GPA outcome, as it reports in Figure~\ref{fig.balance.appendix.tcmsvprecalc.gpa} a SMD well outside the $\pm 0.10$ benchmark for the covariate with the highest partial correlation (0.92). In Figure~\ref{fig.balance.appendix.tcmsvprecalc.enroll}, TFB-K leaves SMDs outside the $\pm 0.10$ benchmark for two of the ten highlighted covariates with the highest partial correlations with the college enrollment outcome. But TFB-K has drastically reduced the SMDs from the starting SMDs for both of these covariates. Thus, we deem TFB-K to be viable for the college enrollment outcome.

\clearpage
\newpage

\begin{figure}[!h]
	\vspace{0.15in}
	\begin{center}
	\caption{Balance plots for TFB-L, TFB-BART, and TFB-K for each outcome when estimating the effects of taking IDS over Precalculus in twelfth grade}\label{fig.balance.appendix.idsvprecalc}
    \vspace{-.60in}
    \begin{subfigure}{.95\textwidth}
    \begin{center}
    \includegraphics[scale=0.60]{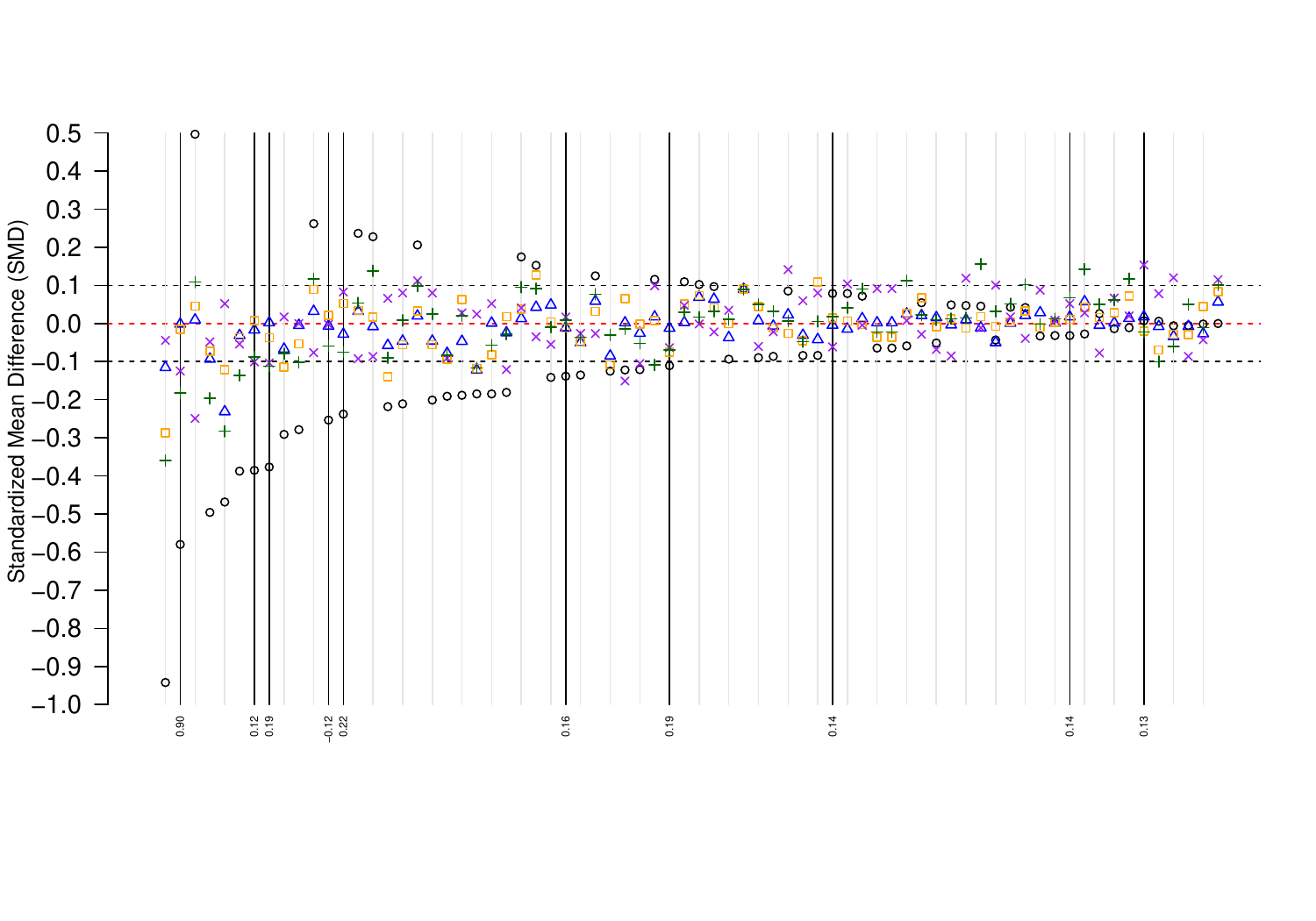} 
    \end{center}
    \vspace{-.90in}
    \subcaption{End-of-high school GPA}\label{fig.balance.appendix.idsvprecalc.gpa}
    \end{subfigure}
    \begin{subfigure}{.95\textwidth}
    \vspace{-1.1in}
    \begin{center}
    \includegraphics[scale=0.60]{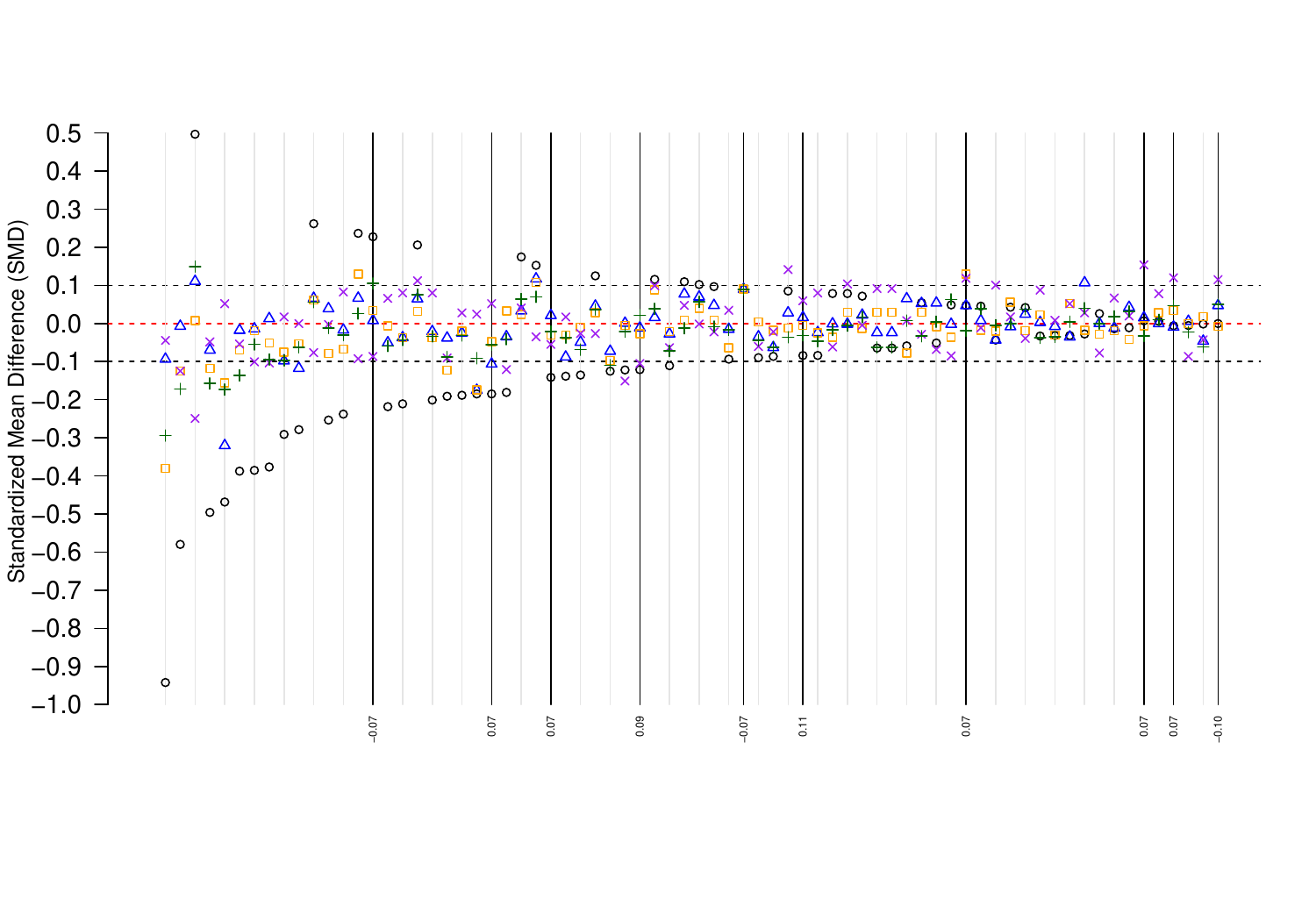} 
    \end{center}
    \vspace{-.90in}
    \subcaption{College enrollment}\label{fig.balance.appendix.idsvprecalc.enroll}
    \end{subfigure}
    \vspace{-0.35in}
    \begin{center}
    \includegraphics[scale=0.30]{Figures_Tables/LAERI_balance_appendix_Legend.pdf} 
    \end{center}
    \vspace{-0.4in}
    \subcaption*{\textit{Note:} SMDs before and after applying TFB to estimate the effects of taking IDS over Precalculus in twelfth grade on the \textit{(a)} end-of-high school GPA outcome, and the \textit{(b)} college enrollment outcome. For each outcome, the ten (non-school indicator) covariates with the highest absolute partial correlations with the outcome are highlighted with black, vertical lines, and the corresponding partial correlation values are listed on the x-axis. Horizontal black dotted lines indicate the common SMD benchmark of $\pm 0.10$  (\citealp{austin2009balance, normand2001validating}). The horizontal red dotted line indicates a SMD of 0, or exact balance. Vertical light gray lines are visual aids help the reader follow the SMD for a covariate across the weighting methods. }
	\end{center}
	\vspace{-0.25in}
	\end{figure}

\paragraph{IDS vs. Precalculus.} Figure~\ref{fig.balance.appendix.idsvprecalc} reports the SMDs before and after weighting when comparing taking IDS to taking Precalculus in twelfth grade. As with comparing TCMS to Precalculus, TFB-L, TFB-BART, and TFB-K all achieve acceptable balance on the covariates when applied to the college enrollment outcome. TFB-L and TFB-BART are again viable when applied to the end-of-high school GPA outcome, but TFB-K is unreliable for the end-of-high school GPA outcome and is omitted from Table~\ref{tab.idsvprecalc.estimates}. 

TFB-L and TFB-BART for both outcomes obtain acceptable balance on the covariates.  In Figure~\ref{fig.balance.appendix.idsvprecalc.gpa}, TFB-L and TFB-BART achieve SMDs within $\pm 0.10$ for all ten of the highlighted covariates with the highest partial correlations with the end-of-high school GPA outcome. In Figure~\ref{fig.balance.appendix.idsvprecalc.enroll}, TFB-L and TFB-BART both leave a SMD outside the $\pm 0.10$ benchmark for one of the ten highlighted covariates with the highest partial correlations with the college enrollment outcome. But the SMDs for those covariates are just barely outside of the benchmark range. Thus, we deem TFB-L and TFB-BART to be viable for both of these outcomes in this course-taking comparison.

TFB-K is again unreliable for the end-of-high school GPA outcome. In Figure~\ref{fig.balance.appendix.idsvprecalc.gpa}, it reports a SMD well outside of the $\pm 0.10$ benchmark for the covariate with the highest partial correlation (0.90). In constrast, in Figure~\ref{fig.balance.appendix.idsvprecalc.enroll}, TFB-K leaves a SMD outside the $\pm 0.10$ benchmark for just one of the ten highlighted covariates with the highest partial correlations with the college enrollment outcome. But the SMD value is just barely outside the benchmark. Thus, TFB-K is viable for the college enrollment outcome.

\clearpage
\newpage

\begin{figure}[!h]
	\vspace{0.15in}
	\begin{center}
	\caption{Balance plots for TFB-L, TFB-BART, and TFB-K for each outcome when estimating the effects of taking TCMS over Statistics in twelfth grade}\label{fig.balance.appendix.tcmsvregstat}
    \vspace{-.60in}
    \begin{subfigure}{.95\textwidth}
    \begin{center}
    \includegraphics[scale=0.60]{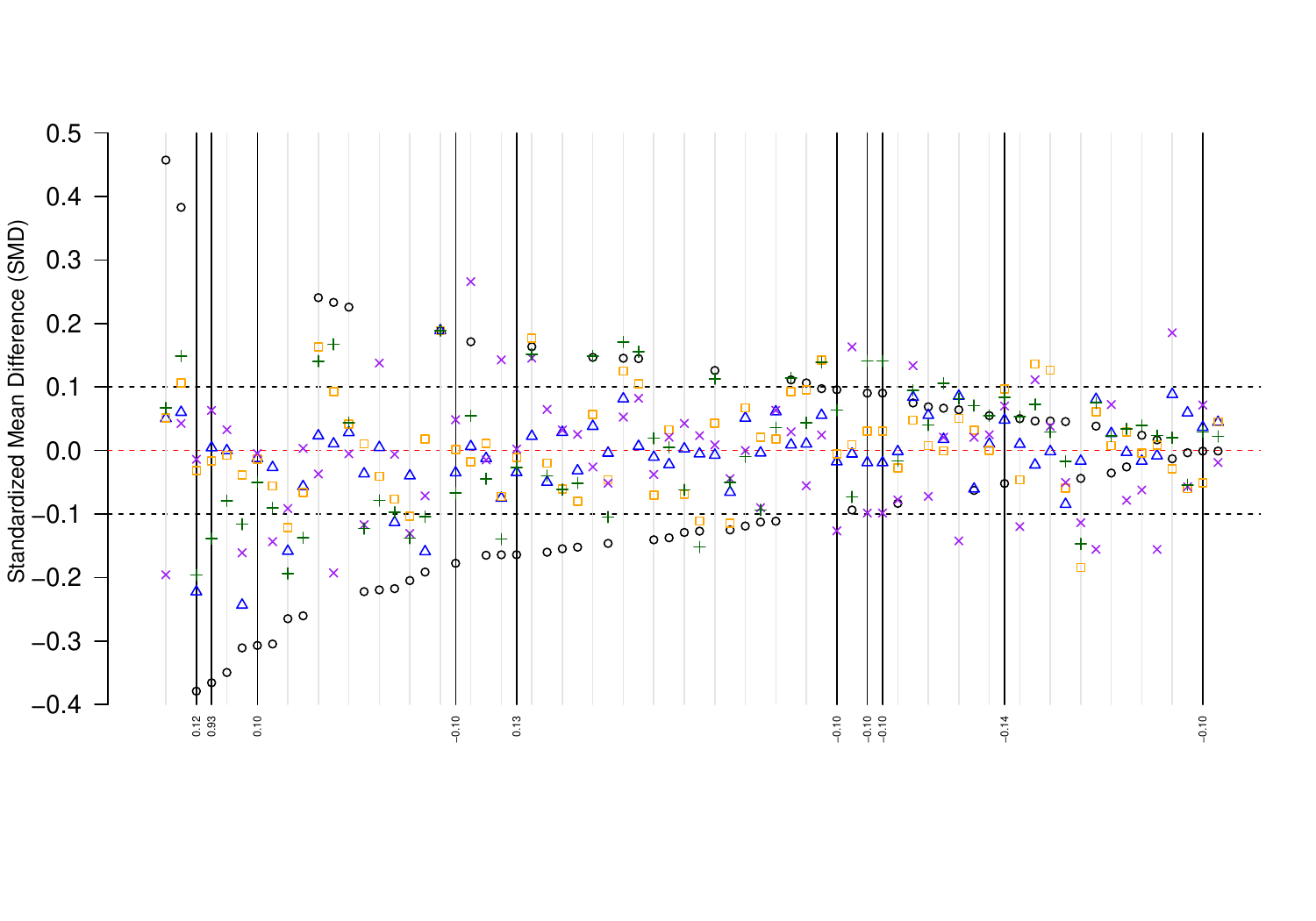} 
    \end{center}
    \vspace{-.90in}
    \subcaption{End-of-high school GPA}\label{fig.balance.appendix.tcmsvregstat.gpa}
    \end{subfigure}
    \begin{subfigure}{.95\textwidth}
    \vspace{-1.1in}
    \begin{center}
    \includegraphics[scale=0.60]{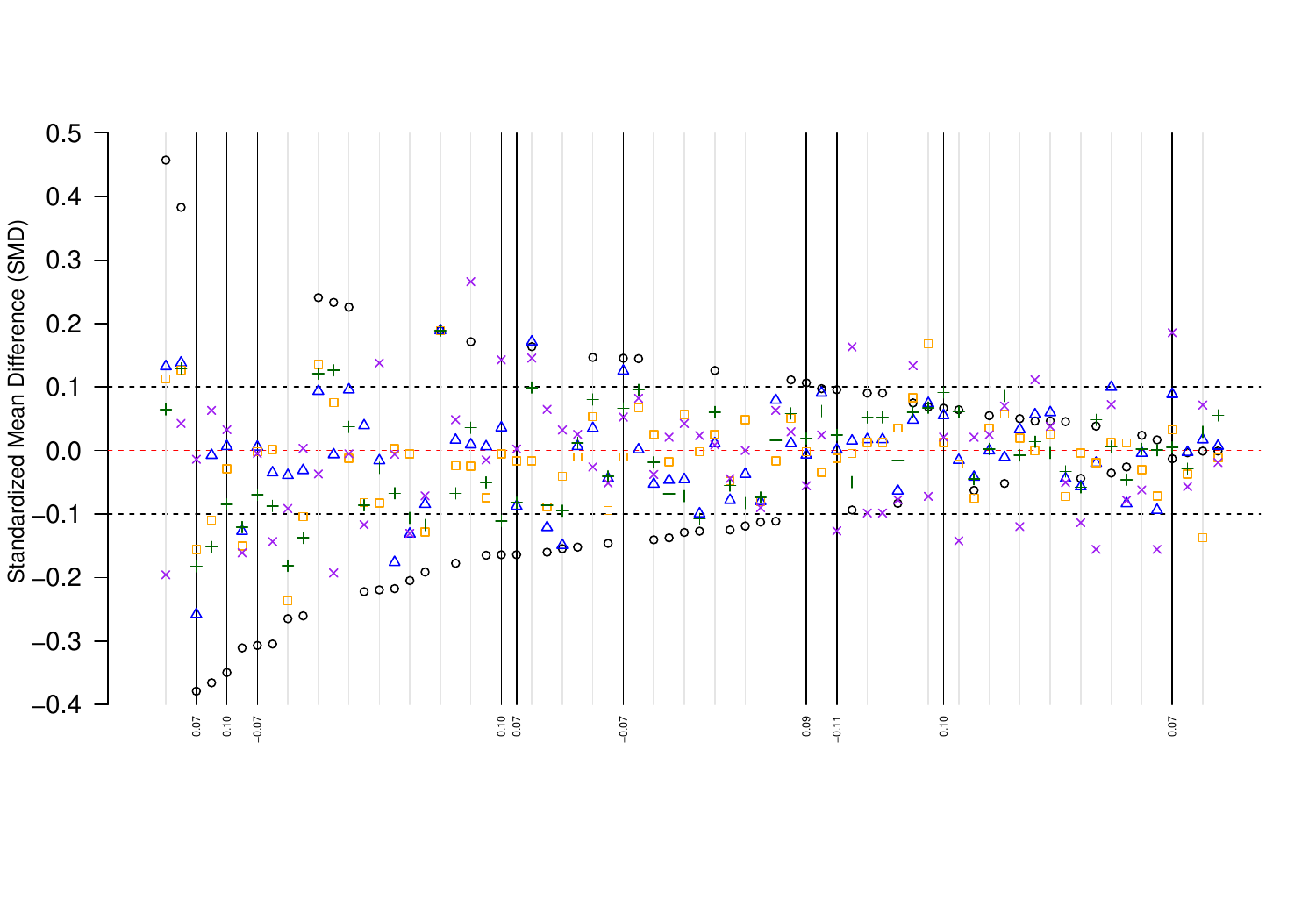} 
    \end{center}
    \vspace{-.90in}
    \subcaption{College enrollment}\label{fig.balance.appendix.tcmsvregstat.enroll}
    \end{subfigure}
    \vspace{-0.35in}
    \begin{center}
    \includegraphics[scale=0.30]{Figures_Tables/LAERI_balance_appendix_Legend.pdf} 
    \end{center}
    \vspace{-0.4in}
    \subcaption*{\textit{Note:} SMDs before and after applying TFB to estimate the effects of taking TCMS over Statistics in twelfth grade on the \textit{(a)} end-of-high school GPA outcome, and the \textit{(b)} college enrollment outcome. For each outcome, the ten (non-school indicator) covariates with the highest absolute partial correlations with the outcome are highlighted with black, vertical lines, and the corresponding partial correlation values are listed on the x-axis. Horizontal black dotted lines indicate the common SMD benchmark of $\pm 0.10$  (\citealp{austin2009balance, normand2001validating}). The horizontal red dotted line indicates a SMD of 0, or exact balance. Vertical light gray lines are visual aids help the reader follow the SMD for a covariate across the weighting methods. }
	\end{center}
	\vspace{-0.25in}
	\end{figure}

\paragraph{TCMS vs. Statistics.} Figure~\ref{fig.balance.appendix.tcmsvregstat} reports the SMDs before and after weighting when comparing taking TCMS to taking Statistics in twelfth grade. In summary, all of TFB-L, TFB-BART, and TFB-K attain acceptable balance on the covariates when applied to the college enrollment outcome. TFB-L and TFB-BART are also viable when applied to the end-of-high school GPA outcome. However, TFB-K is deemed unreliable for the end-of-high school GPA outcome, and is thus omitted in Table~\ref{tab.tcmsvregstat.estimates}. 

TFB-L obtains acceptable balance on the covariates for both outcomes.  In Figure~\ref{fig.balance.appendix.tcmsvregstat.gpa}, TFB-L reports SMDs within $\pm 0.10$ for nine of the ten highlighted covariates with the highest partial correlations with the end-of-high school GPA outcome. The one SMD that is outside the benchmark range is below -0.20, which is admittedly concerning, but TFB-L does meaningfully reduce the starting SMD for that variable. Additionally, TFB-L attains very good balance on the covariate with the highest partial correlation (0.93). In Figure~\ref{fig.balance.appendix.tcmsvregstat.enroll}, TFB-L leaves SMDs outside the $\pm 0.10$ benchmark for two of the ten highlighted covariates with the highest partial correlations with the college enrollment outcome. But one of those SMDs is not far outside of the benchmark range. For the other covariate, TFB-L is successful in reducing the starting SMD, though the remaining SMD below -0.20 is concerning. Nevertheless, we believe TFB-L is still viable for both of these outcomes in this course-taking comparison.

TFB-BART also obtains acceptable balance on the covariates for both outcomes.  In Figure~\ref{fig.balance.appendix.tcmsvregstat.gpa}, TFB-BART achieves SMDs within $\pm 0.10$ for all ten of the highlighted covariates with the highest partial correlations with the end-of-high school GPA outcome. In Figure~\ref{fig.balance.appendix.tcmsvregstat.enroll}, TFB-BART leaves a SMD outside the $\pm 0.10$ benchmark for just one of the ten highlighted covariates with the highest partial correlations with the college enrollment outcome. But TFB-BART significantly reduces the starting SMD for that variable, and the remaining SMD is not far from the benchmark (roughly -0.15). Thus, we deem TFB-BART to be viable for both outcomes in this course-taking comparison.

TFB-K is again unreliable for the end-of-high school GPA outcome here, as in Figure~\ref{fig.balance.appendix.tcmsvregstat.gpa} it leaves a SMD outside of the $\pm 0.10$ benchmark for the covariate with the highest partial correlation (0.93). In Figure~\ref{fig.balance.appendix.tcmsvregstat.enroll}, TFB-K leaves SMDs outside the $\pm 0.10$ benchmark for two of the ten highlighted covariates with the highest partial correlations with the college enrollment outcome. But one of those SMDs is just barely outside the benchmark range, and TFB-K has drastically reduced the SMD from the starting SMD for the other covariate. Therefore, we deem TFB-K's balance to be acceptable enough for the college enrollment outcome.

\clearpage
\newpage

\begin{figure}[!h]
	\vspace{0.15in}
	\begin{center}
	\caption{Balance plots for TFB-L, TFB-BART, and TFB-K for each outcome when estimating the effects of taking IDS over Statistics in twelfth grade}\label{fig.balance.appendix.idsvregstat}
    \vspace{-.60in}
    \begin{subfigure}{.95\textwidth}
    \begin{center}
    \includegraphics[scale=0.60]{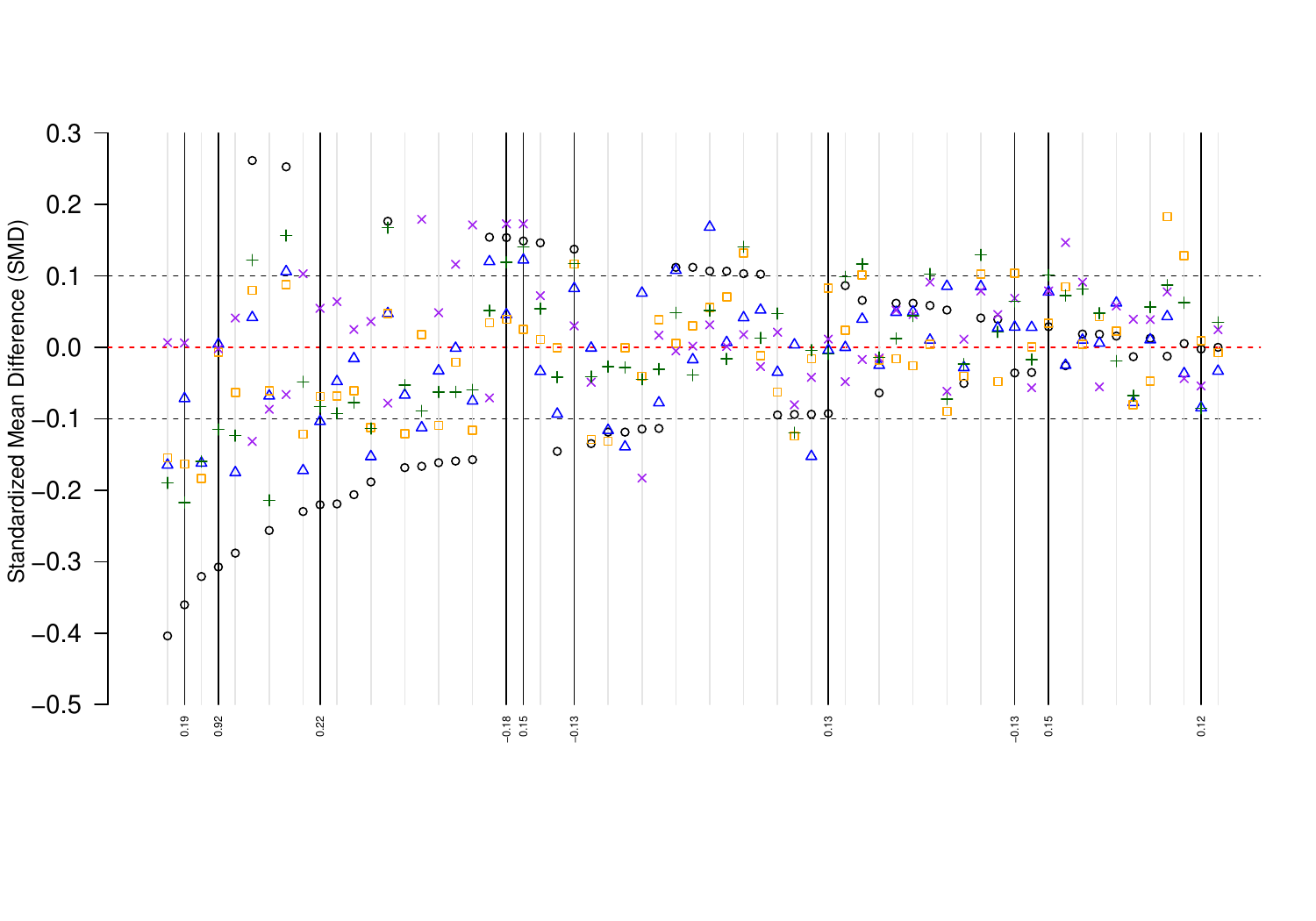} 
    \end{center}
    \vspace{-.90in}
    \subcaption{End-of-high school GPA}\label{fig.balance.appendix.idsvregstat.gpa}
    \end{subfigure}
    \begin{subfigure}{.95\textwidth}
    \vspace{-1.1in}
    \begin{center}
    \includegraphics[scale=0.60]{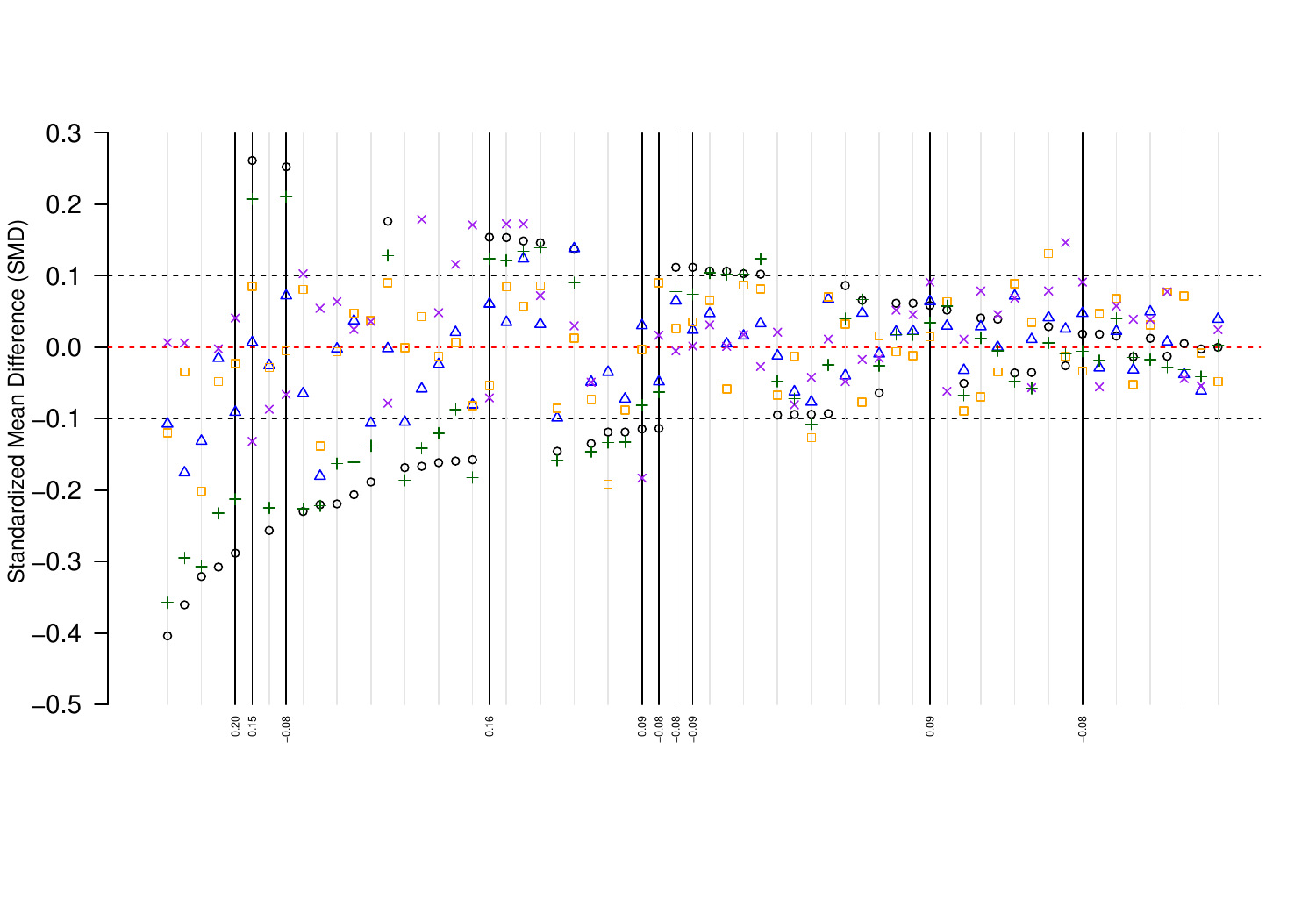} 
    \end{center}
    \vspace{-.90in}
    \subcaption{College enrollment}\label{fig.balance.appendix.idsvregstat.enroll}
    \end{subfigure}
    \vspace{-0.35in}
    \begin{center}
    \includegraphics[scale=0.30]{Figures_Tables/LAERI_balance_appendix_Legend.pdf} 
    \end{center}
    \vspace{-0.4in}
    \subcaption*{\textit{Note:} SMDs before and after applying TFB to estimate the effects of taking IDS over Statistics in twelfth grade on the \textit{(a)} end-of-high school GPA outcome, and the \textit{(b)} college enrollment outcome. For each outcome, the ten (non-school indicator) covariates with the highest absolute partial correlations with the outcome are highlighted with black, vertical lines, and the corresponding partial correlation values are listed on the x-axis. Horizontal black dotted lines indicate the common SMD benchmark of $\pm 0.10$  (\citealp{austin2009balance, normand2001validating}). The horizontal red dotted line indicates a SMD of 0, or exact balance. Vertical light gray lines are visual aids help the reader follow the SMD for a covariate across the weighting methods. }
	\end{center}
	\vspace{-0.25in}
	\end{figure}

\paragraph{IDS vs. Statistics.} Figure~\ref{fig.balance.appendix.idsvregstat} reports the SMDs before and after weighting when comparing taking IDS to taking Statistics in twelfth grade. In summary,  TFB-L and TFB-BART report acceptable balance on the covariates when applied to both outcomes. However, we deem TFB-K to be unreliable for both outcomes, and thus omit all of TFB-K's estimates in Table~\ref{tab.idsvregstat.estimates}. 

TFB-L reports acceptable balance on the covariates for both outcomes.  In Figure~\ref{fig.balance.appendix.idsvregstat.gpa}, TFB-L reports SMDs outside the $\pm 0.10$ benchmark for two of the ten highlighted covariates with the highest partial correlations with the end-of-high school GPA outcome. But both of those SMDs are just barely outside the benchmark range. Furthermore, TFB-L attains near perfect balance on the covariate with the highest partial correlation (0.92). In Figure~\ref{fig.balance.appendix.idsvregstat.enroll}, TFB-L achieves SMDs within the $\pm 0.10$ benchmark for all of the ten highlighted covariates with the highest partial correlations with the college enrollment outcome. Thus, TFB-L is deemed viable for both of these outcomes in this course-taking comparison.

TFB-BART also has acceptable balance on the covariates for both outcomes.  In Figure~\ref{fig.balance.appendix.idsvregstat.gpa}, TFB-BART leaves SMDs outside the $\pm 0.10$ benchmark for three of the ten highlighted covariates with the highest partial correlations with the end-of-high school GPA outcome. However, two of those SMDs are just barely above 0.10. For the third variable, TFB-BART significantly reduces the starting SMD for that variable, and the remaining SMD is not too far from the benchmark (roughly -0.15). Additionally, TFB-BART achieves near perfect balance on the covariate with the highest partial correlation (0.92). In Figure~\ref{fig.balance.appendix.idsvregstat.enroll}, TFB-BART reports SMDs inside the $\pm 0.10$ benchmark for all ten of the highlighted covariates with the highest partial correlations with the college enrollment outcome. Therefore, we consider TFB-BART to be viable for both outcomes in this course-taking comparison.

TFB-K struggles in this course comparison. In Figure~\ref{fig.balance.appendix.idsvregstat.gpa}, it leaves a SMD outside of the $\pm 0.10$ benchmark for the covariate with the highest partial correlation with the end-of-high school GPA outcome (0.92). Then, in Figure~\ref{fig.balance.appendix.idsvregstat.enroll}, TFB-K leaves SMDs outside the $\pm 0.10$ benchmark for the three covariates with the highest partial correlations with the college enrollment outcome (0.20, 0.16, and 0.15). As a result, we deem TFB-K to be unreliable for both outcomes considered here.

\clearpage
\newpage

\subsection{Results: All estimates when comparing different math courses}\label{app:application1_additional_estimates}

Finally, this section reports and discusses all of the TFB estimates when comparing different types of math courses. We also report here the comparison estimates from The LAERI Report. 

As a reminder from Appendix~\ref{app:application1_additional_cvresults}, among the TFB-related models, the LASSO was, overall, the most predictive modeling method for the end-of-high school GPA outcome, and BART was the most predictive for the college enrollment outcome. Then, in Appendix~\ref{app:application1_additional_balance}, TFB-L and TFB-BART report acceptable levels of balance on the covariates for both outcomes across all of the course comparisons. For these reasons, we consider TFB-L to be the primary estimation approach for the end-of-high school GPA outcome, and consider TFB-BART to be the primary estimation approach for the college enrollment outcome, which is why we only report those estimates in Table~\ref{tab.othercourses.estimates} from Section~\ref{subsec:lausd_tcmsids} of the main text. Nevertheless, we report and discuss all viable TFB and comparison methods here to assess any differences between the methods in their overall conclusions, or differences in the magnitudes and signs of their estimates.

\paragraph{Statistics vs. Precalculus.} Table~\ref{tab.regstatvprecalc.estimates} reports all estimated effects of taking Statistics over Precalculus on the two outcomes examined. In summary, the TFB estimates tell a slightly stronger story than do the comparison estimates from The LAERI Report. The TFB estimates imply that taking Statistics over Precalculus in twelfth grade causes a small increase in one's end-of-high school GPA by about 0.01 GPA points, but there is no discernible effect on college enrollment. 
 \begin{table}[!h]
\vspace{0.05in}
\caption{Estimated effects of taking Statistics over Precalculus in twelfth grade}\label{tab.regstatvprecalc.estimates}

\vspace{-0.15in}

	\begin{center}
	\footnotesize
     \begin{tabular}{ |m{2.1cm} | m{1.38cm} m{1.2cm} m{1.3cm} m{1.3cm} | m{1.38cm} m{1.2cm} m{1.3cm} m{1.3cm} |}
     \hline
     \multicolumn{1}{ |c| }{ \textbf{Method} } & \multicolumn{4}{ c| }{\textbf{End-of-High School GPA}} & \multicolumn{4}{ c| }{\textbf{College Enrollment}} \\
      & Estimate & (SE) & RESS: & RESS:  & Estimate & (SE) & RESS: & RESS: \\
      &  & & Control & Treated  & & & Control & Treated \\
     \hline
     Unadjusted & -0.100* & (0.044) & 1.000 & 1.000 & 0.001 & (0.017) & 1.000 & 1.000 \\
     \hline
     WLS: CMatch & 0.009 & (0.005) & 0.686 & 0.498 & -0.019 & (0.020) & 0.686 & 0.498 \\
     WLS: KBAL & 0.009 & (0.005) & 0.786 & 0.649 & -0.003 & (0.018) & 0.786 & 0.649 \\
     WLS: KMatch & 0.010* & (0.005) & 0.735 & 0.558 & -0.009 & (0.021) & 0.735 & 0.558 \\
     OLS & 0.010* & (0.005) & 1.000 & 1.000 & -0.006 & (0.016) & 1.000 & 1.000 \\
     \hline
     TFB-L & 0.013** & (0.005) & 0.668 & 0.395 & 0.001 & (0.017) & 0.666 & 0.490 \\
     TFB-BART & 0.011** & (0.004) & 0.831 & 0.703 & 0.001 & (0.019) & 0.561 & 0.471 \\
     TFB-K & 0.006 & (0.009) & 0.662 & 0.459 & -0.015 & (0.016) & 0.735 & 0.549 \\
     \hline
     \end{tabular}

     \normalsize
	\end{center}
    \subcaption*{\textit{Note:} The ``Unadjusted" estimates are unweighted differences in means. WLS = Weighted Least Squares. CMatch = Preferential Within-Cluster Matching, which was the primary estimation approach in The LAERI Report. KBAL = Kernel Balancing. KMatch = Kernel Matching. OLS = Unweighted Least Squares. SE = Standard Error. The non-TFB estimates and associated standard errors are pulled directly from The LAERI Report. Standard errors for TFB are estimated per Appendix~\ref{subsec:tfb.asymptotics}. * $p<0.05$, ** $p<0.01$, *** $p<0.001$.} 
\vspace{-0.15in}
\end{table}

TFB-L and TFB-BART both report positive estimated effects of taking Statistics over Precalculus on end-of-high school GPA that are statistically significant at the 0.01 level. While TFB-K's estimate is not statistically significant at even the 0.05 level, it should be noted that TFB-L's and TFB-BART's outcome models are much more predictive for this outcome than is TFB-K's outcome model (see Appendix~\ref{app:application1_additional_cvresults}), so TFB-L and TFB-BART should be trusted over TFB-K here. Two of the comparison methods from The LAERI Report (KMatch and OLS) also report positive and statistically significant estimates for this outcome, but the statistical significance is at a lower threshold (the 0.05 level, as opposed to the 0.01 level for TFB-L and TFB-BART). 

None of the TFB estimates are statistically significant for the college enrollment outcome, nor are any of the four comparison methods from The LAERI Report. While TFB-L and TFB-BART report point estimates that are of the opposite sign as are those from The LAERI Report's comparison methods, the estimates are very small, and are nowhere near statistical significance.

\paragraph{TCMS vs. Precalculus.} Table~\ref{tab.tcmsvprecalc.estimates} reports all estimated effects of taking TCMS over Precalculus on the two outcomes examined. The TFB estimates and the comparison estimates from The LAERI Report are very aligned: there is no discernible effect of taking TCMS over Precalculus in twelfth grade on either end-of-high school GPA or college enrollment. While the estimates for end-of-high school GPA are mostly positive, and the estimates for college enrollment are mostly negative, none of them are near statistical significance at the 0.05 level.
 \begin{table}[!h]
\vspace{0.05in}
\caption{Estimated effects of taking TCMS over Precalculus in twelfth grade}\label{tab.tcmsvprecalc.estimates}

\vspace{-0.15in}

	\begin{center}
	\footnotesize
     \begin{tabular}{ |m{2.1cm} | m{1.38cm} m{1.2cm} m{1.3cm} m{1.3cm} | m{1.38cm} m{1.2cm} m{1.3cm} m{1.3cm} |}
     \hline
     \multicolumn{1}{ |c| }{ \textbf{Method} } & \multicolumn{4}{ c| }{\textbf{End-of-High School GPA}} & \multicolumn{4}{ c| }{\textbf{College Enrollment}} \\
      & Estimate & (SE) & RESS: & RESS:  & Estimate & (SE) & RESS: & RESS: \\
      &  & & Control & Treated  & & & Control & Treated \\
     \hline
     Unadjusted & -0.390*** & (0.048) & 1.000 & 1.000 & -0.055* & (0.027) & 1.000 & 1.000 \\
     \hline
     WLS: CMatch & -0.001 & (0.015) & 0.283 & 0.576 & -0.027 & (0.035) & 0.283 & 0.576 \\
     WLS: KBAL & 0.021 & (0.014) & 0.124 & 1.000 & -0.001 & (0.039) & 0.124 & 1.000 \\
     WLS: KMatch & 0.010 & (0.016) & 0.170 & 0.901 & 0.019 & (0.030) & 0.170 & 0.901 \\
     OLS & 0.003 & (0.015) & 1.000 & 1.000 & -0.008 & (0.028) & 1.000 & 1.000 \\
     \hline
     TFB-L & 0.015 & (0.012) & 0.109 & 1.000 & -0.019 & (0.029) & 0.368 & 1.000 \\
     TFB-BART & 0.009 & (0.012) & 0.117 & 1.000 & -0.021 & (0.048) & 0.072 & 1.000 \\
     TFB-K & - & - & - & - & -0.039 & (0.029) & 0.234 & 1.000 \\
     \hline
     \end{tabular}

     \normalsize
	\end{center}
    \subcaption*{\textit{Note:} The ``Unadjusted" estimates are unweighted differences in means. WLS = Weighted Least Squares. CMatch = Preferential Within-Cluster Matching, which was the primary estimation approach in The LAERI Report. KBAL = Kernel Balancing. KMatch = Kernel Matching. OLS = Unweighted Least Squares. SE = Standard Error. The non-TFB estimates and associated standard errors are pulled directly from The LAERI Report. Standard errors for TFB are estimated per Appendix~\ref{subsec:tfb.asymptotics}. TFB-K estimates for the end-of-high school GPA outcome are omitted because they are deemed unreliable (see Appendices~\ref{app:application1_additional_cvresults} and~\ref{app:application1_additional_balance}). * $p<0.05$, ** $p<0.01$, *** $p<0.001$.} 
\vspace{-0.15in}
\end{table}

\paragraph{IDS vs. Precalculus.} Table~\ref{tab.idsvprecalc.estimates} reports all estimated effects of taking IDS over Precalculus on the two outcomes examined. To summarize, the TFB estimates tell the same story as do the comparison estimates from The LAERI Report: taking IDS over Precalculus in twelfth grade causes a small increase in one's end-of-high school GPA (of around 0.03-0.05 GPA points), but there is no discernible effect on college enrollment. 
 \begin{table}[!h]
\vspace{0.05in}
\caption{Estimated effects of taking IDS over Precalculus in twelfth grade}\label{tab.idsvprecalc.estimates}

\vspace{-0.15in}

	\begin{center}
	\footnotesize
     \begin{tabular}{ |m{2.1cm} | m{1.38cm} m{1.2cm} m{1.3cm} m{1.3cm} | m{1.38cm} m{1.2cm} m{1.3cm} m{1.3cm} |}
     \hline
     \multicolumn{1}{ |c| }{ \textbf{Method} } & \multicolumn{4}{ c| }{\textbf{End-of-High School GPA}} & \multicolumn{4}{ c| }{\textbf{College Enrollment}} \\
      & Estimate & (SE) & RESS: & RESS:  & Estimate & (SE) & RESS: & RESS: \\
      &  & & Control & Treated  & & & Control & Treated \\
     \hline
     Unadjusted & -0.243*** & (0.071) & 1.000 & 1.000 & -0.056 & (0.033) & 1.000 & 1.000 \\
     \hline
     WLS: CMatch & 0.054*** & (0.010) & 0.211 & 0.605 & -0.020 & (0.051) & 0.211 & 0.605 \\
     WLS: KBAL & - & - & - & - & - & - & - & - \\
     WLS: KMatch & 0.042*** & (0.011) & 0.166 & 0.931 & -0.041 & (0.056) & 0.166 & 0.931 \\
     OLS & 0.043*** & (0.009) & 1.000 & 1.000 & -0.030 & (0.044) & 1.000 & 1.000 \\
     \hline
     TFB-L & 0.030** & (0.011) & 0.147 & 1.000 & -0.021 & (0.043) & 0.265 & 1.000 \\
     TFB-BART & 0.038*** & (0.011) & 0.128 & 1.000 & -0.047 & (0.044) & 0.158 & 1.000 \\
     TFB-K & - & - & - & - & -0.068 & (0.038) & 0.267 & 1.000 \\
     \hline
     \end{tabular}

     \normalsize
	\end{center}
    \subcaption*{\textit{Note:} The ``Unadjusted" estimates are unweighted differences in means. WLS = Weighted Least Squares. CMatch = Preferential Within-Cluster Matching, which was the primary estimation approach in The LAERI Report. KBAL = Kernel Balancing. KMatch = Kernel Matching. OLS = Unweighted Least Squares. SE = Standard Error. The non-TFB estimates and associated standard errors are pulled directly from The LAERI Report. KBAL estimates are omitted because they were deemed unreliable in The LAERI Report. Standard errors for TFB are estimated per Appendix~\ref{subsec:tfb.asymptotics}. TFB-K estimates for the end-of-high school GPA outcome are omitted because they are deemed unreliable (see Appendices~\ref{app:application1_additional_cvresults} and~\ref{app:application1_additional_balance}). * $p<0.05$, ** $p<0.01$, *** $p<0.001$.} 
\vspace{-0.15in}
\end{table}

All of the shown methods report positive estimated effects on end-of-high school GPA of taking IDS over Precalculus that are statistically significant at the 0.01 level. Most of the estimates, with the exception of TFB-L, are also statistically significant at the 0.001 level. 
	
None of the estimates are statistically significant for the college enrollment outcome. However, they are all negative, and TFB-K's estimate is approaching statistical significance at the 0.05 level. 

\paragraph{TCMS vs. Statistics.} Table~\ref{tab.tcmsvregstat.estimates} reports all estimated effects of taking TCMS over Statistics on the two outcomes examined. The TFB estimates and the The LAERI Report's estimates are very aligned: there is no discernible effect of taking TCMS over Statistics in twelfth grade on either end-of-high school GPA or college enrollment. While the estimates for college enrollment are all negative, none of them are near statistical significance at the 0.05 level.
 \begin{table}[!h]
\vspace{0.05in}
\caption{Estimated effects of taking TCMS over Statistics in twelfth grade}\label{tab.tcmsvregstat.estimates}

\vspace{-0.15in}

	\begin{center}
	\footnotesize
     \begin{tabular}{ |m{2.1cm} | m{1.38cm} m{1.2cm} m{1.3cm} m{1.3cm} | m{1.38cm} m{1.2cm} m{1.3cm} m{1.3cm} |}
     \hline
     \multicolumn{1}{ |c| }{ \textbf{Method} } & \multicolumn{4}{ c| }{\textbf{End-of-High School GPA}} & \multicolumn{4}{ c| }{\textbf{College Enrollment}} \\
      & Estimate & (SE) & RESS: & RESS:  & Estimate & (SE) & RESS: & RESS: \\
      &  & & Control & Treated  & & & Control & Treated \\
     \hline
     Unadjusted & -0.205* & (0.089) & 1.000 & 1.000 & -0.039 & (0.038) & 1.000 & 1.000 \\
     \hline
     WLS: CMatch & - & - & - & - & - & - & - & - \\
     WLS: KBAL & - & - & - & - & - & - & - & - \\
     WLS: KMatch & -0.010 & (0.012) & 0.257 & 0.902 & -0.007 & (0.034) & 0.257 & 0.902 \\
     OLS & -0.007 & (0.013) & 1.000 & 1.000 & -0.003 & (0.034) & 1.000 & 1.000 \\
     \hline
     TFB-L & 0.018 & (0.014) & 0.262 & 1.000 & -0.005 & (0.040) & 0.400 & 1.000 \\
     TFB-BART & -0.003 & (0.010) & 0.243 & 1.000 & -0.021 & (0.037) & 0.277 & 1.000 \\
     TFB-K & - & - & - & - & -0.014 & (0.035) & 0.481 & 1.000 \\
     \hline
     \end{tabular}

     \normalsize
	\end{center}
    \subcaption*{\textit{Note:} The ``Unadjusted" estimates are unweighted differences in means. WLS = Weighted Least Squares. CMatch = Preferential Within-Cluster Matching, which was the primary estimation approach in The LAERI Report. KBAL = Kernel Balancing. KMatch = Kernel Matching. OLS = Unweighted Least Squares. SE = Standard Error. The non-TFB estimates and associated standard errors are pulled directly from The LAERI Report. CMatch and KBAL estimates are omitted because they were deemed unreliable in The LAERI Report. Standard errors for TFB are estimated per Appendix~\ref{subsec:tfb.asymptotics}. TFB-K estimates for the end-of-high school GPA outcome are omitted because they are deemed unreliable (see Appendices~\ref{app:application1_additional_cvresults} and~\ref{app:application1_additional_balance}). * $p<0.05$, ** $p<0.01$, *** $p<0.001$.} 
\vspace{-0.15in}
\end{table}

\paragraph{IDS vs. Statistics.} Table~\ref{tab.idsvregstat.estimates} reports all estimated effects of taking IDS over Statistics on the two outcomes examined. In summary, TFB estimates are slightly more conservative than are the comparison estimates from The LAERI Report, but generally tell the same story. TFB's estimates imply that taking IDS over Statistics in twelfth grade causes a small increase in one's end-of-high school GPA by about 0.03 GPA points, but there is no discernible effect on college enrollment. 
 \begin{table}[!h]
\vspace{0.05in}
\caption{Estimated effects of taking IDS over Statistics in twelfth grade}\label{tab.idsvregstat.estimates}

\vspace{-0.15in}

	\begin{center}
	\footnotesize
     \begin{tabular}{ |m{2.1cm} | m{1.38cm} m{1.2cm} m{1.3cm} m{1.3cm} | m{1.38cm} m{1.2cm} m{1.3cm} m{1.3cm} |}
     \hline
     \multicolumn{1}{ |c| }{ \textbf{Method} } & \multicolumn{4}{ c| }{\textbf{End-of-High School GPA}} & \multicolumn{4}{ c| }{\textbf{College Enrollment}} \\
      & Estimate & (SE) & RESS: & RESS:  & Estimate & (SE) & RESS: & RESS: \\
      &  & & Control & Treated  & & & Control & Treated \\
     \hline
     Unadjusted & -0.133** & (0.047) & 1.000 & 1.000 & -0.091** & (0.034) & 1.000 & 1.000 \\
     \hline
     WLS: CMatch & 0.017 & (0.010) & 0.278 & 0.654 & -0.025 & (0.084) & 0.278 & 0.654 \\
     WLS: KBAL & 0.059*** & (0.009) & 0.210 & 1.000 & 0.028 & (0.040) & 0.210 & 1.000 \\
     WLS: KMatch & 0.037*** & (0.006) & 0.399 & 0.817 & -0.012 & (0.054) & 0.399 & 0.817 \\
     OLS & 0.037*** & (0.009) & 1.000 & 1.000 & -0.014 & (0.053) & 1.000 & 1.000 \\
     \hline
     TFB-L & 0.030* & (0.015) & 0.516 & 1.000 & -0.037 & (0.059) & 0.433 & 1.000 \\
     TFB-BART & 0.026* & (0.011) & 0.237 & 1.000 & -0.038 & (0.055) & 0.278 & 1.000 \\
     TFB-K & - & - & - & - & - & - & - & - \\
     \hline
     \end{tabular}

     \normalsize
	\end{center}
    \subcaption*{\textit{Note:} The ``Unadjusted" estimates are unweighted differences in means. WLS = Weighted Least Squares. CMatch = Preferential Within-Cluster Matching, which was the primary estimation approach in The LAERI Report. KBAL = Kernel Balancing. KMatch = Kernel Matching. OLS = Unweighted Least Squares. SE = Standard Error. The non-TFB estimates and associated standard errors are pulled directly from The LAERI Report. Standard errors for TFB are estimated per Appendix~\ref{subsec:tfb.asymptotics}. TFB-K estimates for both outcomes are omitted because they are deemed unreliable (see Appendices~\ref{app:application1_additional_cvresults} and~\ref{app:application1_additional_balance}). * $p<0.05$, ** $p<0.01$, *** $p<0.001$.} 
\vspace{-0.15in}
\end{table}

TFB-L and TFB-BART both report positive estimated effects on end-of-high school GPA of taking IDS over Statistics that are statistically significant at the 0.05 level. Meanwhile, three of the comparison methods from the LAERI Report (KBAL, KMatch, and OLS) report positive estimates that are statistically significant at the 0.001 level. The difference with TFB-L and TFB-BART is that their point estimates (roughly 0.03) are smaller than these three comparison methods' point estimates (0.037-0.059), and the TFB estimates have larger estimated standard errors. 

None of the TFB estimates are statistically significant for the college enrollment outcome, nor are any of the four comparison methods from The LAERI Report. However, TFB-L's and TFB-BART's estimates are noticeably more negative than are the estimates from the comparison methods. Although, none of the estimates are close to statistical significance at the 0.05 level. 

\newpage
\clearpage

\section{Simulations}\label{app:simulations}

\subsection{Extension to Demonstration 1}\label{app:dgp1_extension}

This section investigates how TFB-K performs  after varying important aspects of DGP 1 from Section~\ref{sec:dgp1}. As in DGP 1, $X$ is drawn from a bivariate normal distribution around $C$, one of four equally likely (cluster) centers that are unknown to the researcher,
	\begin{align}
		& X_i \overset{iid}{\sim} \mathcal{N} (C_i , \mathbf{I}_2) \ \ \text{where} \ \ C \in \{c_1= (0, 0), \ c_2= (0, 5), \ c_3 = (5, 0), \ c_4= (5, 5) \} \nonumber \\
  & \text{with} \ \ P(C = c_{\ell}) = \frac{1}{4}   \ \ \text{for} \ \ \ell = 1, \dots, 4
	\end{align}
Though $D$ and $Y$ again depend on $Z^{(\ell)}$ for $\ell = 1, \dots, 4$,
	\begin{align}
		Z^{(\ell)} = \frac{1}{ ||X - c_{\ell} ||_2 + 1} 
	\end{align}
which are unknown to the researcher. 

Several aspects of the probability of treatment and model for the outcome are varied. The (log odds of the) probability of treatment is:
	\begin{align}\label{eq:dgp1_extend_ps}
		\mathrm{log} \frac{\pi(X)}{1 - \pi(X)} &=  \psi \times \1 \{C = c_1 \} ( Z^{(1)} - 0.47)  +  \sum_{\ell = 2}^{4}   \1 \{C = c_{\ell} \}  (20) (Z^{(\ell)} - 0.47)
	\end{align}
where $\psi$, the coefficient on $Z^{(1)}$, takes values  $\psi \in (2, 4, 8)$. $\psi$ controls the extent of overlap in the distributions of $Z^{(1)}$ between the treated and control groups --- $\psi = 2$ yields ``high" overlap, $\psi = 8$ yields ``low" overlap, and $\psi = 4$ yields ``medium" overlap. The level of overlap for $Z^{(2)}$, $Z^{(3)}$, and $Z^{(4)}$ is not varied here because the goal is to see if TFB still leaves imbalance in $Z^{(1)}$ to overcome low overlap in the other $Z^{(\ell)}$ for different levels of overlap in $Z^{(1)}$.

The model for the outcome is:
	\begin{align}\label{eq:dgp1_extend_y}
		Y = 10 Z^{(1)} +\sum_{\ell = 2}^4 Z^{(\ell)} + \epsilon, \ \ \ \epsilon_i \overset{iid}{\sim} N(0, \sigma^2) 
	\end{align}
where $\sigma^2$, the variance of the error term $\epsilon$, is varied so that the model $R^2 = \frac{\var(Y - \epsilon)}{\var(Y)}$ takes values $R^2 \in \{ 0.30, 0.60, 0.90 \}$ in (\ref{eq:dgp1_extend_y}). 
The $R^2$ is varied to see how the maximum predictive power of a $\hat{f}_0$ influences TFB-K's performance. Finally, two sample sizes are tried: $n \in \{500, 1000\}.$ To summarize, the following aspects are varied:
\begin{itemize}
    \item $\psi \in \{2, 4, 8\}$ in (\ref{eq:dgp1_extend_ps})
    \item $R^2 \in \{ 0.30, 0.60, 0.90 \}$ in (\ref{eq:dgp1_extend_y})\item $n \in \{500, 1000\}$ 
\end{itemize}
Note that $\psi = 4$, $R^2 = 0.60$, and $n=1000$ recreates DGP 1 in Section~\ref{sec:dgp1}. 

\begin{figure}
    \caption{RMSE in Extended DGP 1}\label{fig.app.dgp1.rmse}
        
    \vspace{-0.35in}
    
    \begin{subfigure}{.45\textwidth}
    \begin{center}
    \includegraphics[scale=.375]{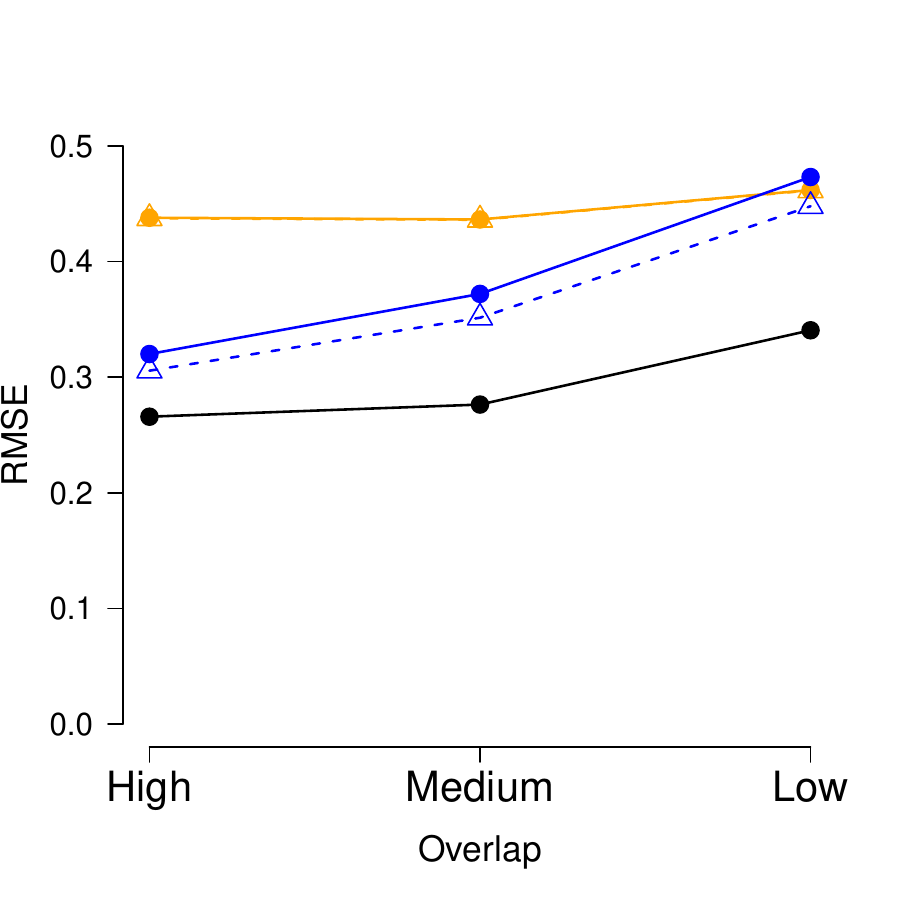}
    \subcaption{$R^2=0.30$, $n=500$}
    \end{center}
    \end{subfigure} 
    \begin{subfigure}{.45\textwidth}
    \begin{center}
    \includegraphics[scale=.375]{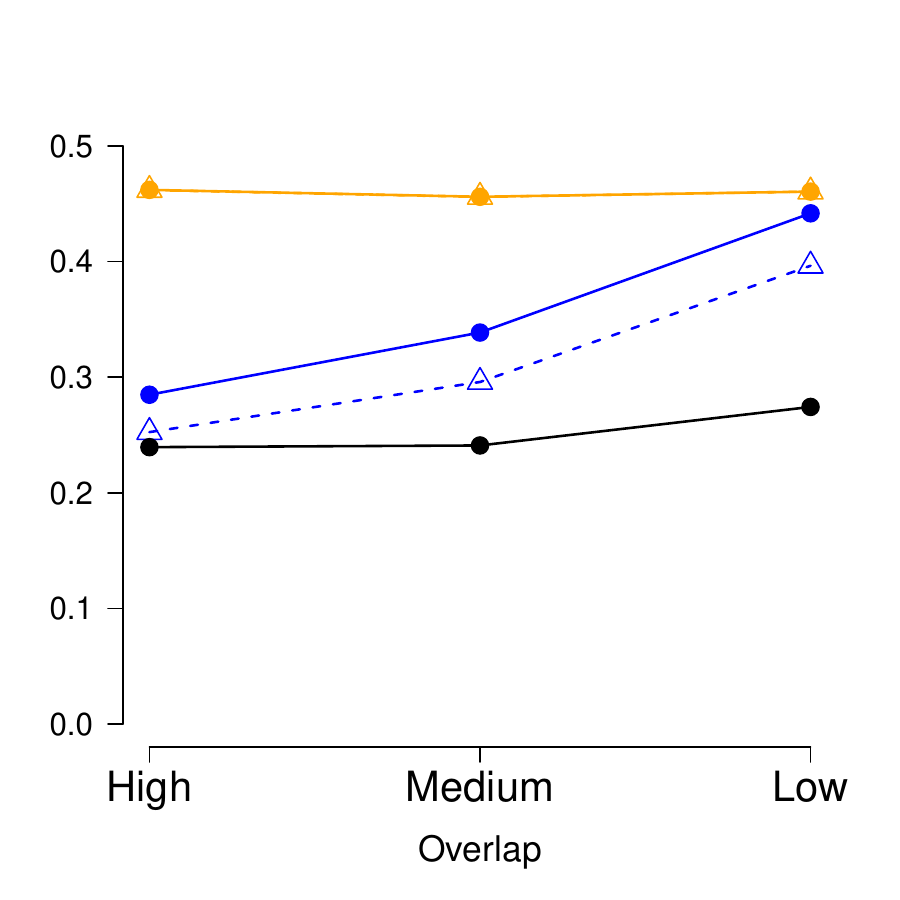}
    \subcaption{$R^2=0.30$, $n=1000$}
    \end{center}
    \end{subfigure}

    \vspace{-0.20in}

    \begin{subfigure}{.45\textwidth}
    \begin{center}
    \includegraphics[scale=.375]{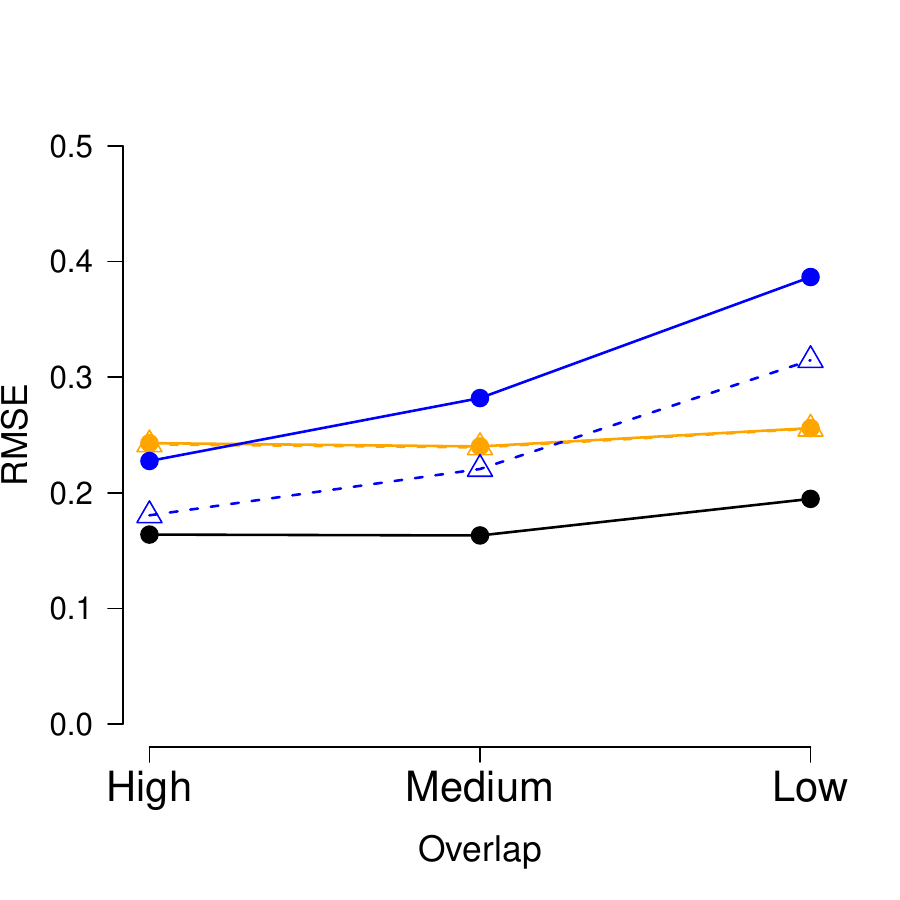}
    \subcaption{$R^2=0.60$, $n=500$}
    \end{center}
    \end{subfigure} 
    \begin{subfigure}{.45\textwidth}
    \begin{center}
    \includegraphics[scale=.375]{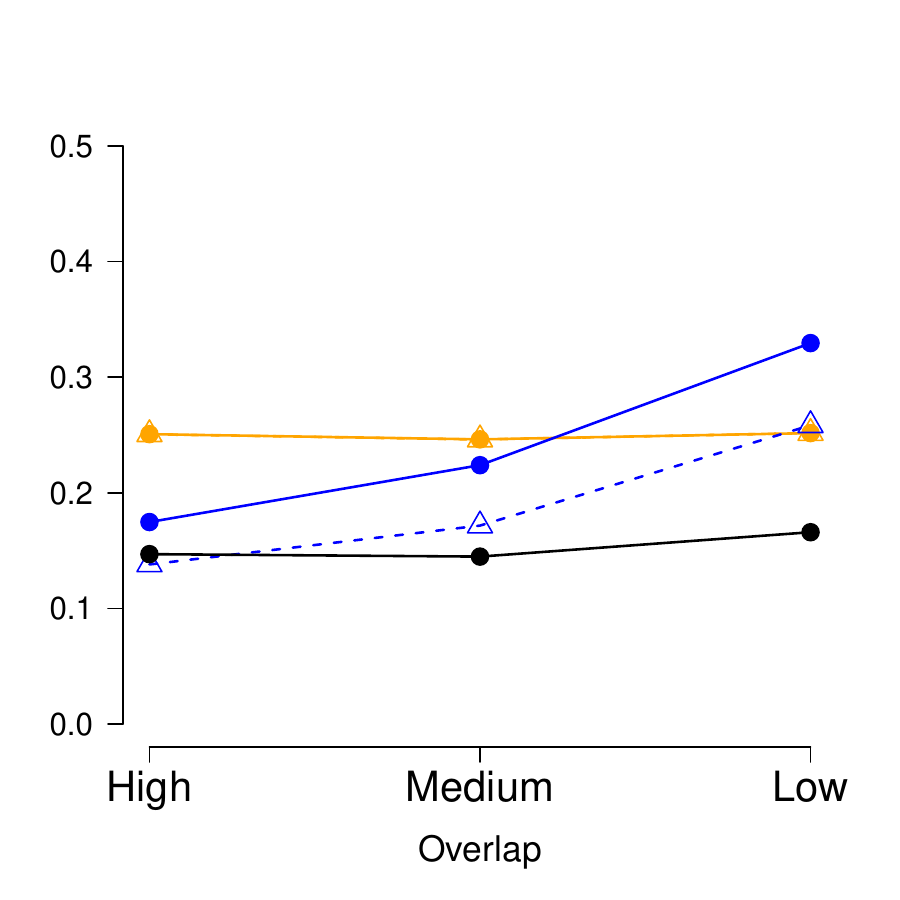}
    \subcaption{$R^2=0.60$, $n=1000$}
    \end{center}
    \end{subfigure}

    \vspace{-0.20in}

    \begin{subfigure}{.45\textwidth}
    \begin{center}
    \includegraphics[scale=.375]{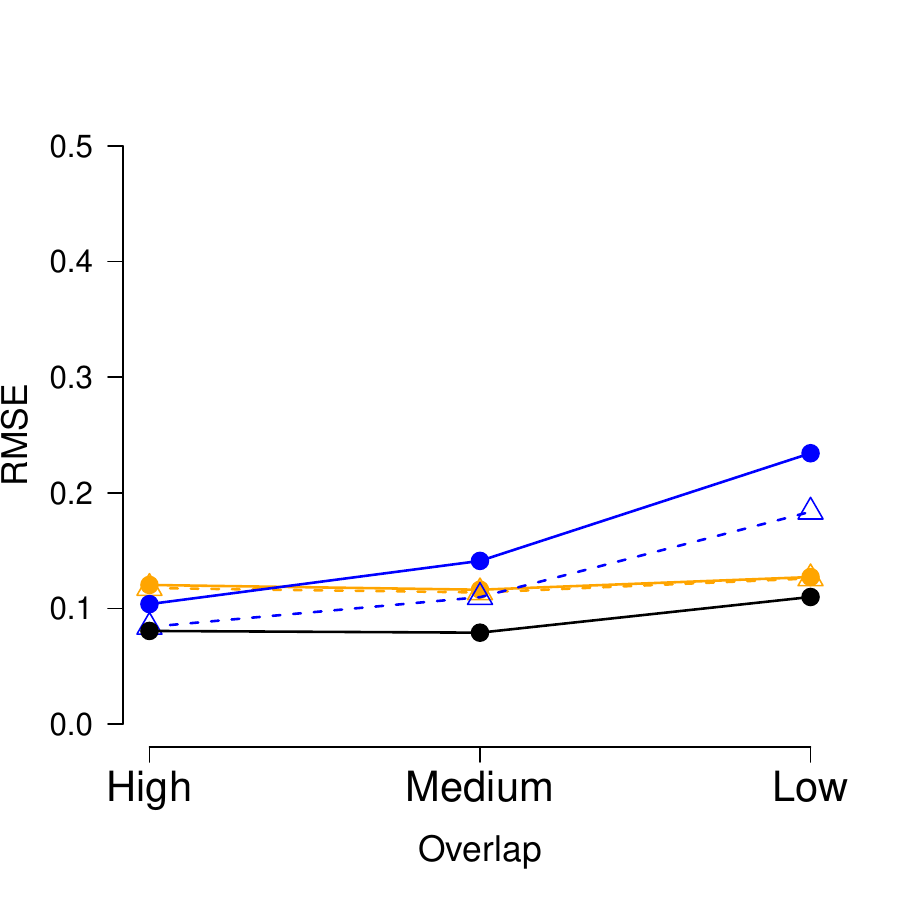}
    \subcaption{$R^2=0.90$, $n=500$}
    \end{center}
    \end{subfigure} 
    \begin{subfigure}{.45\textwidth}
    \begin{center}
    \includegraphics[scale=.375]{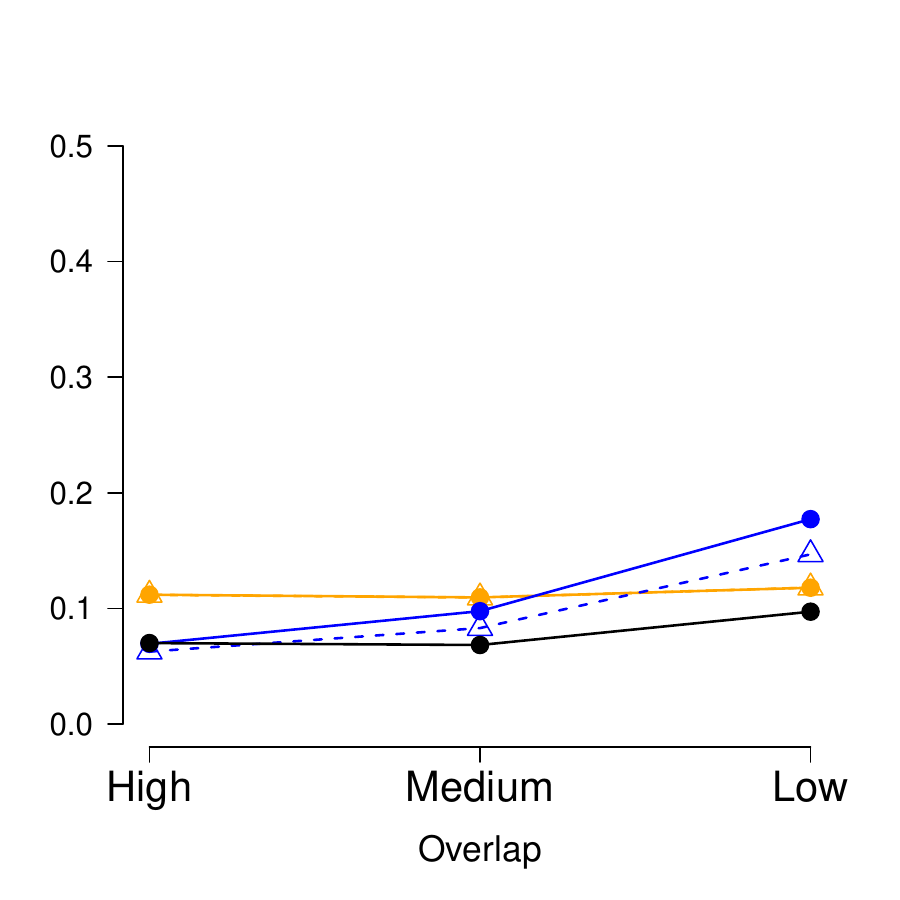}
    \subcaption{$R^2=0.90$, $n=1000$}
    \end{center}
    \end{subfigure}

    \vspace{-0.35in}
    
    \begin{center}
    \includegraphics[scale=.40]{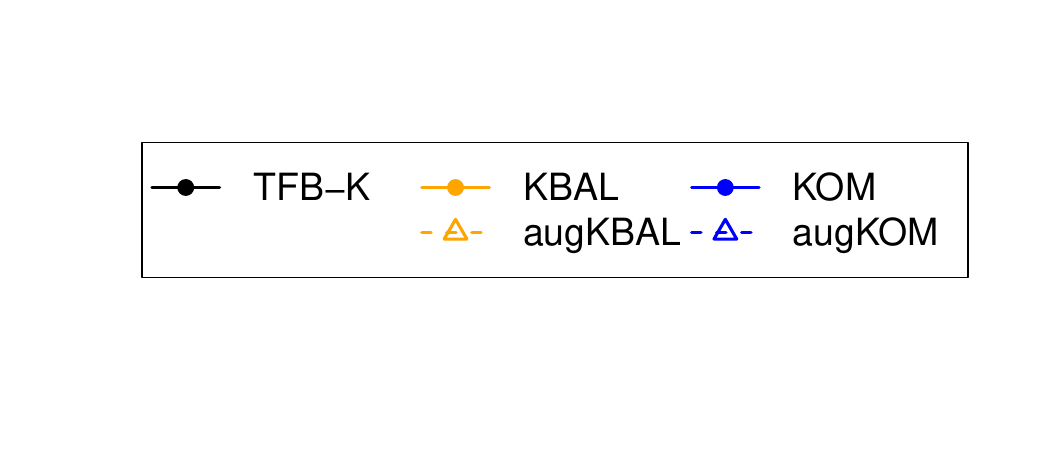}
    \end{center}

    \vspace{-0.35in}
    
    \subcaption*{\textit{Note:} Results across 200 draws from each setting. High Overlap corresponds to $\psi=2$, Medium Overlap corresponds to $\psi=4$, and Low Overlap corresponds to $\psi=8$ in (\ref{eq:dgp1_extend_ps}).}

     \end{figure}

\begin{figure}
    \caption{Bias in Extended DGP 1}\label{fig.app.dgp1.bias}
        
    \vspace{-0.35in}
    
    \begin{subfigure}{.45\textwidth}
    \begin{center}
    \includegraphics[scale=.375]{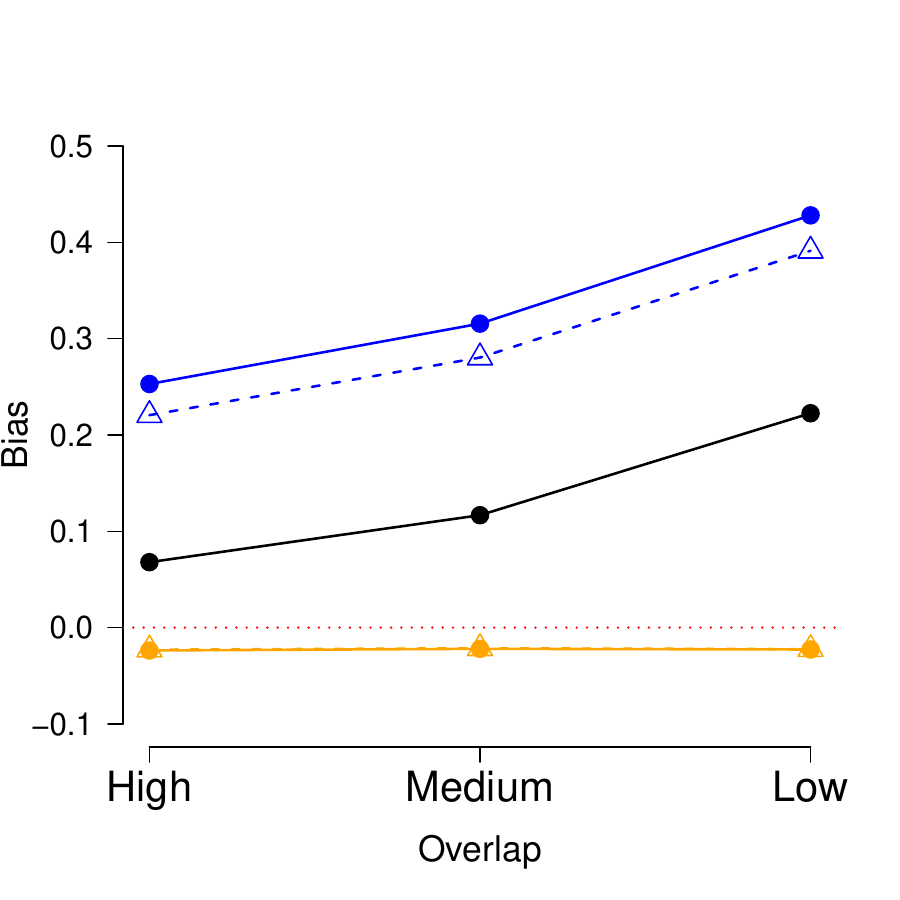}
    \subcaption{$R^2=0.30$, $n=500$}
    \end{center}
    \end{subfigure} 
    \begin{subfigure}{.45\textwidth}
    \begin{center}
    \includegraphics[scale=.375]{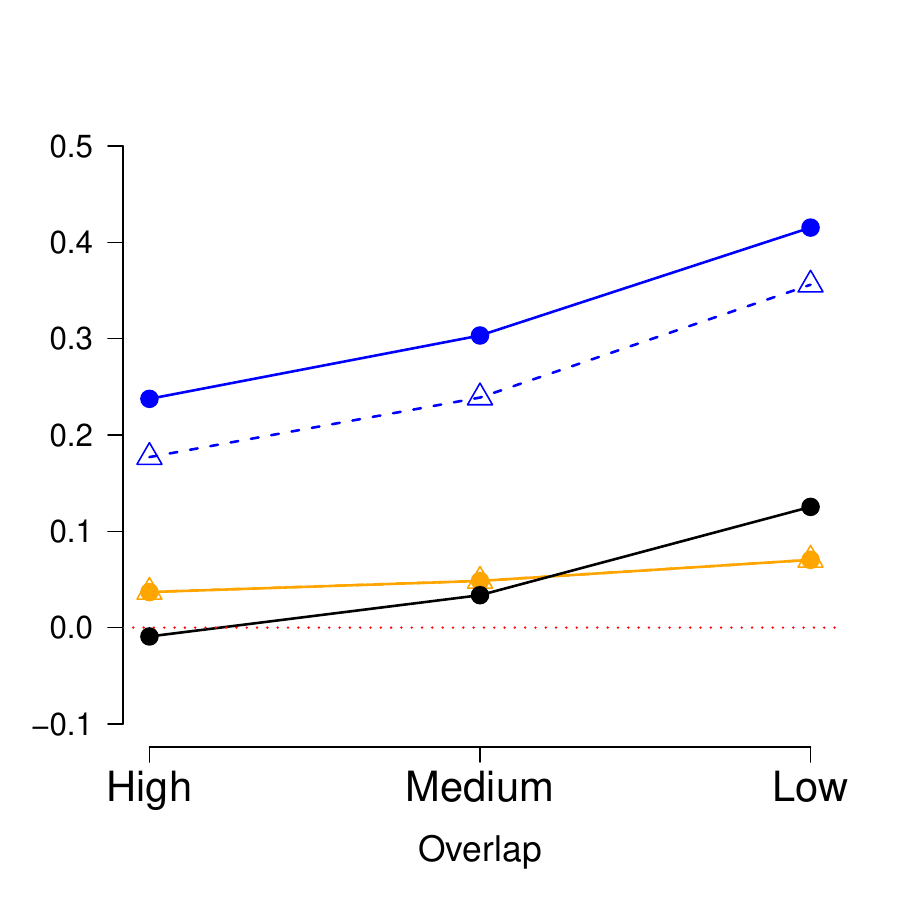}
    \subcaption{$R^2=0.30$, $n=1000$}
    \end{center}
    \end{subfigure}

    \vspace{-0.20in}

    \begin{subfigure}{.45\textwidth}
    \begin{center}
    \includegraphics[scale=.375]{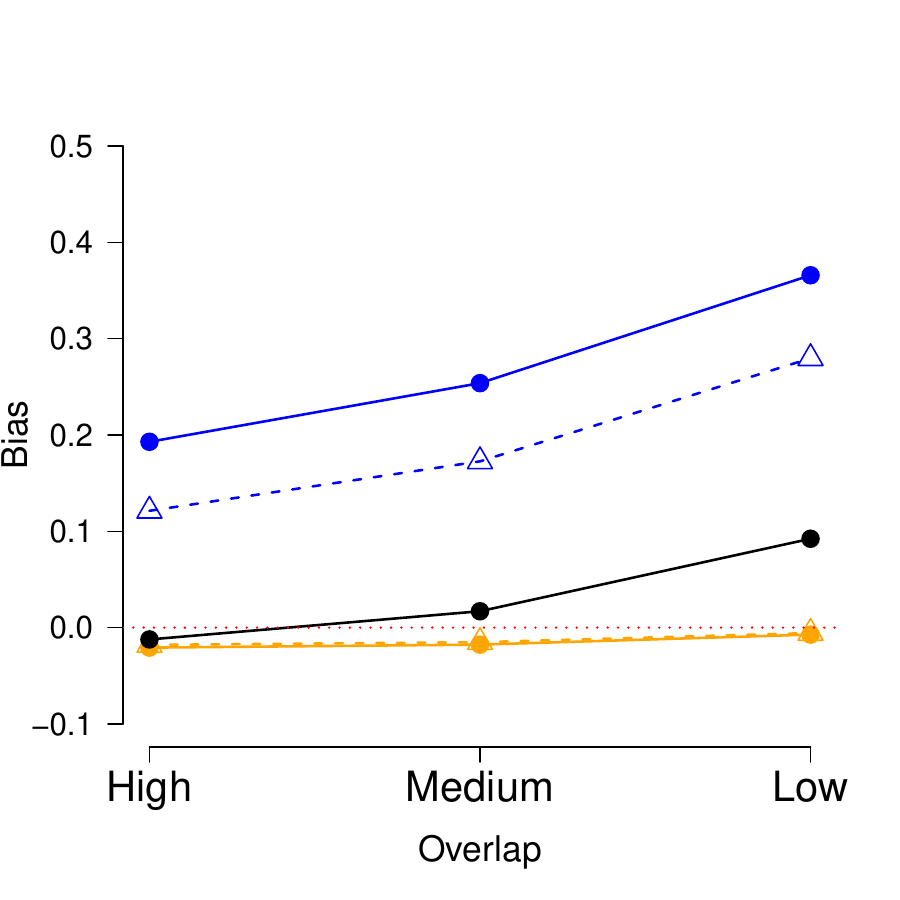}
    \subcaption{$R^2 = 0.60$, $n=500$}
    \end{center}
    \end{subfigure} 
    \begin{subfigure}{.45\textwidth}
    \begin{center}
    \includegraphics[scale=.375]{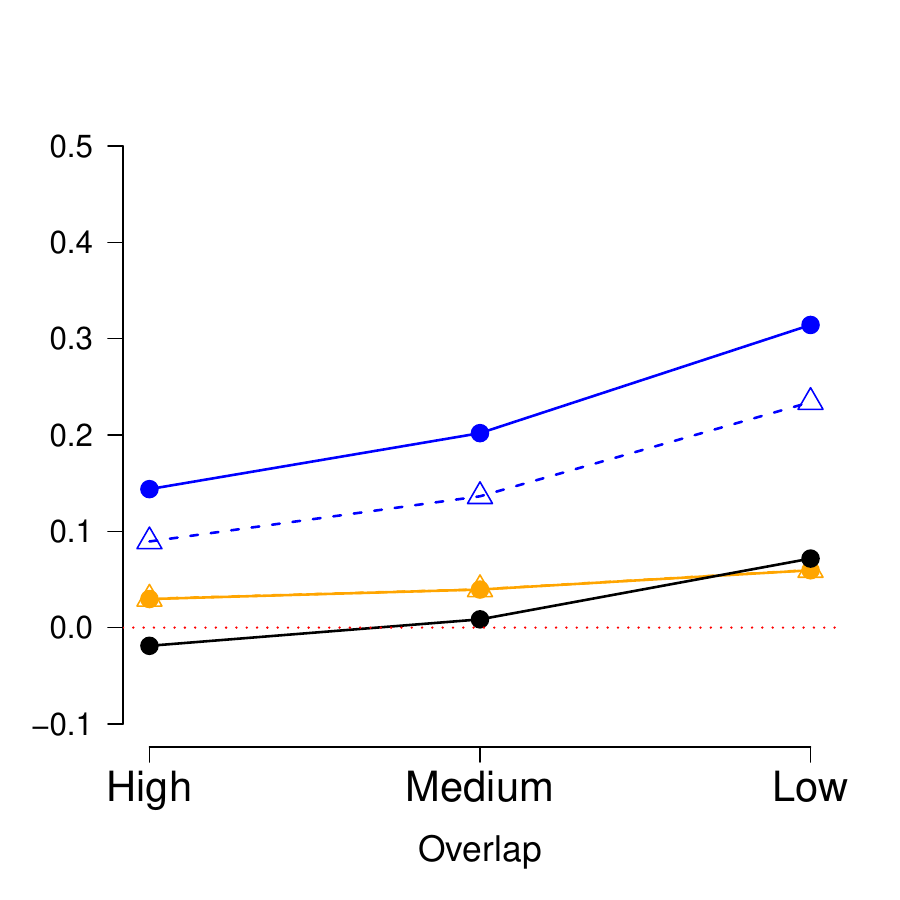}
    \subcaption{$R^2 = 0.60$, $n=1000$}
    \end{center}
    \end{subfigure}

    \vspace{-0.20in}

    \begin{subfigure}{.45\textwidth}
    \begin{center}
    \includegraphics[scale=.375]{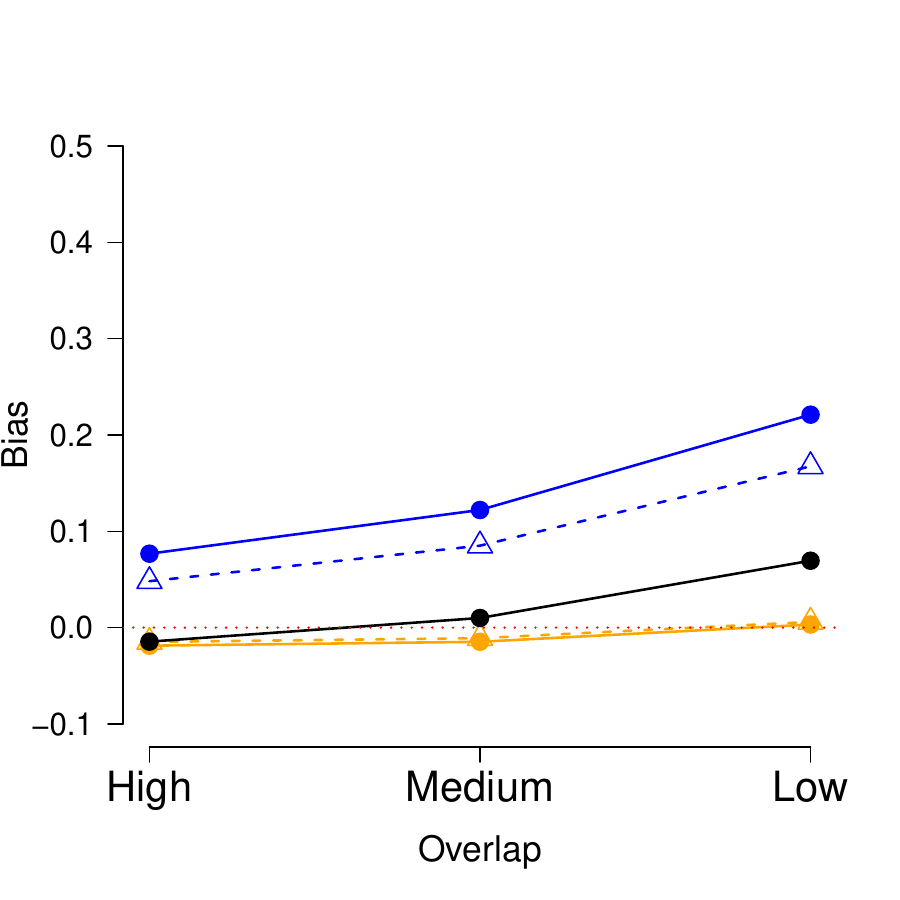}
    \subcaption{$R^2 = 0.90$, $n=500$}
    \end{center}
    \end{subfigure} 
    \begin{subfigure}{.45\textwidth}
    \begin{center}
    \includegraphics[scale=.375]{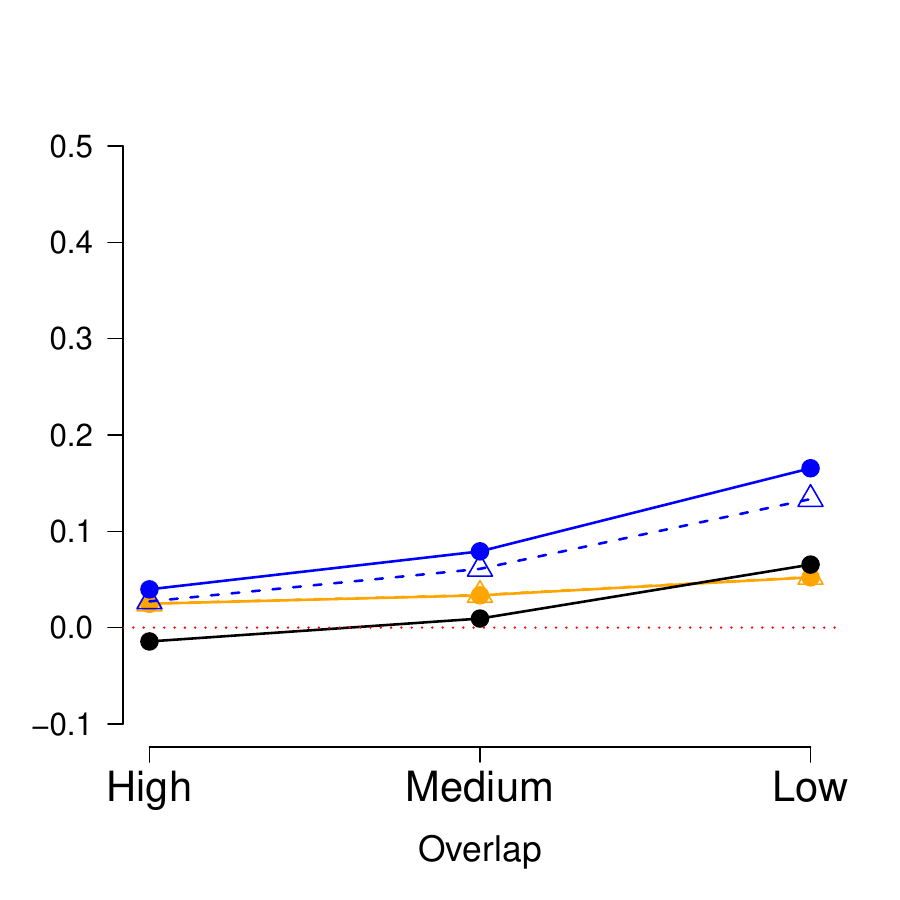}
    \subcaption{$R^2 = 0.90$, $n=1000$}
    \end{center}
    \end{subfigure}

    \vspace{-0.35in}
    
    \begin{center}
    \includegraphics[scale=.40]{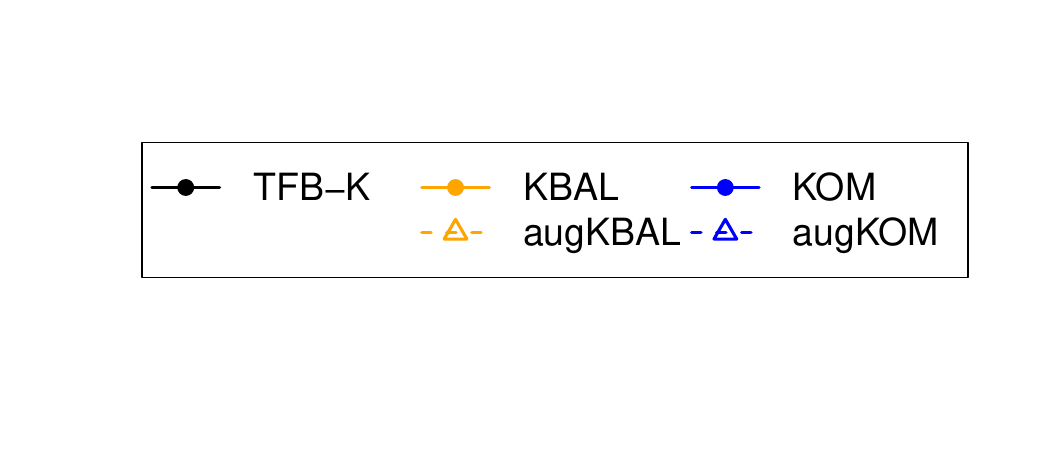}
    \end{center}

    \vspace{-0.35in}
    
    \subcaption*{\textit{Note:} Results across 200 draws from each setting. High Overlap corresponds to $\psi=2$, Medium Overlap corresponds to $\psi=4$, and Low Overlap corresponds to $\psi=8$ in (\ref{eq:dgp1_extend_ps}). The red dotted line indicates zero bias. }

     \end{figure}

\begin{figure}
    \caption{Median Imbalance in $Z_i^{(1)}$ in Extended DGP 1}\label{fig.app.dgp1.imbalance}
        
    \vspace{-0.35in}
    
    \begin{subfigure}{.45\textwidth}
    \begin{center}
    \includegraphics[scale=.375]{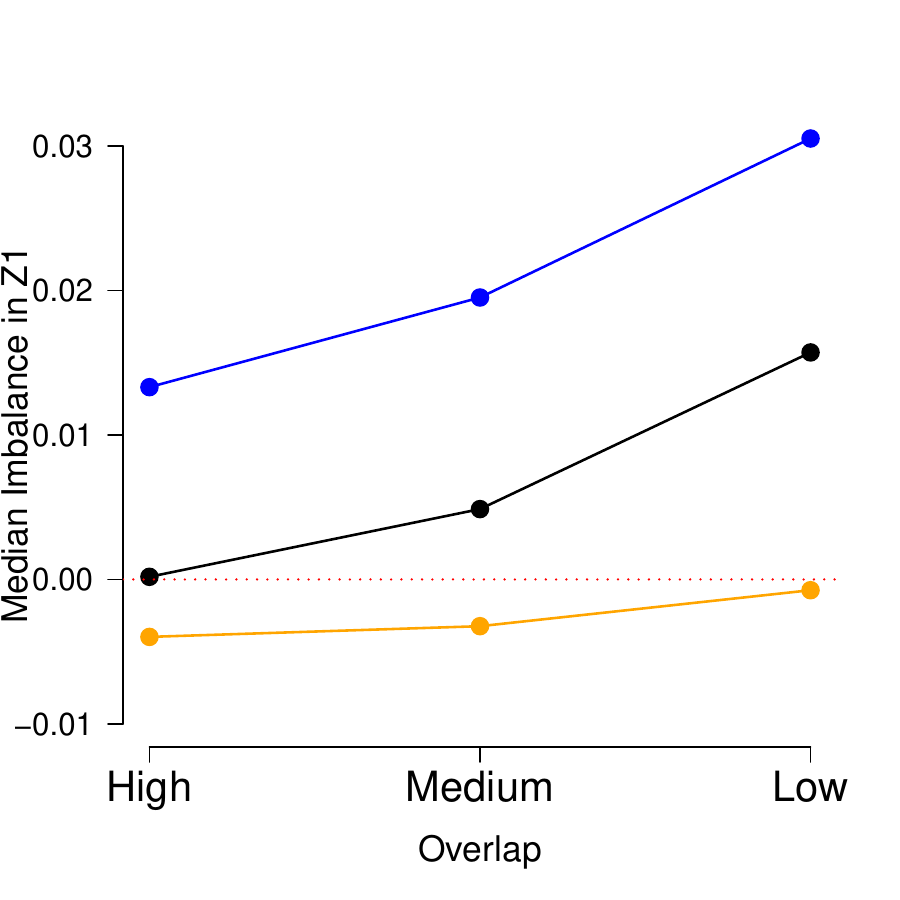}
    \subcaption{$R^2=0.30$, $n=500$}
    \end{center}
    \end{subfigure} 
    \begin{subfigure}{.45\textwidth}
    \begin{center}
    \includegraphics[scale=.375]{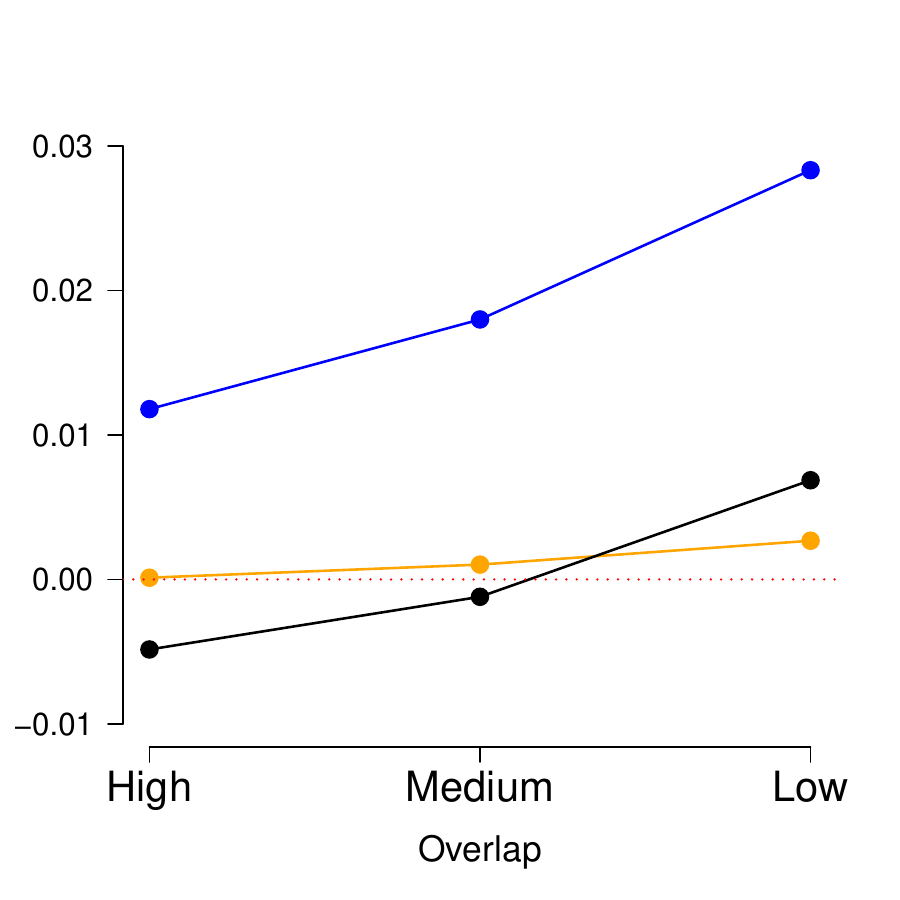}
    \subcaption{$R^2=0.30$, $n=1000$}
    \end{center}
    \end{subfigure}

    \vspace{-0.20in}

    \begin{subfigure}{.45\textwidth}
    \begin{center}
    \includegraphics[scale=.375]{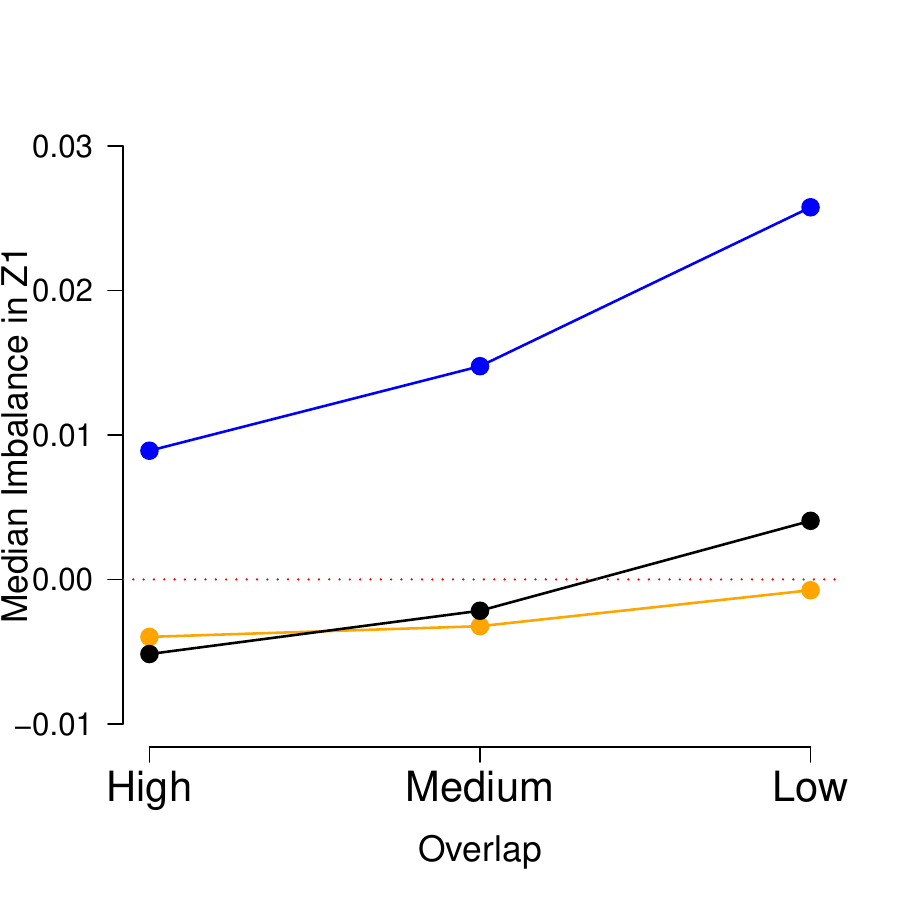}
    \subcaption{$R^2 = 0.60$, $n=500$}
    \end{center}
    \end{subfigure} 
    \begin{subfigure}{.45\textwidth}
    \begin{center}
    \includegraphics[scale=.375]{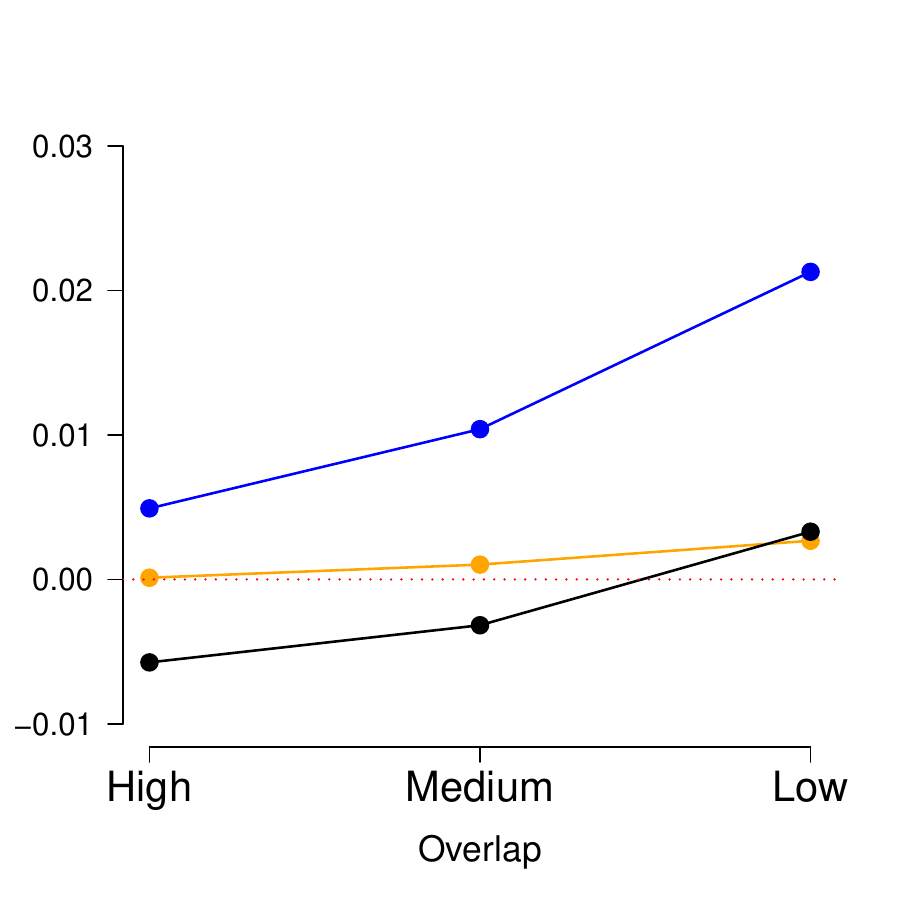}
    \subcaption{$R^2 = 0.60$, $n=1000$}
    \end{center}
    \end{subfigure}

    \vspace{-0.20in}

    \begin{subfigure}{.45\textwidth}
    \begin{center}
    \includegraphics[scale=.375]{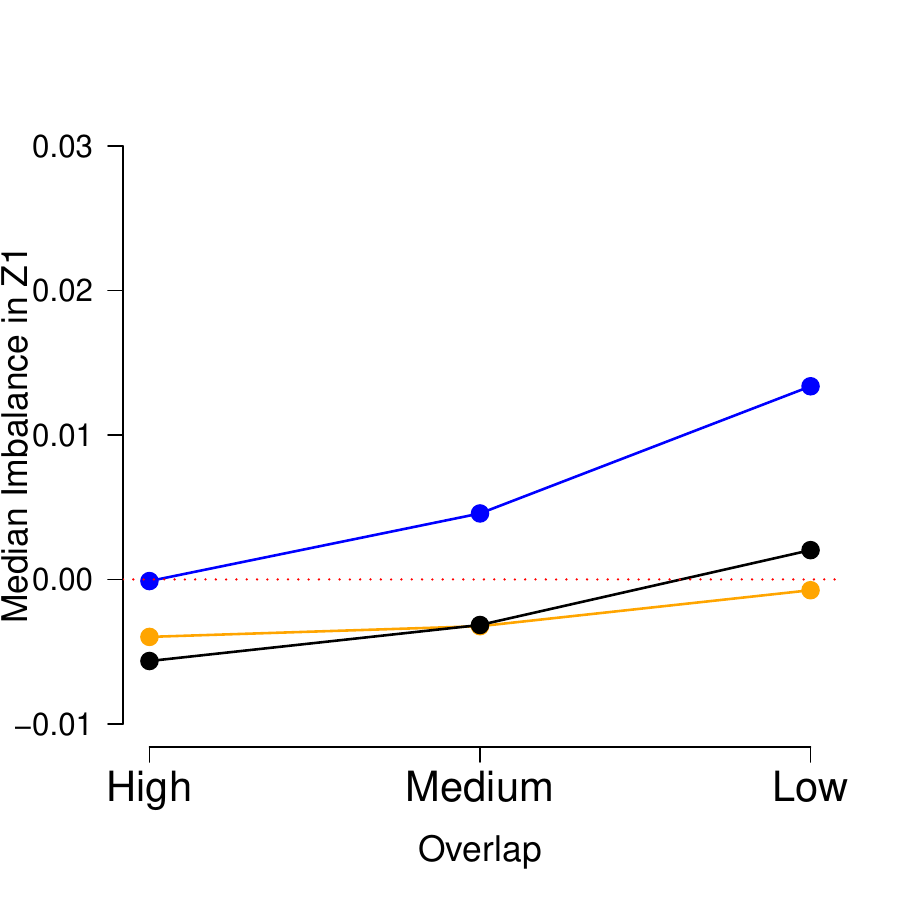}
    \subcaption{$R^2 = 0.90$, $n=500$}
    \end{center}
    \end{subfigure} 
    \begin{subfigure}{.45\textwidth}
    \begin{center}
    \includegraphics[scale=.375]{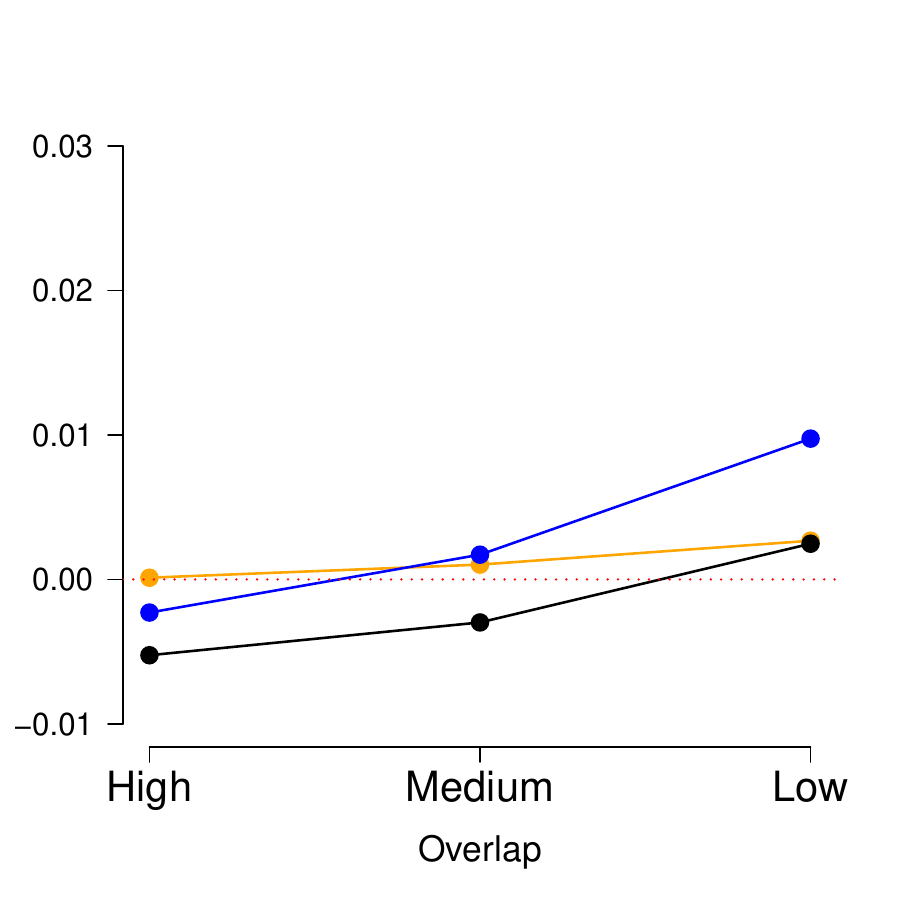}
    \subcaption{$R^2 = 0.90$, $n=1000$}
    \end{center}
    \end{subfigure}

    \vspace{-0.35in}
    
    \begin{center}
    \includegraphics[scale=.40]{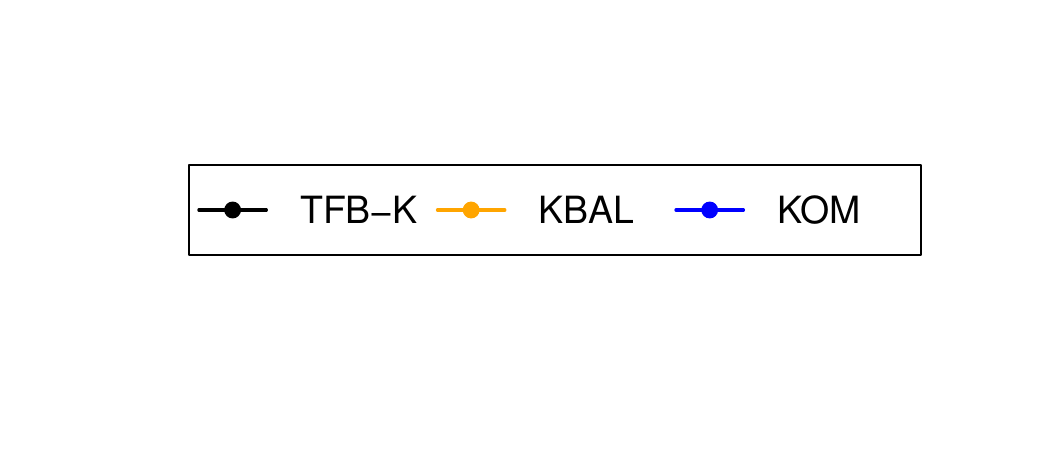}
    \end{center}

    \vspace{-0.35in}
    
    \subcaption*{\textit{Note:} Results across 200 draws from each setting. High Overlap corresponds to $\psi=2$, Medium Overlap corresponds to $\psi=4$, and Low Overlap corresponds to $\psi=8$ in (\ref{eq:dgp1_extend_ps}). The red dotted line indicates zero mean imbalance. }

     \end{figure}

\begin{figure}
    \caption{Coverage of 95\% Confidence Intervals for TFB-K in Extended DGP 1}\label{fig.app.dgp1.coverage}
        
    \vspace{-0.35in}
    
    \begin{subfigure}{.45\textwidth}
    \begin{center}
    \includegraphics[scale=.375]{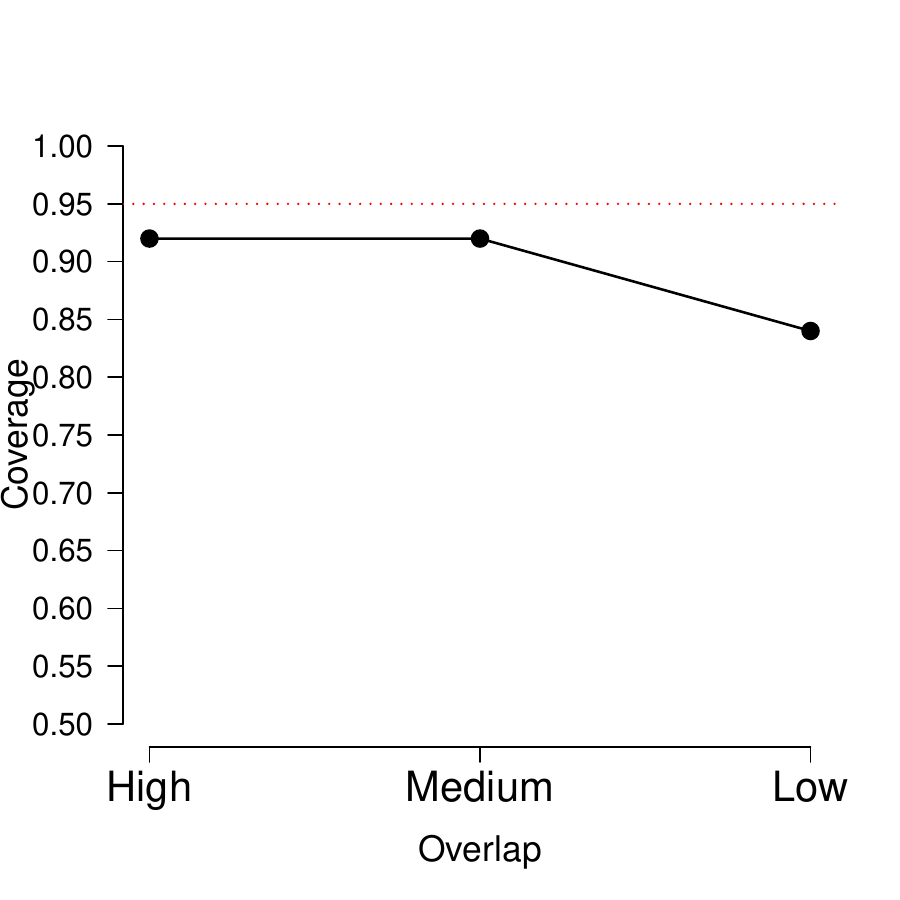}
    \subcaption{$R^2=0.30$, $n=500$}
    \end{center}
    \end{subfigure} 
    \begin{subfigure}{.45\textwidth}
    \begin{center}
    \includegraphics[scale=.375]{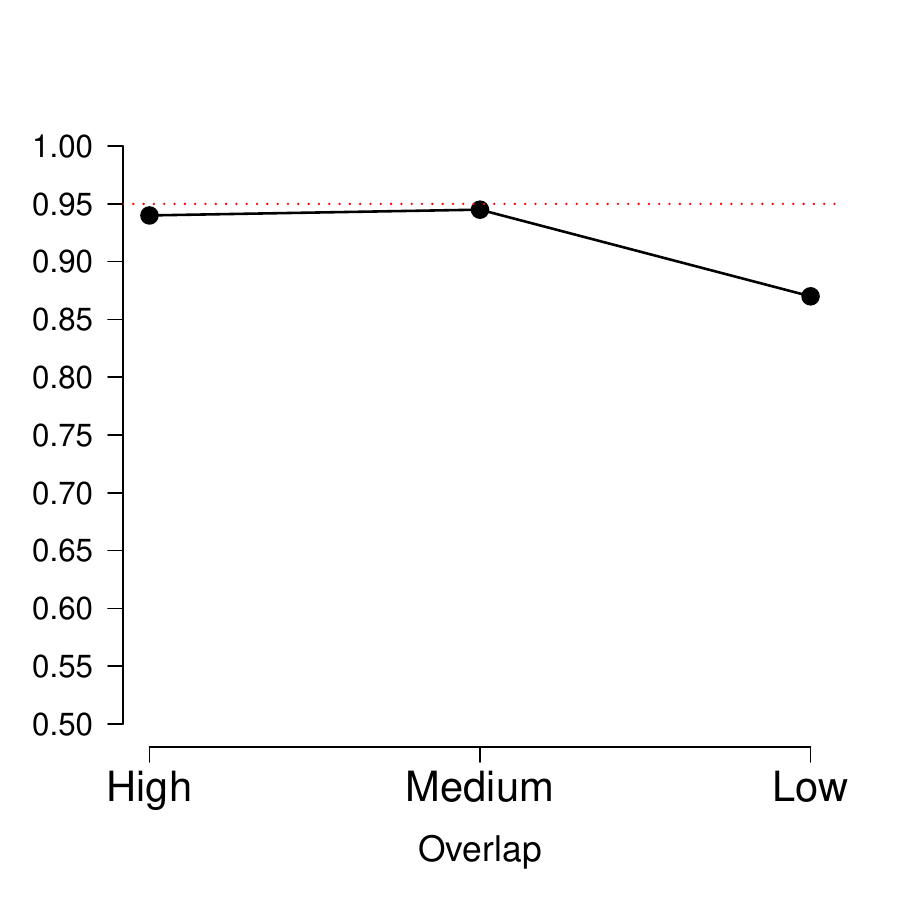}
    \subcaption{$R^2=0.30$, $n=1000$}
    \end{center}
    \end{subfigure}

    \vspace{-0.20in}

    \begin{subfigure}{.45\textwidth}
    \begin{center}
    \includegraphics[scale=.375]{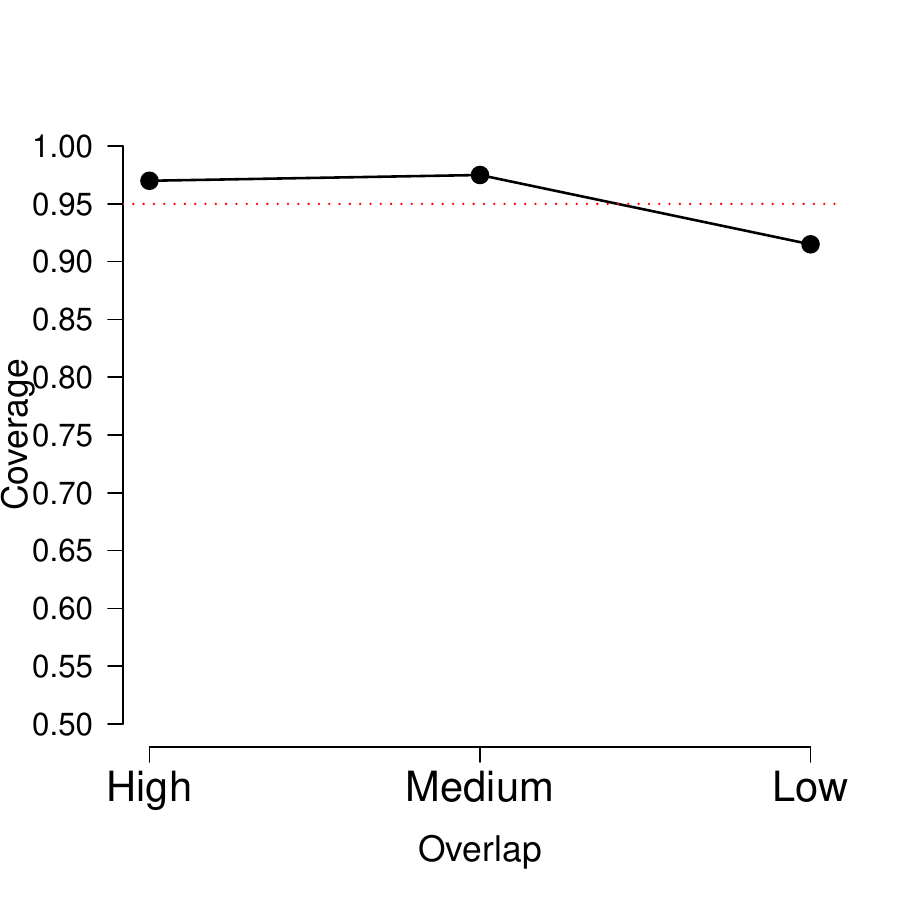}
    \subcaption{$R^2 = 0.60$, $n=500$}
    \end{center}
    \end{subfigure} 
    \begin{subfigure}{.45\textwidth}
    \begin{center}
    \includegraphics[scale=.375]{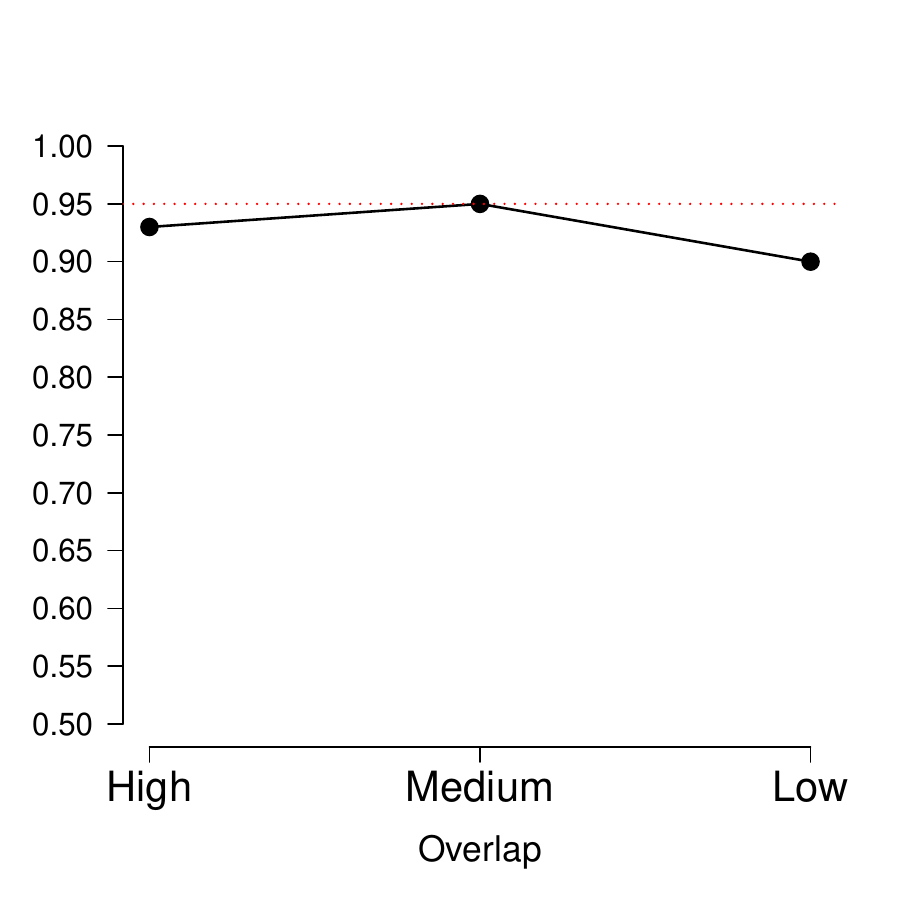}
    \subcaption{$R^2 = 0.60$, $n=1000$}
    \end{center}
    \end{subfigure}

    \vspace{-0.20in}

    \begin{subfigure}{.45\textwidth}
    \begin{center}
    \includegraphics[scale=.375]{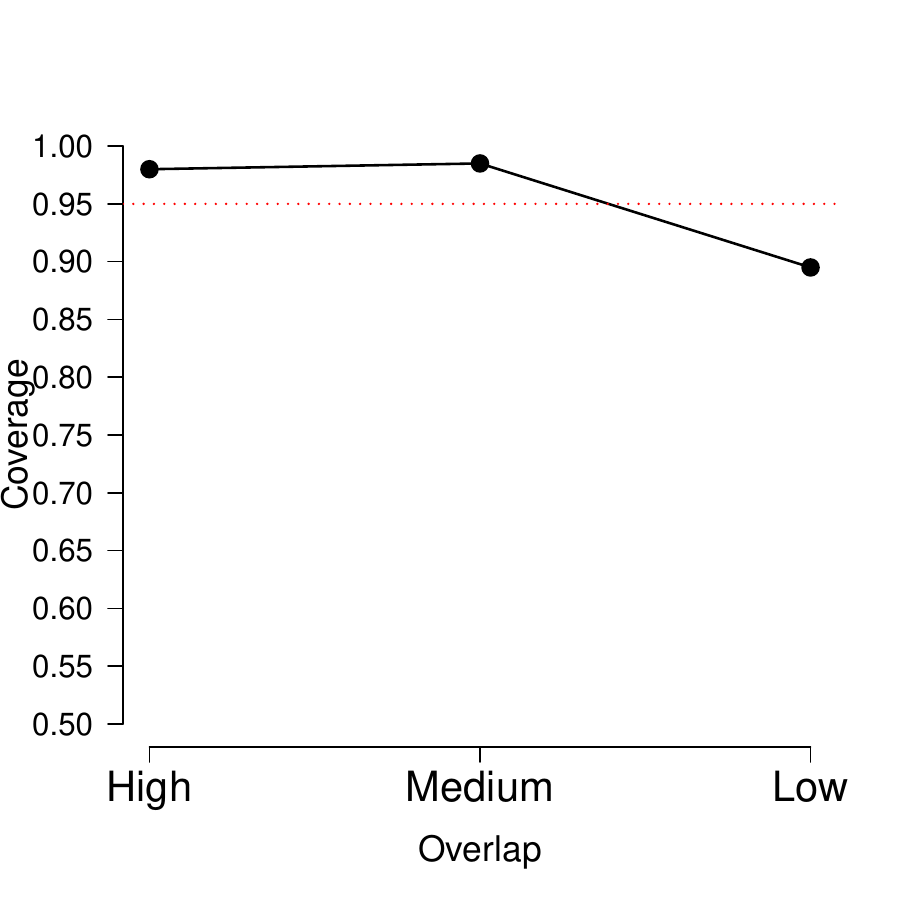}
    \subcaption{$R^2 = 0.90$, $n=500$}
    \end{center}
    \end{subfigure} 
    \begin{subfigure}{.45\textwidth}
    \begin{center}
    \includegraphics[scale=.375]{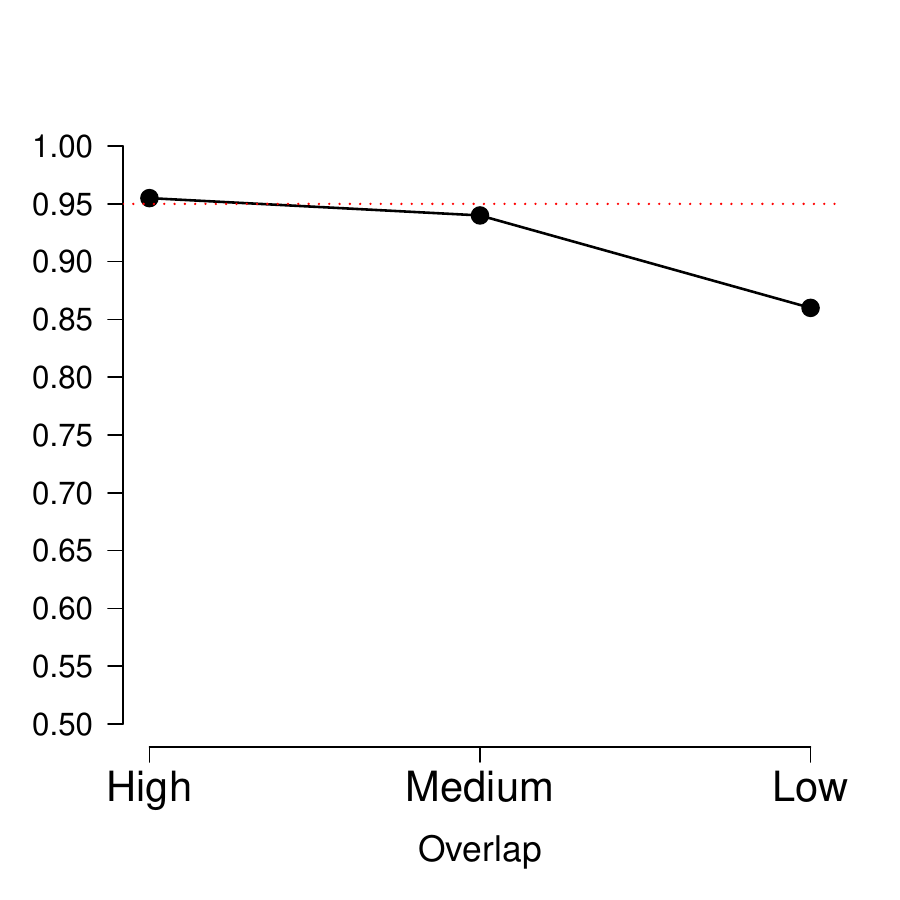}
    \subcaption{$R^2 = 0.90$, $n=1000$}
    \end{center}
    \end{subfigure}


    \vspace{0.2in}
    
    \subcaption*{\textit{Note:} Results across 200 draws from each setting. Confidence intervals are found by the proposal in Section~\ref{subsec:tfb.asymptotics}. Overlap corresponds to $\psi=2$, Medium Overlap corresponds to $\psi=4$, and Low Overlap corresponds to $\psi=8$ in (\ref{eq:dgp1_extend_ps}). The red dotted line indicates the target coverage rate of 0.95. }

     \end{figure}

As in Section~\ref{sec:dgp1}, TFB-K is compared to Kernel Balancing (KBAL; \citealp{hazlett2018kernel}) and Kernel Optimal Matching (KOM; \citealp{kallus2020generalized}). All methods are implemented in the same way as in Section~\ref{subsec:tfb.demonstrations}. Also included in the comparison are the augmented forms of KBAL (augKBAL) and KOM (augKOM), both using KRLS as their regression function. Figure~\ref{fig.app.dgp1.rmse} reports the RMSE of each estimator, Figure~\ref{fig.app.dgp1.bias} reports the bias, Figure~\ref{fig.app.dgp1.imbalance} reports the median imbalance in $Z^{(1)}$, and Figure~\ref{fig.app.dgp1.coverage} reports the coverage rate of 95\% confidence intervals for TFB found through the proposal in Appendix~\ref{subsec:tfb.asymptotics} across the 200 iterations from each setting.

Starting with RMSE in Figure~\ref{fig.app.dgp1.rmse}, as expected, the RMSE for each method tends to increase as overlap in $Z^{(1)}$ decreases (i.e., $\psi$ increases), and RMSE tends to decrease as $R^2$ and $n$ increase. The exceptions are KBAL and augKBAL, whose RMSEs are fairly consistent as $n$ and overlap change. In the vast majority of settings, TFB-K has the lowest RMSE. augKOM has the second lowest RMSE in the large majority of settings.

Figure~\ref{fig.app.dgp1.bias} shows that TFB-K does not always have the least bias in every setting --- KBAL tends to have the least bias when $n=500$ or overlap is low. However, TFB-K does have either the lowest or second lowest bias across all of the settings. Note also that the settings in which TFB-K shows little to no bias --- when there is high or medium overlap combined with $R^2 \geq 0.60$, or there is high overlap combined with $R^2=0.30$ and $n=1000$ --- are all settings in which TFB-K's median imbalance for $Z^{(1)}$ is negative (see Figure~\ref{fig.app.dgp1.imbalance}), i.e., the opposite sign of the original imbalance. Recall that in the original DGP 1 this allowed TFB-K to be effectively unbiased. That this typically occurs when the $R^2$ is high and there is high or medium overlap makes sense. When the $R^2$ is high, $\hat{f}_0$ is more predictive of the outcome and thus has lower variance, meaning TFB-K can more confidently leave imbalance in $Z^{(1)}$. Further, when there is high or medium overlap, leaving negative imbalance in $Z^{(1)}$ does not require outlandishly high variance weights. Finally, Figure~\ref{fig.app.dgp1.coverage} shows that the proposed confidence intervals for TFB-K from Appendix~\ref{subsec:tfb.asymptotics} reach, or are very near, the target coverage rate of 0.95 in the same settings in which TFB-K shows little bias (i.e., when when there is high or medium overlap combined with $R^2 \geq 0.60$, and high overlap with $n=1000$ and $R^2 = 0.30$).

In summary, TFB-K performs very well even after varying important aspects of DGP 1 --- it consistently has the lowest RMSE among the comparison methods tried. It also shows little bias when the model $R^2$ is higher, and there is acceptable overlap in key confounders. When the model $R^2$ and the overlap are lower, TFB-K appears to prioritize lower variance weights, leaving imbalance in confounders that induces bias. However, even in these settings, TFB-K often still has the lowest RMSE. Finally, the proposed confidence intervals from Section~\ref{subsec:tfb.asymptotics} perform well when TFB-K shows little bias.


\clearpage

\newpage

\subsection{Extension to Demonstration 2}\label{app:acic_extension}

In this appendix section, we provide a deeper inspection into the differences between TFB-BART, TFB-K, and the g-computation estimators applied to the ACIC DGPs in Section~\ref{sec:acic}. Specifically, we model the differences between the TFB estimators and the g-computation estimators in each of the 77 DGPs as a function of the DGP ``knobs" (see \citealp{dorie2019automated}), for example levels of treatment effect heterogeneity and degree of overlap. 

To construct the 77 DGPs, \cite{dorie2019automated} varied six knobs when simulating the datasets:
\begin{enumerate}
    \item \textbf{Treatment Model:} Functional form of the assignment mechanism. Grouped into ``Polynomial" (39 DGPs), ``Linear" (6), and ``Step(-Function)" (32). We use ``Polynomial" as the reference group in our models. 
    \item \textbf{Treatment Percent:} Percentage of observations that are treated. Grouped into ``Low" (40) and ``High" (37). We use ``Low" as the reference group in our models.
    \item \textbf{Overlap:} Degree of overlap in the covariates. Grouped into ``Full" (32) and ``Penalize" (45). We use ``Full" as the reference group in our models. 
    \item \textbf{Response Model:} Functional form of the outcome variable, as a function of the covariates (and the treatment). Grouped into ``Exponential" (39), ``Linear" (6), and ``Step" (32). We use ``Exponential" as the reference group in our models.
    \item \textbf{Alignment:} Correlation between the (logit of the) propensity score and the outcome. This knob is intended to control the alignment between the functions of the covariates that contribute to propensity score, and the functions that contribute to the outcome. Grouped into ``High" (39), ``Low" (36), and  ``None" (2). We use ``High" as the reference group in our models. 
    \item \textbf{Effect Heterogeneity:} Level of heterogeneity in the treatment effects (conditional on $X$). Grouped into  ``High" (43), ``Low" (32), and ``None" (2). We use ``Low" as the reference group in our models.
\end{enumerate}
We summarize these knobs above, and point readers to \cite{dorie2019automated} for more detail. See also Table 2 in Appendix A.1 of \cite{dorie2019automated} for the knob groups, and the knobs that are used in each of the 77 DGPs. 

\subsubsection{TFB vs. g-computation}\label{app:acic_extension_tfbvgcomp}

We first consider the differences between each TFB estimator and its analogous g-computation estimator. We focus on the standardized RMSEs in each DGP from the TFB and the g-computation estimators. We calculate the difference between the RMSEs of the TFB and g-computation estimators in each DGP: $\text{RMSE}_{\text{g-comp}} - \text{RMSE}_{\text{TFB}}$, where positive values indicate that TFB is outperforming the g-computation estimator (i.e., has lower RMSE), and negative values indicate that the g-computation estimator is outperforming TFB. We then model these differences as a function of dummy variables that indicate the DGP knob groups described above in a linear OLS regression. We calculate robust standard errors (\citealp{white1980heteroskedasticity}) for these models. Table~\ref{tab.acic.reg.bart} reports the coefficients from the model that compares TFB-BART and the BART g-computation estimators.  Table~\ref{tab.acic.reg.krls} reports the coefficients from the model that compares TFB-K and the KRLS g-computation estimators.

\begin{table}[!h]
\vspace{0.05in}
\caption{Coefficients for differences in standardized RMSE between TFB-BART and g-computation BART estimators in the 77 ACIC DGPs, modeled by DGP knob}\label{tab.acic.reg.bart}

\vspace{-0.15in}

\normalsize
\begin{center}

     \begin{tabular}{ m{5cm} | c c c c }
     \hline
     \hline
     & \textbf{Coefficient} & \textbf{SE} & \textbf{Z} & \textbf{p-value} \\
     \hline
     Intercept & 0.0016** & 0.0004 & 3.6373 & 0.0003 \\
     \hline
     Treatment Model: Linear & 0.0012 & 0.0011 & 1.1766 & 0.2393 \\
     \hline
     Treatment Model: Step & 0.0010** & 0.0003 & 3.0744 & 0.0021 \\
     \hline
     Treatment Percent: High & 0.0012** & 0.0003 & 3.7354 & 0.0002 \\
     \hline
     Overlap: Penalize & 0.0004 & 0.0003 & 1.2993 & 0.1938 \\
     \hline
     Response Model: Linear & -0.0003 & 0.0012 & -0.2163 & 0.8287 \\
     \hline
     Response Model: Step & -0.0006** & 0.0003 & -1.9873 & 0.0469 \\
     \hline
     Alignment: Low & -0.0009** & 0.0003 & -2.6603 & 0.0078 \\
     \hline
     Alignment: None & -0.0039** & 0.0009 & -4.3617 & 0.0000 \\
     \hline
     Effect Heterogeneity: None & -0.0016 & 0.0014 & -1.1621 & 0.2452 \\
     \hline
     Effect Heterogeneity: High & 0.0022** & 0.0003 & 6.7541 & 0.0000 \\
     \hline
     \end{tabular}

    \vspace{0.1in}

\subcaption*{\textit{Note:} Differences in standardized RMSEs between TFB-BART and g-computation BART estimators ($\text{RMSE}_{\text{BART}} - \text{RMSE}_{\text{TFB-BART}}$) in the 77 ACIC DGPs are regressed by dummy variables that indicate the DGP knobs in a linear OLS model. Because the Intercept coefficient is positive, that means that in the reference group DGP, TFB-BART has lower RMSE than does BART. A positive value for a (non-intercept) coefficient indicates that TFB-BART is (on average) improving even more over BART for that knob. A negative value for a coefficient indicates that the gap between TFB-BART and BART is closing. For the Treatment Model knob, the reference group is ``Polynomial". For the Treatment Percent knob, the reference group is ``Low". For the Overlap knob, the reference group is ``Full". For the Response Model knob, the reference group is ``Exponential". For the Alignment knob, the reference group is ``High". For the Effect Heterogeneity knob, the reference group is ``Low". Robust standard errors are used (\citealp{white1980heteroskedasticity}). Starred (**) coefficients indicate statistical significance at the 0.05 level. }

\end{center}

\end{table}

\begin{table}[!h]
\vspace{0.05in}
\caption{Coefficients for differences in standardized RMSE between TFB-K and g-computation KRLS estimators in the 77 ACIC DGPs, modeled by DGP knob}\label{tab.acic.reg.krls}

\vspace{-0.15in}

\normalsize
\begin{center}

     \begin{tabular}{ m{5cm} | c c c c }
     \hline
     \hline
     & \textbf{Coefficient} & \textbf{SE} & \textbf{Z} & \textbf{p-value} \\
     \hline
     Intercept & 0.0277** & 0.0013 & 20.7578 & 0.0000 \\
     \hline
     Treatment Model: Linear & -0.0109** & 0.0045 & -2.4163 & 0.0157 \\
     \hline
     Treatment Model: Step & -0.0057** & 0.0014 & -4.0722 & 0.0000 \\
     \hline
     Treatment Percent: High & -0.0106** & 0.0013 & -8.3998 & 0.0000 \\
     \hline
     Overlap: Penalize & 0.0004 & 0.0014 & 0.3249 & 0.7452 \\
     \hline
     Response Model: Linear & 0.0034 & 0.0047 & 0.7189 & 0.4722 \\
     \hline
     Response Model: Step & 0.0000 & 0.0014 & 0.0268 & 0.9786 \\
     \hline
     Alignment: Low & -0.0052** & 0.0013 & -4.0662 & 0.0000 \\
     \hline
     Alignment: None & -0.0268** & 0.0051 & -5.2447 & 0.0000 \\
     \hline
     Effect Heterogeneity: None & -0.0040 & 0.0130 & -0.3055 & 0.7600 \\
     \hline
     Effect Heterogeneity: High & 0.0111** & 0.0014 & 7.8551 & 0.0000 \\
     \hline
     \end{tabular}

    \vspace{0.1in}

\subcaption*{\textit{Note:} Differences in standardized RMSEs between TFB-K and g-computation KRLS estimators ($\text{RMSE}_{\text{KRLS}} - \text{RMSE}_{\text{TFB-K}}$) in the 77 ACIC DGPs are regressed by dummy variables that indicate the DGP knobs in a linear OLS model. Because the Intercept coefficient is positive, that means that in the reference group DGP, TFB-K has lower RMSE than does KRLS. A positive value for a (non-intercept) coefficient indicates that TFB-K is (on average) improving even more over KRLS for that knob. A negative value for a coefficient indicates that the gap between TFB-K and KRLS is closing. For the Treatment Model knob, the reference group is ``Polynomial". For the Treatment Percent knob, the reference group is ``Low". For the Overlap knob, the reference group is ``Full". For the Response Model knob, the reference group is ``Exponential". For the Alignment knob, the reference group is ``High". For the Effect Heterogeneity knob, the reference group is ``Low". Robust standard errors are used (\citealp{white1980heteroskedasticity}). Starred (**) coefficients indicate statistical significance at the 0.05 level. }

\end{center}

\end{table}

We focus on the similarities between the two forms of TFB. In both Tables~\ref{tab.acic.reg.bart} and~\ref{tab.acic.reg.krls}, the intercept is positive and statistically significant (at the 0.05 level). This indicates that in the reference DGP, TFB clearly outperforms g-computation. The two knobs that have statistically significant differences (over the reference group) in the same direction for both tables are the Alignment and Effect Heterogeneity knobs. Starting with Effect Heterogeneity, both tables show a statistically significant and positive coefficient for the ``High" setting when compared to ``Low" setting (the reference group here). This indicates that TFB tends to outperform g-computation when there are higher levels of treatment effect heterogeneity. 

The negative and statistically significant estimates for the Alignment knob are the most interesting finding. As a reminder, the reference group for this knob is ``High". In both tables, the coefficients for the ``Low" and ``None" settings are negative and statistically significant. For the ``None" setting, this is despite there only being two DGPs (out of 77) with Alignments of ``None". Furthermore, the ``None" coefficients are the largest in effect size of all of the (non-intercept) coefficients in their respective tables. 
These coefficients' being negative indicates that when Alignment is ``High", TFB outperforms g-computation. This is further evidence of TFB's double-robustness (noted in Appendix~\ref{app:tfb.dr}), as when Alignment is ``High", the propensity score $\pi$ is (mostly) a function of the conditional expectation function $f_0$. Because TFB's weights are propensity score weights where $\pi$ is modeled as a function of $\hat{f}_0$ (and $X)$, TFB is implying an (approximately) correctly specified $\pi$.

\subsubsection{TFB-K vs. TFB-BART} 

We now consider the difference between TFB-K and TFB-BART. Similar to Appendix~\ref{app:acic_extension_tfbvgcomp}, we calculate the difference between the RMSEs of TFB-K and TFB-BART in each DGP: $\text{RMSE}_{\text{TFB-K}} - \text{RMSE}_{\text{TFB-BART}}$, where positive values indicate that TFB-BART is outperforming TFB-K (i.e., has lower RMSE), and negative values indicate that the TFB-K is outperforming TFB-BART. We again model these differences as a function of dummy variables that indicate the DGP knob groups described above in a linear OLS regression, and calculate robust standard errors (\citealp{white1980heteroskedasticity}). Table~\ref{tab.acic.tfbkvstfbbart} reports the coefficients from the model that compares TFB-K and TFB-BART. 
\begin{table}[!h]
\vspace{0.05in}
\caption{Coefficients for differences in standardized RMSE between TFB-K and TFB-BART in the 77 ACIC DGPs, modeled by DGP knob}\label{tab.acic.tfbkvstfbbart}

\vspace{-0.15in}

\normalsize
\begin{center}

     \begin{tabular}{ m{5cm} | c c c c }
     \hline
     \hline
     & \textbf{Coefficient} & \textbf{SE} & \textbf{Z} & \textbf{p-value} \\
     \hline
     Intercept & 0.0527** & 0.0040 & 13.2183 & 0.0000 \\
     \hline
     Treatment Model: Linear & -0.0208** & 0.0089 & -2.3416 & 0.0192 \\
     \hline
     Treatment Model: Step & 0.0249** & 0.0030 & 8.4116 & 0.0000 \\
     \hline
     Treatment Percent: High & -0.0069** & 0.0027 & -2.5053 & 0.0122 \\
     \hline
     Overlap: Penalize & -0.0127** & 0.0030 & -4.2858 & 0.0000 \\
     \hline
     Response Model: Linear & 0.0110 & 0.0086 & 1.2771 & 0.2016 \\
     \hline
     Response Model: Step & 0.0074** & 0.0030 & 2.5092 & 0.0121 \\
     \hline
     Alignment: Low & -0.0198** & 0.0028 & -6.9989 & 0.0000 \\
     \hline
     Alignment: None & -0.0481** & 0.0058 & -8.2349 & 0.0000 \\
     \hline
     Effect Heterogeneity: None & 0.0021 & 0.0203 & 0.1031 & 0.9179 \\
     \hline
     Effect Heterogeneity: High & 0.0301** & 0.0030 & 10.1566 & 0.0000 \\
     \hline
     \end{tabular}

    \vspace{0.1in}

\subcaption*{\textit{Note:} Differences in standardized RMSEs between TFB-K and TFB-BART ($\text{RMSE}_{\text{TFB-K}} - \text{RMSE}_{\text{TFB-BART}}$) in the 77 ACIC DGPs are regressed by dummy variables that indicate the DGP knobs in a linear OLS model. Because the Intercept coefficient is positive, that means that in the reference group DGP, TFB-BART has lower RMSE than does TFB-K. A positive value for a (non-intercept) coefficient indicates that TFB-BART is (on average) improving even more over TFB-K for that knob. A negative value for a coefficient indicates that the gap between TFB-K and TFB-BART is closing. For the Treatment Model knob, the reference group is ``Polynomial". For the Treatment Percent knob, the reference group is ``Low". For the Overlap knob, the reference group is ``Full". For the Response Model knob, the reference group is ``Exponential". For the Alignment knob, the reference group is ``High". For the Effect Heterogeneity knob, the reference group is ``Low". Robust standard errors are used (\citealp{white1980heteroskedasticity}). Starred (**) coefficients indicate statistical significance at the 0.05 level. }

\end{center}

\end{table}

In Table~\ref{tab.acic.tfbkvstfbbart}, the intercept is positive and statistically significant (at the 0.05 level). This indicates that in the reference DGP, TFB-BART clearly outperforms TFB-K. Many of the coefficients in Table~\ref{tab.acic.tfbkvstfbbart} are statistically significant, and several are even negative, which would imply a smaller difference between TFB-K's and TFB-BART's RMSEs. However, the magnitude of the intercept estimate is also the largest of all of the coefficient estimates, meaning that the OLS model estimates that there is not any one single setting that, if changed from the reference group, can entirely close the gap between TFB-K and TFB-BART.  

The most intuitive finding is that the ``Treatment Model: Step" and ``Response Model: Step" coefficients are both positive and statistically significant, indicating that TFB-K struggles, in relation to TFB-K, when step functions are introduced into the DGP.  This is to be expected, as BART models non-smooth functions better than does KRLS. Similarly, there is a negative estimate for ``Treatment Model: Linear", meaning when the treatment model is easier to model, TFB-K and TFB-BART perform more similarly. The fact that knobs relating to the treatment model show statistically significant estimates gives evidence to the importance of the implied propensity score that TFB estimates through its choice of regression estimator (see Appendix~\ref{app:tfb.dr}).

\clearpage

\newpage

\subsection{Demonstration 3: Role of $\hat{V}_{\beta_0}$
}\label{sec:dgp2}

In this demonstration, we illustrate the role of $\hat{V}_{\beta_0}$ in allowing TFB to prioritize balance in certain dimensions of $X$. In particular, we show how  estimating $f_0$ with sparsity helps prioritize the dimensions of $X$ that are most influential on $Y$. We apply TFB-L, which estimates $f_0$ with LASSO regression (modified, per \citealp{chatterjee2011bootstrapping}), in a high-dimensional simulated setting.

Consider a DGP, henceforth ``DGP 2", in which $X$ contains three types of variables: (i) ``confounders", $Z$, that are related to both $D$ and $Y$; (ii) ``distractors", $A$, that are related to $D$, and are difficult to balance, but are unrelated to $Y$; and (iii) ``extraneous" variables, $U$, that are independent of both $D$ and $Y$. The outcome and the log odds of treatment are both linear in $Z = [Z^{(1)} \ \cdots \ Z^{(4)}]$, 
	\begin{align}\label{eq:dgp2_d}
		\mathrm{log} \frac{\pi(X)}{1 - \pi(X)} = \frac{1}{5} (Z^{(1)}  + Z^{(2)}  + Z^{(3)}  + Z^{(4)}) \ \ \ \ \ \ \ \ \ \ \ \ \ \ \ \ \ \ \ \ \ \ \ \ \ \ \ \ \ \ \ \ \ \ \ \ 
    \end{align}
    \vspace{-0.4in}
    \begin{align}\label{eq:dgp2_y}
		\ \ \ \ \ \ \ \ Y = 8 Z^{(1)}  + 4 Z^{(2)}  + 2 Z^{(3)}  + 1 Z^{(4)} + \epsilon, \ \epsilon_i\overset{iid}{\sim} N(0, 9.21^2)
	\end{align}
where the  $Z^{(\ell)}$ are equally imbalanced, and have varying levels of influence on $Y$. Furthermore, the ATT is 0, and $\epsilon_i\overset{iid}{\sim} N(0, 9.21^2)$ allows the true $R^2$ to be approximately 0.50. The $Z$, the distractors ($A$), and the extraneous variables ($U$) are then generated as 
	\begin{align}\label{eq:dgp2_covariates}
		Z_i  \overset{iid}{\sim} \mathcal{N} (0, \mathbf{I}_4) \  , \ A_i \ | \ D_i \overset{iid}{\sim} \mathcal{N} ( D_i \vec{\1}_5, \mathbf{I}_5) \ , \ \text{and} \ U_i \overset{iid}{\sim} \mathcal{N} ( 0, \mathbf{I}_{10})
	\end{align}
Note that we do not vary aspects of DGP 2 here for clarity, but we do so in Appendix~\ref{app:dgp2_extension}. We summarize the results from those more extended simulations after inspecting the focused setting laid out above.
    
\begin{figure}[!h]
	\vspace{0.15in}
	\begin{center}
	\caption{Performance of TFB in DGP 2}\label{fig.dgp2}
    \vspace{-.25in}
    \begin{subfigure}{.55\textwidth}
    \begin{center}
    \includegraphics[scale=0.43]{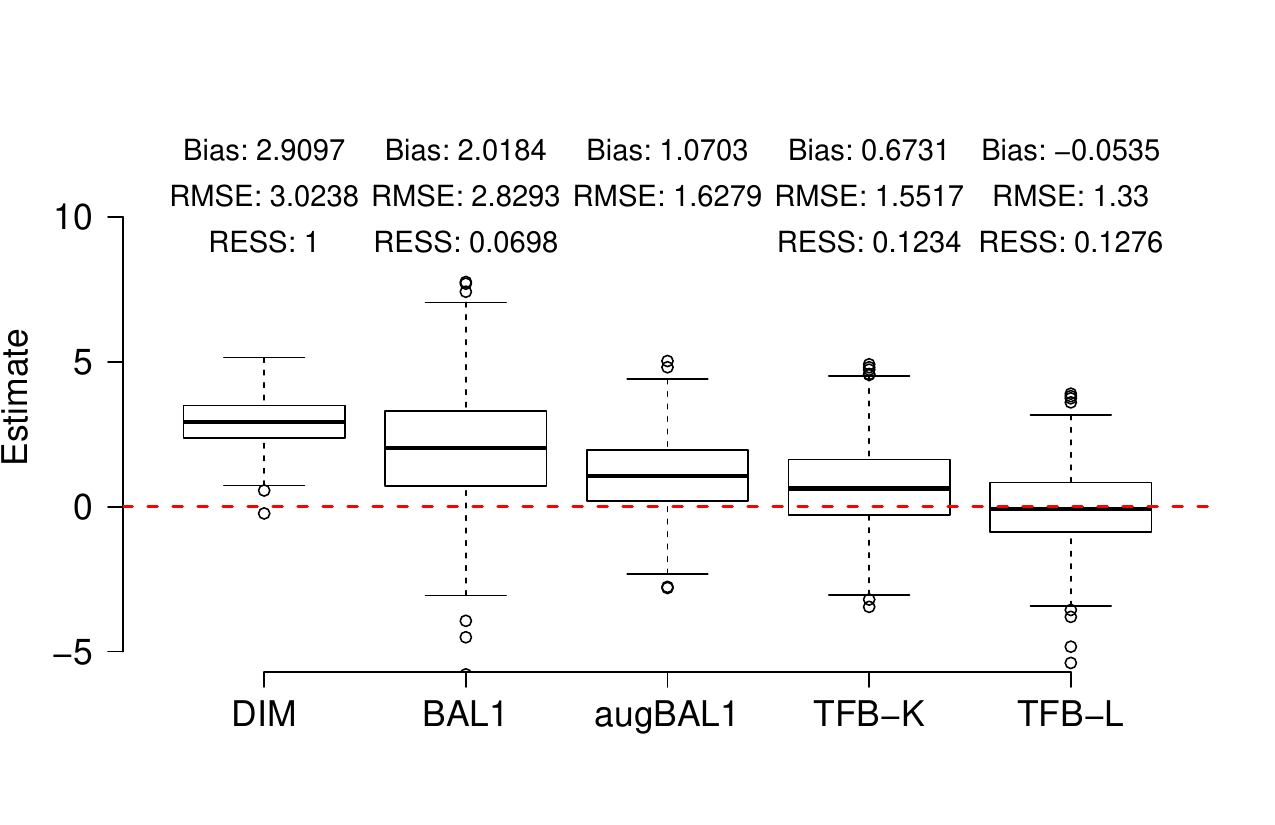} 
    \end{center}
    \vspace{-.25in}
    \subcaption{Bias, RMSE, and RESS}\label{fig.dgp2.bias}
    \end{subfigure}
    \begin{subfigure}{.40\textwidth}
    \begin{center}
    \includegraphics[scale=0.43]{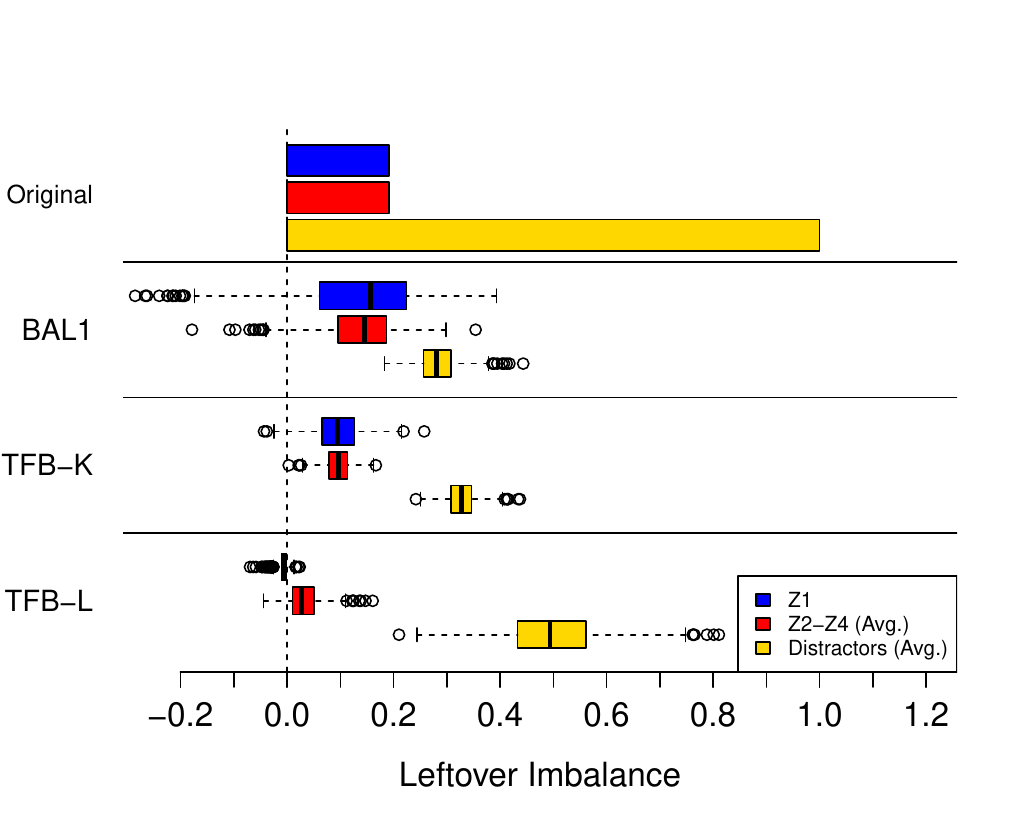} 
    \end{center}
    \vspace{-.25in}
    \subcaption{Leftover imbalance}\label{fig.dgp2.balance}
    \end{subfigure}
    \subcaption*{\textit{Note:} Comparison of TFB-L, TFB-K, \ref{eq:approxbal_w1} where the $\delta^{(\ell)}$ are all equal, the augmented form of BAL1 (augBAL1), and a difference in means (DIM) across 1000 draws from DGP 2 with $n=1000$. Augmented forms of TFB estimators are not included because they are numerically identical to the TFB estimators shown. \textit{(a)} Distributions of the estimates from each method, as well as their bias, RMSE, and RESS. \textit{(b)} Leftover imbalance from each method after weighting, where imbalance in (i) $Z^{(2)}$, $Z^{(3)}$, and $Z^{(4)}$ and (ii) the distractors is averaged at each iteration. }
	\end{center}
	\vspace{-0.25in}
	\end{figure}

Balance in $Z$ is sufficient to unbiasedly estimate the ATT. However, $A$ and $U$, particularly $A$ which is significantly more imbalanced than is $Z$, complicate balancing the entirety of $X$. However, TFB-L performs well by applying LASSO regression. Figure~\ref{fig.dgp2} compares TFB-L with \ref{eq:approxbal_w1} where the $\delta^{(\ell)}$ are all equal, and its augmented form (augABAL1). Further, to make the task of balancing $Z$ more challenging for each method, TFB-L and BAL1 expand on $X$, additionally (and needlessly) seeking balance in its squares and pairwise interactions, a total of 209 variables. Figure~\ref{fig.dgp2} also includes TFB-K to further illustrate the effect of TFB-L's estimating $f_0$ with sparsity.\footnote{BAL1 here is the method from \cite{zubizarreta2015stable} with $\delta$ implied by the \texttt{approx.balance} function in \texttt{R} (from the \texttt{balanceHD} package) with default parameters. We use the \texttt{residualBalance.ate} function (from the same package) to implement augBAL1, with the LASSO as the regression estimator. augBAL1 is a special case of the estimator studied by \cite{athey2018approximate}. TFB-K uses the Gaussian kernel with bandwidth $b = P = 19$.} 
TFB-L shows minimal bias, while BAL1 is more comparable to a difference in means. TFB-L also boasts a lower RMSE than BAL1 by over a factor of two, and a higher RESS. TFB-L outperforms augBAL1 too, though augBAL1 reduces the gap between BAL1 and TFB-L.  
TFB-K also performs well here, and is a marked improvement over (aug)BAL1 in both bias and RMSE, but falls short of TFB-L. 

\begin{figure}[!h]
\vspace{0.15in}
\begin{center}
\caption{Diagonal elements of $\hat{V}_{\beta_0}^{\frac{1}{2}}$ for TFB-L (see Table~\ref{tab:abbreviations}) in DGP 2}\label{fig.dgp2.vdiag}
	\vspace{-0.25in}
    \begin{center}
    \includegraphics[scale=0.50]{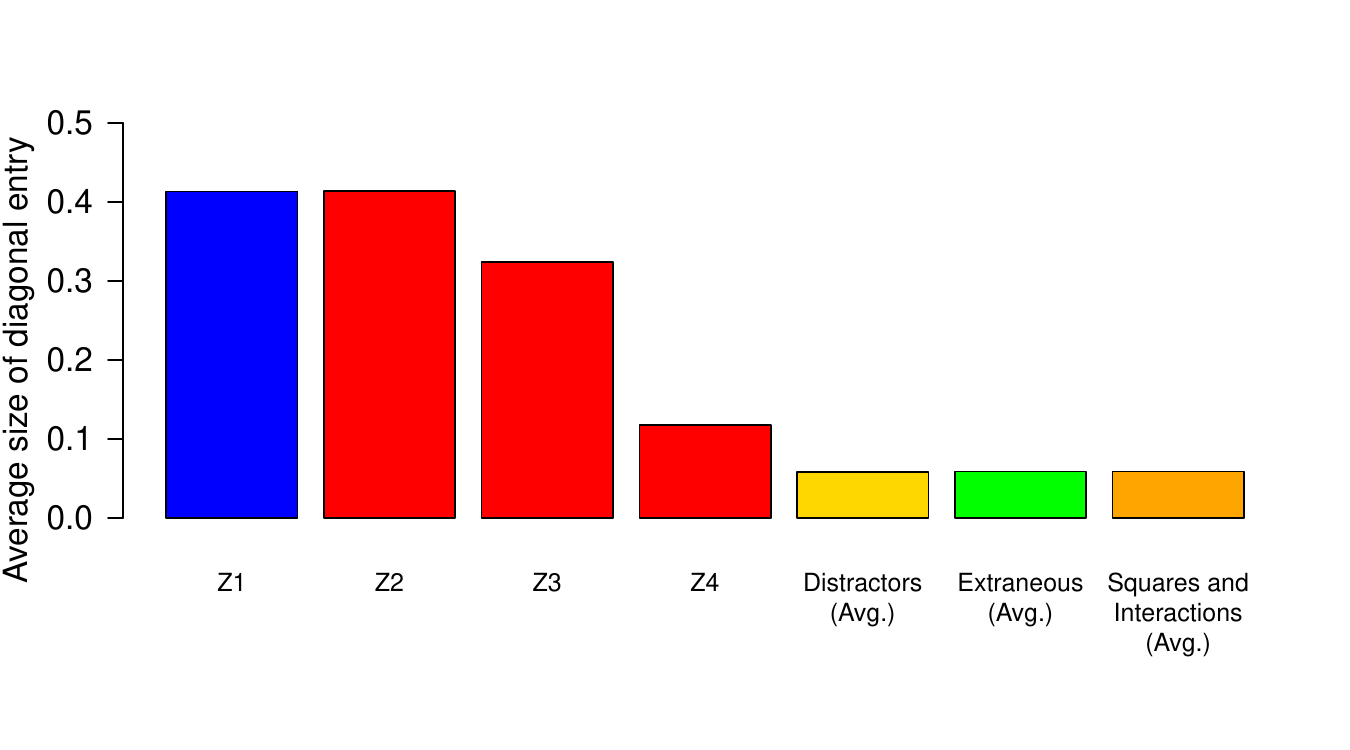} 
    \end{center}
    \vspace{-0.15in}
    \subcaption*{\textit{Note:} Each variable's diagonal entry in $\hat{V}_{\beta_0}^{\frac{1}{2}}$ is averaged across 1000 draws of DGP 2 with $n=1000$. The average entries of (i) the distractors, (ii) the extraneous variables, and (iii) the squares and pairwise interactions of $X_i$ are then averaged across their 5, 10, and 190 dimensions, respectively.}
\end{center}
\vspace{-0.25in}
\end{figure}

As in DGP 1, the leftover imbalance in Figure~\ref{fig.dgp2.balance} reveals why TFB-L excels. Balancing the distractors in $A$ keeps BAL1 from properly balancing $Z$ and reduces its RESS. Meanwhile, TFB-L largely disregards balance in $A$, and obtains good balance in the $Z$, particularly in $Z^{(1)}$, the most influential variable on $Y$. This is a byproduct of how TFB-L models $f_0$ --- LASSO regression forces coefficients in $\hat{\beta}_0$ to 0 when their respective variables are weakly correlated with $Y$, deflating their variances. Accordingly, Figure~\ref{fig.dgp2.vdiag} shows that the distractors', extraneous variables', and polynomial terms' diagonal elements in $\hat{V}_{\beta_0}$ are small, illuminating why TFB-L devalues balance in those variables. Additionally, the diagonal elements for the $Z^{(\ell)}$ in $\hat{V}_{\beta_0}$ increase with the size of their coefficients in the model for $Y$. Meanwhile, this variable selection does not carry over to TFB-K --- it obtains balance on the distractors at almost the level of BAL1, while falling short of TFB-L's balance in $Z^{(1)}$. This is because KRLS does not estimate coefficients for $K_{i}$ with sparsity, so the diagonal entries of the corresponding $\hat{V}_{\beta_0}$ depend more on the variance of the elements of $K_{i}$ than on their contribution to $Y$. Nevertheless, TFB-K still obtains far better balance on $Z$ than does BAL1, which explains its improved performance. 

As with DGP 1, we extend our analysis of DGP 2 by varying its aspects in Appendix \ref{app:dgp2_extension}. Across 18 settings, we vary (i) the sample size ($n$); (ii) the number of distractor and extraneous variables ($A$ and $U$ in (\ref{eq:dgp2_covariates})); and (iii) the true $R^2$ for the model (through the variance of $\epsilon$ in (\ref{eq:dgp2_y})). In short, TFB-L's strong performance persists across all of the variations --- it has the lowest bias and RMSE across all of the 18 settings examined.

\newpage

\subsection{Extension to Demonstration 3}\label{app:dgp2_extension}

This section investigates how TFB performs  after varying important aspects of DGP 2 from Appendix~\ref{sec:dgp2}. As in DGP 2, $X$ contains three types of variables: (i) ``confounders", $Z$, that are related to both $D$ and $Y$; (ii) ``distractors", $A$, that are related to $D$, and are difficult to balance, but are unrelated to $Y$; and (iii) ``extraneous" variables, $U$, that are independent of both $D$ and $Y$. The $Z$ are again generated as
	\begin{align}
		Z_i  \overset{iid}{\sim} \mathcal{N} (0, \mathbf{I}_4)
    \end{align}
and the log odds of treatment is again linear in $Z$ , 
	\begin{align}
		\mathrm{log} \frac{\pi(X)}{1 - \pi(X)} = \frac{1}{5} (Z^{(1)}  + Z^{(2)}  + Z^{(3)}  + Z^{(4)}) 
    \end{align}

Here, the $R^2$ of the model for the outcome, and the dimensions of the distractor and extraneous variables are varied. First, the outcome is linear in $Z$, 
	\begin{align}\label{eq:dgp2_extend_y}
		Y = 8 Z^{(1)}  + 4 Z^{(2)}  + 2 Z^{(3)}  + 1 Z^{(4)} + \epsilon, \ \epsilon_i\overset{iid}{\sim} N(0, \sigma^2)
	\end{align}
where $\sigma^2$, the variance of the error term $\epsilon$, is varied so that the $R^2 = \frac{\var(Y - \epsilon)}{\var(Y)}$ takes values $R^2 \in \{ 0.20, 0.50, 0.80 \}$. 
The $R^2$ is varied to examine how the maximum predictive power of a $\hat{f}_0$ influences TFB's performance.

Then, the distractors ($A$) and the extraneous variables ($U$) are generated as 
	\begin{align}\label{eq:dgp2_extend_au}
		A_i \ | \ D_i \overset{iid}{\sim} \mathcal{N} ( D_i \vec{\1}_{5k}, \mathbf{I}_{5k}) \ \ \text{and} \ \ U_i \overset{iid}{\sim} \mathcal{N} ( 0, \mathbf{I}_{10k})
	\end{align}
where $k \in \{1, 2, 3\}$. In other words, we consider settings with $5$ distractors and $10$ extraneous variables ($k=1$); $10$ distractors and $20$ extraneous variables ($k=2$); and $15$ distractors and $30$ extraneous variables ($k=3$). Finally, two samples sizes are tried: $n \in \{500, 1000\}.$ To summarize, the following aspects are varied:
\begin{itemize}
    \item $R^2 \in \{ 0.20, 0.50, 0.80 \}$ 
    in (\ref{eq:dgp2_extend_y})
    \item $k \in \{1, 2, 3 \}$  in (\ref{eq:dgp2_extend_au})
    \item $n \in \{500, 1000\}$
\end{itemize}
Note that $k = 1$, $R^2 = 0.50$ and $n=1000$ essentially recreates DGP 2 in Appendix~\ref{sec:dgp2}. 

\begin{figure}
    \caption{RMSE in Extended DGP 2}\label{fig.app.dgp2.rmse}
        
    \vspace{-0.35in}
    
    \begin{subfigure}{.45\textwidth}
    \begin{center}
    \includegraphics[scale=.375]{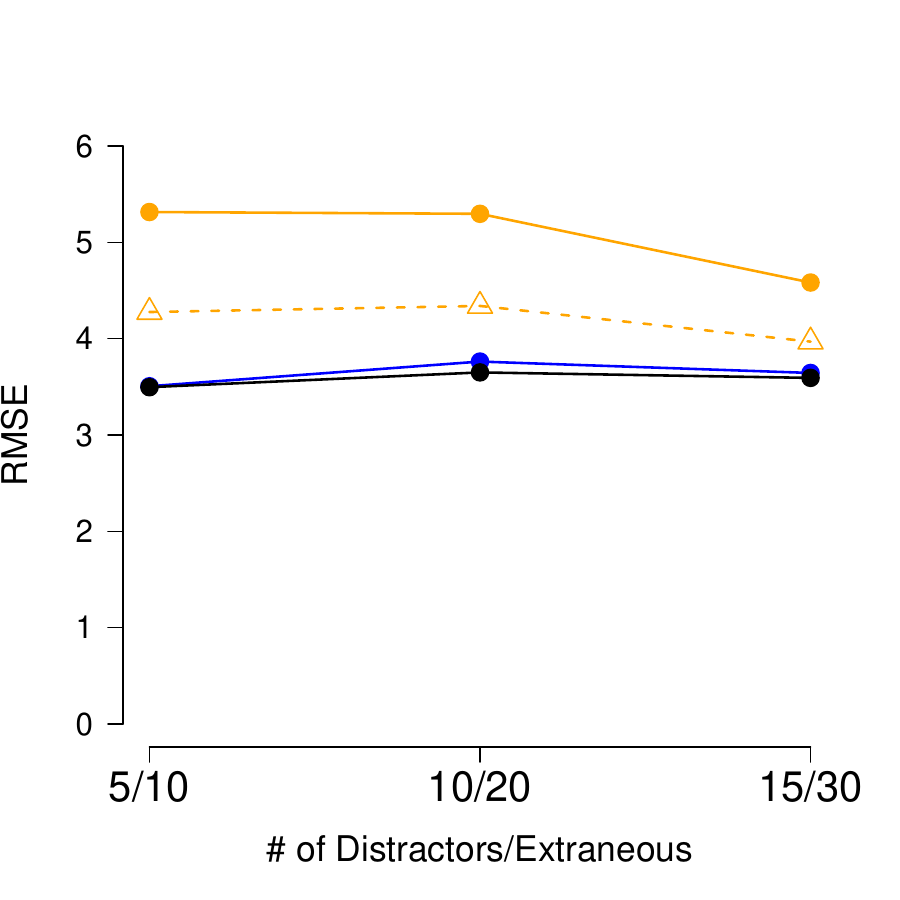}
    \subcaption{$R^2 = 0.20$, $n=500$}
    \end{center}
    \end{subfigure} 
    \begin{subfigure}{.45\textwidth}
    \begin{center}
    \includegraphics[scale=.375]{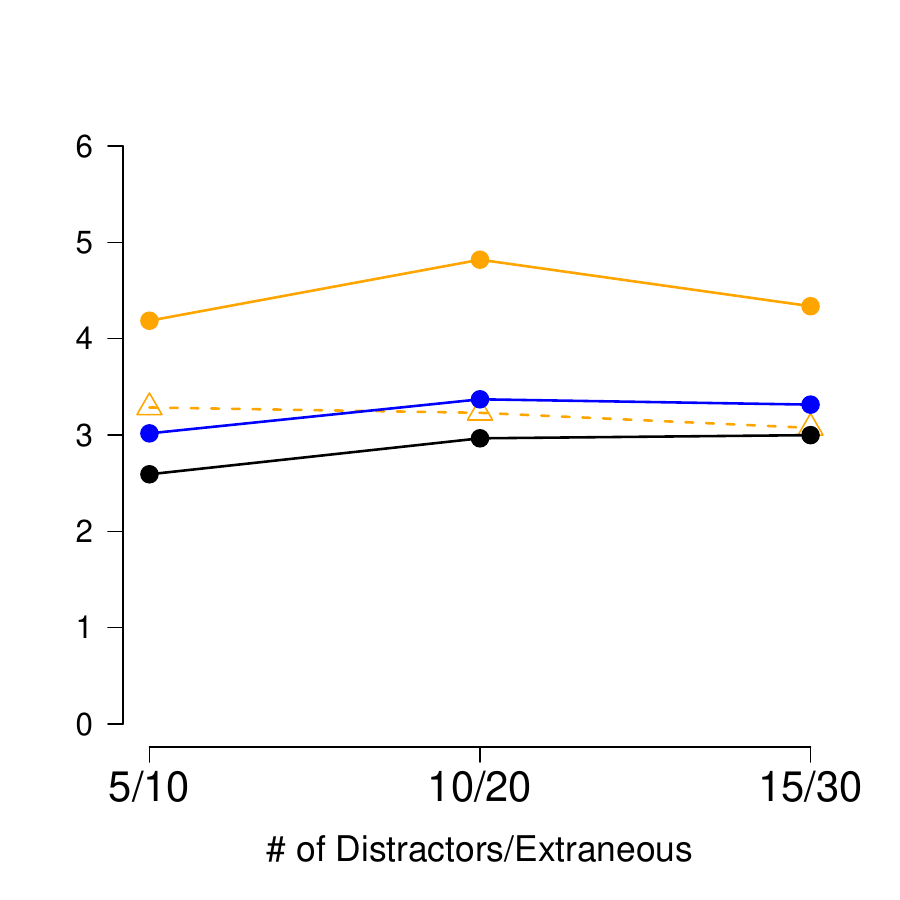}
    \subcaption{$R^2 = 0.20$, $n=1000$}
    \end{center}
    \end{subfigure}

    \vspace{-0.20in}

    \begin{subfigure}{.45\textwidth}
    \begin{center}
    \includegraphics[scale=.375]{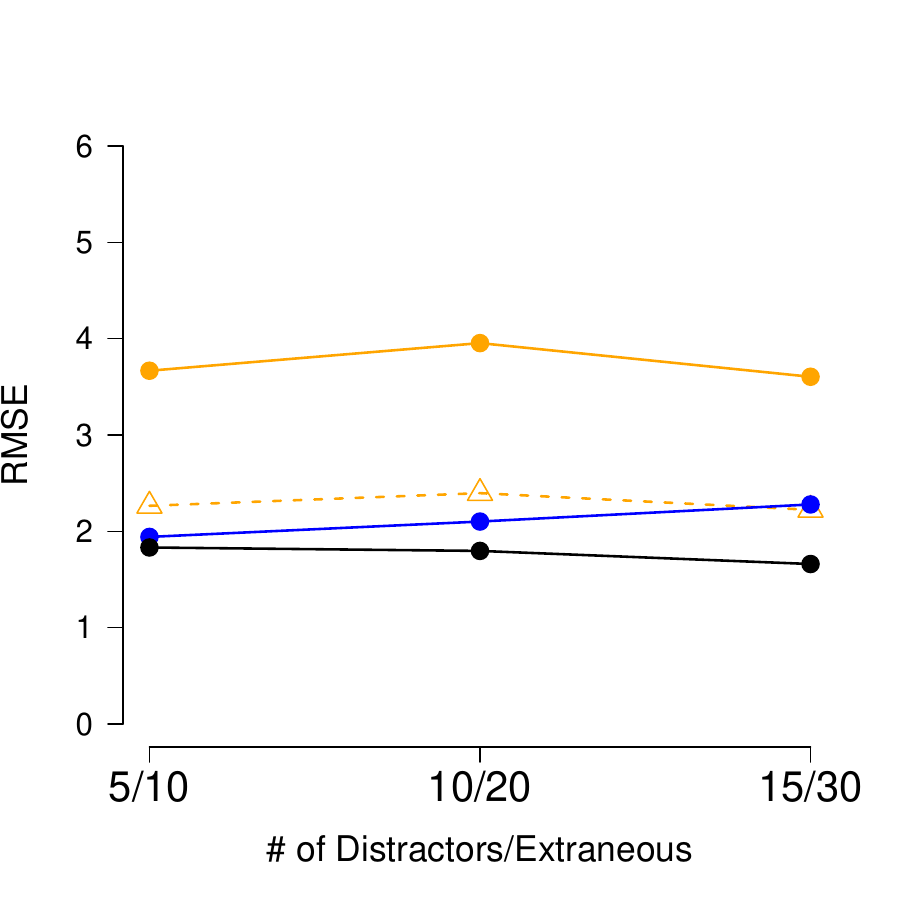}
    \subcaption{$R^2 = 0.50$, $n=500$}
    \end{center}
    \end{subfigure} 
    \begin{subfigure}{.45\textwidth}
    \begin{center}
    \includegraphics[scale=.375]{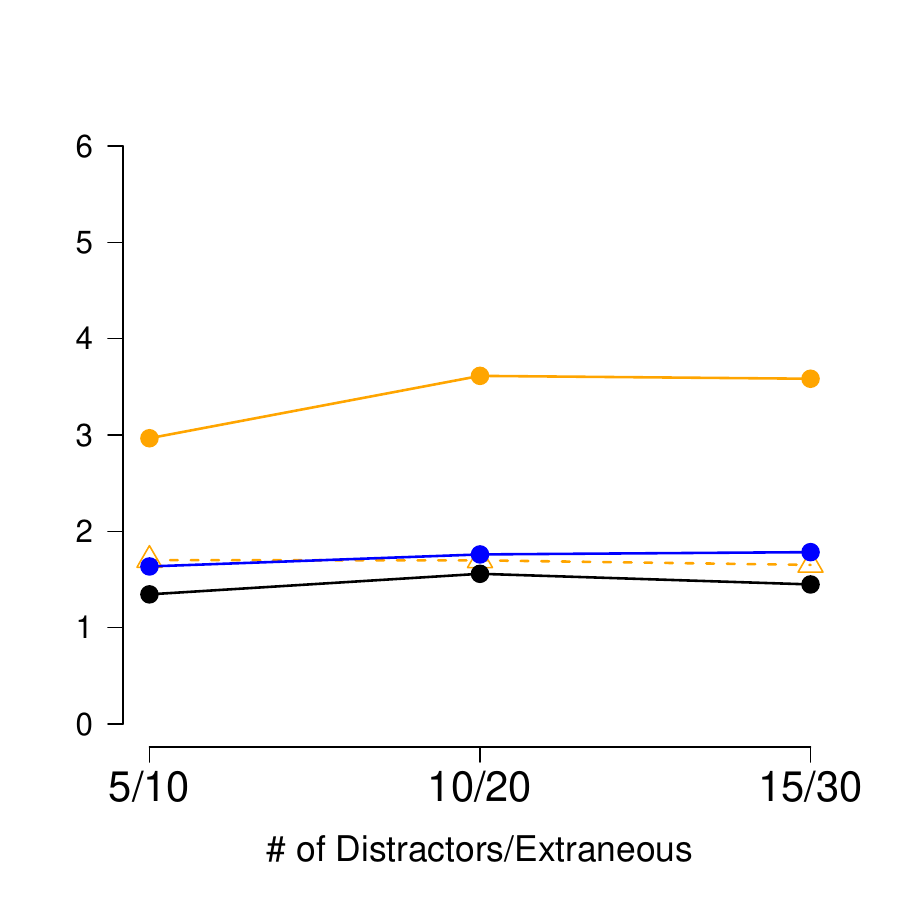}
    \subcaption{$R^2 = 0.50$, $n=1000$}
    \end{center}
    \end{subfigure}

    \vspace{-0.20in}

    \begin{subfigure}{.45\textwidth}
    \begin{center}
    \includegraphics[scale=.375]{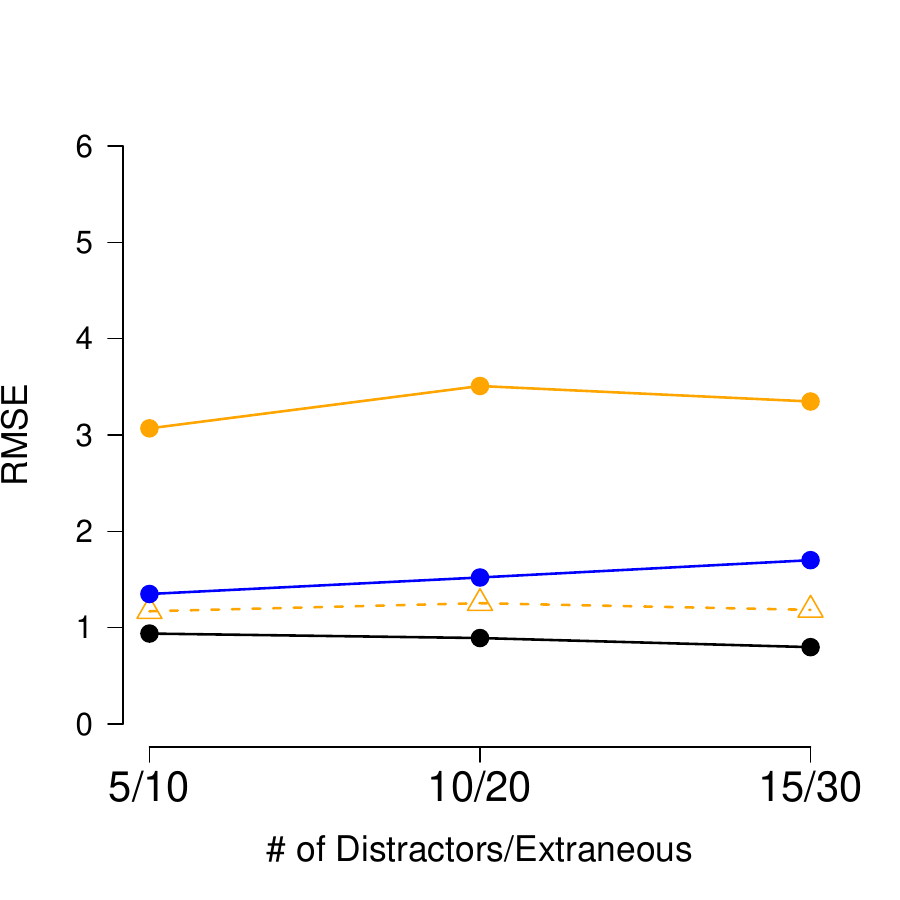}
    \subcaption{$R^2 = 0.80$, $n=500$}
    \end{center}
    \end{subfigure} 
    \begin{subfigure}{.45\textwidth}
    \begin{center}
    \includegraphics[scale=.375]{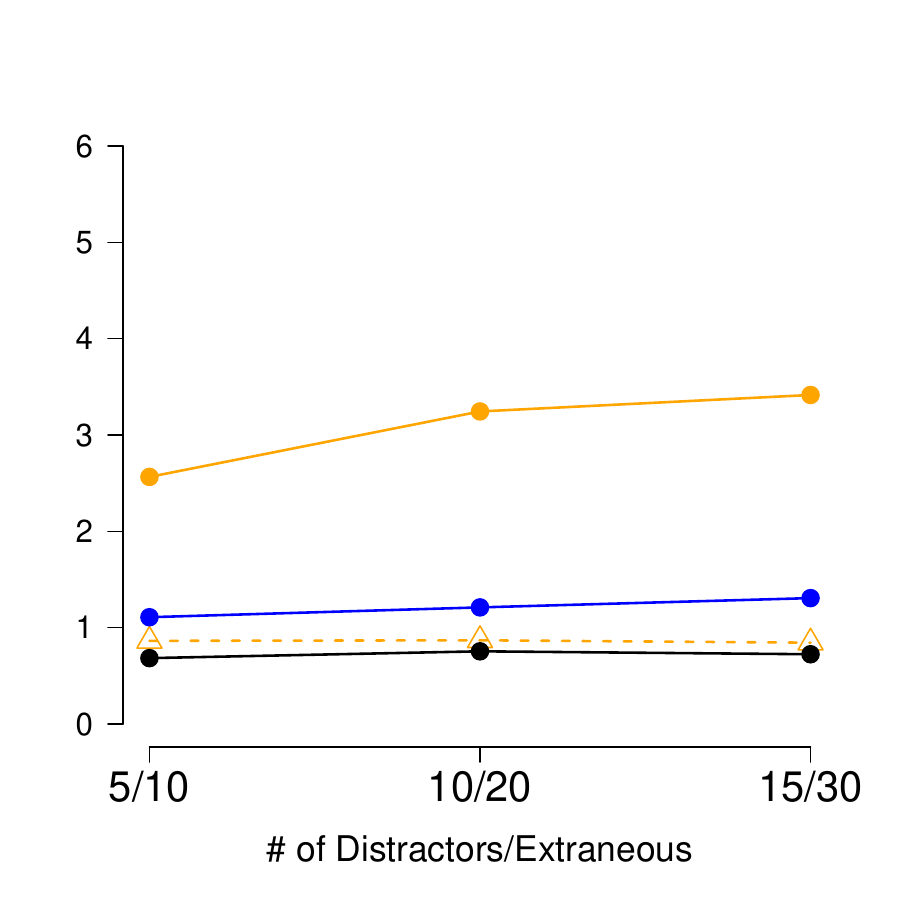}
    \subcaption{$R^2 = 0.80$, $n=1000$}
    \end{center}
    \end{subfigure}

    \vspace{-0.35in}
    
    \begin{center}
    \includegraphics[scale=.40]{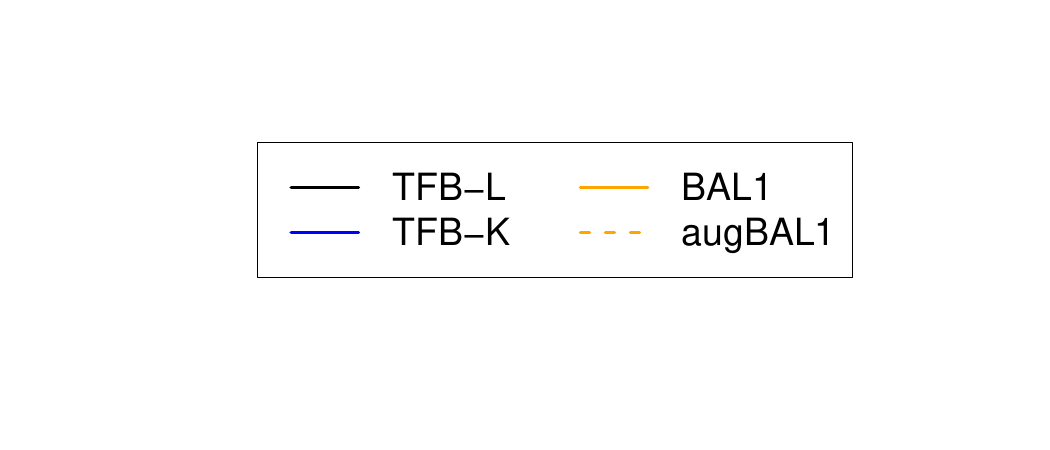}
    \end{center}

    \vspace{-0.35in}
    
    \subcaption*{\textit{Note:} Results across 200 draws from each setting. 5 distractors and 10 extraneous variables corresponds to $k=1$ in (\ref{eq:dgp2_extend_au}); 10 distractors and 20 extraneous variables corresponds to $k=2$; and 15 distractors and 30 extraneous variables corresponds to $k=3$.}

     \end{figure}

\begin{figure}
    \caption{Bias in Extended DGP 2}\label{fig.app.dgp2.bias}
        
    \vspace{-0.35in}
    
    \begin{subfigure}{.45\textwidth}
    \begin{center}
    \includegraphics[scale=.375]{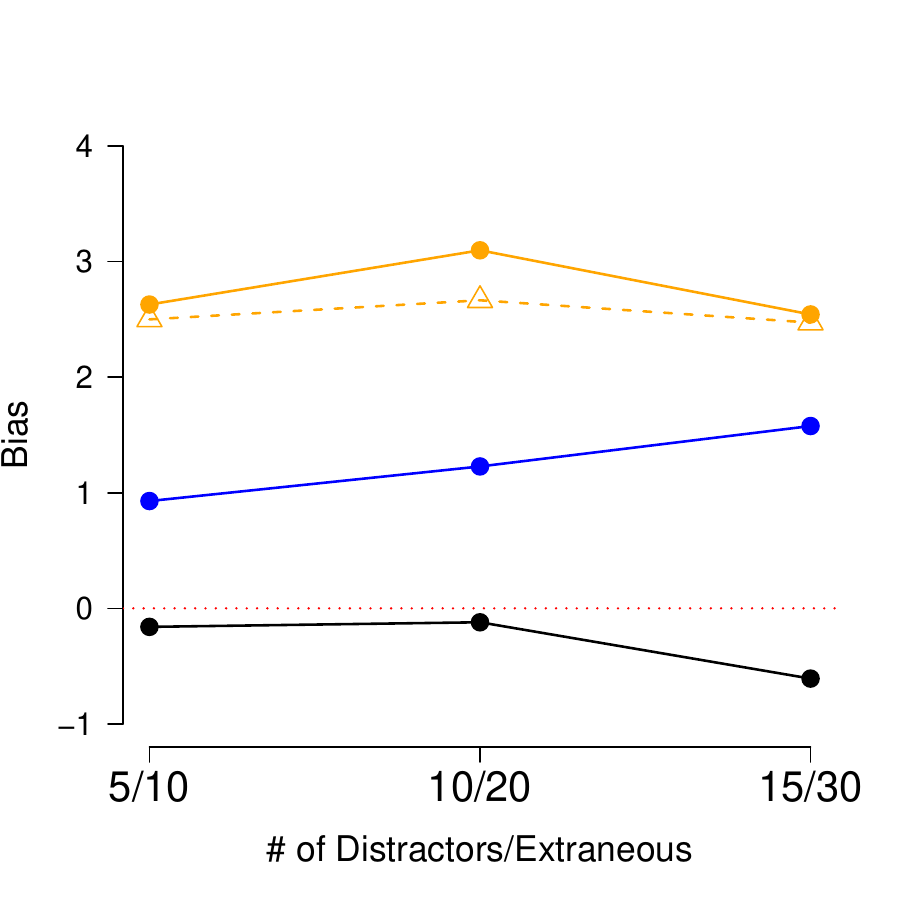}
    \subcaption{$R^2 = 0.20$, $n=500$}
    \end{center}
    \end{subfigure} 
    \begin{subfigure}{.45\textwidth}
    \begin{center}
    \includegraphics[scale=.375]{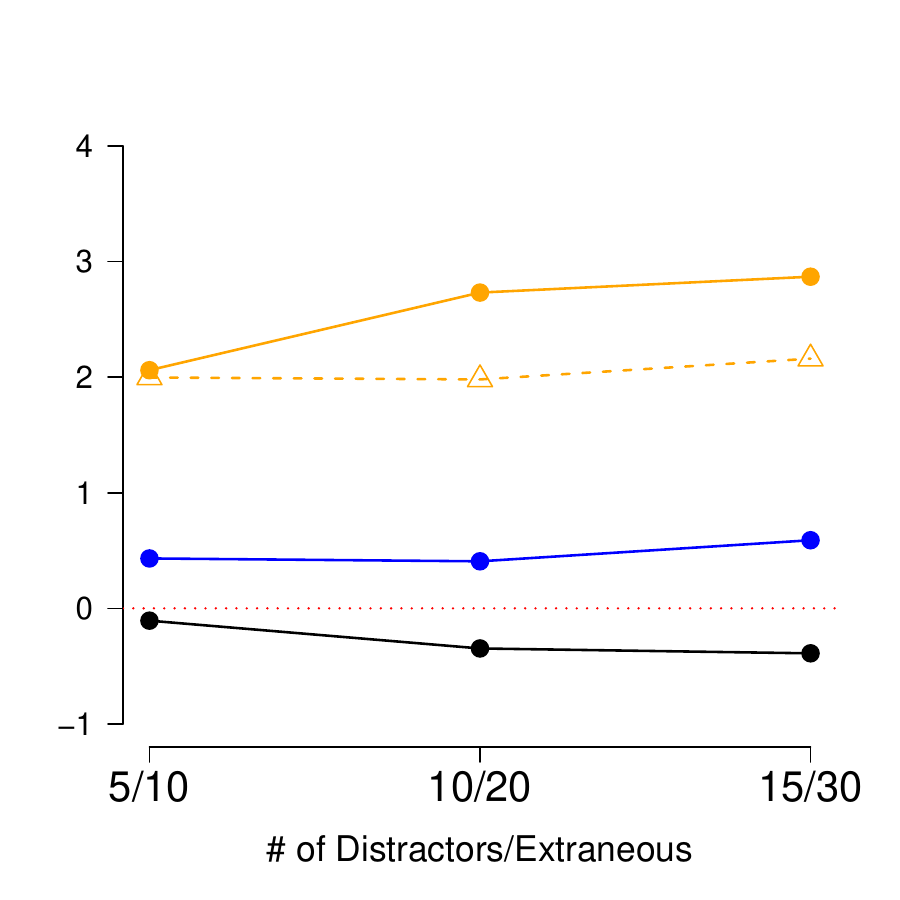}
    \subcaption{$R^2 = 0.20$, $n=1000$}
    \end{center}
    \end{subfigure}

    \vspace{-0.20in}

    \begin{subfigure}{.45\textwidth}
    \begin{center}
    \includegraphics[scale=.375]{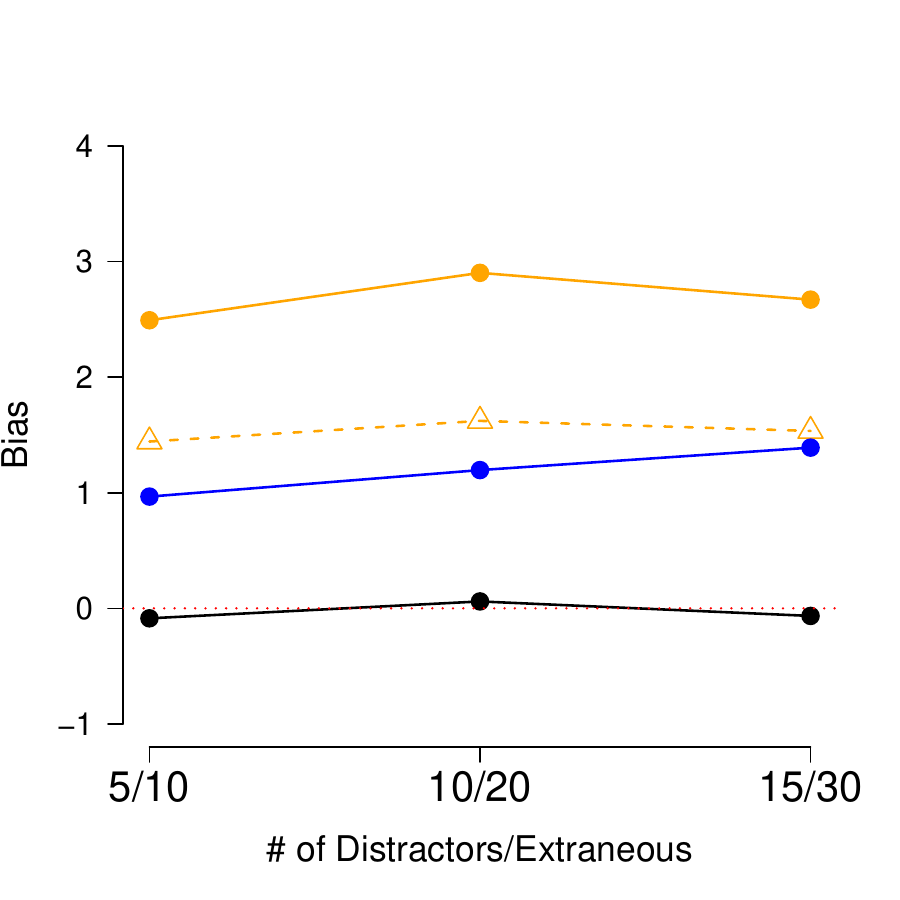}
    \subcaption{$R^2 = 0.50$, $n=500$}
    \end{center}
    \end{subfigure} 
    \begin{subfigure}{.45\textwidth}
    \begin{center}
    \includegraphics[scale=.375]{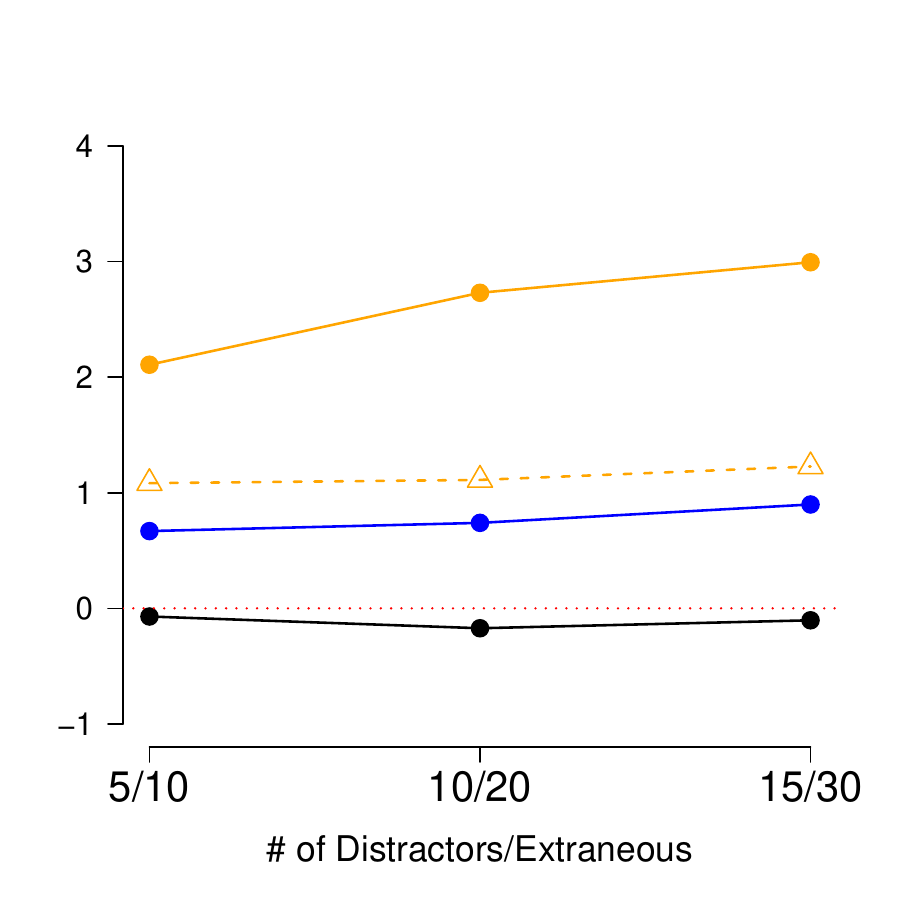}
    \subcaption{$R^2 = 0.50$, $n=1000$}
    \end{center}
    \end{subfigure}

    \vspace{-0.20in}

    \begin{subfigure}{.45\textwidth}
    \begin{center}
    \includegraphics[scale=.375]{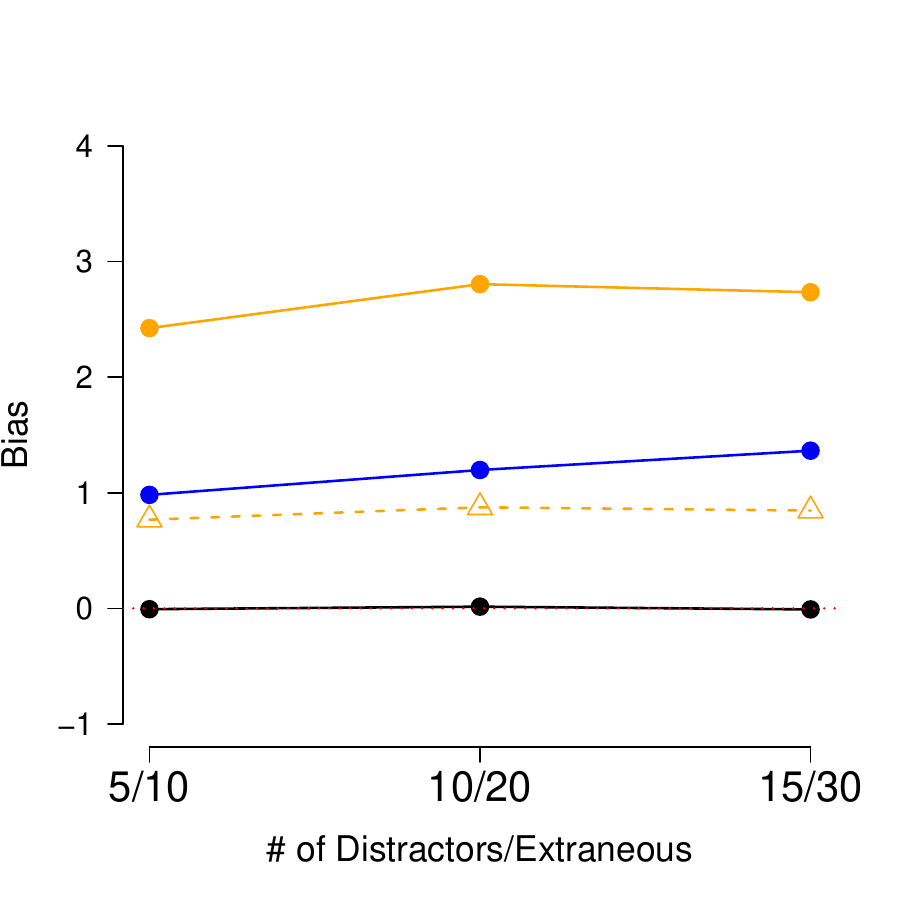}
    \subcaption{$R^2 = 0.80$, $n=500$}
    \end{center}
    \end{subfigure} 
    \begin{subfigure}{.45\textwidth}
    \begin{center}
    \includegraphics[scale=.375]{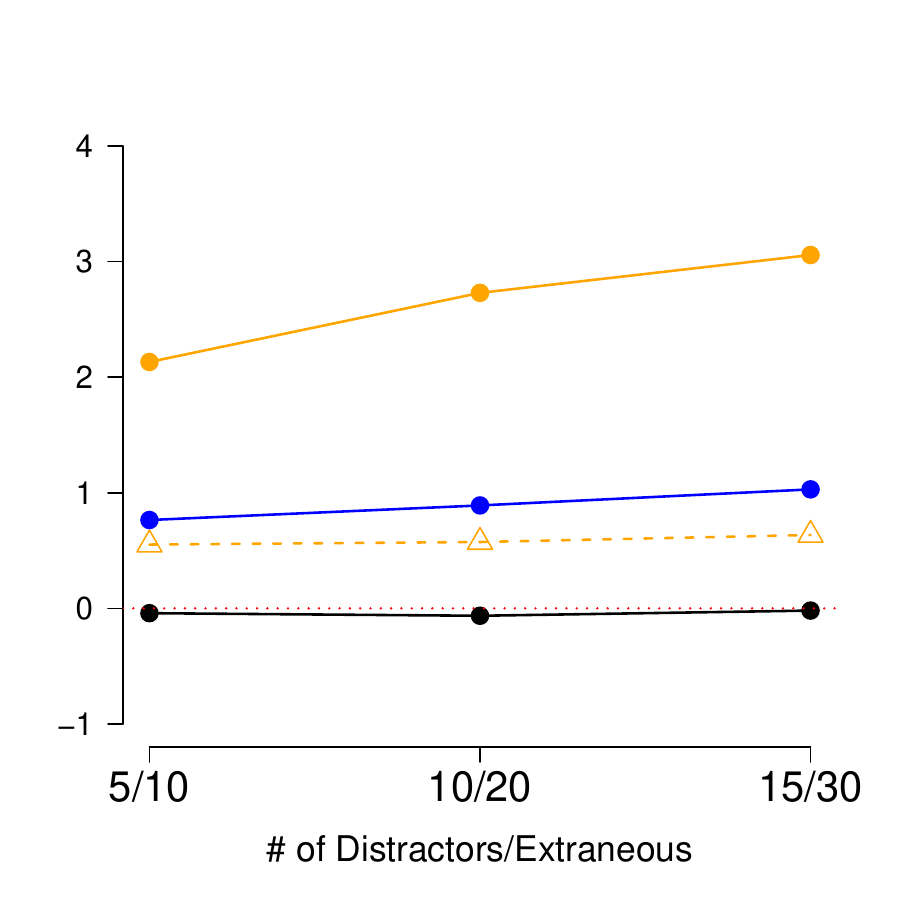}
    \subcaption{$R^2 = 0.80$, $n=1000$}
    \end{center}
    \end{subfigure}

    \vspace{-0.35in}
    
    \begin{center}
    \includegraphics[scale=.40]{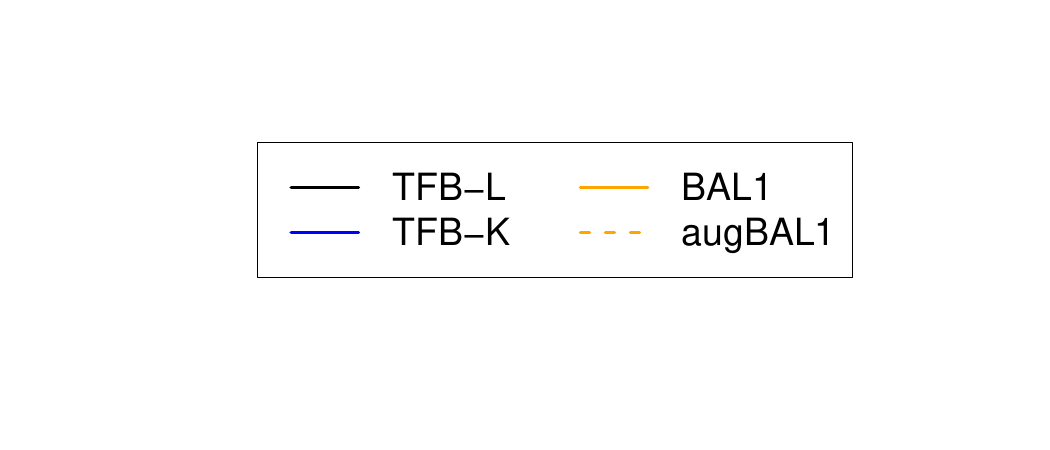}
    \end{center}

    \vspace{-0.35in}
    
    \subcaption*{\textit{Note:} Results across 200 draws from each setting. The red dotted line indicates zero bias. 5 distractors and 10 extraneous variables corresponds to $k=1$ in (\ref{eq:dgp2_extend_au}); 10 distractors and 20 extraneous variables corresponds to $k=2$; and 15 distractors and 30 extraneous variables corresponds to $k=3$.}

     \end{figure}

\begin{figure}
    \caption{Coverage of 95\% Confidence Intervals for TFB-L in Extended DGP 2}\label{fig.app.dgp2.coverage}
        
    \vspace{-0.35in}
    
    \begin{subfigure}{.45\textwidth}
    \begin{center}
    \includegraphics[scale=.375]{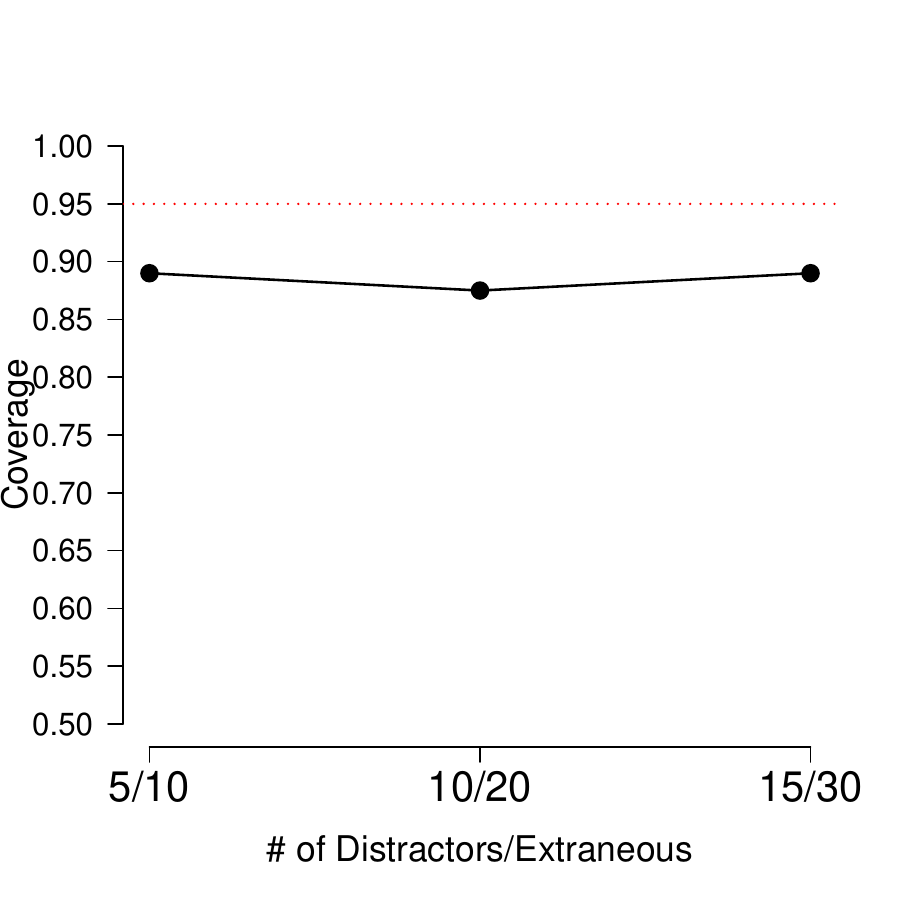}
    \subcaption{$R^2 = 0.20$, $n=500$}
    \end{center}
    \end{subfigure} 
    \begin{subfigure}{.45\textwidth}
    \begin{center}
    \includegraphics[scale=.375]{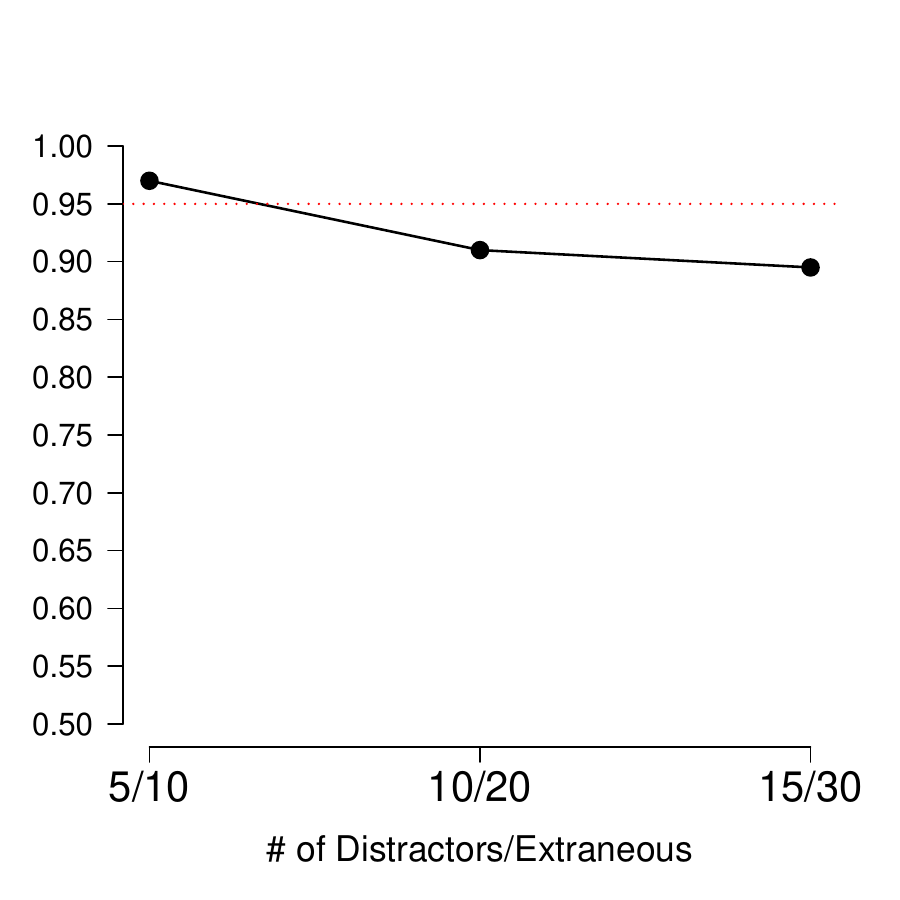}
    \subcaption{$R^2 = 0.20$, $n=1000$}
    \end{center}
    \end{subfigure}

    \vspace{-0.20in}

    \begin{subfigure}{.45\textwidth}
    \begin{center}
    \includegraphics[scale=.375]{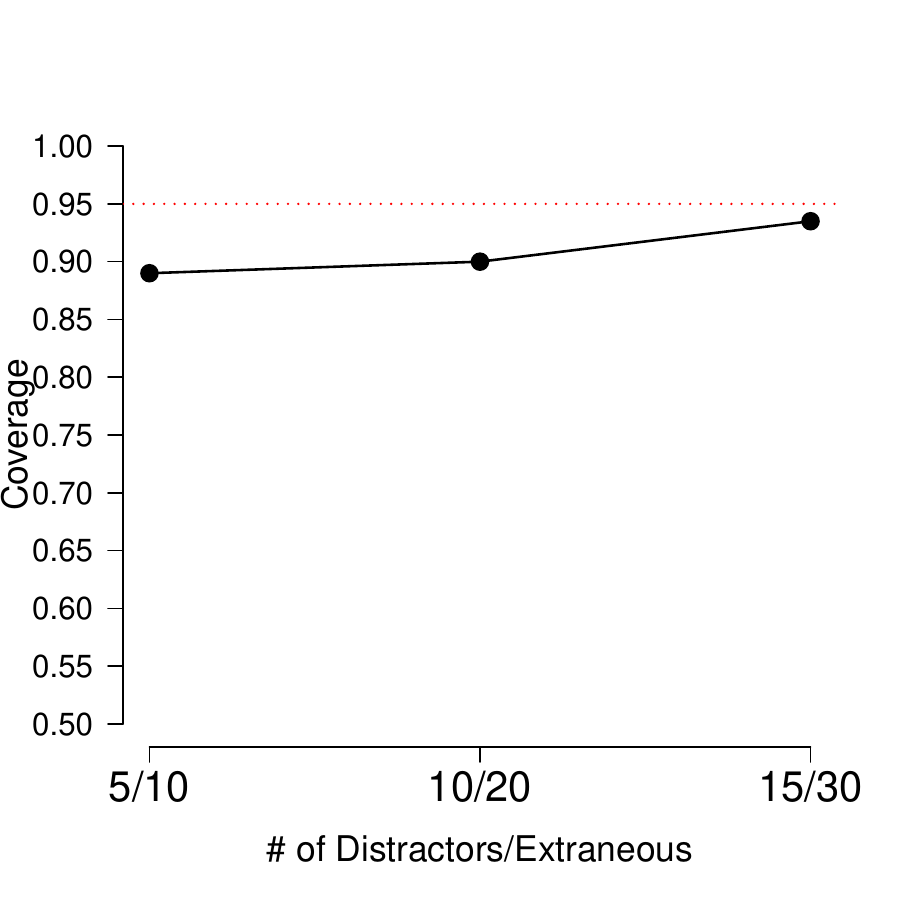}
    \subcaption{$R^2 = 0.50$, $n=500$}
    \end{center}
    \end{subfigure} 
    \begin{subfigure}{.45\textwidth}
    \begin{center}
    \includegraphics[scale=.375]{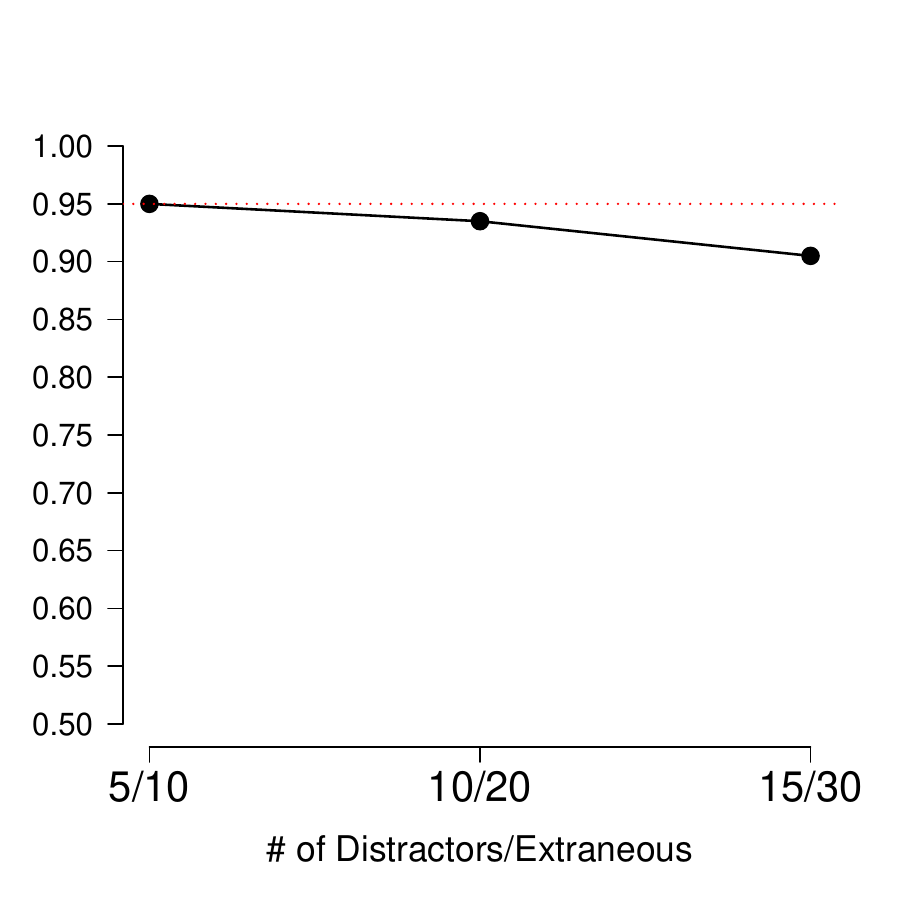}
    \subcaption{$R^2 = 0.50$, $n=1000$}
    \end{center}
    \end{subfigure}

    \vspace{-0.20in}

    \begin{subfigure}{.45\textwidth}
    \begin{center}
    \includegraphics[scale=.375]{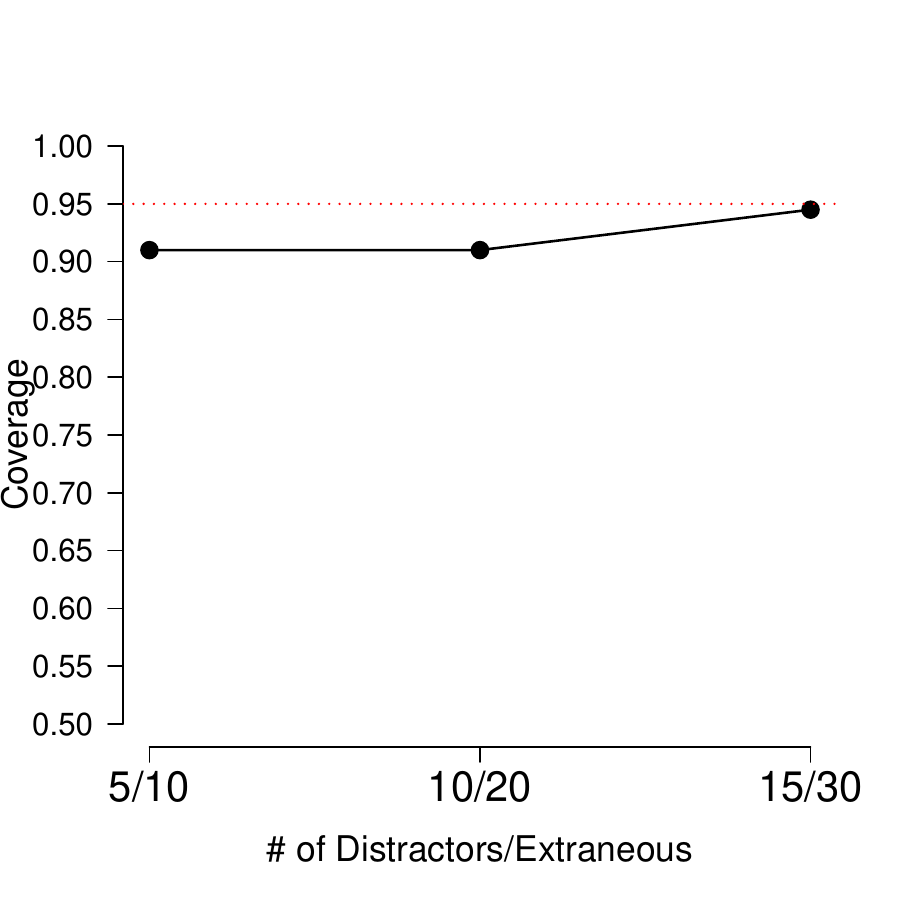}
    \subcaption{$R^2 = 0.80$, $n=500$}
    \end{center}
    \end{subfigure} 
    \begin{subfigure}{.45\textwidth}
    \begin{center}
    \includegraphics[scale=.375]{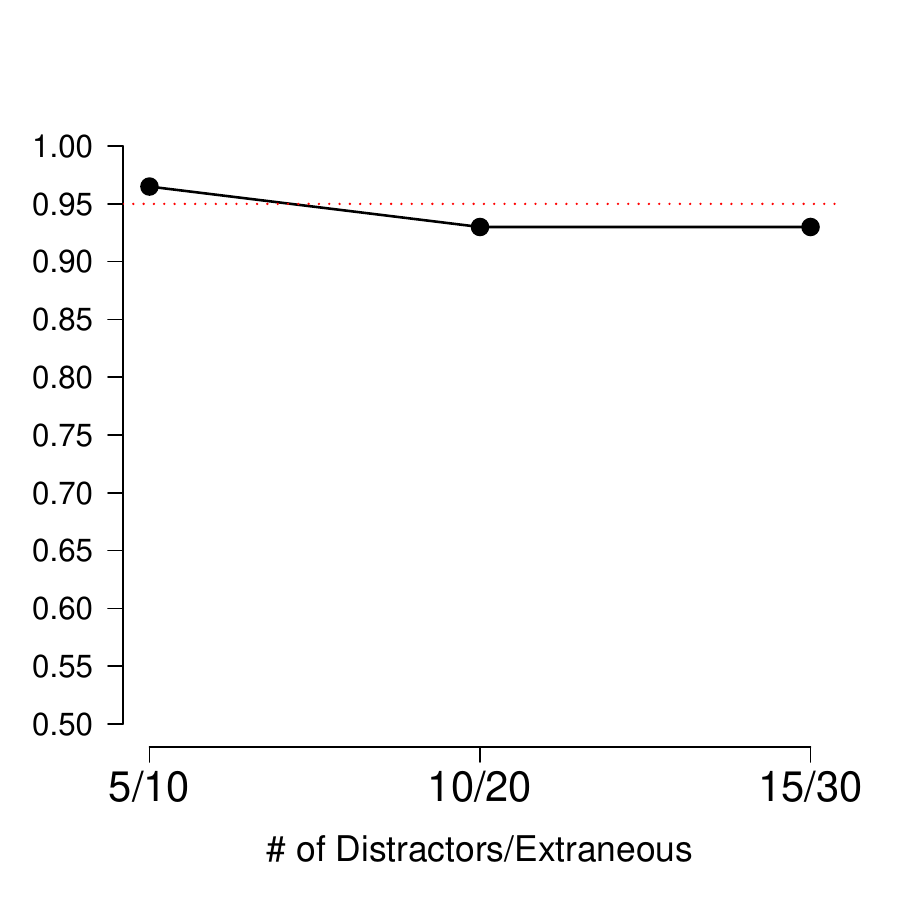}
    \subcaption{$R^2 = 0.80$, $n=1000$}
    \end{center}
    \end{subfigure}

    \vspace{0.2in}
    
    \subcaption*{\textit{Note:} Results across 200 draws from each setting. Confidence intervals are found by the proposal in Section~\ref{subsec:tfb.asymptotics}. 5 distractors and 10 extraneous variables corresponds to $k=1$ in (\ref{eq:dgp2_extend_au}); 10 distractors and 20 extraneous variables corresponds to $k=2$; and 15 distractors and 30 extraneous variables corresponds to $k=3$. The red dotted line indicates the target coverage rate of 0.95. }

     \end{figure}

As in Appendix~\ref{sec:dgp2}, TFB-L and TFB-K are compared to an application of \ref{eq:approxbal_w1} that does not prioritize balance in any dimensions of $X$. All methods are implemented in the same way as in Appendix~\ref{sec:dgp2}. Note that for TFB-L and BAL1, this involves adding to $X$ all its squares and pairwise interactions --- when $k=1$ (5 distractors and 10 extraneous variables) this totals 209 variables, when $k=2$ (10 distractors and 20 extraneous variables) this totals 629 variables, and when $k=3$ (15 distractors and 30 extraneous variables) this totals 1274 variables.
Also included in the comparison is the augmented form of  BAL1 (augBAL1), which uses the LASSO for its regression function. 
Figure~\ref{fig.app.dgp2.rmse} reports the RMSE of each estimator, Figure~\ref{fig.app.dgp2.bias} reports the bias, and Figure~\ref{fig.app.dgp2.coverage} reports the coverage rate of 95\% confidence intervals for TFB-L found through the proposal in Appendix~\ref{subsec:tfb.asymptotics} across 200 iterations from each setting.

Starting with RMSE in Figure~\ref{fig.app.dgp2.rmse}, TFB-L has the lowest RMSE in every setting. 
Figure~\ref{fig.app.dgp2.bias} then shows that TFB-L consistently has the lowest bias of all of the estimators tried here. 
However, TFB-L does show some bias when $R^2 = 0.20$. 
Finally, Figure~\ref{fig.app.dgp2.coverage} shows that the proposed confidence intervals for TFB-L from Appendix~\ref{subsec:tfb.asymptotics} consistently reach, or fall just below, the target coverage rate of 0.95 when $n=1000$ and $R^2 \geq 0.50$.


\clearpage

\newpage

\subsection{Double robustness of TFB}\label{app:tfb.dr}

In this section, we show that TFB can exhibit double robustness due to the equivalence between balancing weights and propensity score weights (\citealp{zhao2017entropy, wang2020minimal}). This equivalence was first noted by  \cite{zhao2017entropy}, who proved that Entropy Balancing is doubly robust because the resulting $\hat{\tau}_{\mathrm{wdim}}$ is unbiased when $f_0$ is linear in $X$, and the resulting weights are propensity score weights with $\pi$ estimated by logistic regression (i.e., log-odds linear in $X$) and parameters that minimize a loss function different from the traditional negative log-likelihood from maximum likelihood estimation. \cite{wang2020minimal} then proved that \ref{eq:approxbal_w1} weights where $B(\cdot)$ is a convex function are equivalent to propensity score weights where $\pi$ has been estimated with $L_1$ shrinkage and the appropriate link and loss functions. \cite{wang2020minimal} also note that certain \ref{eq:approxbal_w2} weights can be rewritten as \ref{eq:approxbal_w1} weights by using a Lagrange relaxation of the balance constraints. 

We first show that TFB weights can be rewritten in this way (i.e., as \ref{eq:approxbal_w1} with a convex $B(\cdot)$). Let $\tilde{\mathbf{X}} = [ \mathbf{X} \hat{V}_{\beta_0}^{\frac{1}{2}} \ \mathbf{X} \hat{\beta}_0]$ be the block matrix that binds together $\mathbf{X} \hat{V}_{\beta_0}^{\frac{1}{2}}$ and $\mathbf{X} \hat{\beta}_0$ column-wise. Then, given TFB weights $\hat{\mathbf{w}}_{\mathrm{TFB}}$, consider  \ref{eq:approxbal_w1}  weights $\hat{\mathbf{w}}_{\mathrm{altTFB}}$ found by:
\begin{align}
    & \hat{\mathbf{w}}_{\mathrm{altTFB}} = \underset{\mathbf{w}}{\text{argmin}} \ || \textbf{w} ||_2^2  \nonumber \\
    \text{where \ \ } & | \mathrm{imbal}( \textbf{w}, \tilde{\mathbf{X}}, \mathbf{D}) | \leq | \mathrm{imbal}( \hat{\mathbf{w}}_{\mathrm{TFB}}, \tilde{\mathbf{X}}, \mathbf{D}) | , \nonumber \\
    & \textbf{w} \geq 0, \ \text{and} \ \frac{1}{n_0} \sum_{i:D_i=0}w_i = 1
\end{align}
Because TFI (in (\ref{eq:linTFImbal})) is strictly increasing in the absolute value of the individual components of $\mathrm{imbal}( \textbf{w}, \tilde{\mathbf{X}}, \mathbf{D})$, it is straightforward to show that $\hat{\mathbf{w}}_{\mathrm{altTFB}} = \hat{\mathbf{w}}_{\mathrm{TFB}}$.\footnote{In this footnote, we omit $\mathbf{D}$ from $\mathrm{imbal}(\cdot)$ and $(\mathbf{D}, \hat{\beta}_0, \hat{V}_{\beta_0}, q)$ from $\mathrm{TFI}(\cdot)$ for clarity. 
Assume that $\hat{\mathbf{w}}_{\mathrm{altTFB}} \neq \hat{\mathbf{w}}_{\mathrm{TFB}}$. 
Then, by definition,  $|| \hat{\mathbf{w}}_{\mathrm{altTFB}} ||_2^2 < || \hat{\mathbf{w}}_{\mathrm{TFB}} ||_2^2$.
It is then worthwhile to compare $\mathrm{TFI}(\mathbf{\hat{w}}_{\mathrm{altTFB}}, \mathbf{X})$ and $\mathrm{TFI}(\mathbf{\hat{w}}_{\mathrm{TFB}}, \mathbf{X})$, as TFI and $|| \cdot ||_2^2$ are the two components of TFB's objective function. For each $\tilde{\mathbf{X}}^{(\ell)}$, the $\ell$th column of $\tilde{\mathbf{X}}$, there are then two possibilities: (i) $|\mathrm{imbal}( \hat{\mathbf{w}}_{\mathrm{altTFB}}, \tilde{\mathbf{X}}^{(\ell)})| < |\mathrm{imbal}( \hat{\mathbf{w}}_{\mathrm{TFB}}, \tilde{\mathbf{X}}^{(\ell)})|$ or (ii) 
$| \mathrm{imbal}( \hat{\mathbf{w}}_{\mathrm{altTFB}}, \tilde{\mathbf{X}}^{(\ell)}) | = | \mathrm{imbal}( \hat{\mathbf{w}}_{\mathrm{TFB}}, \tilde{\mathbf{X}}^{(\ell)}) |$. In either case, $\mathrm{TFI}(\mathbf{\hat{w}}_{\mathrm{altTFB}}, \mathbf{X}) \leq \mathrm{TFI}(\mathbf{\hat{w}}_{\mathrm{TFB}}, \mathbf{X})$, because TFI is strictly increasing in the absolute value of the individual components of $\mathrm{imbal}( \textbf{w}, \tilde{\mathbf{X}})$. This fact, combined with that $|| \hat{\mathbf{w}}_{\mathrm{altTFB}} ||_2 < || \hat{\mathbf{w}}_{\mathrm{TFB}} ||_2$ by necessity, means TFB's objection function evaluated at $\hat{\mathbf{w}}_{\mathrm{altTFB}}$ is strictly less than $\hat{\mathbf{w}}_{\mathrm{TFB}}$. This is a contradiction, because $\hat{\mathbf{w}}_{\mathrm{TFB}}$ is the minimizer of TFB's objective function. Therefore, it must be the case that $\hat{\mathbf{w}}_{\mathrm{altTFB}} = \hat{\mathbf{w}}_{\mathrm{TFB}}$.
}
In other words, TFB weights can be rewritten as \ref{eq:approxbal_w1} weights with $B(\mathbf{w}) = || \textbf{w} ||_2^2$, which is convex, $X$ has been exchanged with $\tilde{X} = [ (\hat{V}_{\beta_0}^{\frac{1}{2}} X)^{\top} \ X^{\top} \hat{\beta}_0]^{\top}$, and $\delta = | \mathrm{imbal}( \hat{\mathbf{w}}_{\mathrm{TFB}}, \tilde{\mathbf{X}}, \mathbf{D}) |$. Thus, TFB weights can be rewritten as the \ref{eq:approxbal_w1} weights that \cite{wang2020minimal} study, and thus the equivalence with propensity score weights applies.

Because of this fact, we can apply Theorem 1 from \cite{wang2020minimal} to show that \textit{when estimating the ATT}:\footnote{Note that the assumed model for $\pi$ is different when estimating the ATE.}
\begin{align}
    \hat{w}_{\mathrm{TFB}} \propto  \frac{\hat{\pi} (X)}{1 - \hat{\pi} (X)} \ \ \text{where} \ \ \hat{\pi} (X) = \frac{\tilde{X}^{\top} \tilde{\alpha}}{1 + \tilde{X}^{\top} \tilde{\alpha}}
\end{align}
where $\hat{\pi}$ is an estimated propensity score with $\tilde{\alpha}$ being the associated coefficient vector, estimated with $L_1$ shrinkage and the appropriate link and loss functions. Thus, TFB weights are propensity score weights that model the propensity score as:
\begin{align}\label{eq:tfb_pscore}
    \pi(X) = \frac{ X^{\top} \eta}{1 + X \eta}
\end{align}
and estimate $\eta$ (with shrinkage, and the appropriate link and loss functions). As a result, TFB should be consistent whenever the true propensity score follows the model in $(\ref{eq:tfb_pscore})$ above, regardless of the consistency of the $\hat{f}_0$ that TFB uses. Hence, TFB should be doubly robust.

We now demonstrate this double robustness through simulation. We focus on TFB's performance when it has and has not correctly specified $\pi$, rather than when it has or has not correctly specified $f_0$. This is because we already demonstrate in multiple places TFB's consistency when $f_0$ is correctly specified but TFB's implied $\pi$ (in (\ref{eq:tfb_pscore})) is not (e.g., Appendices~\ref{sec:dgp2}, ~\ref{app:glm} and~\ref{app:overlap}). We consider two DGPs where $X$ is bivariate, and each dimension is generated as an independent uniform random variable:
\begin{align}
    X^{(\ell)} \sim \mathrm{Unif}(-1, 1)
\end{align}
However, $f_0$ is not linear in $X$. Instead, it is linear in the bivariate $Z$ where:
\begin{align}
    Z^{(\ell)} = ( X^{(\ell)} + 0.25 )^2
\end{align}
The outcome is then generated as:
\begin{align}
    Y = 10 Z^{(1)} + Z^{(2)} + \epsilon, \ \ \epsilon \sim N(0, 5.86^2)
\end{align}
where $\var (\epsilon) = 5.86^2$ allows $R^2 = \frac{\var(Y - \epsilon)}{\var(Y)} \approx \frac{1}{3}$. Additionally, there is no treatment effect (i.e., $\mathrm{ATT}=0$). We then consider two settings for the propensity score. The first is:
\begin{align}\label{eq:sim_dr_ps1}
    \pi_1 (X) = \frac{\exp( Z^{(1)} + Z^{(2)} - 0.8)}{1 + \exp( Z^{(1)} + Z^{(2)} - 0.8 )}
\end{align}
The second is:
\begin{align}\label{eq:sim_dr_ps2}
    \pi_2 (X) = \frac{(1.1 + 0.2X^{(1)} + 0.9X^{(2)})}{1 + (1.1 + 0.2X^{(1)} + 0.9X^{(2)})}
\end{align}
In other words, the PS model implied by TFB (in (\ref{eq:tfb_pscore})) is misspecified in the first setting (in (\ref{eq:sim_dr_ps1})), and is correctly specified in the second setting (in (\ref{eq:sim_dr_ps2})).

We apply the linear form of TFB here (i.e., we use the untransformed $X$, and $\hat{f}_0 (X) = X^{\top} \hat{\beta}_0$) with a $\hat{f}_0$ that is guaranteed to be biased. We first perform an OLS regression of $Y$ on $X$ in the control group to obtain the starting coefficient estimate $\hat{\beta}_{\mathrm{ols}}$. We then add a constant, biasing term to this starting estimate to obtain $\hat{\beta}_{0}$:
\begin{align}
    \hat{\beta}_{0} = \hat{\beta}_{\mathrm{ols}} + [0.25 \ 0.25 \ 0.25]^{\top}
\end{align}
where an intercept term is included in the above coefficients. Because the biasing term for $\hat{\beta}_{0}$ above is a constant vector, we use the sandwich variance estimator (\citealp{white1980heteroskedasticity}) for $\hat{V}_{\beta_0}$. We also apply g-computation (\citealp{robins1986new}) with the same $\hat{f}_0$ as TFB, and \ref{eq:approxbal_w1} with $\delta=0$ (i.e., exact balance on $X$) and $B(\mathbf{w}) = || \mathbf{w} ||_2^2$ (i.e., the method studied by \citealp{zubizarreta2015stable}). We should expect all of these estimators to show bias in the DGP with the first propensity score model (in (\ref{eq:sim_dr_ps1})), because all of the estimators have misspecified $f_0$ and $\pi$. However, we should expect TFB and \ref{eq:approxbal_w1} to be consistent in the DGP with the second propensity score model (in (\ref{eq:sim_dr_ps2})), because they have correctly specified $\pi$. We should expect the g-computation estimator to show bias in this second DGP because it still misspecifies, and poorly estimates, $f_0$.

\begin{figure}[!h]
	\vspace{0.15in}
	\begin{center}
	\caption{Demonstration of TFB's double robustness in DGPs from Appendix~\ref{app:tfb.dr}}\label{fig.dr}
    \vspace{-.40in}
    \begin{subfigure}{.48\textwidth}
    \begin{center}
    \includegraphics[scale=0.47]{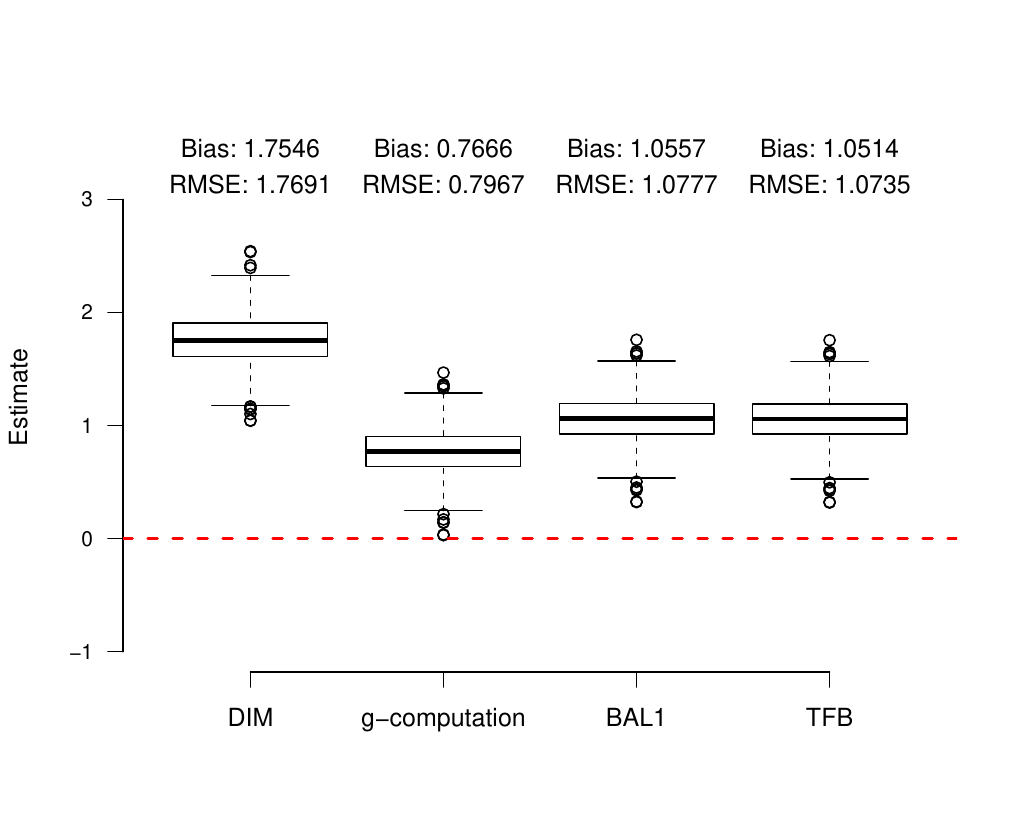} 
    \end{center}
    \vspace{-.25in}
    \subcaption{Misspecified $\pi$ (in (\ref{eq:sim_dr_ps1}))}\label{fig.dr.misspecify}
    \end{subfigure}
    \begin{subfigure}{.48\textwidth}
    \begin{center}
    \includegraphics[scale=0.47]{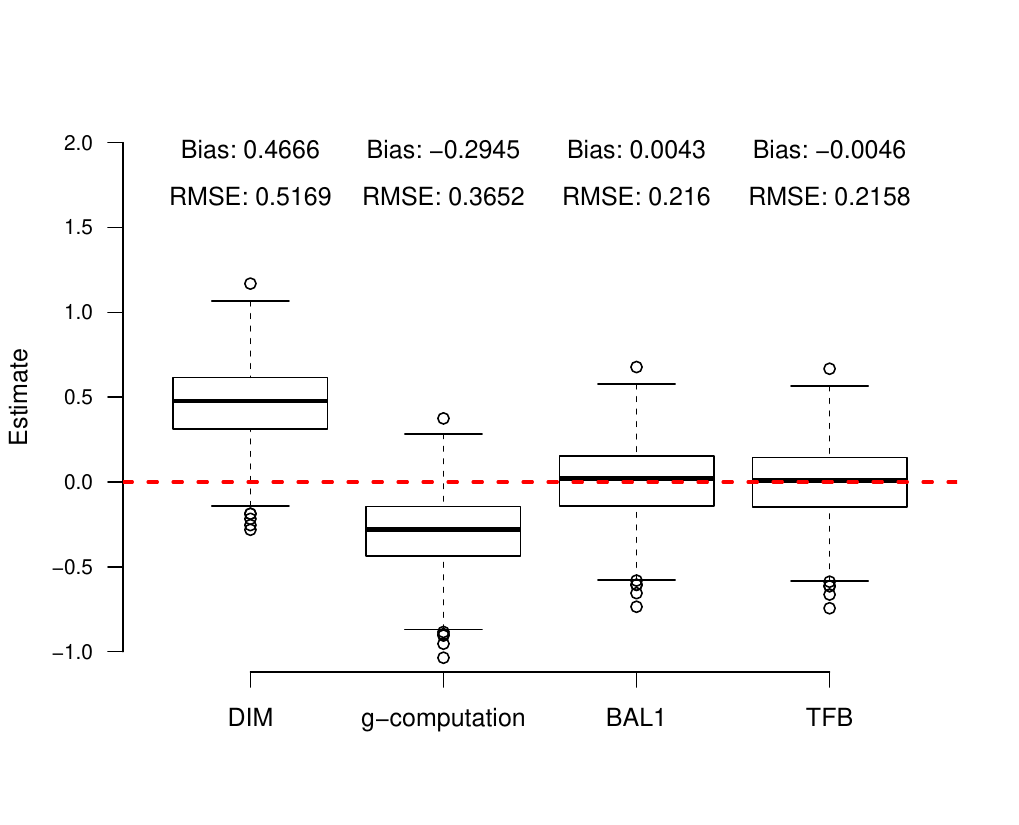} 
    \end{center}
    \vspace{-.25in}
    \subcaption{Correctly specified $\pi$ (in (\ref{eq:sim_dr_ps2}))}\label{fig.dr.correct}
    \end{subfigure}
    \subcaption*{\textit{Note:} Results across 1000 draws with $n=4000$. Distributions of the estimates from each method, as well as their bias and RMSE. \textit{(a)} TFB misspecifies $\pi$, which is truly modeled as (\ref{eq:sim_dr_ps1}). \textit{(b)} TFB correctly specifies $\pi$, which is truly modeled as (\ref{eq:sim_dr_ps2}).}
	\end{center}
	\vspace{-0.25in}
	\end{figure}

These expectations are confirmed in Figure~\ref{fig.dr}, which reports the distributions of these estimators in the DGPs described above when $n=4000$. When TFB misspecifies $\pi$ (in Figure~\ref{fig.dr.misspecify}), TFB shows clear bias. However, when TFB correctly specifies $\pi$ (in Figure~\ref{fig.dr.correct}), TFB is effectively unbiased, despite its $\hat{f}_0$ producing a biased g-computation estimator. TFB and our implementation of \ref{eq:approxbal_w1} here perform very similarly in both settings. 

Finally, note that although the model for $\pi$ assumed by TFB in (\ref{eq:tfb_pscore}) may be restrictive when using an untransformed $X$, it becomes very flexible after exchanging $X$ with a high-dimensional transformation (e.g., exchanging each $X_i$ with $K_i$ for TFB-K). Relatedly, because TFB behaves very similarly to an augmented estimator, and its weights can be written as propensity score weights, it seems very plausible that TFB has the ``rate double robustness" (e.g., \citealp{smucler2019unifying}) property, as in \cite{farrell2015robust} and \cite{chernozhukov2018double}, where an estimator is $\sqrt{n}$-consistent as long as the product of the rates of convergence of $\hat{f}_0$ and $\hat{\pi}$ is above a certain threshold. This provides further motivation for using flexible machine learning models to obtain $\hat{f}_0$, as \cite{chernozhukov2018double} suggest with their doubly robust, augmented estimator. 



\subsection{Benefits of TFB over classical doubly robust estimation}\label{app:tfb_versus_dr}

In this appendix, we demonstrate that TFB can be more efficient, in finite samples, than doubly robust estimators that employ what we refer to as ``classical" propensity score weights, where $\hat{\pi}$ is found by explicitly modeling $D$ as a function of $X$ with a modeling technique that prioritizes predictive accuracy for $D$ (e.g., traditional logistic regression, or BART), and without any input from $Y$. This includes the canonical doubly robust estimators from \cite{robins1994estimation} and \cite{robins1995semiparametric}, and Double Machine Learning from \cite{chernozhukov2018double}. We show that TFB can be more efficient that these doubly robust estimators even when the doubly robust method has perfectly specified both $\pi$ and $f_0$. 

Consider a DGP where $X$ is three-dimensional, and each dimension is generated independently as a standard normal random variable:
\begin{align}
    X \sim \mathcal{N}(0, \mathbf{I}_3)
\end{align}
$f_0$ is then linear in $X$, with $X^{(1)}$ and $X^{(2)}$ being far more influential for $Y$ than is  $X^{(3)}$:
\begin{align}
    Y = X^{(1)} + 2 X^{(2)} + 0.05 X^{(3)} + \epsilon, \ \ \epsilon \sim N(0, 3.15^2)
\end{align}
where $\var (\epsilon) = 3.15^2$ allows $R^2 = \frac{\var(Y - \epsilon)}{\var(Y)} \approx \frac{1}{3}$. Additionally, there is no treatment effect (i.e., $\mathrm{ATT}=0$). The propensity score is then:
\begin{align}
    \pi (X) = \frac{\exp( 0.5 X^{(1)} + 0.5 X^{(2)} + 3.5 X^{(3)}) }{1 + \exp( 0.5 X^{(1)} + 0.5 X^{(2)} + 3.5 X^{(3)}) }
\end{align}
where $X^{(3)}$ shows far more imbalance before weighting than do $X^{(1)}$ and $X^{(2)}$.

We apply a form of TFB that uses a simple OLS regression of $Y$ on $X$ in the control group for $\hat{f}_0$, and the corresponding sandwich variance estimator (\citealp{white1980heteroskedasticity}) for $\hat{V}_{\beta_0}$. Note that this outcome model perfectly specifies $f_0$. We compare TFB with a $\hat{\tau}_{\mathrm{wdim}}$ and employs (inverse) propensity score weights (IPW) and its augmented form (augIPW), where $\hat{f}_0$ is the same as for the TFB estimator. The  propensity score weights for IPW and augIPW apply traditional logistic regression to predict $D$ with $X$ for $\hat{\pi}$, which perfectly specifies $\pi$. We also apply g-computation (\citealp{robins1986new}) with the same $\hat{f}_0$ as TFB.

The key challenge in this DGP is the role of $X^{(3)}$, which is highly imbalanced to start, due to its large coefficient in the propensity score model, but has very limited influence on $Y$. Any $\hat{\pi}$ that is fit directly (i.e., by modeling $D$ as a function of $X$ without any input from $Y$) will produce weights that vary greatly with $X^{(3)}$. However, this variation will do little to decrease the bias in the resulting estimator, because of the small influence of $X^{(3)}$ on $Y$, but will greatly increase the estimator's variance. Therefore, while all of the estimators applied here correctly specify $f_0$ and/or $\pi$, and thus should show little to no bias, one should expect the IPW and augIPW estimators to have unnecessarily  high variances.

\begin{figure}[!h]
	\vspace{0.15in}
	\begin{center}
	\caption{Comparing TFB with classical doubly robust estimation}\label{fig.tfb_versus_dr}
    \vspace{-.40in}
    \begin{center}
    \includegraphics[scale=0.55]{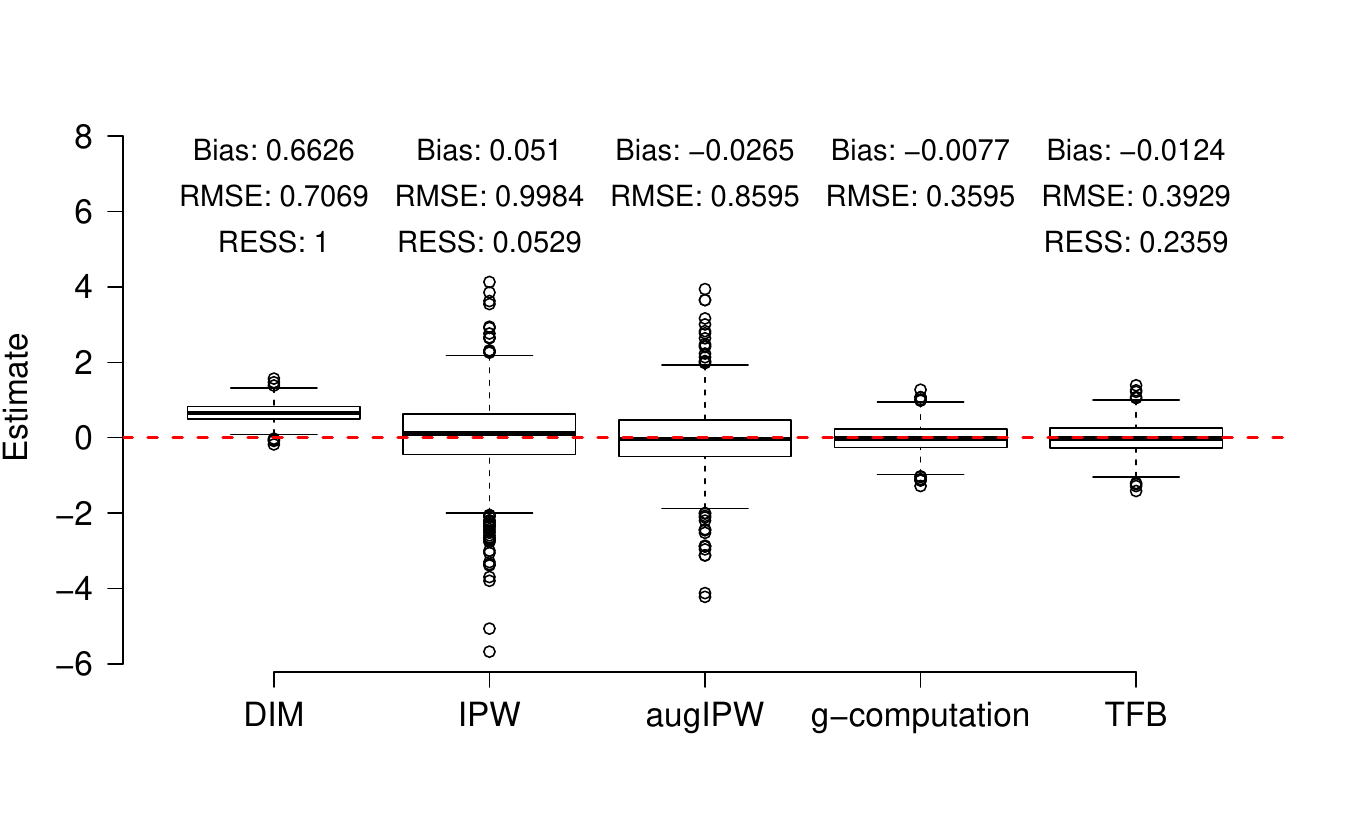} 
    \end{center}
      \subcaption*{\textit{Note:} Results across 1000 draws of the DGP described in Appendix~\ref{app:tfb_versus_dr} with $n=1000$. Distributions of the estimates from each method, as well as their bias, RMSE, and RESS.}
	\end{center}
	\vspace{-0.25in}
	\end{figure}

Figure~\ref{fig.tfb_versus_dr} confirms this hypothesis, applying all of the estimators to the above DGP with $n=1000$. While all estimators shown are effectively unbiased, the IPW and augIPW estimators show extremely high variance. These variances are so high that the RMSEs of IPW and augIPW are actually higher than that of the unweighted difference in means (DIM). The reason for these excessively high variances is the high variance of the propensity score weights these estimators employ: the RESS of IPW is minuscule (0.0529), and is much lower than the RESS of TFB's weights (0.2359). Accordingly, TFB's RMSE (0.3929) is less than half of the RMSEs of IPW (0.9984) and augIPW (0.8595). The g-computation and TFB estimators perform similarly, with g-computation having a slightly lower RMSE (0.3595). The increase in RMSE from the g-computation estimator to augIPW in particular shows how the inclusion of the high variance propensity score weights has hindered estimation.

Note that while we have applied estimators here with very simple $\hat{f}_0$ and $\hat{\pi}$, the lesson of this demonstration carries over to estimators that apply $\hat{f}_0$ and $\hat{\pi}$ found through, for example, machine learning, as in Double Machine Learning (\citealp{chernozhukov2018double}) or with TFB-L, TFB-K, and TFB-BART. No matter what regression technique is used for $\hat{\pi}$, as long as $\hat{\pi}$ is fit (i) by only prioritizing predictive accuracy for $D$, and (ii) without any input from $Y$, the resulting propensity score weights run the risk of needlessly increasing the variance of the resulting doubly robust estimator because the $\hat{\pi}$ will be highly sensitive to any covariates that have strong association with treatment, regardless of their associations with the outcome. Meanwhile, even though TFB does imply a $\hat{\pi}$ with corresponding propensity score weights (see Appendix~\ref{app:tfb.dr}), the $\hat{\pi}$ is heavily influenced by TFB's $\hat{f}_0$, and will thus be less sensitive to covariates that are less predictive of $Y$.

\subsection{Choosing $q$}\label{subsec:tfb.q}

In this appendix, we consider $q$, the probability threshold in TFB. Note that $q$ influences TFB's weights through the constant $Q_q (\mathcal{X}^2_P)$ in (\ref{eq:linTFImbal}). Thus, when $q$ is closer to 1, TFB prioritizes reducing the imbalance in $X$, which could yield less bias but a higher variance  estimator. When $q$ is closer to 0, reducing imbalance becomes less important, and lowering the variance of the weights is more important. Thus $q$ presents a bias-variance trade-off. Nevertheless, we recommend $q=0.95$ generally, and this is used in all demonstrations here.

\begin{figure}[!h]
\begin{center}
\caption{Demonstration of TFB-K (see Table~\ref{tab:abbreviations}) in DGP 1 for varying $q$}\label{fig.TFB.q1}
    \begin{subfigure}{.45\textwidth}
    \begin{center}
    \includegraphics[scale=0.5]{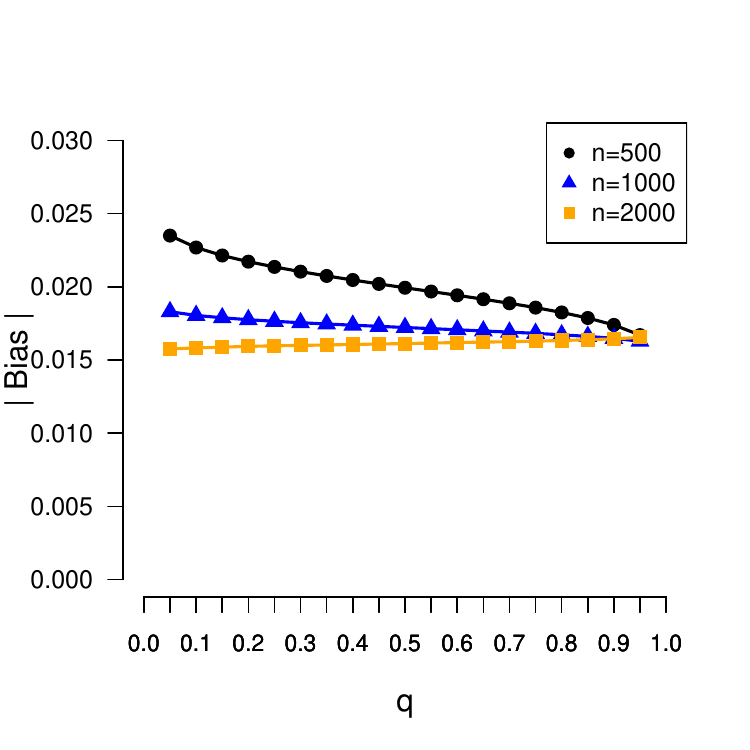} 
    \end{center}
    \vspace{-.25in}
    \subcaption{$|$Bias$|$}
    \end{subfigure}
    \begin{subfigure}{.45\textwidth}
    \begin{center}
    \includegraphics[scale=0.5]{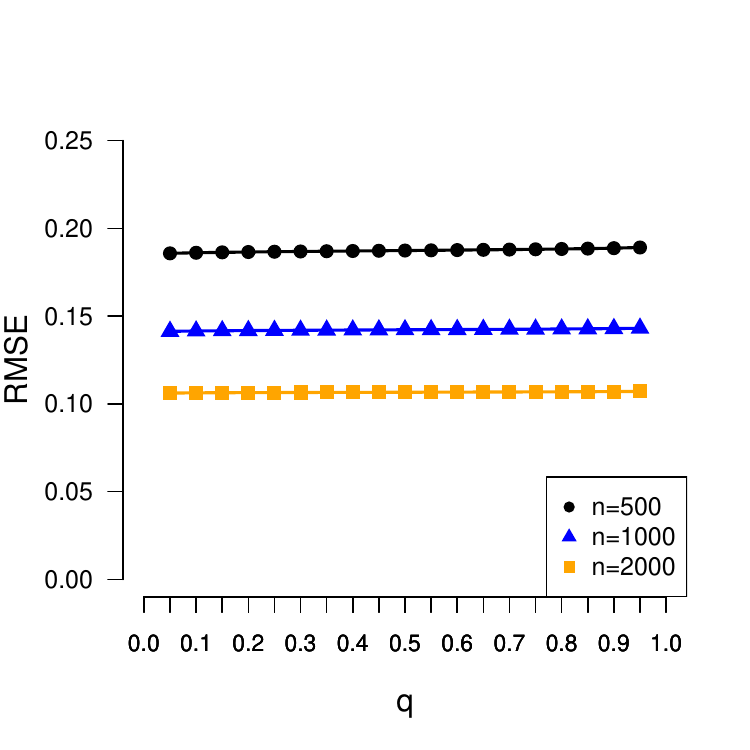} 
    \end{center}
    \vspace{-.25in}
    \subcaption{RMSE}
    \end{subfigure}
    \subcaption*{\textit{Note:} Results across 1000 draws of DGP 1. $q$ takes values from 0.05 to 0.95 in intervals of 0.05. TFB-K, where KRLS estimates $f_0$, is applied.  }
\end{center}
\vspace{-0.25in}
\end{figure}

\begin{figure}
\begin{center}
\caption{Demonstration of TFB-K in DGP 2  for varying $q$}\label{fig.TFB.q.dgp2}
    \begin{subfigure}{.45\textwidth}
    \begin{center}
    \includegraphics[scale=0.5]{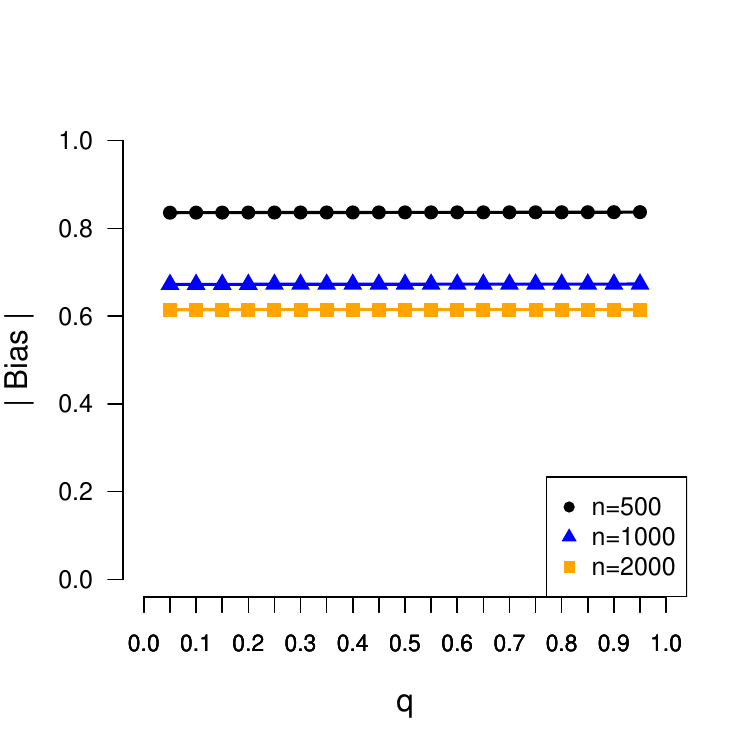} 
    \end{center}
    \vspace{-.25in}
    \subcaption{$|$Bias$|$}
    \end{subfigure}
    \begin{subfigure}{.45\textwidth}
    \begin{center}
    \includegraphics[scale=0.5]{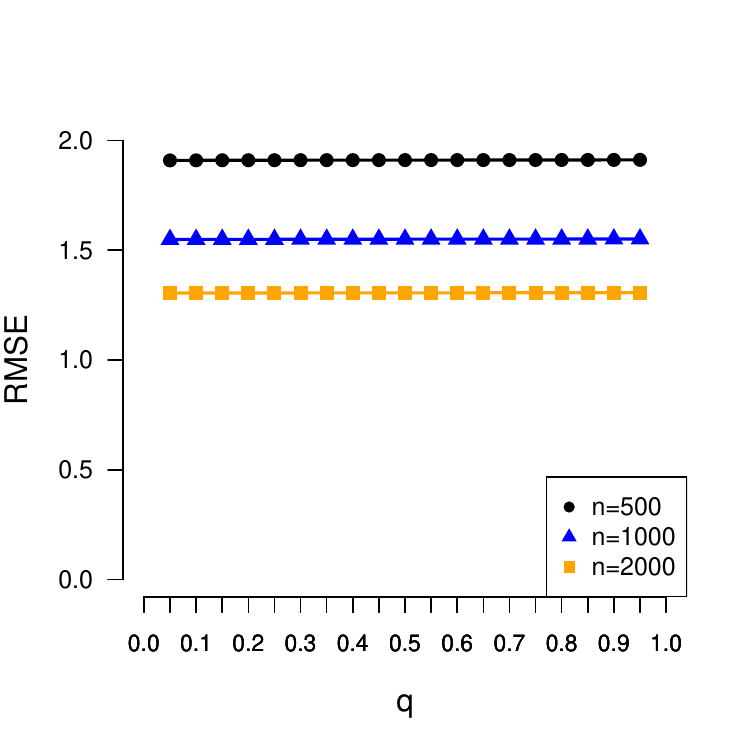} 
    \end{center}
    \vspace{-.25in}
    \subcaption{RMSE}
    \end{subfigure}
    \subcaption*{\textit{Note:} Results across 1000 draws of DGP 2. $q$ takes values from 0.05 to 0.95 in intervals of 0.05. TFB-K (see Table~\ref{tab:abbreviations}) is applied.  }
\end{center}
\vspace{-0.25in}
\end{figure}

To justify this recommendation, we consider two scenarios that deserve separate discussions: (i) when $P \rightarrow \infty$ as $n \rightarrow \infty$ and (ii) when $P$ is fixed. Starting with the first scenario, notice that this applies to TFB-K, which replaces the $X_i$ with the $K_i$ and $P = n$, and TFB-BART, which replaces $\mathbf{X}$ with $\mathbf{I}_n$ and thus $P=n$. This also applies to TFB-L in settings where one would add more non-linear transformations to $X$ as $n$ grows. Note here that:
	\begin{align}\label{eq:chisq_quantile}
		Q_q (\mathcal{X}^2_P) &= \sqrt{2P} Q_q \biggr[ \sqrt{ \frac{P}{2} } \biggr( \frac{\sum_{\ell=1}^{P} Z_{\ell}^2}{P} - 1 \biggr) \biggr] + P  \sim \sqrt{2P} \Phi^{-1} (q) + P
	\end{align}
where $Z_\ell$ are iid standard normal variables, $\Phi^{-1} (q)$ is the $q$th quantile of a $N(0, 1)$, and $\sqrt{ \frac{P}{2} } ( \frac{\sum_{\ell=1}^{P} Z_{\ell}^2}{P} - 1) \overset{d}{\rightarrow} N(0, 1)$ by the Central Limit Theorem. By (\ref{eq:chisq_quantile}) above, $\frac{ \sqrt{Q_{q_1} (\mathcal{X}^2_P)} }{ \sqrt{Q_{q_2} (\mathcal{X}^2_P)} } \rightarrow 1 $ for fixed $q_1$ and $q_2$ and as $P \rightarrow \infty$, meaning that the specific choice of $q$ becomes inconsequential when $n$, and thus $P$, is large.  Figure~\ref{fig.TFB.q1} shows this empirically --- TFB-K's bias and RMSE in DGP 1 (from Section~\ref{sec:dgp1}) become stable across $q$ as $n$ increases. However, with smaller $n$, there is a bias-variance trade-off --- bias decreases as $q$ increases, due to the higher emphasis on reducing imbalance in $K_i$, but the RMSE is mostly unchanged, implying a higher variance estimator.\footnote{We see a similar trend in Appendix~\ref{app:hetero}, which explores an extension to DGP 1 that allows for heteroscedastic error.} In DGP 2 (from Appendix~\ref{sec:dgp2}), we see similar results. Figure~\ref{fig.TFB.q.dgp2} shows that at every sample size tried for DGP 2, TFB-K's bias and RMSE are stable across all values of $q$ tried. 

Now consider the setting where $P$ is fixed. This is never the case for TFB-K or TFB-BART, but applies to TFB-L in cases where one would \textit{not} expand $X$ were the sample size to grow, and would like to take advantage of the LASSO's variable selection. In this setting, $\hat{V}_{\beta_0} \overset{p}{\rightarrow} 0$ quite generally, so TFB will devalue overall balance in $X$ as $n$ grows. A $q$ near 1 may thus be preferable in this case because it will yield better balance on $X$, which may be valuable if $f_0$ has been misspecified (i.e., is not linear in $X$) --- for example, see Appendix~\ref{app:tfb.dr}. Therefore, because $q$ appears to be inconsequential in the case where $P \rightarrow \infty$, and $q$ near 1 is advisable when $P$ is fixed, $q=0.95$ seems a reasonable choice.



\subsection{Generalized linear models with TFB }\label{app:glm}

In this section, we consider applying a generalized linear model (GLM) to the non-linear extension of TFB. Assuming that $f_0$ can be modeled as a GLM, we can write  
    \begin{align}
        f_0(X) = g(X^{\top} \beta_0)
    \end{align}
where $g$ is the inverse link function of the GLM. We then estimate $f_0$ as $\hat{f}_0 (X) = g(X^{\top} \hat{\beta}_0)$ for some estimate $\hat{\beta}_0$ obtained through maximum likelihood estimation (MLE). Then let $\hat{V}_{\hat{\beta}_0}$ be a consistent estimator of the variance of $\hat{\beta}_0$.

The non-linear extension of TFB requires a $\hat{V}_{\hat{f}_0 (\mathbf{X})}$. Because $\hat{\beta}_0$ is obtained through MLE, under the GLM model for $f_0$ and iid data,
    \begin{align*}
        \hat{V}_{\hat{\beta}_0}^{-\frac{1}{2}} ( \hat{\beta}_0 - \beta_0) \overset{d}{\rightarrow} N(0, \mathbf{I}_P)
    \end{align*}
Applying the Delta Method with a general function $h(\cdot)$ then gives
    \begin{align}
         \biggr( \frac{\partial h(\beta_0)}{\partial \beta_0} \hat{V}_{\hat{\beta}_0} \frac{\partial h(\beta_0)}{\partial \beta_0}^{\top} \biggr)^{-\frac{1}{2}} \biggr( h(\hat{\beta}_0) - h(\beta_0) \biggr) \xrightarrow{d} N(0, \mathbf{I}_P)
    \end{align}
Finally, letting $h(\beta_0) = \hat{f}_0 (\mathbf{X}) = [ g(X_1^{\top} \hat{\beta}_0) \ \cdots \ g(X_n^{\top} \hat{\beta}_0)]^{\top}$ to the above yields a plausible choice for $\hat{V}_{\hat{f}_0 (\mathbf{X})}$ in the GLM setting:
    \begin{align}\label{eq:vhat_glm}
        \hat{V}_{\hat{f}_0 (\mathbf{X})} &= [g'] \mathbf{X} \hat{V}_{\hat{\beta}_0} \mathbf{X}^\top [g'] \nonumber \\
        \text{where} \ \ \ [g'] &= \begin{bmatrix} 
            g'(X_1^\top\hat{\beta}_0) & 0 & \cdots & 0 \\
            0 & g'(X_2^\top\hat{\beta}_0) & \cdots & 0 \\
            \vdots & \vdots & \ddots & \vdots \\
            0 & 0 & \cdots & g'(X_n^\top\hat{\beta}_0) \end{bmatrix}
    \end{align}
Given this choice of $\hat{V}_{\hat{f}_0 (\mathbf{X})}$, the scaled imbalance term in the objective function for the non-linear extension of TFB becomes:
    \begin{align}\label{eq:tfb_glm_imbal}
        || \hat{V}_{\hat{f}_0 (\mathbf{X})}^{\frac{1}{2}} \mathrm{imbal}(\mathbf{w}, \mathbf{I}_n, \mathbf{D}) ||_2 &= || \hat{V}_{\hat{\beta}_0}^{\frac{1}{2}} \mathrm{imbal}(\mathbf{w}, [g']\mathbf{X}, \mathbf{D}) ||_2 
        \ \ \ 
        \text{where} 
        \ \ \ 
        [g'] \mathbf{X} = \begin{bmatrix} 
            g'(X_1^\top\hat{\beta}_0) X_1^{\top} \\ \vdots \\ g'(X_n^\top\hat{\beta}_0) X_n^{\top}
        \end{bmatrix}
    \end{align}

To summarize, this version of TFB seeks balance in $g'(X_i^\top\hat{\beta}_0) X$. We can understand this choice by examining a first-order Taylor approximation of $g(X^{\top} \beta_0)$:
    \begin{align}
        g(X^{\top} \beta_0) &\approx g(\tilde{X}^{\top}\beta_0) + g'(\tilde{X}^{\top}\beta_0)(X-\tilde{X})^{\top}\beta_0 \nonumber \\
        &= \biggr( g(\tilde{X}^{\top} \beta_0) - g'(\tilde{X}^{\top} \beta_0) \tilde{X}^{\top} \beta_0 \biggr) + g'(\tilde{X}^{\top}\beta_0)X^{\top}\beta_0
    \end{align}
for some $\tilde{X} \in \mathbb{R}^{P}$ near $X$. Thus, $f_0$ (under a GLM model) can be approximated as linear in $g'(\tilde{X}^{\top} \beta_0)X$. Since the Taylor approximation requires $\tilde{X}$ to be near $X$, we may substitute $\tilde{X}$ for $X$, further approximating  $f_0$ as a linear function of $g'(X^{\top}\beta_0)X$. Therefore, balance in $g'(X^{\top} \hat{\beta}_0)X$ should translate to balance in $f_0$ when $\hat{\beta}_0$ is a good estimate of $\beta_0$, and $g'(X^{\top} \hat{\beta}_0)X$ is precisely what this GLM version of TFB seeks balance in, per (\ref{eq:tfb_glm_imbal}). 

To demonstrate, consider a DGP where $P=5$ with $X^{(5)}=1$ being an intercept term, and each other dimension of $X$ is simulated independently from a standard normal distribution:
    \begin{align}
        X_i^{(\ell)} \overset{iid}{\sim} N(0, 1) \ \ \text{for} \ \ \ell=1, \dots, 4
    \end{align}
Next, we generate the propensity scores according to a probit model,
    \begin{align}
        \pi(X) &= \Phi(\psi(X)) \nonumber \\
    \text{where} \ \ \ \psi(X) &= \frac{2.2\biggr( 0.25[X^{(1)}]^3 + [X^{(2)}]^3 + [X^{(3)}]^3 + [X^{(4)}]^3 \biggr)}{[X^{(1)}]^2 + 2}
    \end{align}
where $\Phi (\cdot)$ is the CDF of the standard normal distribution. Finally, we generate $Y$ as a binary outcome variable that also comes from a probit model:
    \begin{align}
        p(Y = 1 | X, D) = \Phi \biggr(X^{(1)} + \frac{1}{10}\sum_{\ell=2}^4 X^{(\ell)} - 0.8 \biggr)
\end{align}
Note that $X^{(1)}$ is the most influential variable in determining $Y$, and the ATT is 0.

\begin{figure}[!h]
	\vspace{0.15in}
	\begin{center}
	\caption{Performance of TFB with probit regression in the DGP from Appendix~\ref{app:glm} }\label{fig.glm}
    \vspace{-.25in}
    \begin{subfigure}{.45\textwidth}
    \begin{center}
    \includegraphics[scale=0.43]{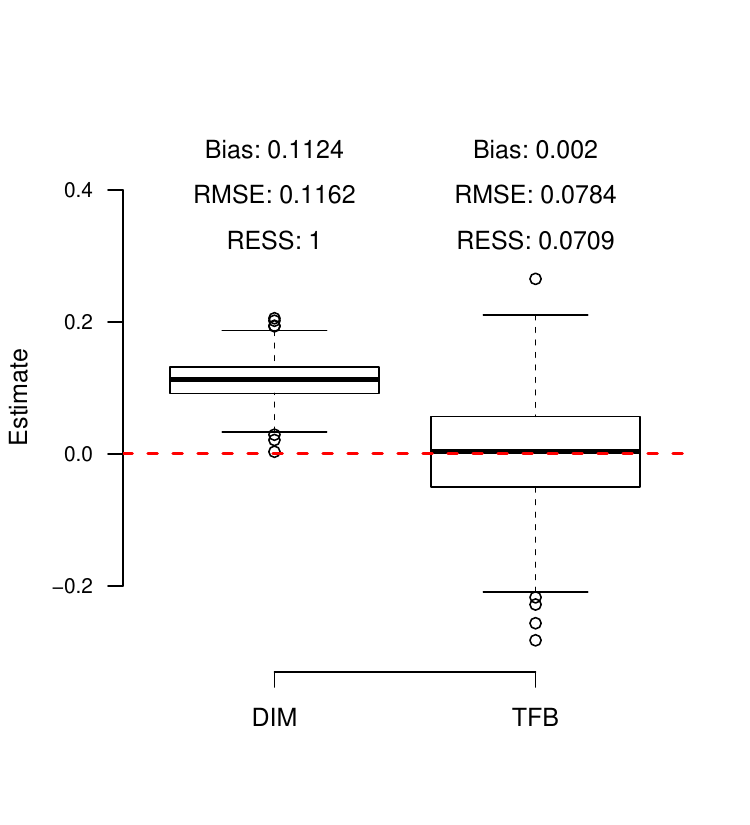} 
    \end{center}
    \vspace{-.25in}
    \subcaption{Bias, RMSE, and RESS}\label{fig.glm.bias}
    \end{subfigure}
    \begin{subfigure}{.45\textwidth}
    \begin{center}
    \includegraphics[scale=0.43]{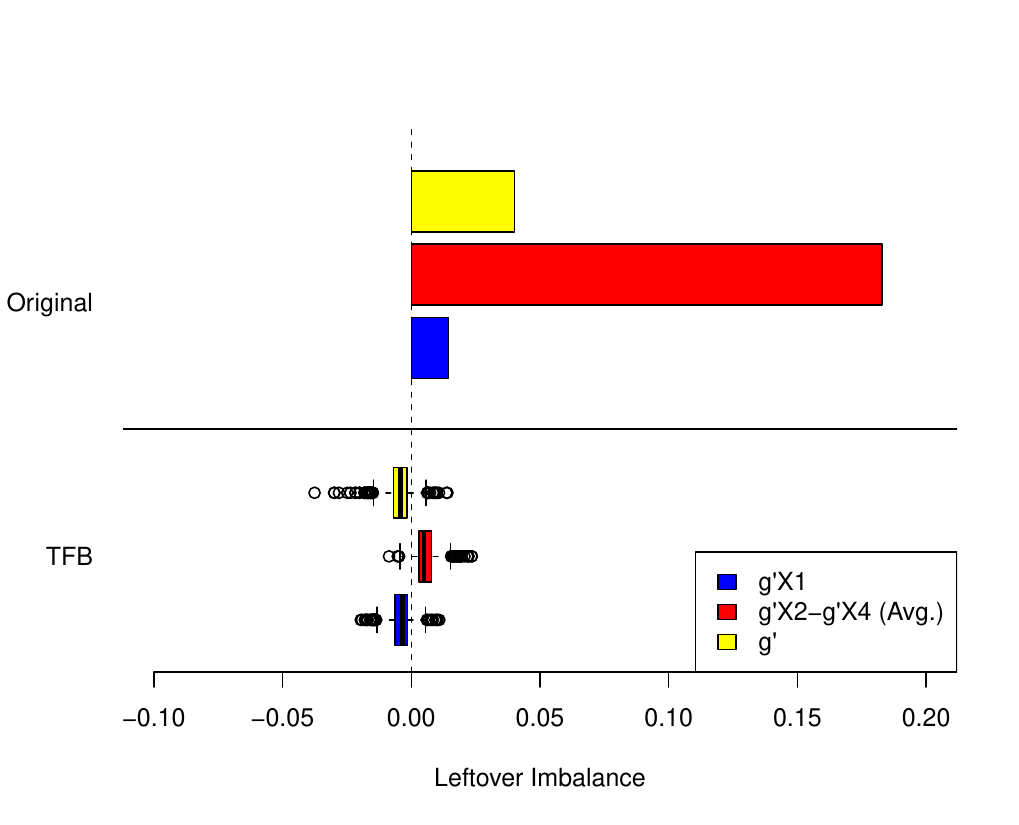} 
    \end{center}
    \vspace{-.25in}
    \subcaption{Leftover imbalance}\label{fig.glm.balance}
    \end{subfigure}
    \subcaption*{\textit{Note:} Results across 1000 draw of DGP from Appendix~\ref{app:glm} with $n=1000$. TFB with $f_0$ estimated by a probit regression of $Y$ on $X$ is applied. \textit{(a)} Distributions of the estimates from each method, as well as their bias, RMSE, and RESS. \textit{(b)} Leftover imbalance after weighting in  $g'(X^{\top} \beta_0)$ and the $g'(X^{\top} \beta_0)X^{(\ell)}$.}
	\end{center}
	\vspace{-0.25in}
	\end{figure}

We apply the non-linear extension of TFB with GLM probit regression as the regression estimator. Note in this setting that the inverse link function $g(\cdot)$ is the standard normal CDF $\Phi (\cdot)$, and 
    \begin{align}
        X^{\top} \beta_0 = X^{(1)} + \frac{1}{10}\sum_{\ell=2}^4 X^{(\ell)} - 0.8
    \end{align}
Thus, this application of TFB should seek balance in
    \begin{align}
        g'(X^{\top} \beta_0)X &= \phi (X^{\top} \beta_0) X \nonumber \\
        &= \phi \biggr(X^{(1)} + \frac{1}{10}\sum_{\ell=2}^4 X^{(\ell)} - 0.8\biggr) X
    \end{align}
where $\phi (\cdot)$ is the probability density function of a standard normal random variable. Figure~\ref{fig.glm} reports how this application of TFB performs in 1000 iterations of this DGP with $n=1000$, and investigates TFB's average leftover imbalance in $g'(X^{\top} \beta_0)X$ after weighting. 

Because TFB here has correctly specified $f_0$, it is not surprising that the TFB estimator is effectively unbiased in Figure~\ref{fig.glm.bias}. More interesting is how TFB chooses to leave imbalance in certain dimensions of $g'(X^{\top} \beta_0)X$, shown in Figure~\ref{fig.glm.balance}. As does TFB-K in DGP 1, TFB here chooses to leave negative imbalances in  $g'(X^{\top} \beta_0)$ and $g'(X^{\top} \beta_0)X^{(1)}$, which are of the opposite sign of the original imbalance in these components before weighting. TFB does this to counteract leftover imbalance $g'(X^{\top} \beta_0)X^{(\ell)}$ for $\ell=$2, 3, and 4, because those components have a much higher starting imbalance than do the other dimensions of $g'(X^{\top} \beta_0)X$, and are thus harder to get good balance in.

\subsection{Danger of misspecifying $f_0$ in TFB}\label{app:miss}

In DGP 1 from Section~\ref{sec:dgp1} and DGP 2 from Appendix~\ref{sec:dgp2}, the applications of TFB either correctly specify $f_0$ or (in the case of TFB-K) $f_0$ is well-approximated by a $\hat{f}_0$ from KRLS. This section demonstrates the bias that can occur when TFB-L  incorrectly specifies $f_0$ (and its implied propensity score model is also incorrect, per Appendix~\ref{app:tfb.dr}). 

An extension to DGP 2 is considered here. As in DGP 2, $X$ contains three types of variables: (i) ``confounders", $Z$, that are related to both $D$ and $Y$; (ii) ``distractors", $A$, that are related to $D$, and are difficult to balance, but are unrelated to $Y$; and (iii) ``extraneous" variables, $U$, that are independent of both $D$ and $Y$. The $Z$, $A$, and $U$ are again generated as
	\begin{align}
		Z_i  \overset{iid}{\sim} \mathcal{N} (0, \mathbf{I}_4) \  , \ A_i \ | \ D_i \overset{iid}{\sim} \mathcal{N} ( D_i \vec{\1}_5, \mathbf{I}_5) \ , \ \text{and} \ U_i \overset{iid}{\sim} \mathcal{N} ( 0, \mathbf{I}_{10})
	\end{align}
The difference from DGP 2 comes in the propensity store and model for the outcome. Instead of both the log odds of treatment and $f_0$ being linear in $Z$, the key transformation of $Z$,
	\begin{align}
        \phi(Z) = \frac{ (Z^{(1)})^3 + 1 }{ (Z^{(2)})^2 + 1 }
    \end{align}
is very influential in both models:
	\begin{align}
		\mathrm{log} \frac{\pi(X)}{1 - \pi(X)} &= \frac{1}{5} (Z^{(1)}  + Z^{(2)}  + Z^{(3)}  + Z^{(4)}) + \frac{3}{10} ( \phi(Z)  - 0.66 )  \\
        Y &= 8 Z^{(1)}  + 4 Z^{(2)}  + 2 Z^{(3)}  + 1 Z^{(4)} + 5 \phi(Z) + \epsilon, \ \ \ \epsilon_i\overset{iid}{\sim} N(0, 20.7^2)
    \end{align}
Centering $\phi(Z)$ by 0.66 in the propensity score allows $p(D=1) \approx 0.50$, and choosing $\var(\epsilon) = 20.7^2$ allows the model $R^2 = \frac{\var(Y - \epsilon)}{\var(Y)}$ to be approximately 0.50. 

\begin{figure}[!h]
	\vspace{0.15in}
	\begin{center}
	\caption{Comparison of estimators that misspecify $f_0$ in the DGP from Appendix~\ref{app:miss}}\label{fig.miss}
    \vspace{-.25in}
    \begin{subfigure}{.45\textwidth}
    \begin{center}
    \includegraphics[scale=0.42]{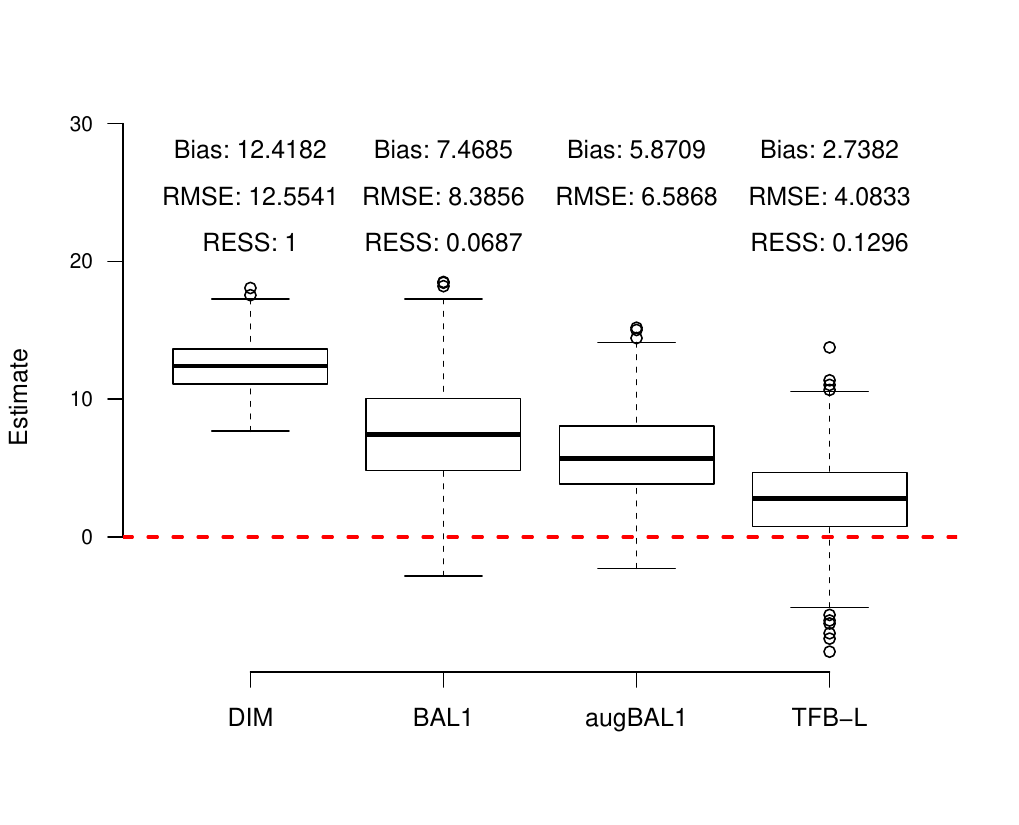} 
    \end{center}
    \vspace{-.25in}
    \subcaption{$n=1000$}\label{fig.miss.n1}
    \end{subfigure}
    \begin{subfigure}{.45\textwidth}
    \begin{center}
    \includegraphics[scale=0.42]{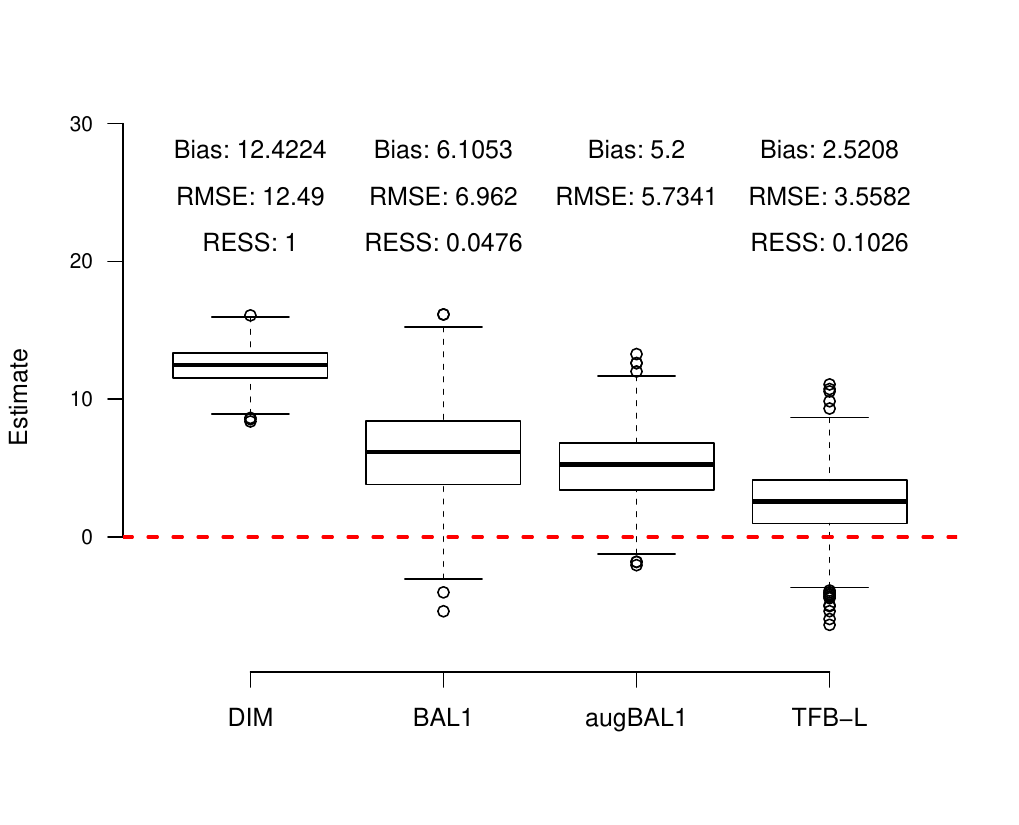} 
    \end{center}
    \vspace{-.25in}
    \subcaption{$n=2000$}\label{fig.miss.n2}
    \end{subfigure}
    \subcaption*{\textit{Note:} Results across 1000 draws of the DGP from Appendix~\ref{app:miss} for $n=1000$ and $n=2000$. Distributions of the estimates from each method, as well as their bias, RMSE, and RESS.}
	\end{center}
	\vspace{-0.25in}
	\end{figure}

As in Appendix~\ref{sec:dgp2}, TFB-L is compared to an application of \ref{eq:approxbal_w1} that does not prioritize balance in any dimensions of $X$. Also included in the comparison is the augmented form of  BAL1 (augBAL1), which uses the LASSO for its regression function. All methods are implemented in the same way as in Appendix~\ref{sec:dgp2}. 
Note that this involves adding to $X$ all its squares and pairwise interactions, which totals 209 variables. However, this does not include $\phi(Z)$, meaning that all methods should expect bias. This bias is confirmed in Figure~\ref{fig.miss} --- all methods, including TFB-L, show noticeable bias at $n=1000$ and $n=2000$.

\subsection{Obviating complete overlap}\label{app:overlap}

Section~\ref{subsec:wprops} discusses how the traditional assumption of \textit{complete} overlap in the conditional distributions of $X$, often written as (\ref{eq:psoverlap}), is not required for \ref{eq:ewc}. In fact all that is needed is for the average $f_0 (X)$ in the treated group to be in the range of
the $f_0 (X)$ in the control group. For example, consider the following DGP:
    \begin{align}\label{eq:dgp_overlap}
        Y &= X^2 - X + 0.25 + \epsilon \nonumber \\
        \text{where} \ \ & X_i \overset{iid}{\sim} N(0, 1), \ \  \epsilon_i \overset{iid}{\sim} N(0, 1) \nonumber \\
        \text{and} \ \ & D = \1 \{X > 0 \}
    \end{align}
Here this no overlap in the distribution of $X$ --- the control group only has values less than 0, and the treated group only has values greater than 0. Thus, both propensity score methods and any balancing method that requires exact balance on $X$ would be impossible. However, $\E[ f_0(X) \ | \ D=1] \approx 0.45$, and the $f_0(X) $ among the control group have a range of $(0.25, \infty)$. Thus, \ref{eq:ewc} is often possible.

\begin{figure}[!h]
	\vspace{0.15in}
	\begin{center}
	\caption{Performance of TFB in the DGP in (\ref{eq:dgp_overlap}) }\label{fig.overlap}
    \vspace{-.25in}
    \begin{subfigure}{.45\textwidth}
    \begin{center}
    \includegraphics[scale=0.43]{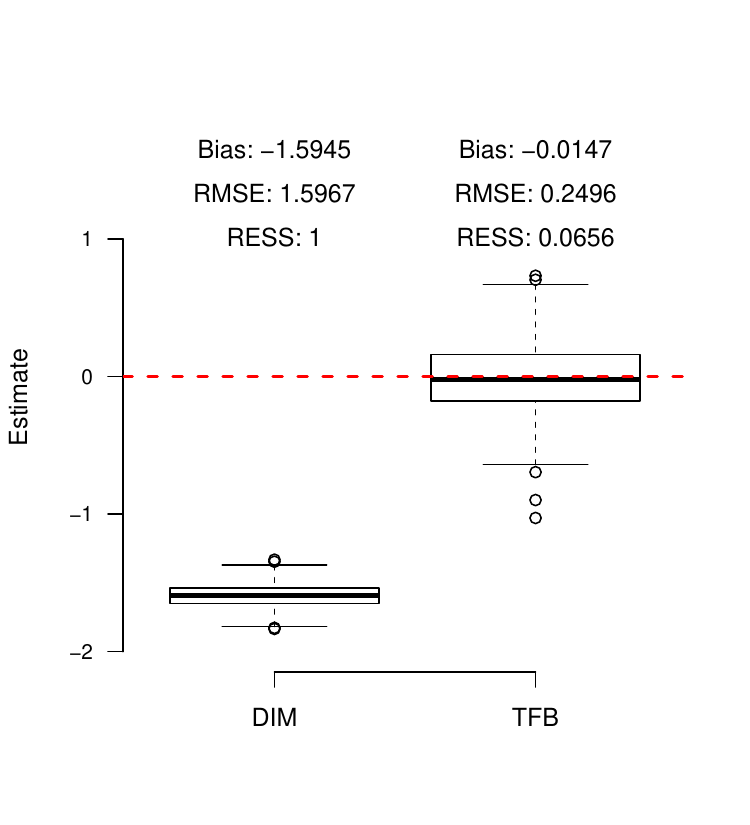} 
    \end{center}
    \vspace{-.25in}
    \subcaption{Bias, RMSE, and RESS}\label{fig.overlap.bias}
    \end{subfigure}
    \begin{subfigure}{.45\textwidth}
    \begin{center}
    \includegraphics[scale=0.43]{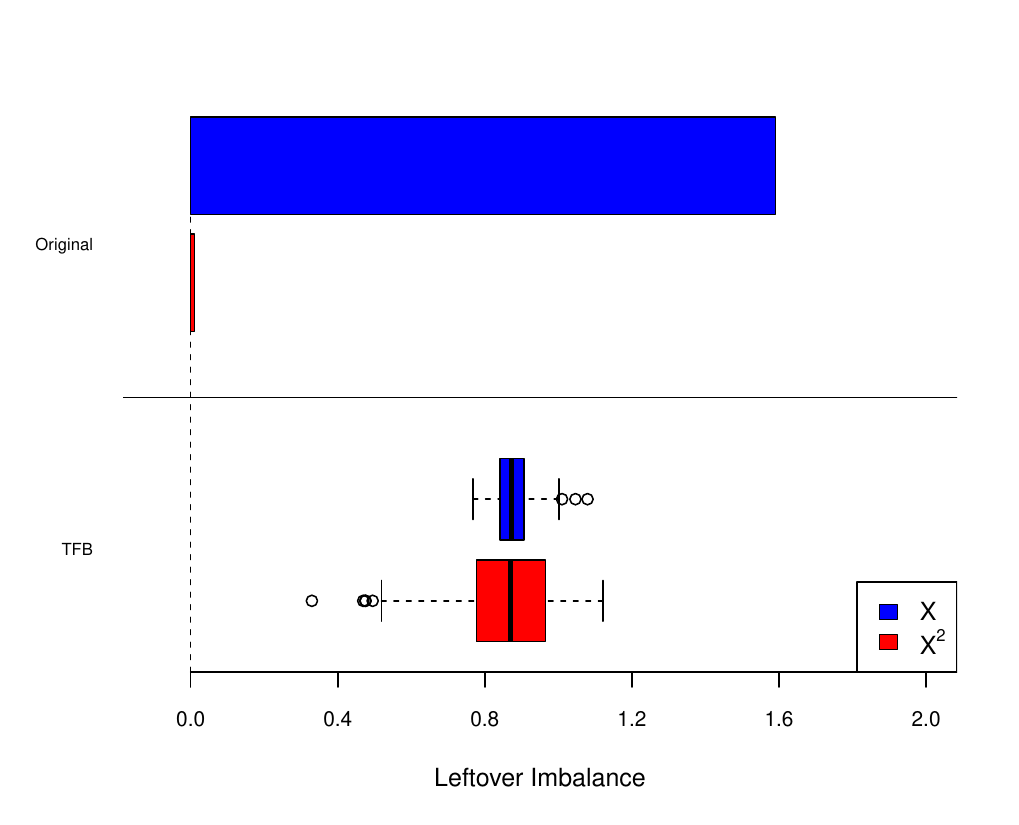} 
    \end{center}
    \vspace{-.25in}
    \subcaption{Leftover imbalance}\label{fig.overlap.balance}
    \end{subfigure}
    \subcaption*{\textit{Note:} Results across 1000 draw of (\ref{eq:dgp_overlap}) with $n=2000$. TFB with $f_0$ estimated by an OLS of $Y$ on $X$ and $X^2$ is applied. \textit{(a)} Distributions of the estimates from each method, as well as their bias, RMSE, and RESS. \textit{(b)} Leftover imbalance after weighting. The initial imbalance in $X^2$ is exactly 0.}
	\end{center}
	\vspace{-0.25in}
	\end{figure}

Consider now applying TFB to this DGP. Here, for simplicity, TFB will use an OLS that regresses $Y$ on $(X, X^2)$ to estimate $f_0$, and the sandwich variance estimator (\citealp{white1980heteroskedasticity}) for $\hat{V}_{\beta_0}$. Figure~\ref{fig.overlap} reports TFB's performance in this DGP, and investigates the leftover imbalance in $X$ and $X^2$ after weighting. Despite the lack of overlap in $X$, TFB is effectively unbiased at $n=2000$. Per Figure~\ref{fig.overlap.balance}, TFB achieves this by \textit{inducing} imbalance in $X^2$ to offset imbalance in $X$. 

We make two key disclaimers about this simulated example. First, even though OLS with a small number of regressors was applied here for TFB, we still do not recommend this practice in general. In this example, the OLS perfectly specifies $f_0$, which is unlikely in practice. Second, this simulated setting is even more unrealistic in that TFB's outcome model correctly extrapolates in regions of the $X$ distribution where there is no overlap. As discussed in the main text, we largely recommend applying nonparametric modeling techniques with TFB, which often do not extrapolate by design. For example, KRLS with a Gaussian kernel (\citealp{hainmueller2014kernel})  tends to predict the average $Y$ for  values of $X$ that are beyond the data, so this modeling technique would struggle with TFB in settings where traditional overlap is violated. Nevertheless, despite these disclaimers, this simulated example still has an important result that has implications for the assumptions required for TFB's asymptotic properties. It is clear that, depending on the consistency of the regression estimator $\hat{f}_0$, one may only need a very weak overlap assumption in order for TFB to be asymptotically unbiased. Note also that there are nonparametric techniques that can extrapolate, such as Gaussian processes with non-stationary kernels (e.g., see \citealp{cho2026inference}). We also recommend LASSO regression with TFB, after adding to $X$ numerous non-linear transformations, which can also extrapolate.


\subsection{Heteroscedastic error in DGP 1}\label{app:hetero}

This section considers an extension to DGP 1 from Section~\ref{sec:dgp1} in which the error in the model for $Y$ in  (\ref{eq:dgp1_y}) is heteroscedastic. Here,  $X$, $Z$, and $D$ are generated exactly as they are in DGP 1. However the model for $Y$ is now:
	\begin{align}\label{eq:dgp1_y_hetero}
		Y = 10 Z^{(1)} +\sum_{\ell = 2}^4 Z^{(\ell)} + \epsilon, \ \ \ \epsilon_i \ | \ X_i \overset{iid}{\sim} N \biggr( 0, \frac{f_0 (X_i)^2}{8} \biggr)
	\end{align}
where the conditional variance of $\epsilon$ is now a function of $X$. Furthermore, dividing $f_0 (X)^2$ by 8 in the variance of $\epsilon$ results in a true $R^2 = \frac{\var(Y - \epsilon)}{\var(Y)}$ of approximately 0.60, just like in the original DGP 1. 

We consider the same estimators in this extension as the demonstrations on DGP 1 in Section~\ref{sec:dgp1}. The goal of this experiment is to see if TFB-K's performance, which uses $\hat{V}_{\hat{\beta}_0}$ as described in \cite{hainmueller2014kernel}, is sensitive to heteroscedastic error. As mentioned previously, \cite{hainmueller2014kernel} show this $\hat{V}_{\hat{\beta}_0}$ is only valid \textit{for inference on the coefficient vector} under homoscedastic error. 

\begin{figure}[!h]
	\vspace{0.15in}
	\begin{center}
	\caption{Performance of TFB in DGP 1 with Heteroscedastic Error}\label{fig.dgp1.hetero}
    \vspace{-.25in}
    \begin{subfigure}{.90\textwidth}
    \begin{center}
    \includegraphics[scale=0.55]{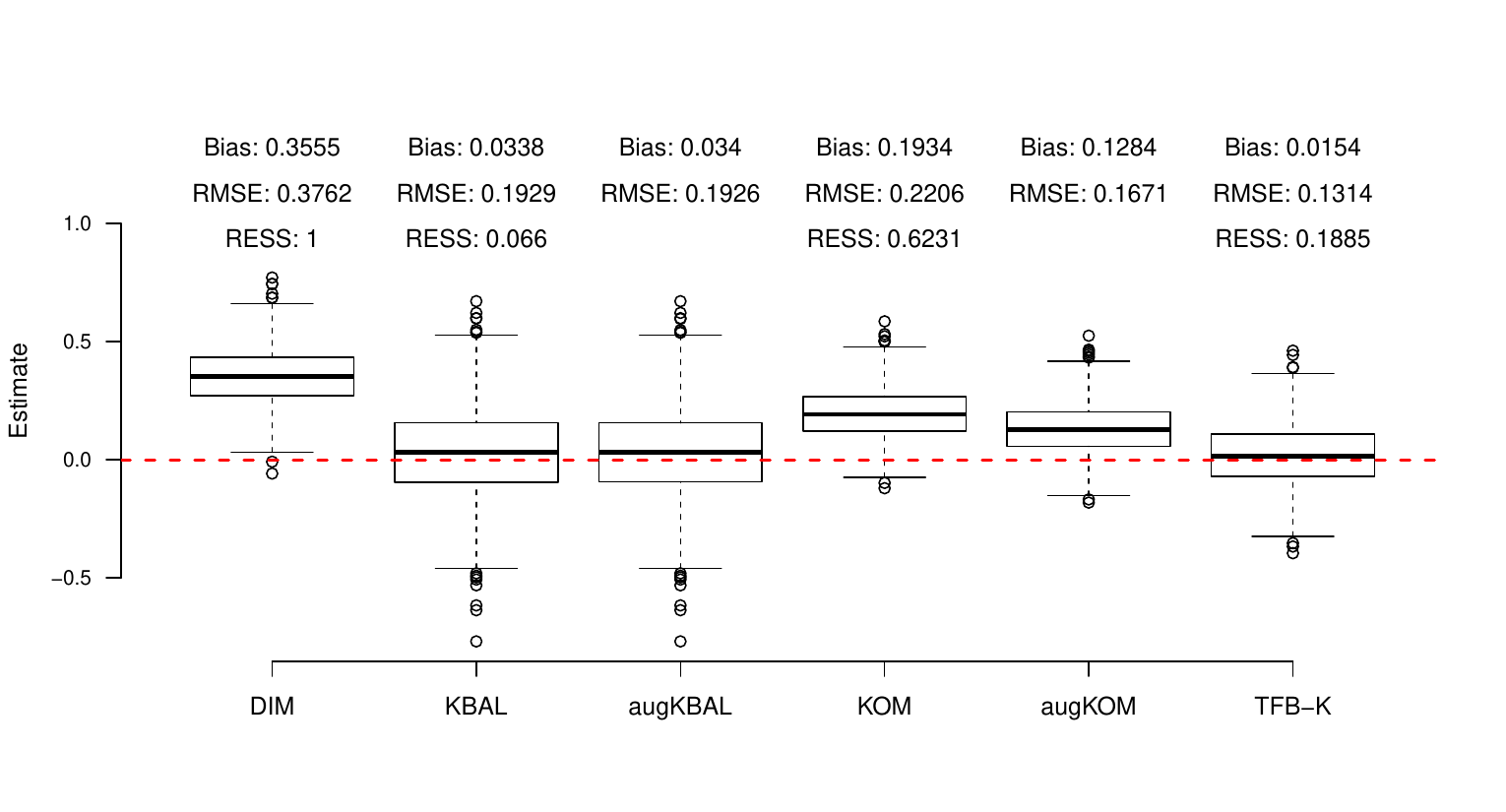} 
    \end{center}
    \vspace{-.25in}
    \subcaption{Bias, RMSE, and RESS}\label{fig.dgp1.hetero.bias}
    \end{subfigure}
    \begin{subfigure}{.90\textwidth}
    \begin{center}
    \includegraphics[scale=0.55]{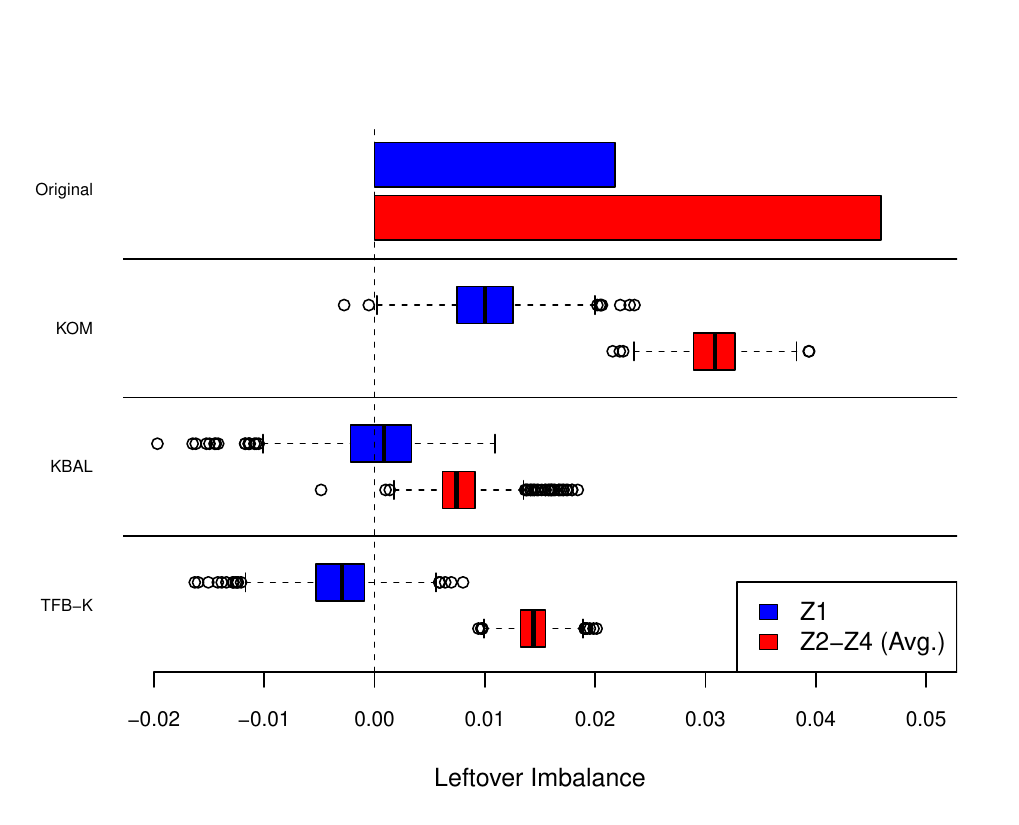} 
    \end{center}
    \vspace{-.20in}
    \subcaption{Leftover imbalance}\label{fig.dgp1.hetero.balance}
    \end{subfigure}
    \vspace{0.15in}
    
    \subcaption*{\textit{Note:} Comparison of TFB-K (see Table~\ref{tab:abbreviations}), Kernel Optimal Matching (KOM) and its augmented form (augKOM), Kernel Balancing (KBAL) and its augmented form (augKBAL), and a difference in means (DIM) across 1000 draws from DGP 1 with heteroscedastic error with $n=1000$. augKOM and augKBAL use KRLS for $\hat{f}_0$. \textit{(a)} Distributions of the estimates from each method, as well as their bias, RMSE, and RESS. \textit{(b)} Leftover imbalance from each method after weighting, where imbalance for $Z^{(2)}$, $Z^{(3)}$, and $Z^{(4)}$ is averaged at each iteration. }
	\end{center}
	\vspace{-0.25in}
	\end{figure}

\begin{figure}
\vspace{0.15in}
\begin{center}
\caption{Demonstration of TFB-K in DGP 1 with Heteroscedastic Error for varying $q$}\label{fig.TFB.q1.dgp1hetero}
    \vspace{-.25in}
    \begin{subfigure}{.45\textwidth}
    \begin{center}
    \includegraphics[scale=0.5]{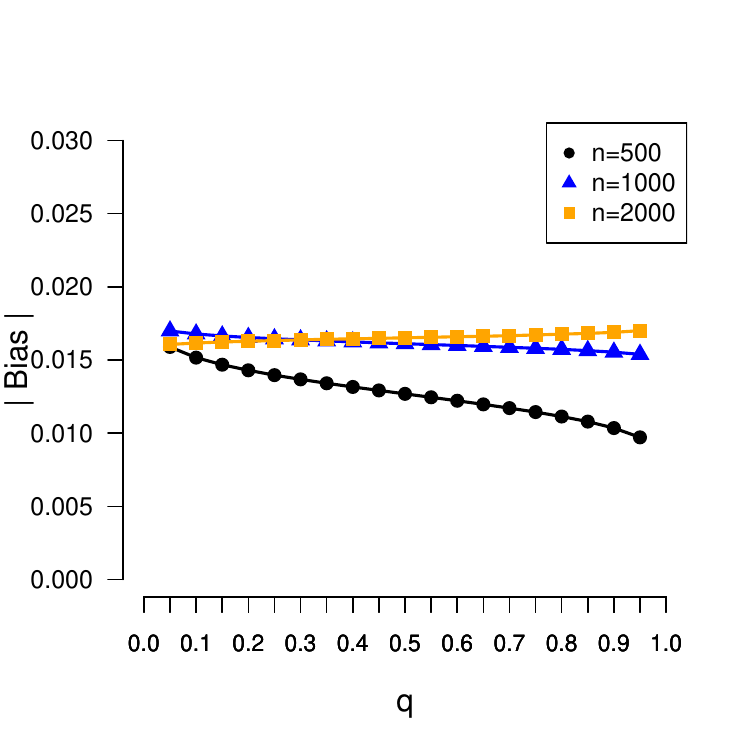} 
    \end{center}
    \vspace{-.25in}
    \subcaption{$|$Bias$|$}
    \end{subfigure}
    \begin{subfigure}{.45\textwidth}
    \begin{center}
    \includegraphics[scale=0.5]{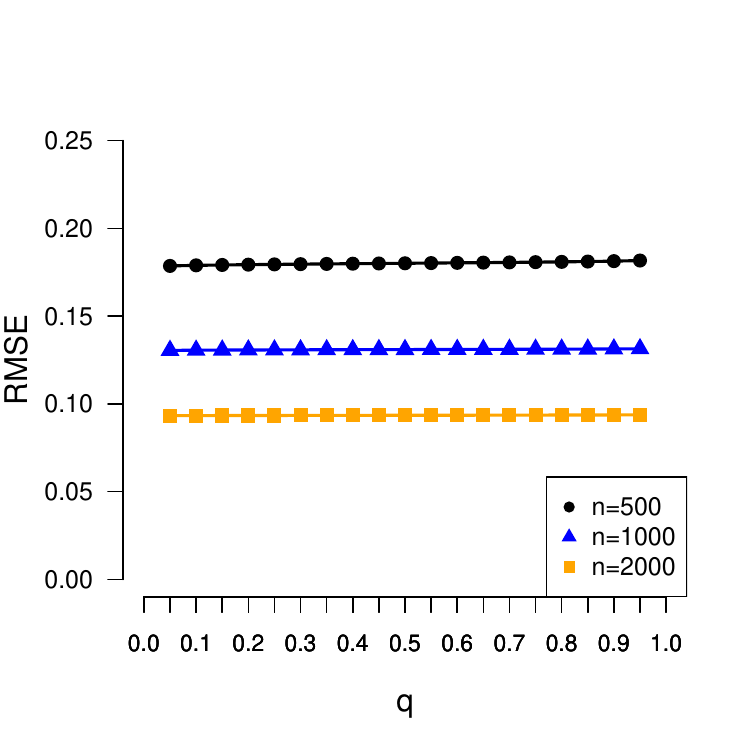} 
    \end{center}
    \vspace{-.25in}
    \subcaption{RMSE}
    \end{subfigure}
    \subcaption*{\textit{Note:} Results across 1000 draws of DGP 1 with heteroscedastic error. $q$ takes values from 0.05 to 0.95 in intervals of 0.05. TFB-K (see Table~\ref{tab:abbreviations}) is applied.  }
\end{center}
\vspace{-0.25in}
\end{figure}

Figure~\ref{fig.dgp1.hetero} compares the estimators' performance in this extension, and investigates the leftover imbalance in the $Z^{(\ell)}$ after weighting. As in Section~\ref{sec:dgp1}, TFB-K still has the lowest RMSE and shows little bias. Further, comparing Figure~\ref{fig.dgp1.hetero.balance} to Figure~\ref{fig.dgp1.balance}, we see that TFB-K shows similar tendencies in the original DGP 1 and this extension --- it leaves imbalance in $Z^{(1)}$ of the opposite sign to offset imbalance in the other $Z^{(\ell)}$. Figure~\ref{fig.TFB.q1.dgp1hetero} then examines the bias and RMSE of TFB-K in this extension for varying $q$. Comparing Figure~\ref{fig.TFB.q1.dgp1hetero} to Figure~\ref{fig.TFB.q1}, we again see very similar trends. With smaller $n$, there a bias-variance trade-off in that the bias decreases as $q$ increases, but the RMSE is relatively stable. When $n=2000$, both the bias and RMSE are quite stable across all values of $q$.

\clearpage

\newpage

\subsection{Performance of the standard LASSO with TFB-L in DGP 2}\label{app:lasso_dgp2}

\begin{figure}[!h]
	\vspace{0.15in}
	\begin{center}
	\caption{Comparison of standard and modified LASSO with TFB-L in DGP 2 }\label{fig.mlasso}
    \vspace{-.25in}
    \begin{subfigure}{.32\textwidth}
    \begin{center}
    \includegraphics[scale=0.50]{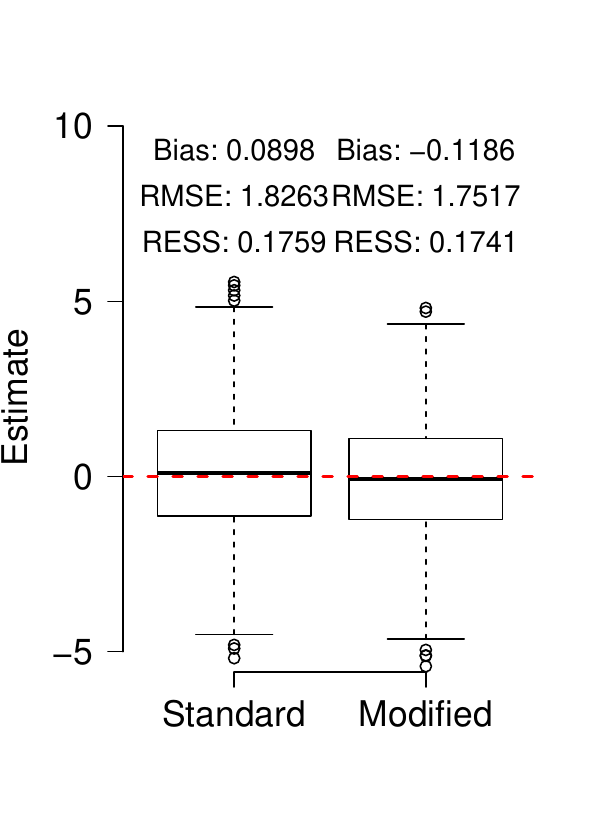} 
    \end{center}
    \vspace{-.25in}
    \subcaption{$n=500$}
    \end{subfigure}
    \begin{subfigure}{.32\textwidth}
    \begin{center}
    \includegraphics[scale=0.50]{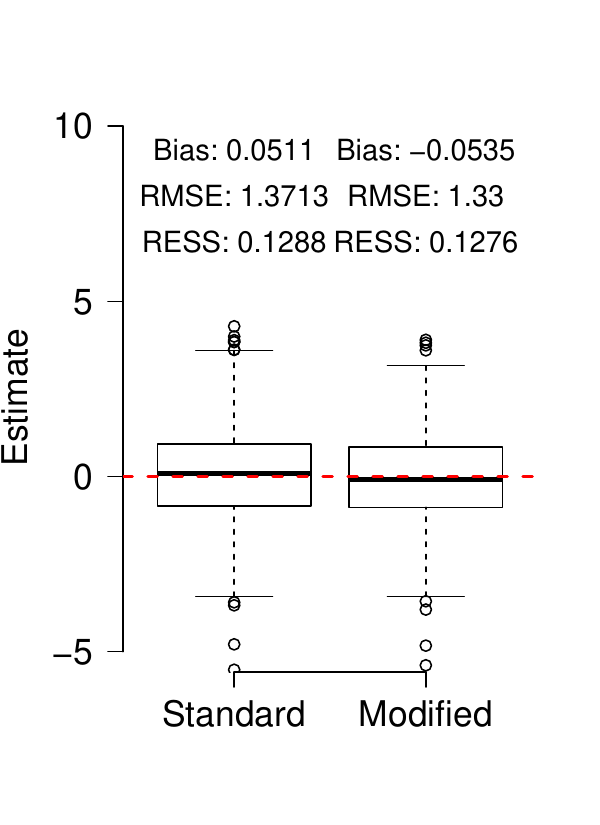} 
    \end{center}
    \vspace{-.25in}
    \subcaption{$n=1000$}
    \end{subfigure}
    \begin{subfigure}{.32\textwidth}
    \begin{center}
    \includegraphics[scale=0.50]{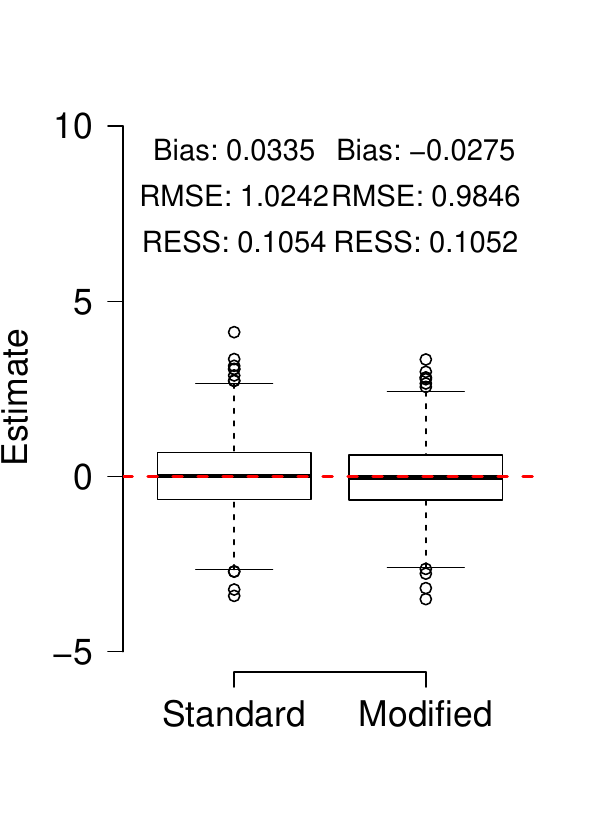} 
    \end{center}
    \vspace{-.25in}
    \subcaption{$n=2000$}
    \end{subfigure}
    \subcaption*{\textit{Note:} Results across 1000 draws of DGP 2 with \textit{(a)} $n=500$, \textit{(b)} $n=1000$, and \textit{(c)} $n=2000$. Distributions of TFB-L estimates when the standard (``Standard") and modified (``Modified") LASSO are used, along with the corresponding bias, RMSE, and RESS.}
	\end{center}
	\vspace{-0.25in}
	\end{figure}

In this appendix, we compare the performance of the modified LASSO (\citealp{chatterjee2010asymptotic}) and the standard LASSO with TFB-L in DGP 2 (from Appendix~\ref{sec:dgp2}). Figure~\ref{fig.mlasso} reports the distributions of the estimates for each LASSO type for $n \in \{500, 1000, 2000\}$. In general, both the standard and the modified LASSO perform very well in DGP 2, and show few differences on average.

\begin{figure}[!h]
\vspace{0.15in}
\begin{center}
\caption{Diagonal elements of $\hat{V}_{\beta_0}^{\frac{1}{2}}$ for TFB-L with the standard LASSO in DGP 2}\label{fig.dgp2.vdiag.mlasso}
	\vspace{-0.25in}
    \begin{center}
    \includegraphics[scale=0.50]{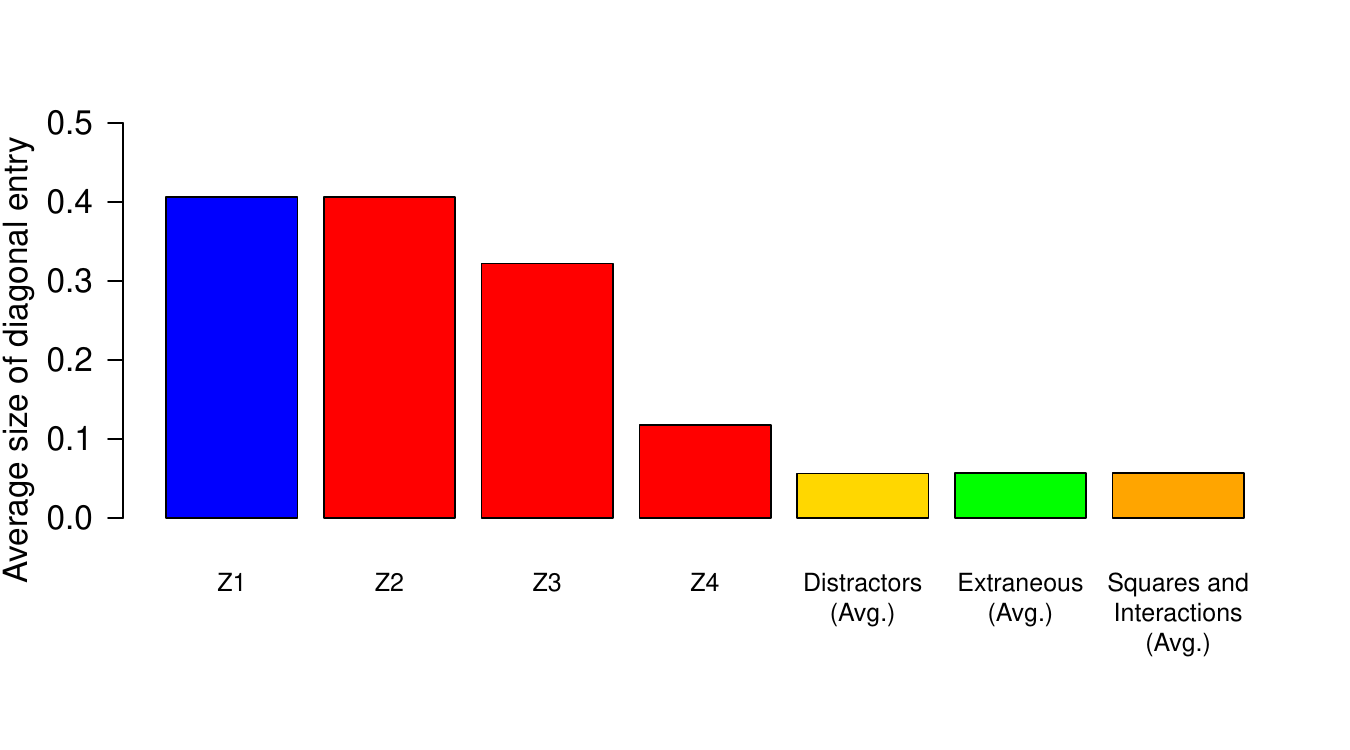} 
    \end{center}
    \vspace{-0.15in}
    \subcaption*{\textit{Note:} Each variable's diagonal entry in $\hat{V}_{\beta_0}^{\frac{1}{2}}$ is averaged across 1000 draws of DGP 2 with $n=1000$. The average entries of (i) the distractors, (ii) the extraneous variables, and (iii) the squares and pairwise interactions of $X$ are then averaged across their 5, 10, and 190 dimensions, respectively.}
\end{center}
\vspace{-0.25in}
\end{figure}

As with the modified LASSO, Figure~\ref{fig.dgp2.vdiag.mlasso} shows that the distractors', extraneous variables', and polynomial terms' diagonal elements in $\hat{V}_{\beta_0}$ are small even when using the standard LASSO. As a result, TFB-L again devalues balance in those variables. The diagonal elements for the $Z^{(\ell)}$ in $\hat{V}_{\beta_0}$ again increase with the size of their coefficients in the model for $Y$. 

\clearpage
\newpage


\section{Sample splitting}\label{subsec:samplesplitting}

Recall that the central motivation for TFB is to minimize (an approximation of) the EWC Bias. Further, Proposition~\ref{prop:ewc} states that the bias of $\hat{\tau}_{\mathrm{wdim}}$ is exactly the expectation of the EWC Bias. However, this result requires that each pair of $w_i$ and $Y_i$ be independent of each other, given $\mathbf{X}$ and $\mathbf{D}$. This is violated by TFB as defined in Section~\ref{subsec:tfb.weights}, but can be remedied by sample splitting (as in \citealp{chernozhukov2018double}), if desired. Further, because TFB can be written as \ref{eq:approxbal_w1} (see Appendix~\ref{app:tfb.dr}), sample splitting allows TFB to fit into the ``cross-balancing" framework studied by \cite{jin2025cross}. 
The cross-balancing framework has several desirable asymptotic guarantees: for example, under certain conditions, $\sqrt{n}$-consistency; double-robustness, given the connection between balancing weights and propensity score weights; and a property similar to ``rate double robustness" (e.g., \citealp{smucler2019unifying}), as in \cite{farrell2015robust} and \cite{chernozhukov2018double}, where an estimator is $\sqrt{n}$-consistent as long as the product of the rates of convergence of $\hat{f}_0$ and $\hat{\pi}$ is above a certain threshold. Though, \cite{chen2022debiased} show that, under certain conditions, sample splitting is not required for the ``double/debiased machine learning" estimators studied by \cite{chernozhukov2018double} to be $\sqrt{n}$-consistent and asymptotically normal. Thus, it seems plausible that, under certain conditions, TFB does not require sample splitting to obtain the same proven asymptotic guarantees given by \cite{jin2025cross} for cross-balancing.

Nevertheless, because of any potential theoretical guarantees that may come from sample splitting, we explore in this appendix the possibility of sample splitting in TFB. In Section~\ref{sec:samplesplitting_sim}, we compare TFB with and without sample splitting. We then demonstrate sample splitting with TFB on real data in Appendix~\ref{sec:applications}. Our \texttt{R} package \href{https://github.com/bleth/tfb}{\texttt{tfb}} can implement sample splitting with TFB, if desired.

Sample splitting with TFB requires randomly splitting the full sample into two non-overlapping (sub-)samples: Samples 1 and 2. Let the superscripts $(s1)$ and $(s2)$ denote values that come from Samples 1 and 2, respectively (e.g., $\hat{\textbf{w}}_{\mathrm{TFB}}^{(s1)})$. First, use Sample 2 to obtain $(\hat{\beta}^{(s2)}_0, \hat{V}^{(s2)}_{\beta_0})$, and apply them to TFB in Sample 1 to form $\hat{\textbf{w}}^{(s1)}_{\mathrm{TFB}}$ and $\hat{\tau}_{\mathrm{wdim}}^{(s1)} (\hat{\textbf{w}}_{\mathrm{TFB}}^{(s1)})$. Next, switch the roles of the two sub-samples to obtain $(\hat{\beta}^{(s1)}_0, \hat{V}^{(s1)}_{\beta_0})$ from Sample 1, and $\hat{\textbf{w}}_{\mathrm{TFB}}^{(s2)}$ and  $\hat{\tau}_{\mathrm{wdim}}^{(s2)} (\hat{\textbf{w}}_{\mathrm{TFB}}^{(s2)})$ from Sample 2. The final estimate is then the average of the estimates from each sub-sample:
    \begin{align}\label{eq:tfb_ss_est}
        \hat{\tau}_{\mathrm{wdim}}^{(\mathrm{TFB})} = \frac{ \hat{\tau}_{\mathrm{wdim}}^{(s1)} (\hat{\textbf{w}}_{\mathrm{TFB}}^{(s1)}) + \hat{\tau}_{\mathrm{wdim}}^{(s2)} (\hat{\textbf{w}}_{\mathrm{TFB}}^{(s2)}) }{2}
    \end{align}
Note that while TFB's weights with sample splitting still depend on the observed outcomes, the weights in Sample 1 are independent of the outcomes \textit{in Sample 1} (given $\textbf{X}$ and $\textbf{D}$), and the weights in Sample 2 are independent of the outcomes \textit{in Sample 2} (given $\textbf{X}$ and $\textbf{D}$). Thus, each pair of $w_i$ and $Y_i$ is independent of each other (given $\textbf{X}$ and $\textbf{D}$), as required in the conditions for Proposition~\ref{prop:ewc}. 

Finally, when sample-splitting in real-data applications, it is prudent to examine estimates from many random splits of the sample. However, if a single estimate is required for reporting, \cite{chernozhukov2018double} suggest the median. We demonstrate this procedure in Section~\ref{sec:applications}.


\subsection{Comparing TFB with and without sample splitting}\label{sec:samplesplitting_sim}

\begin{figure}[!h]
	\vspace{0.15in}
	\begin{center}
	\caption{Standardized RMSE of TFB with and without sample splitting in ACIC DGPs}\label{fig.SS}
    \vspace{-.25in}
    \begin{subfigure}{.45\textwidth}
    \begin{center}
    \includegraphics[scale=0.50]{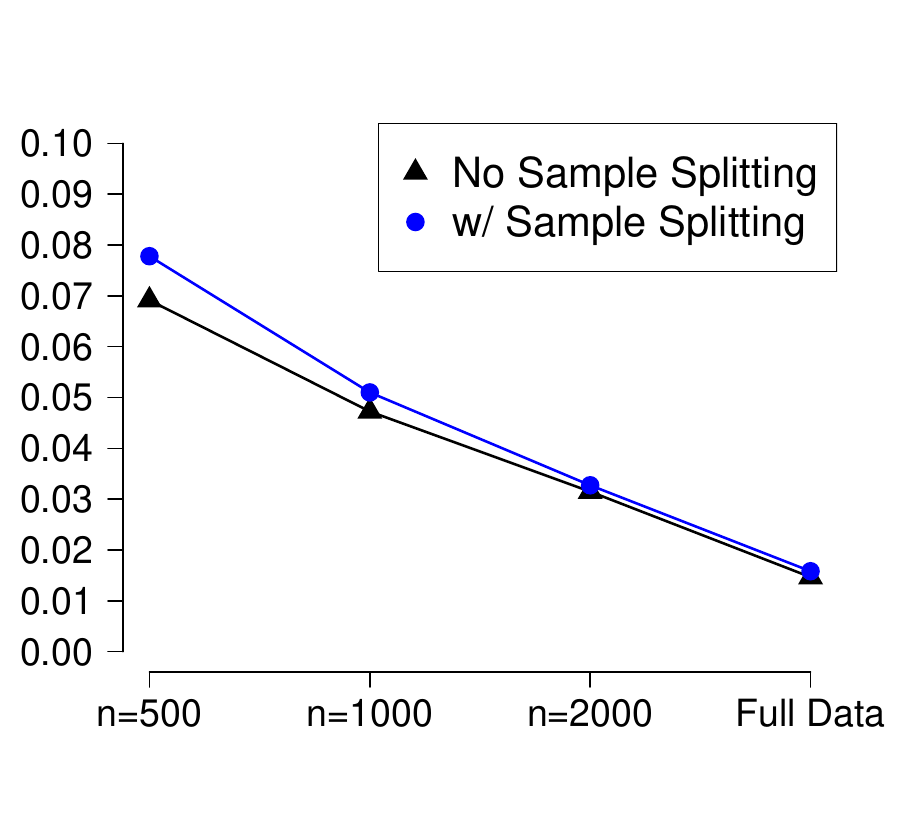} 
    \end{center}
    \vspace{-.25in}
    \subcaption{TFB-BART}\label{fig.SS.bart}
    \end{subfigure}
    \begin{subfigure}{.45\textwidth}
    \begin{center}
    \includegraphics[scale=0.50]{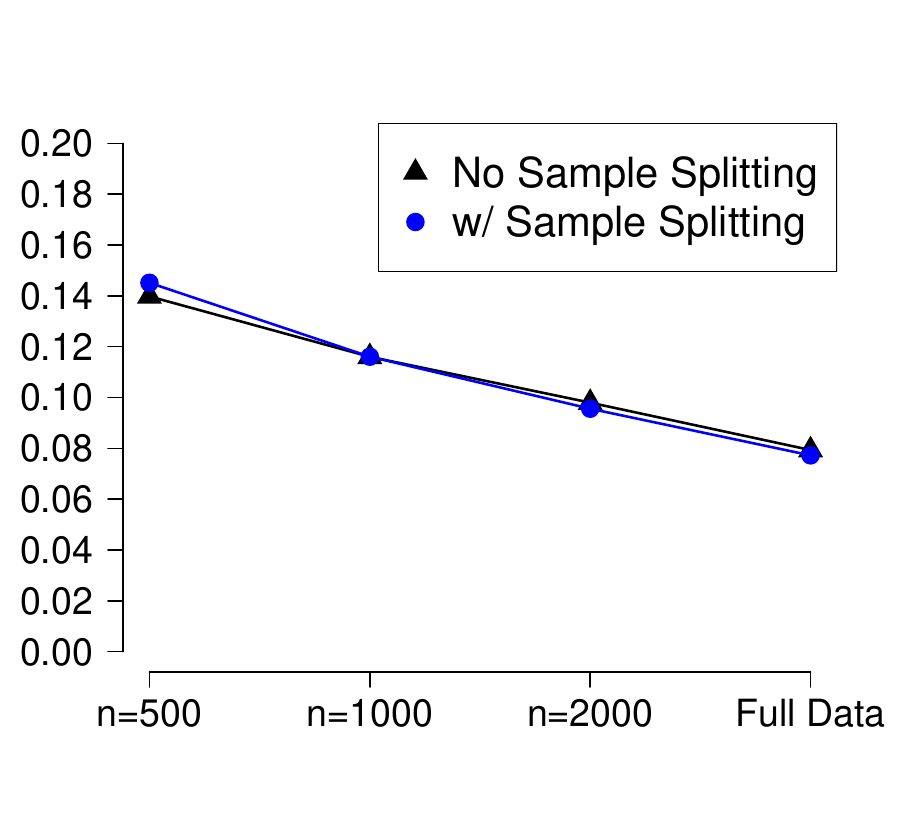} 
   \end{center}
    \vspace{-.25in}
    \subcaption{TFB-K}\label{fig.SS.krls}
    \end{subfigure}
    \subcaption*{\textit{Note:} Standardized RMSE over the 7700 ACIC challenge datasets for TFB without sample splitting (black triangles) and with sample splitting (blue circles) across random samples of size $n \in \{500, 1000, 2000, 4802\}$. The full datasets have $n=4802$. Note that the scale of the $y$-axis is larger when using TFB-K (Figure~\ref{fig.SS.krls}).}
	\end{center}
	\vspace{-0.25in}
	\end{figure}

In this section, we compare TFB with and without sampling splitting in the ACIC challenge DGPs from Section~\ref{sec:acic}. We again apply TFB-K and TFB-BART. Further, in addition to using all $n=4802$ observations, we also examine TFB's performance on random subsets of each dataset of size $n \in \{500, 1000, 2000\}$. Figure~\ref{fig.SS} reports the standardized RMSE for TFB with and without sampling splitting at each sample size.

Figure~\ref{fig.SS} shows that TFB's performance with and without sample splitting converges as the sample size increases. However, TFB with sample splitting has higher RMSE in smaller samples. This is because with sample splitting the required $\hat{f}_0$ are based on half of the data, and thus are often meaningfully worse estimates of $f_0$ when $n$ is small (e.g., $n=500$). 

Additionally, note that we have applied TFB with sample splitting to all of the other demonstrations and simulations in Section~\ref{subsec:tfb.demonstrations} and Appendix~\ref{app:simulations}. We do not see any meaningful differences between TFB with and without sample splitting in these applications. We do not include these results here, but they are available upon request.

Putting all of this evidence together, in terms of finite-sample performance, we do not find a compelling reason to implement sample splitting with TFB. Further, sample splitting complicates analysis. While there could be theoretical, asymptotic reasons for implementing sample splitting (e.g., \citealp{jin2025cross}), we leave deeper investigations of this to future work. At the same time, when sample sizes are large, there does not appear to be much risk in sample splitting. One potential way of checking whether sample splitting is risky in a given dataset would be to compare the $\hat{f}_0$ from the split samples with a $\hat{f}_0$ fit to the full dataset. If the estimates are similar, then sample splitting likely poses minimal risk. If the estimates are very different, then sample splitting is likely risky.


\subsection{Application 2: National Supported Work Demonstration}\label{sec:applications}

This section assesses TFB's performance on the National Supported Work (NSW) Demonstration  data, accessible from \cite{nswweb}. We also demonstrate sample splitting with TFB here. 

NSW was a labor training program in the mid-1970s that randomly selected its participants from a group of candidates. In addition to experimental treated and control groups drawn from the program candidates, data on comparison groups of  non-candidates are available. \cite{lalonde1986evaluating} and \cite{dehejia1999causal} use these data to compare non-experimental estimates of NSW's impact on 1978 earnings (the outcome here) to the unbiased experimental estimate. Defining the treated (185 units), experimental control (260), and non-candidate comparison (2,490) groups as in \cite{dehejia1999causal} yields an experimental estimate of \$1,794 and a naive observational estimate (i.e., the unweighted difference in means between the experimental treated and non-candidate comparison groups) of -\$15,205.

\begin{figure}[!h]
	\vspace{0.15in}
	\begin{center}
	\caption{Performance on NSW data}\label{fig.nsw1}
    \vspace{-.25in}
    \begin{subfigure}{.45\textwidth}
    \begin{center}
    \includegraphics[scale=0.55]{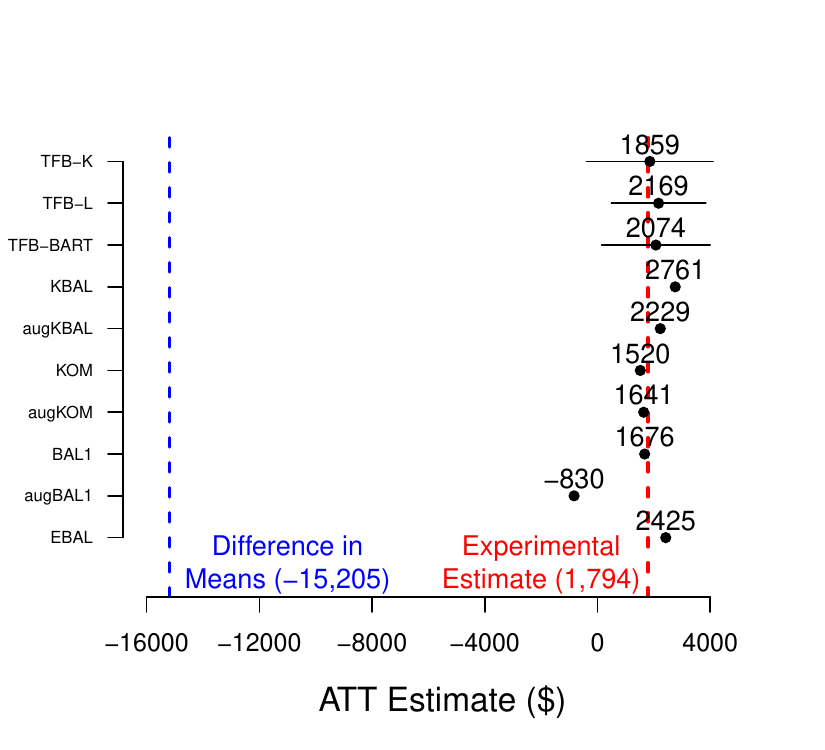} 
    \end{center}
    \vspace{-.25in}
    \subcaption{Comparison of estimates}\label{fig.nsw1a}
    \end{subfigure}
    \begin{subfigure}{.45\textwidth}
    \begin{center}
    \includegraphics[scale=0.55]{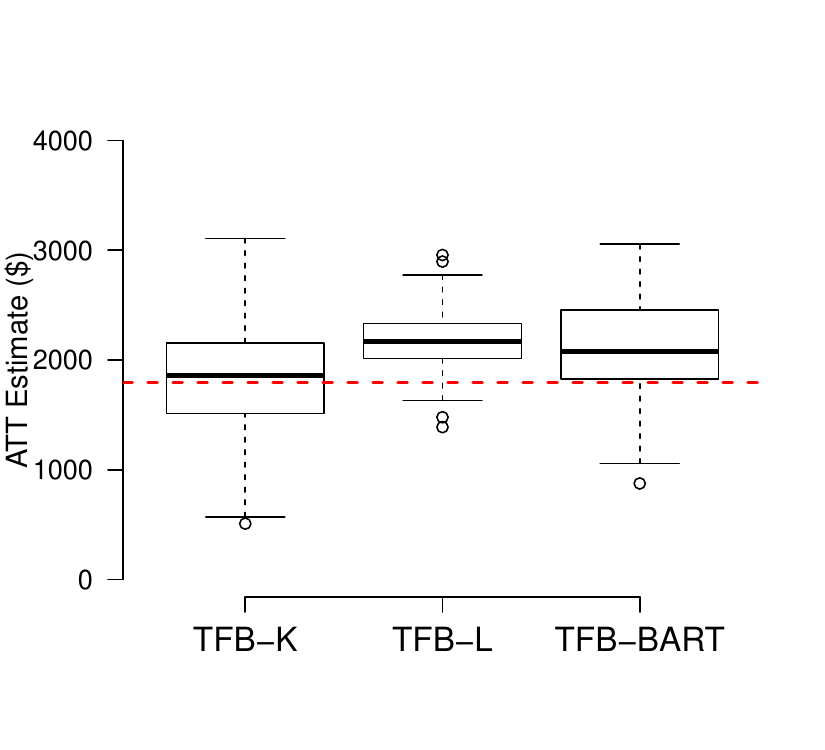} 
    \end{center}
    \vspace{-.25in}
    \subcaption{TFB estimates across 100 sample splits}\label{fig.nsw1b}
    \end{subfigure}
        \subcaption*{\textit{Note:} \textit{(a)} Comparison of TFB-K, TFB-L, TFB-BART, EBAL, (aug)BAL1, (aug)KBAL, and (aug)KOM. TFB estimates are the median estimates from 100 random splits of the sample. 95\% confidence intervals are provided for TFB estimators, with confidence intervals for the median estimate per Appendix~\ref{subsec:tfb.asymptotics}. \textit{(b)} Boxplots of TFB estimates over 100 sample splits. The dashed line is the experimental benchmark (\$1,794). }
   \end{center}
\vspace{-0.25in}
\end{figure}

The available covariates are age, years of education, an indicator of having a high school degree, 1974 and 1975 earnings, indicators of unemployment in 1974 and 1975, marital status, and several indicators of race/ethnicity. Figure~\ref{fig.nsw1} compares TFB-L, TFB-K, and TFB-BART to seven other estimators. 
The first of these estimators is Entropy Balancing (EBAL; \citealp{hainmueller2012entropy}), which is an exact balancing method (see Section~\ref{subsec:litreview}). The next three estimators are the same approximate balancing methods that were used in the demonstrations from Section~\ref{subsec:tfb.demonstrations} and Appendix~\ref{sec:dgp2}: \ref{eq:approxbal_w1} where all the $\delta^{(\ell)}$ are equal (see Section~\ref{subsec:litreview}), Kernel Balancing (KBAL; \citealp{hazlett2018kernel}), and Kernel Optimal Matching (KOM; \citealp{kallus2020generalized}). The final three estimators are augmented forms (see Section~\ref{subsec:tfb.weights}) of the three approximate balancing methods.\footnote{The \texttt{ebalance} function in \texttt{R} (from the \texttt{ebal} package) was used to implement EBAL. Each other method's specifics are consistent with Section~\ref{subsec:tfb.demonstrations} and Appendix~\ref{sec:dgp2}.} TFB-K, TFB-BART, EBAL, (aug)KBAL, and (aug)KOM are limited to the 10 untransformed covariates, while (aug)BAL1 and TFB-L additionally use the covariates' squared terms and pairwise interactions, a total of 52 covariates.\footnote{Following \cite{hainmueller2012entropy}, we exclude various squares and interactions. These are the interaction between the indicator for high school degree and years of education, the interactions of the race/ethnicity indicators, and the squares and interactions of real earnings in 1974 and 1975. }  All TFB estimators employ sample splitting here, so estimates across 100 random splits of the sample are provided in Figure~\ref{fig.nsw1b}, and Figure~\ref{fig.nsw1a} reports the median estimate.

As are the other estimators, the TFB estimators are large improvements over the naive estimate of -\$15,205. TFB-K is the closest to the experimental benchmark, and TFB-BART and TFB-L are the fifth and sixth closest (of ten). However, as this is just one realization of the data, drawing conclusions from these minute differences is challenging. It is more affirming that TFB performs comparably to other well-known estimators and that the experimental estimate falls within all of the TFB estimators' 95\% confidence intervals (as do the other estimators excluding augBAL1). 

Finally, Table~\ref{tab.nsw1} reports the standardized imbalance in the covariates for each weighting method. 
\begin{table}[!h]
\vspace{0.05in}
\caption{Leftover imbalance in NSW data}\label{tab.nsw1}

\vspace{-0.15in}

	\begin{center}
	\scriptsize
     \begin{tabular}{ l | c | c |  c  c  c | c  c  c  c  c }
      & Part. Corr. & Init. Imbal. & TFB-K & TFB-L & TFB-BART & EBAL & BAL1 & KBAL & KOM \\
     \hline
     Age & -0.08 & -0.86 & -0.13 & -0.33 & -0.21 & 0 & -0.03 & -0.10 & -0.11 \\
     \hline
     Years of Education & 0.11 & -0.58 & -0.05 & 0.04 & -0.02 & 0 & -0.05 & 0 & -0.03 \\
     \hline
     Ethnicity (Black) & -0.02 & 1.30 & 0.08 & 0.17 & 0.15 & 0 & 0.05 & 0.03 & 0.05 \\
     \hline
     Ethnicity (Hispanic) & 0.04 & 0.15 & 0.04 & 0.06 & 0.06 & 0 & 0.05 & -0.06 & 0.02 \\
     \hline
     Married & 0.04 & -1.76 & -0.08 & -0.23 & -0.15 & 0 & -0.06 & -0.04 & -0.06 \\
     \hline
     No HS Degree & 0.02 & 0.85 & 0.06 & 0.05 & 0.05 & 0 & 0.06 & 0.06 & 0.05 \\
     \hline
     Earnings (1974) & 0.18 & -1.26 & -0.01 & 0.01 & -0.01 & 0 & -0.02 & 0 & -0.05 \\
     \hline
     Earnings (1975) & 0.32 & -1.26 & -0.03 & -0.01 & -0.03 & 0 & -0.06 & 0.01 & -0.07 \\
     \hline
     Unemployed (1974) & 0.02 & 1.85 & 0.02 & 0.06 & 0.05 & 0 & 0.06 & 0 & 0.08 \\
     \hline
     Unemployed (1975) & -0.03 & 1.46 & -0.02 & -0.06 & -0.04 & 0 & 0.06 & -0.01 & 0.02 \\
     \end{tabular}

     \normalsize
	\end{center}
	\vspace{0.05in}
    \subcaption*{\textit{Note:} The ``Part. Corr." column reports the partial correlation (given the other variables) of each variable and real earnings in 1978 (i.e., the outcome)  
    within the non-candidate comparison group, and the ``Init. Imbal." column reports the  
    imbalance before weighting. The remaining columns report the 
    imbalance after weighting for each method. All covariates are standardized prior to calculating imbalances. Imbalances for TFB are averaged across 100 splits of the sample. See Table~\ref{tab:abbreviations} for descriptions of TFB-K, TFB-L, and TFB-BART.}
\vspace{-0.15in}
\end{table}
TFB shows similar tendencies here as in previous demonstrations. TFB estimators obtain very good balance on 1974 and 1975 earnings, which have the highest partial correlations with the outcome. Furthermore, because TFB-K does not estimate $f_0$ with sparsity, TFB-K obtains good balance in all of the covariates, even those with low partial correlations with the outcome. Meanwhile, TFB-L and TFB-BART leave imbalance in age, the indicator of being Black, and the indicator of being married, the latter two of which show low partial correlations with the outcome.

\newpage
\clearpage

\section{Extensions to the ATC and the ATE}\label{app:ate_atc}

Although this paper describes TFB in the context of estimating the ATT, it is easily extended to estimate the ATC and the ATE. This appendix details how to do so, and our \texttt{R} package \href{https://github.com/bleth/tfb}{\texttt{tfb}} can implement TFB to estimate the ATT, the ATC, and the ATE. 

Although it is a slight abuse of notation, let $\textbf{w} \geq 0$ be a general vector of non-negative weights. In the context of the ATC, these weights will only be defined for treated units. In the context of the ATE, these weights will be defined for the full sample. 

\subsection{Weighted difference in mean estimators for the ATC and the ATE}\label{app:estimators_ate_atc}

We first define the weighted difference in means estimators that will be used to estimate the ATC and the ATE. 

\subsubsection{ATC}
A weighted difference in means estimator for the ATC takes the form
	\begin{align}\label{eq:tau_atc}
		\hat{\tau}_{\mathrm{wdim}}^{(\mathrm{ATC})} ( \textbf{w} )  = \frac{1}{n_1} \sum_{i : D_i = 1} w_i Y_i - \frac{1}{n_0} \sum_{i:D_i = 0} Y_i
	\end{align}
where $\textbf{w} \geq 0$ and, often, $\frac{1}{n_1} \sum_{D_i=1} w_i = 1$. 

\subsubsection{ATE}
A weighted difference in means estimator for the ATE takes the form
	\begin{align}\label{eq:tau_ate}
		\hat{\tau}_{\mathrm{wdim}}^{(\mathrm{ATE})} ( \textbf{w} ) = \frac{1}{n_1} \sum_{i : D_i = 1} w_i Y_i - \frac{1}{n_0} \sum_{i:D_i = 0} w_i Y_i 
	\end{align}
where $\textbf{w} \geq 0$ and, often, $\frac{1}{n_0} \sum_{D_i=0} w_i = 1$ and $\frac{1}{n_1} \sum_{D_i=1} w_i = 1$. 

\subsection{TFI for the ATC and the ATE}\label{app:tfi_ate_atc}

This section extends TFI to weighted difference in means estimators of the ATC and the ATE. Like when estimating the ATT, these extensions of TFI are approximations of the bias of $\hat{\tau}_{\mathrm{wdim}}^{(\mathrm{ATC})}$ and $\hat{\tau}_{\mathrm{wdim}}^{(\mathrm{ATE})}$ in estimating the ATC and ATE, respectively. 

First, however, we require extra notation. Define 
	\begin{align}\label{eq:imbal_atc}
		\mathrm{imbal}^{(\mathrm{ATC})} ( \textbf{w}, g(\textbf{X}), \textbf{D} ) = \frac{1}{n_1} \sum_{i:D_i = 1} w_i g(X_i) - \frac{1}{n_0} \sum_{i:D_i = 0} g(X_i)
	\end{align} 
where $g$ is an arbitrary vector-valued function. In words, $\mathrm{imbal}^{(\mathrm{ATC})}  ( \textbf{w}, g(\textbf{X}), \textbf{D} )$ is the imbalance in $g(X)$ between the unweighted control group and the weighted treated group. Then, define 
	\begin{align}\label{eq:imbal_ate}
		& \mathrm{imbal}^{(\mathrm{ATE}, 0)}( \textbf{w}, g(\textbf{X}), \textbf{D}) = \frac{1}{n} \sum_{i = 1}^n g(X_i) - \frac{1}{n_0} \sum_{i: D_i = 0} w_i g(X_i) \nonumber \\
		\text{and} \ \ & \mathrm{imbal}^{(\mathrm{ATE}, 1)}( \textbf{w}, g(\textbf{X}), \textbf{D}) = \frac{1}{n} \sum_{i = 1}^n g(X_i) - \frac{1}{n_1} \sum_{i: D_i = 1} w_i g(X_i)
	\end{align}
In words, $\mathrm{imbal}^{(\mathrm{ATE}, d)}(\textbf{w}, g(\textbf{X}), \textbf{D})$ is the imbalance in $g (X_i)$ between the full sample and the weighted group with treatment status $d \in \{0, 1\}$. Finally, let $\hat{\beta}_d$ for $d \in \{0, 1\}$ be the coefficients from regression estimators $\hat{f}_d (X) = X^{\top} \hat{\beta}_d$, and let $\hat{V}_{\hat{\beta}_d}$ be estimates of their variances. 

\subsubsection{ATC}
TFI for a $\hat{\tau}_{\mathrm{wdim}}^{(\mathrm{ATC})} (\textbf{w}) $ is defined as
	\begin{align}\label{eq:tfi_atc}
		& \mathrm{TFI}^{(\mathrm{ATC})} (\textbf{w}, \textbf{X}, \textbf{D}, \hat{\beta}_1, \hat{V}_{\beta_1}, q) = \nonumber \\
		& \ \ \ \ \ \ \ \ \ \ \ \ \ \ \sqrt{Q_q (\mathcal{X}^2_P)} \times || \hat{V}_{\beta_1}^{\frac{1}{2}} \mathrm{imbal}^{(\mathrm{ATC})}(\textbf{w}, \textbf{X}, \textbf{D}) ||_2 +  |\mathrm{imbal}^{(\mathrm{ATC})}(\textbf{w}, \textbf{X}, \textbf{D})^{\top}\hat{\beta}_1|
	\end{align}
As before, $q$ is a probability threshold chosen by the researcher.

\subsubsection{ATE}
Define TFI for a  $\hat{\tau}_{\mathrm{wdim}}^{(\mathrm{ATE})} (\textbf{w}) $ as
	\begin{align}\label{eq:tfi_ate}
		& \mathrm{TFI}^{(\mathrm{ATE})} (\textbf{w}, \textbf{X}, \textbf{D}, \hat{\beta}_0, \hat{V}_{\beta_0}, \hat{\beta}_1, \hat{V}_{\beta_1}, q) = \nonumber \\
		& \ \ \ \ \ \ \ \ \ \ \ \ \ \ \sqrt{Q_q (\mathcal{X}^2_P)} \times || \hat{V}_{\beta_0}^{\frac{1}{2}} \mathrm{imbal}^{(\mathrm{ATE , 0})}(\textbf{w}, \textbf{X}, \textbf{D}) ||_2 +  |\mathrm{imbal}^{(\mathrm{ATE , 0})}(\textbf{w}, \textbf{X}, \textbf{D})^{\top}\hat{\beta}_0| \ + \nonumber \\
		& \ \ \ \ \ \ \ \ \ \ \ \ \ \ \sqrt{Q_q (\mathcal{X}^2_P)} \times || \hat{V}_{\beta_1}^{\frac{1}{2}} \mathrm{imbal}^{(\mathrm{ATE , 1})}(\textbf{w}, \textbf{X}, \textbf{D}) ||_2 +  |\mathrm{imbal}^{(\mathrm{ATE , 1})}(\textbf{w}, \textbf{X}, \textbf{D})^{\top}\hat{\beta}_1| 
	\end{align}
Again, $q$ is a probability threshold chosen by the user.

\subsection{TFB for the ATC and the ATE}\label{app:tfb_ate_atc}

This section details how TFB finds weights to estimate the ATC and the ATE.

\subsubsection{ATC}
Similar to when estimating the ATT, TFB chooses weights to estimate the ATC that minimize the corresponding TFI subject to constraints on the weights' variance. In other words,
	\begin{align}\label{eq:tfb_atc}
		& \hat{\textbf{w}}_{\mathrm{TFB}}^{(\mathrm{ATC})} = \underset{\textbf{w}}{\text{argmin}}\biggr[ [ \mathrm{TFI}^{(\mathrm{ATC})}(\textbf{w}, \textbf{X}, \textbf{D}, \hat{\beta}_1, \hat{V}_{\beta_1}, q) ]^2  + \frac{ \hat{\sigma}^2 (1) }{ n_1^2 } \sum_{i:D_i = 1} w_i^2 \biggr] \nonumber \\
		\text{where} \ \ \ & \textbf{w} \geq 0 \ \text{and} \ \frac{1}{n_1} \sum_{i:D_i=1}w_i = 1
	\end{align}
where $\hat{\sigma}^2 (1) = \frac{1}{n_1} \sum_{i:D_i = 1} (Y_i - X_i^{\top} \hat{\beta}_1)^2$. 

\subsubsection{ATE}
To estimate the ATE, TFB chooses
	\begin{align}\label{eq:tfb_ate}
		& \hat{\textbf{w}}_{\mathrm{TFB}}^{(\mathrm{ATE})} = \underset{\textbf{w}}{\text{argmin}}\biggr[ [  \mathrm{TFI}^{(\mathrm{ATE})} (\textbf{w}, \textbf{X}, \textbf{D}, \hat{\beta}_0, \hat{V}_{\beta_0}, \hat{\beta}_1, \hat{V}_{\beta_1}, q) ]^2  +  \frac{ \hat{\sigma}^2(0) }{ n_0^2 } \sum_{i:D_i = 0} w_i^2 + \frac{ \hat{\sigma}^2 (1) }{n_1^2} \sum_{i:D_i = 1} w_i^2 \biggr] \nonumber \\
		\text{where} \ \ \ & \textbf{w} \geq 0, \  \frac{1}{n_0} \sum_{i:D_i=0}w_i = 1, \ \text{and} \ \frac{1}{n_1} \sum_{i:D_i=1}w_i = 1
	\end{align}
where $\hat{\sigma}^2 (0) = \frac{1}{n_0} \sum_{i:D_i = 0} (Y_i - X_i^{\top} \hat{\beta}_0)^2$ and  $\hat{\sigma}^2 (1) = \frac{1}{n_1} \sum_{i:D_i = 1} (Y_i - X_i^{\top} \hat{\beta}_1)^2$.

\newpage
\clearpage

\section{Variance estimation and confidence intervals}\label{subsec:tfb.asymptotics}

Given that TFB's weights are smooth, a bootstrap may be valid for variance estimation and confidence intervals. However, because high-dimensional regression estimators are recommended for TFB, the computation time required for a bootstrap may be impractical. This section thus proposes a closed-form estimated variance, and resulting confidence interval, for TFB. These are implemented in our \texttt{R} package \href{https://github.com/bleth/tfb}{\texttt{tfb}}. Asymptotic validity is not formally proven, but it is shown to perform well in DGP 1 (from Section~\ref{sec:dgp1}) and DGP 2 (from Appendix~\ref{sec:dgp2}).

\subsection{Variance estimation for the ATT}\label{app:variance_att}

\subsubsection{Conditions for the asymptotic normality of a general $\hat{\tau}_{\mathrm{wdim}}$ and a consistent variance estimator}

First, consider sufficient conditions for the asymptotic normality of  $\hat{\tau}_{\mathrm{wdim}} (\hat{\textbf{w}})$ for general weights $\hat{\textbf{w}}$ estimated from the data:
	\begin{proposition}[Conditions for asymptotic normality of a general $\hat{\tau}_{\mathrm{wdim}}$]\label{thm:an_att} Given Assumption~\ref{asm:ci}, further assume the following conditions: 
\begin{itemize}
    \item[(i)]  $\hat{\textbf{w}} = \textbf{w}_{\mathrm{vec}} (\textbf{X}, \textbf{D})$ for a function $\textbf{w}_{\mathrm{vec}}  ( \cdot )$;  
    \item[(ii)]  $\underset{n \rightarrow \infty}{\mathrm{plim}} \ \frac{1}{n_0} \sum_{i : D_i = 0} \hat{w}_i^2 \sigma^2 (0, X_i)  < \infty$; 
	\item[(iii)]  $\exists \underline{C}, \overline{C}>0$ such that $\underline{C} \leq \sigma^2 (0, X)  \leq \overline{C}$; 
    \item[(iv)]  $\E ( [Y(1) - f_0 (X) - \mathrm{ATT}]^2 \ | \ D=1 ) < \infty $; 
    \item[(v)] $\exists \delta>0$ such that 
		\begin{align*}
			\frac{\sum_{i:D_i=0} \hat{w}_i^{2 + \delta}}{(\sum_{i:D_i=0} \hat{w}_i^2)^{\frac{2+\delta}{2}}} \overset{a.s.}{\rightarrow} 0 \ \ \ \text{and} \ \ \ \E( | \epsilon (0) |^{2 + \delta} \ | \ \textbf{X}) \leq \overline{C}_\delta \ \ \ \text{for \ some} \ \ \ \overline{C}_\delta > 0 \text{;}
		\end{align*}
	and 
    \item[(vi)]  $\mathrm{EWC \ Bias} = \mathrm{imbal} (\hat{\textbf{w}}, f_0 (\textbf{X}), \textbf{D}) = o_p (n^{-1/2})$. 
\end{itemize}
	Then, $V_{\mathrm{ATT}}^{-\frac{1}{2}} (\hat{\textbf{w}})(\hat{\tau}_{\mathrm{wdim}} (\hat{\textbf{w}}) - \mathrm{ATT}) \overset{d}{\rightarrow} N(0, 1)$ where
		\begin{align*}
			V_{\mathrm{ATT}} (\hat{\textbf{w}}) = \frac{1}{n_1} \E \biggr( [Y(1) - f_0 (X) - \mathrm{ATT}]^2 \ \biggr| \ D=1 \biggr) +  \frac{1}{n_0^2}\sum_{i:D_i=0} \hat{w}_i^2 \sigma^2 (0, X_i) 
		\end{align*}
 	\end{proposition}
Proof is given in Appendix~\ref{app:thm1pf}. Condition (i) requires that $\hat{\textbf{w}}$ does not depend on the $Y_i$ given $\textbf{X}$ and $\textbf{D}$. Conditions (ii) and (v) then restrict how extreme the $\hat{\textbf{w}}$ can be in asymptopia. Conditions (iii) and (v) restrict the moments of $\epsilon (0)$, and condition (iv) restricts the variance of the conditional average treatment effects (i.e., $\E[Y (1) - Y (0) \ | \ X]$). Finally, condition (vi) is key to the $\sqrt{n}$-consistency of a $\hat{\tau}_{\mathrm{wdim}} (\hat{\textbf{w}})$, requiring that the bias of $\hat{\tau}_{\mathrm{wdim}}  (\hat{\textbf{w}})$ converges in probability to 0 at a fast enough rate. 

Note again that Proposition~\ref{thm:an_att} does not specify the conditions under which TFB is asymptotically normal, but rather when a $\hat{\tau}_{\mathrm{wdim}}$ with \textit{general} weights is asymptotically normal.  Nevertheless, it is worthwhile discussing conditions (i) and (vi) in the context of TFB. First, note that condition (i) is violated by TFB as defined. However, sample splitting, defined in Appendix~\ref{subsec:samplesplitting}, allows condition (i) to hold for TFB individually within each of the two splits of the sample. We explore applying sample splitting with TFB in the context of variance estimation below.

Next, it should be emphasized that condition (vi) is far from trivial --- for example, \cite{abadie2006large} show it does not hold generally for the matching estimators they examine if $P>1$. However, there is precedent for balancing weights estimators, and particularly augmented estimators, satisfying this condition under additional assumptions (e.g., \citealp{athey2018approximate}; \citealp{hirshberg2017augmented}). In particular, that TFB is often approximately equal to an augmented estimator in practice reveals an relevant connection to the ``augmented minimax linear estimator" explored by \cite{hirshberg2017augmented}. This class of estimator can be expressed as $\hat{\tau}_{\mathrm{aug}}$ with \ref{eq:approxbal_w2} weights with $B(\textbf{w}) \propto || \textbf{w} ||_2^2$ and a general $\mathcal{F}$. 
Recall that TFB chooses \ref{eq:approxbal_w2} weights as well with $B(\textbf{w}) = \frac{\hat{\sigma}^2 (0) }{n_0^2} || \textbf{w} ||_2^2$ and $\mathcal{F} = \{ x^{\top} \beta \ | \ || \hat{V}_{\beta_0}^{-\frac{1}{2}} (\hat{\beta}_0 - \beta) ||_2 \leq \sqrt{Q_q (\mathcal{X}^2_P)} \}$. Thus, it seems plausible that TFB is asymptotically equivalent to an augmented minimax linear estimator with the same $B(\textbf{w})$ and $\mathcal{F}$. This may imply asymptotic normality for TFB under assumptions similar to those proven in \cite{hirshberg2017augmented}, among them that $\hat{f}_0$ has several consistency properties for $f_0$. \cite{hirshberg2017augmented} also make assumptions on the function space $\mathcal{F}$ for the \ref{eq:approxbal_w2} weights they examine, though they describe $\mathcal{F}$ as a function space that contains the \textit{difference} between a $\hat{f}_0$ and $f_0$ (i.e., $f_0 - \hat{f}_0$), rather than containing $f_0$ itself, which is how TFB frames it.  Nevertheless, because $\hat{V}_{\beta_0}$ (along with $\hat{f}_0$) determines the analogous $\mathcal{F}$ for TFB, the properties of $\hat{V}_{\beta_0}$ likely also play a role here. Further, although it does not appear that \textit{complete} overlap in the conditional distributions of $X$ (as in (\ref{eq:psoverlap})) is essential (see Appendix~\ref{app:overlap}), assurance that there exist weights that satisfy \ref{eq:ewc} in asymptopia would likely be required. Further, relaxing the complete overlap assumption likely requires certain consistency properties for $\hat{f}_0$, for example that $\hat{f}_0$ can extrapolate correctly (in asymptopia) into areas of the $X$ distribution that have no overlap. Proving the exact assumptions needed for TFB to satisfy condition (vi), along with conditions (ii) and (v), is left to future work. Along with \cite{hirshberg2017augmented}, \cite{ben2021balancing}, which studies the asymptotic properties of balancing weight estimators, and \cite{jin2025cross}, which describes the ``cross-balancing" framework that TFB fits into with sample splitting (see Appendix~\ref{subsec:samplesplitting}), may also offer insights. 

Nevertheless, Proposition~\ref{thm:an_att}, and Proposition~\ref{thm:Vhat} below, still suggest a closed-form variance estimator. Turning back to general weights $\hat{\textbf{w}}$ that are estimated from the data, assuming that $\hat{f}_0$ is a consistent estimator of $f_0$ suggests the plug-in estimator for $V_{\mathrm{ATT}} (\hat{\textbf{w}})$,
	\begin{align}\label{eq:vhat}
			\hat{V}_{\mathrm{ATT}} (\hat{\textbf{w}}, \hat{f}_0) = \frac{1}{n_1^2} \sum_{i:D_i=1} [Y_i(1) - \hat{f}_0 (X_i) - \hat{\tau}_{\mathrm{wdim}} (\hat{\textbf{w}}) ]^2  +  \frac{1}{n_0^2}\sum_{i:D_i=0} \hat{w}_i^2 \hat{e}_i^2 (0) 
	\end{align}
where $\hat{e}_i (0) = Y_i (0) - \hat{f}_0 (X_i)$. Proposition~\ref{thm:Vhat} describes conditions under which $\hat{V}_{\mathrm{ATT}} (\hat{\textbf{w}}, \hat{f}_0)   \overset{p}{\rightarrow} V_{\mathrm{ATT}} (\hat{\textbf{w}})$.
	\begin{proposition}[Consistency of $\hat{V}_{\mathrm{ATT}}$]\label{thm:Vhat} Given Assumption~\ref{asm:ci}, 
	and conditions (i)-(vi) for Proposition~\ref{thm:an_att}, further assume that:
    \begin{itemize}
    \item[(vii)] $\frac{1}{n} \sum_{i=1}^n [f_0 (X_i) - \hat{f}_0 (X_i)]^2 = o_p (1)$;
    \newline
	\item[(viii)] $\frac{1}{n_0} \sum_{i:D_i = 0} \hat{w}_i^2 \epsilon_i^2 (0) \overset{p}{\rightarrow} \frac{1}{n_0} \sum_{i:D_i = 0} \hat{w}_i^2 \sigma^2 (0, X_i) $; and 
    \item[(ix)] $\frac{1}{n_0}\sum_{i:D_i=0} \hat{w}_i^2 [f_0 (X_i) - \hat{f}_0 (X_i)]^2 = o_p (1)$. 
    \end{itemize}
	Then, $\hat{V}_{\mathrm{ATT}} (\hat{\textbf{w}}, \hat{f}_0)   \overset{p}{\rightarrow} V_{\mathrm{ATT}} (\hat{\textbf{w}})$. 
	\end{proposition}
See Appendix~\ref{app:thm2pf} for proof. Conditions (vii) and (ix) quantify how accurate an estimator of $f_0$ that $\hat{f}_0$ must be in terms of unweighted and $\hat{\textbf{w}}^2$-weighted mean squared error, respectively. Additionally, conditions (viii) and (ix) allow $\hat{V}_{\mathrm{ATT}} (\hat{\textbf{w}}, \hat{f}_0)$ to be robust to heteroscedastic $\epsilon (0) $.

\subsubsection{Proposed variance estimator and confidence interval}

\begin{figure}[!h]
\vspace{0.15in}
\begin{center}
\caption{Coverage of 95\% confidence intervals for TFB in DGPs 1 and 2}\label{fig.cov1.noSS}
	\vspace{-0.25in}
    \begin{subfigure}{.45\textwidth}
    \begin{center}
    \includegraphics[scale=0.4]{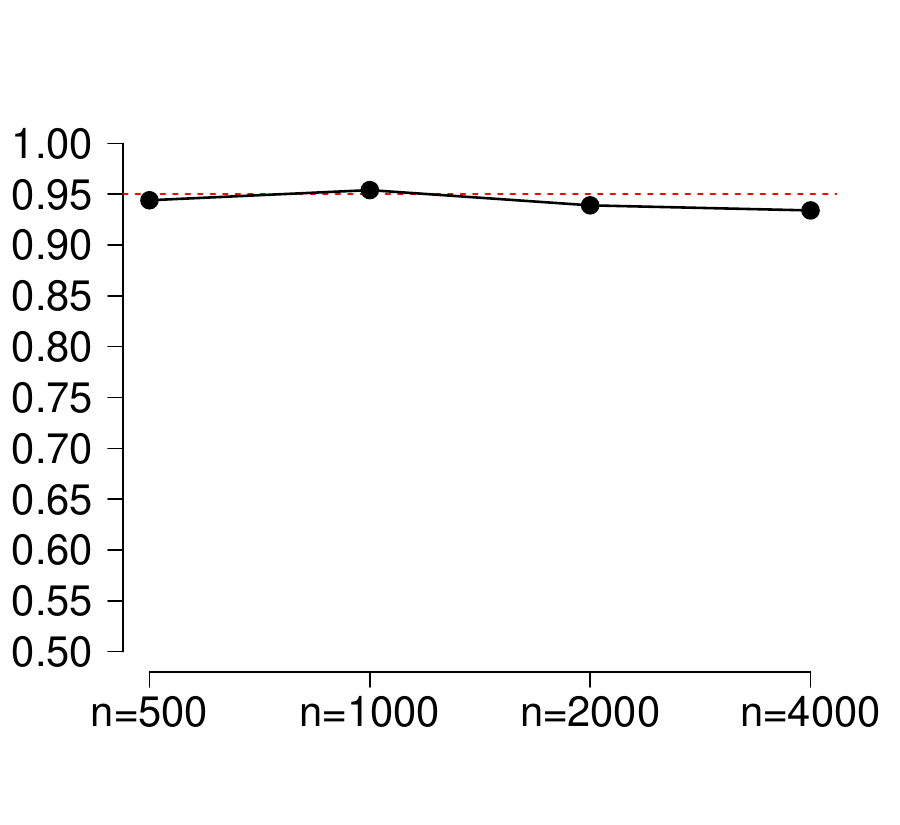} 
    \end{center}
   \vspace{-.25in}
   \subcaption{TFB-K in DGP 1}\label{fig.cov1.noSS.A}
    \end{subfigure}
    \begin{subfigure}{.45\textwidth}
    \begin{center}
    \includegraphics[scale=0.4]{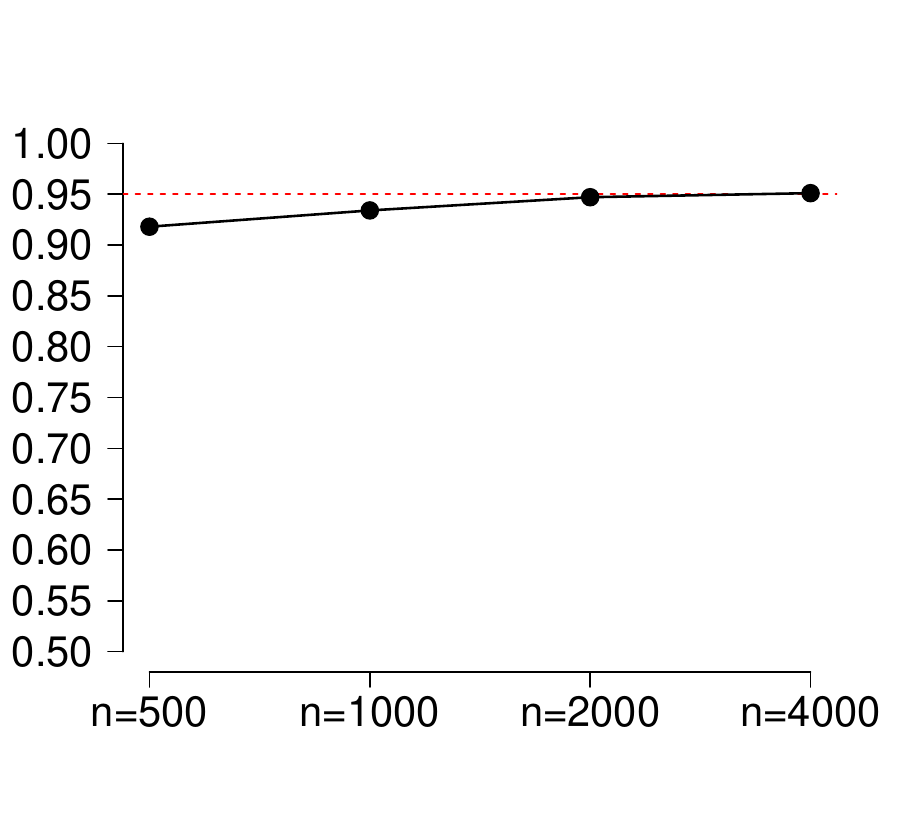} 
    \end{center}
    \vspace{-.25in}
    \subcaption{TFB-L in DGP 2}\label{fig.cov1.noSS.B}
    \end{subfigure}
    \subcaption*{\textit{Note:} Results across 1000 draws from DGPs 1 and 2. TFB-K is applied in DGP 1, and TFB-L is applied in DGP 2.} 
\end{center}
\vspace{-0.25in}
\end{figure}

Propositions~\ref{thm:an_att} and~\ref{thm:Vhat} suggest $\hat{V}_{\mathrm{ATT}} (\hat{\textbf{w}}_{\mathrm{TFB}}, \hat{f}_0)$ as a variance estimator for TFB. $(100 \times \gamma) \%$ confidence intervals can then be obtained using the normal approximation: $\hat{\tau}_{\mathrm{wdim}} (\hat{\textbf{w}}_{\mathrm{TFB}}) \pm \Phi^{-1} ( \frac{1 + \gamma}{2} )  \times \sqrt{\hat{V}_{\mathrm{ATT}} (\hat{\textbf{w}}_{\mathrm{TFB}}, \hat{f}_0)}$. However, TFB violates condition (i) from Proposition~\ref{thm:an_att} as defined. Thus, we might expect that this variance estimator will underestimate the true variance. Nevertheless, in Figure~\ref{fig.cov1.noSS} we see that this variance estimator achieves roughly nominal coverage for TFB-K in DGP 1 (from Section~\ref{sec:dgp1}), and TFB-L in DGP 2 (from Appendix~\ref{sec:dgp2}). Note also that in extensions to DGPs 1 and 2 in Appendices~\ref{app:dgp1_extension} and~\ref{app:dgp2_extension} that vary aspects of the DGPs (e.g., $R^2$, overlap, and sample size), the proposed confidence intervals reach, or fall just below, the target coverage rate of 95\% when TFB's bias is low.

Additionally, sample splitting can correct TFB's violation to condition (i) from Proposition~\ref{thm:an_att}. We first review sample splitting for TFB and its required notation from Appendix~\ref{subsec:samplesplitting}. Sample splitting with TFB requires randomly splitting the full sample into two non-overlapping (sub-)samples: Samples 1 and 2. The superscripts $(s1)$ and $(s2)$ denote values that come from Samples 1 and 2, respectively. First, use Sample 2 to obtain $(\hat{\beta}^{(s2)}_0, \hat{V}^{(s2)}_{\beta_0})$, and apply them to TFB in Sample 1 to form $\hat{\textbf{w}}^{(s1)}_{\mathrm{TFB}}$ and $\hat{\tau}_{\mathrm{wdim}}^{(s1)} (\hat{\textbf{w}}_{\mathrm{TFB}}^{(s1)})$. Next, switch the roles of the two sub-samples to obtain $(\hat{\beta}^{(s1)}_0, \hat{V}^{(s1)}_{\beta_0})$ from Sample 1, and $\hat{\textbf{w}}_{\mathrm{TFB}}^{(s2)}$ and  $\hat{\tau}_{\mathrm{wdim}}^{(s2)} (\hat{\textbf{w}}_{\mathrm{TFB}}^{(s2)})$ from Sample 2. The final estimate is then the average of the estimates from each sub-sample:
    \begin{align}\label{eq:tfb_ss_est}
        \hat{\tau}_{\mathrm{wdim}}^{(\mathrm{TFB})} = \frac{ \hat{\tau}_{\mathrm{wdim}}^{(s1)} (\hat{\textbf{w}}_{\mathrm{TFB}}^{(s1)}) + \hat{\tau}_{\mathrm{wdim}}^{(s2)} (\hat{\textbf{w}}_{\mathrm{TFB}}^{(s2)}) }{2}
    \end{align}

\begin{figure}[!h]
\vspace{0.15in}
\begin{center}
\caption{Coverage of 95\% confidence intervals for TFB with sample splitting in DGPs 1 and 2}\label{fig.cov1}
	\vspace{-0.25in}
    \begin{subfigure}{.45\textwidth}
    \begin{center}
    \includegraphics[scale=0.4]{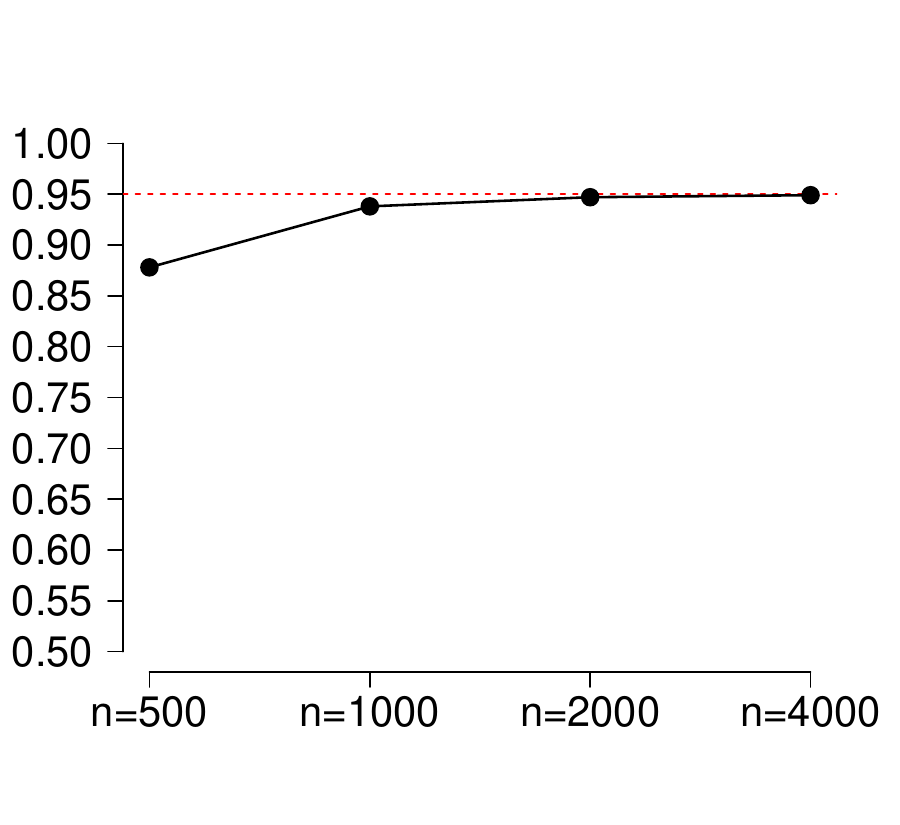} 
    \end{center}
   \vspace{-.25in}
   \subcaption{TFB-K in DGP 1}\label{fig.cov1.A}
    \end{subfigure}
    \begin{subfigure}{.45\textwidth}
    \begin{center}
    \includegraphics[scale=0.4]{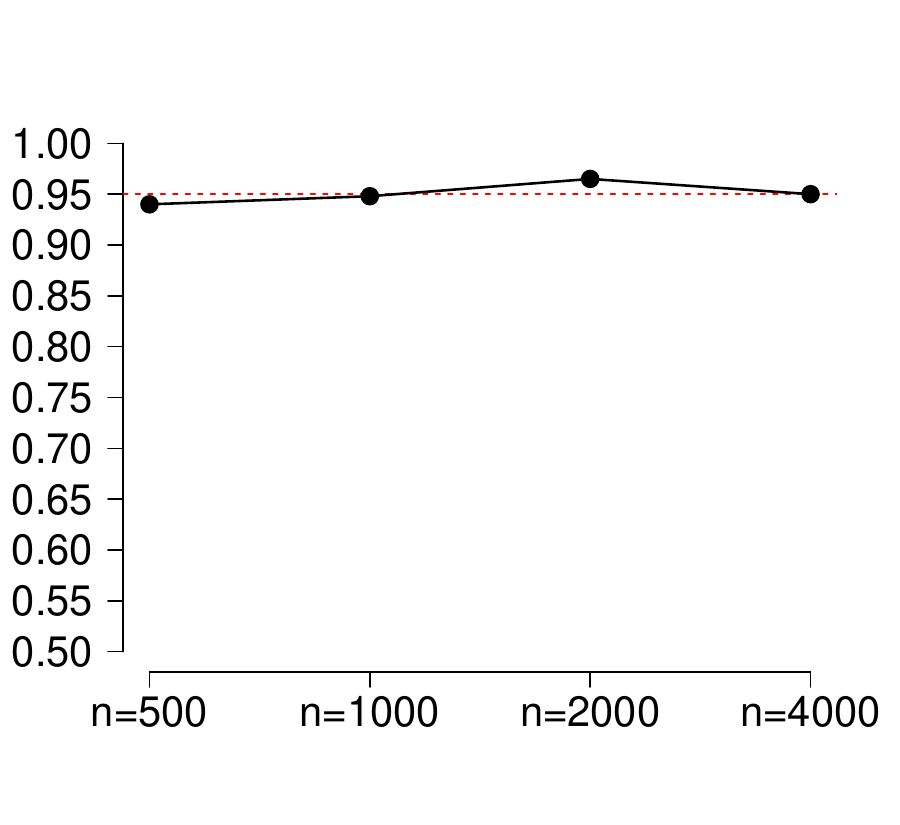} 
    \end{center}
    \vspace{-.25in}
    \subcaption{TFB-L in DGP 2}\label{fig.cov1.B}
    \end{subfigure}
    \subcaption*{\textit{Note:} Results across 1000 draws from DGPs 1 and 2. TFB-K is applied in DGP 1, and TFB-L is applied in DGP 2. Both TFB estimators employ sample splitting.} 
\end{center}
\vspace{-0.25in}
\end{figure}

With sample splitting, condition (i) from Proposition~\ref{thm:an_att} holds for the estimator from each sub-sample. 
Then, following Proposition~\ref{thm:Vhat}, let $\hat{V}_{\mathrm{ATT}}^{(s1)} (\hat{\textbf{w}}_{\mathrm{TFB}}^{(s1)}, \hat{f}_0^{(s2)})$ and $\hat{V}_{\mathrm{ATT}}^{(s2)} (\hat{\textbf{w}}_{\mathrm{TFB}}^{(s2)}, \hat{f}_0^{(s1)})$ be the variance estimators from the two sub-samples. This leads to the proposed variance estimator:
    \begin{align}\label{eq:tfb_ss_var}
        \widehat{\var} (\hat{\tau}_{\mathrm{wdim}}^{(\mathrm{TFB})})  =  \frac{1}{4} \hat{V}_{\mathrm{ATT}}^{(s1)} (\hat{\textbf{w}}_{\mathrm{TFB}}^{(s1)}, \hat{f}_0^{(s2)}) + \frac{1}{4} \hat{V}_{\mathrm{ATT}}^{(s2)} (\hat{\textbf{w}}_{\mathrm{TFB}}^{(s2)}, \hat{f}_0^{(s1)}) 
    \end{align}
$(100 \times \gamma) \%$ confidence intervals can again be obtained using the normal approximation: $\hat{\tau}_{\mathrm{wdim}}^{(\mathrm{TFB})} \pm \Phi^{-1} ( \frac{1 + \gamma}{2} )  \times \sqrt{\widehat{\var} (\hat{\tau}_{\mathrm{wdim}}^{(\mathrm{TFB})})}$. Figure~\ref{fig.cov1} shows that 95\% confidence intervals with the proposed variance estimator for sample splitting achieve essentially 95\% coverage as $n$ increases. The undercoverage for TFB-K when $n=500$ in DGP 1 is because TFB-K still shows bias at that sample size (results available upon request). 

Note that the asymptotic validity of $\widehat{\var} (\hat{\tau}_{\mathrm{wdim}}^{(\mathrm{TFB})})$ in (\ref{eq:tfb_ss_var}) (given the other conditions from Propositions~\ref{thm:an_att} and~\ref{thm:Vhat}) requires that the estimates from each sub-sample be asymptotically uncorrelated. Proving this is left to future work, but Table~\ref{tab1} demonstrates that it is plausible, despite the estimates' being correlated in finite samples (the $Y_i$ in Sample 1 contribute to $\hat{\textbf{w}}_{\mathrm{TFB}}^{(s2)}$ from Sample 2, and vice versa). The correlation between the estimates appears to disappear as $n$ increases in DGPs 1 and 2.
\begin{table}[!h]
\vspace{0.15in}
\begin{center}
\caption{Correlation of $\hat{\tau}_{\mathrm{wdim}}^{(s1)} (\hat{\textbf{w}}_{\mathrm{TFB}}^{(s1)})$ and $\hat{\tau}_{\mathrm{wdim}}^{(s2)} (\hat{\textbf{w}}_{\mathrm{TFB}}^{(s2)})$ from TFB, with sample splitting, in DGPs 1 and 2}\label{tab1}
	\vspace{0.10in}
     \begin{tabular}{ l | c c c c }
     & $n=500$ & $n=1000$ & $n=2000$ & $n=4000$ \\
     \hline
     DGP 1 (TFB-K) & 0.103* & 0.076* & 0.037 & 0.011 \\
     DGP 2 (TFB-L) & 0.104* & 0.07* & -0.008 & 0.005 \\
     \end{tabular}
     \vspace{0.20in}
    \subcaption*{\textit{Note:} Results are across 1000 draws of DGP 1 and DGP 2. Starred (*) values indicate significance at the 0.05 level of the coefficient for $\hat{\tau}_{\mathrm{wdim}}^{(s1)} (\hat{\textbf{w}}_{\mathrm{TFB}}^{(s1)})$ in an OLS regression that predicts $\hat{\tau}_{\mathrm{wdim}}^{(s2)} (\hat{\textbf{w}}_{\mathrm{TFB}}^{(s2)})$, with robust standard errors~\citep{white1980heteroskedasticity}. See Table~\ref{tab:abbreviations} for descriptions of TFB-K and TFB-L. }
\end{center}
\vspace{-0.25in}
\end{table}

Finally, reporting the median estimate across many random splits of the sample, as suggested by \cite{chernozhukov2018double}, requires an adjusted variance. Again, let $\hat{\tau}_{\mathrm{wdim}, s}^{(\mathrm{TFB})}$ be the TFB estimate from the $s$th split of the sample, and let $\widehat{\var} (\hat{\tau}_{\mathrm{wdim}, s}^{(\mathrm{TFB})})$ consistently estimate its variance. Results from \cite{chernozhukov2018double} imply the variance estimator for the median estimate, $\hat{\tau}_{\mathrm{wdim}, \mathrm{med}}^{(\mathrm{TFB})}$:
	\begin{align}\label{eq:samplesplitting_variance}
		\widehat{\var} ( \hat{\tau}_{\mathrm{wdim}, \mathrm{med}}^{(\mathrm{TFB})}  ) = \mathrm{median} \biggr{\{} \widehat{\var} (\hat{\tau}_{\mathrm{wdim}, s}^{(\mathrm{TFB})}) + \frac{(\hat{\tau}_{\mathrm{wdim}, s}^{(\mathrm{TFB})} - \hat{\tau}_{\mathrm{wdim}, \mathrm{med}}^{(\mathrm{TFB})})^2}{n} \biggr{\}}
	\end{align}
$(100 \times \gamma)$\% confidence intervals can be obtained through the normal approximation: $\hat{\tau}_{\mathrm{wdim}, \mathrm{med}}^{(\mathrm{TFB})} \pm \Phi^{-1} ( \frac{1 + \gamma}{2} )  \times \sqrt{\widehat{\var} ( \hat{\tau}_{\mathrm{wdim}, \mathrm{med}}^{(\mathrm{TFB})}  ) }$. We demonstrate this adjusted variance in Appendix~\ref{sec:applications}.

\subsection{Variance estimation for the ATC and the ATE}\label{app:ses_ate_atc}

This section proposes variance estimators for TFB estimators of the ATC and ATE. Although it is a slight abuse of notation, let $\textbf{w} \geq 0$ be a general vector of non-negative weights. In the context of the ATC, these weights will only be defined for treated units. In the context of the ATE, these weights will be defined for the full sample. 

\subsubsection{ATC}

Proposition~\ref{thm:an_atc} describes sufficient conditions to determine the asymptotic normality of a $\hat{\tau}^{(\mathrm{ATC})}_{\mathrm{wdim}} (\hat{\textbf{w}})$ with general weights $\hat{\textbf{w}}$ estimated from the data. Then, Proposition~\ref{thm:Vhat_atc} introduces an estimator of the asymptotic variance of $\hat{\tau}^{(\mathrm{ATC})}_{\mathrm{wdim}} (\hat{\textbf{w}})$ and describes sufficient conditions for the estimator's consistency.

	\begin{proposition}[Conditions for asymptotic normality of a general $\hat{\tau}^{(\mathrm{ATC})}_{\mathrm{wdim}}$]\label{thm:an_atc} Given Assumption~\ref{asm:ci}, 
	further assume the following conditions: 
    \begin{itemize}
        \item[(i)] $\hat{\textbf{w}} = \textbf{w}_{\mathrm{vec}}^{(\mathrm{ATC})} (X, D)$ for a function $\textbf{w}_{\mathrm{vec}}^{(\mathrm{ATC})} (\cdot)$; 
	   \item[(ii)] $\underset{n \rightarrow \infty}{\mathrm{plim}} \ \frac{1}{n_1} \sum_{i : D_i = 1} \hat{w}_i^2 \sigma^2 (1, X_i) < \infty$; 
	   \item[(iii)] $\exists \underline{C}, \overline{C}>0$ such that $\underline{C} \leq \sigma^2 (1, X) \leq \overline{C}$; 
       \item[(iv)] $\E ( [f_1 (X) - Y(0) - \mathrm{ATC}]^2 \ | \ D=0 ) < \infty $; 
       \item[(v)] $\exists \delta>0$ such that 
		\begin{align*}
			\frac{\sum_{i:D_i=1} \hat{w}_i^{2 + \delta}}{(\sum_{i:D_i=1} \hat{w}_i^2)^{\frac{2+\delta}{2}}} \overset{a.s.}{\rightarrow} 0 \ \ \ \text{and} \ \ \ \E( | \epsilon (1) |^{2 + \delta} \ | \ \textbf{X}) \leq \overline{C}_\delta \ \ \ \text{for \ some} \ \ \ \overline{C}_\delta > 0 \text{;}
		\end{align*}
	and 
    \item[(vi)] $\mathrm{imbal}^{(\mathrm{ATC})} (\hat{\textbf{w}}, f_1 (\textbf{X}), \textbf{D}) = o_p (n^{-1/2})$. 
    \end{itemize}
	Then, $V_{\mathrm{ATC}}^{-\frac{1}{2}} (\hat{\textbf{w}}) (\hat{\tau}^{(\mathrm{ATC})}_{\mathrm{wdim}} (\hat{\textbf{w}}) - \mathrm{ATC}) \rightarrow N(0, 1)$ where
		\begin{align*}
			V_{\mathrm{ATC}} (\hat{\textbf{w}}) = \frac{1}{n_0} \E \biggr( [f_1 (X) - Y (0) - \mathrm{ATC}]^2 \ \biggr| \ D=0 \biggr) +  \frac{1}{n_1^2} \sum_{i:D_i=1} \hat{w}_i^2 \sigma^2 (1, X_i)
		\end{align*}
 	\end{proposition}
	
	\begin{proposition}[Estimating $V_{\mathrm{ATC}}$]\label{thm:Vhat_atc} Given Assumption~\ref{asm:ci} and conditions (i)-(vi) for Proposition~\ref{thm:an_atc}, further assume that: 
    \begin{itemize}
       \item[(vii)] $\frac{1}{n} \sum_{i=1}^n [f_1 (X_i) - \hat{f}_1 (X_i)]^2 = o_p (1)$; 
       \item[(viii)] $\frac{1}{n_1} \sum_{i:D_i=1} \hat{w}_i^2 \epsilon_i^2 (1) \overset{p}{\rightarrow} \frac{1}{n_1} \sum_{i : D_i=1} \hat{w}_i^2 \sigma^2 (1, X_i)$; and 
       \item[(ix)] $\frac{1}{n_1}\sum_{i:D_i=1} \hat{w}_i^2 [f_1 (X_i) - \hat{f}_1 (X_i)]^2 = o_p (1)$. 
    \end{itemize}
    Then, $\hat{V}_{\mathrm{ATC}} (\hat{\textbf{w}}, \hat{f}_1) \overset{p}{\rightarrow} V_{\mathrm{ATC}} (\hat{\textbf{w}}) $ where
		\begin{align*}
			\hat{V}_{\mathrm{ATC}} (\hat{\textbf{w}}, \hat{f}_1) = \frac{1}{n_0^2} \sum_{i:D_i=0} [\hat{f}_1 (X_i) - Y_i (0) - \hat{\tau}^{(\mathrm{ATC})}_{\mathrm{wdim}} (\hat{\textbf{w}}) ]^2  +  \frac{1}{n_1^2}\sum_{i:D_i=1} \hat{w}_i^2 [Y_i (1) - \hat{f}_1 (X_i)]^2
		\end{align*}
	\end{proposition}
The proofs for Propositions~\ref{thm:an_atc} and~\ref{thm:Vhat_atc} are omitted here because the ATC is simply the negative of the ATT for the reversed treatment, $\tilde{D} = 1 - D$, and potential outcomes, $\tilde{Y} (\tilde{d}) = Y (1 - \tilde{d})$  (i.e., $\E [ Y (1) - Y (0) \ | \ D = 0] = - \E [ \tilde{Y} (1) - \tilde{Y} (0) \ | \ \tilde{D} = 1])$. Therefore, the proofs for Propositions~\ref{thm:an_atc} and~\ref{thm:Vhat_atc} follow nearly identical steps to those for Propositions~\ref{thm:an_att} and~\ref{thm:Vhat} (in Appendices~\ref{app:thm1pf} and~\ref{app:thm2pf}), respectively. 

We emphasize that Propositions~\ref{thm:an_atc} and~\ref{thm:Vhat_atc} are for an estimator with \textit{general} weights, not TFB specifically. However, as with the ATT, these propositions provide motivation for a closed-form variance estimator for TFB. $(100 \times \gamma) \%$ confidence intervals can then be obtained using the normal approximation.

They also motivate a closed-form variance for TFB with sample splitting. First, randomly split the full sample into two non-overlapping (sub-)samples: Samples 1 and 2. Use Sample 2 to obtain $(\hat{\beta}^{(s2)}_1, \hat{V}^{(s2)}_{\beta_1})$, and apply them to TFB in Sample 1 to form $\hat{\textbf{w}}^{(\mathrm{ATC}),(s1)}_{\mathrm{TFB}}$ and $\hat{\tau}_{\mathrm{wdim}}^{(\mathrm{ATC}),(s1)} (\hat{\textbf{w}}_{\mathrm{TFB}}^{(\mathrm{ATC}),(s1)})$. Next, switch the roles of the two sub-samples to obtain $(\hat{\beta}_1^{(s1)}, \hat{V}^{(s1)}_{\beta_1})$ from Sample 1, and $\hat{\textbf{w}}_{\mathrm{TFB}}^{(\mathrm{ATC}),(s2)}$ and $\hat{\tau}_{\mathrm{wdim}}^{(\mathrm{ATC}),(s2)} (\hat{\textbf{w}}_{\mathrm{TFB}}^{(\mathrm{ATC}),(s2)})$ from Sample 2. The final estimate when sampling splitting is then average of the estimates from each sub-sample:
    \begin{align}
        \hat{\tau}_{\mathrm{wdim}}^{(\mathrm{ATC}),(\mathrm{TFB})} = \frac{ \hat{\tau}_{\mathrm{wdim}}^{(\mathrm{ATC}),(s1)} (\hat{\textbf{w}}_{\mathrm{TFB}}^{(\mathrm{ATC}),(s1)}) + \hat{\tau}_{\mathrm{wdim}}^{(\mathrm{ATC}),(s2)} (\hat{\textbf{w}}_{\mathrm{TFB}}^{(\mathrm{ATC}),(s2)}) }{2}
    \end{align}
The proposed variance estimator when sample splitting is then:
    \begin{align}
        \widehat{\var} ( \hat{\tau}_{\mathrm{wdim}}^{(\mathrm{ATC}),(\mathrm{TFB})})  =  \frac{1}{4} \hat{V}_{\mathrm{ATC}}^{(s1)} (\hat{\textbf{w}}_{\mathrm{TFB}}^{(\mathrm{ATC}),(s1)}, \hat{f}_1^{(s2)}) + \frac{1}{4} \hat{V}_{\mathrm{ATC}}^{(s2)} (\hat{\textbf{w}}_{\mathrm{TFB}}^{(\mathrm{ATC}),(s2)}, \hat{f}_1^{(s1)}) 
    \end{align}
$(100 \times \gamma) \%$ confidence intervals can again be obtained using the normal approximation: $\hat{\tau}_{\mathrm{wdim}}^{(\mathrm{ATC}),(\mathrm{TFB})} \pm \Phi^{-1} ( \frac{1 + \gamma}{2} )  \times \sqrt{\widehat{\var} ( \hat{\tau}_{\mathrm{wdim}}^{(\mathrm{ATC}),(\mathrm{TFB})})}$. 

\subsubsection{ATE}

Proposition~\ref{thm:an_ate} describes sufficient conditions to determine the asymptotic normality of a $\hat{\tau}^{(\mathrm{ATE})}_{\mathrm{wdim}} (\hat{\textbf{w}})$ with general weights $\hat{\textbf{w}}$ estimated from the data. Then, Proposition~\ref{thm:Vhat_ate} introduces an estimator of the asymptotic variance of $\hat{\tau}^{(\mathrm{ATE})}_{\mathrm{wdim}} (\hat{\textbf{w}})$ and describes sufficient conditions for the estimator's consistency.

	\begin{proposition}[Conditions for asymptotic normality of a general $\hat{\tau}^{(\mathrm{ATE})}_{\mathrm{wdim}}$]\label{thm:an_ate} Given Assumption~\ref{asm:ci}, further assume the following conditions:
    \begin{itemize}
	\item[(i)] $\hat{\textbf{w}} = \textbf{w}_{\mathrm{vec}}^{(\mathrm{ATE})} (\textbf{X}, \textbf{D})$ for a function $\textbf{w}_{\mathrm{vec}}^{(\mathrm{ATE})} (\cdot)$; 
	\item[(ii)] $\underset{n \rightarrow \infty}{\mathrm{plim}} \ \frac{1}{n_0} \sum_{i : D_i = 0} \hat{w}_i^2 \sigma^2 (0, X_i) < \infty$ and $\underset{n \rightarrow \infty}{\mathrm{plim}} \ \frac{1}{n_1} \sum_{i : D_i = 1} \hat{w}_i^2 \sigma^2 (1, X_i) < \infty$; 
	\item[(iii)] $\exists \underline{C}, \overline{C}>0$ such that $\underline{C} \leq \sigma^2 (0, X) \leq \overline{C}$ and $\underline{C} \leq \sigma^2 (1, X) \leq \overline{C}$; 
	\item[(iv)] $\E ( [f_1 (X) - f_0 (X) - \mathrm{ATE}]^2 ) < \infty $; 
    \item[(v)] $\exists \delta>0$ such that 
		\begin{align*}
			\text{for some} \ \ \overline{C}_\delta > 0, \ \ \   & \frac{\sum_{i:D_i=0} \hat{w}_i^{2 + \delta}}{(\sum_{i:D_i=0} \hat{w}_i^2)^{\frac{2+\delta}{2}}} \overset{a.s.}{\rightarrow} 0, \ \ \frac{\sum_{i:D_i=1} \hat{w}_i^{2 + \delta}}{(\sum_{i:D_i=1} \hat{w}_i^2)^{\frac{2+\delta}{2}}} \overset{a.s.}{\rightarrow} 0, \ \ \text{and} \\
    & \E( | \epsilon (d) |^{2 + \delta} \ | \ \textbf{X}) \leq \overline{C}_\delta \ \ \text{for} \ \ d \in \{0, 1\} \text{;}
		\end{align*}
	and 
    \item[(vi)] $\mathrm{imbal}^{(\mathrm{ATE, 0})} (\hat{\textbf{w}}, f_0 (\textbf{X}), \textbf{D}) = o_p (n^{-1/2})$ and $\mathrm{imbal}^{(\mathrm{ATE, 1})} (\hat{\textbf{w}}, f_1 (\textbf{X}), \textbf{D})= o_p (n^{-1/2})$. 
    \end{itemize}
    Then, $V_{\mathrm{ATE}}^{-\frac{1}{2}} (\hat{\textbf{w}}) (\hat{\tau}^{(\mathrm{ATE})}_{\mathrm{wdim}} - \mathrm{ATE}) \rightarrow N(0, 1)$ where
		\begin{align*}
			V_{\mathrm{ATE}} (\hat{\textbf{w}}) = \frac{1}{n} \E \biggr( [f_1 (X) - f_0 (X) - \mathrm{ATE}]^2 \biggr) +  \frac{1}{n_0^2} \sum_{i: D_i = 0} \hat{w}_i^2 \sigma^2 (0, X_i) + \frac{1}{n_1^2} \sum_{i: D_i = 1} \hat{w}_i^2 \sigma^2 (1, X_i)
		\end{align*}
 	\end{proposition}
		
	\begin{proposition}[Estimating $V_{\mathrm{ATE}}$]\label{thm:Vhat_ate} Given Assumption~\ref{asm:ci}  
	and conditions (i)-(vi) for Proposition~\ref{thm:an_ate}, further assume that: 
    \begin{itemize}
    \item[(vii)] $\frac{1}{n} \sum_{i=1}^n [f_0 (X_i) - \hat{f}_0 (X_i)]^2 = o_p (1)$ and $\frac{1}{n} \sum_{i=1}^n [f_1 (X_i) - \hat{f}_1 (X_i)]^2 = o_p (1)$; 
    \item[(viii)] $\frac{1}{n_0} \sum_{i:D_i=0} \hat{w}_i^2 \epsilon_i^2 (0) \overset{p}{\rightarrow} \frac{1}{n_0} \sum_{i : D_i=0} \hat{w}_i^2 \sigma^2 (0, X_i) $ and $\frac{1}{n_1} \sum_{i:D_i=1} \hat{w}_i^2 \epsilon_i^2 (1) \overset{p}{\rightarrow} \frac{1}{n_1} \sum_{i : D_i=1} \hat{w}_i^2 \sigma^2 (1, X_i) $; and 
    \item[(ix)] $\frac{1}{n_0} \sum_{i:D_i=0} \hat{w}_i^2 [f_0 (X_i) - \hat{f}_0 (X_i)]^2 = o_p (1)$ and $\frac{1}{n_1} \sum_{i:D_i=1} \hat{w}_i^2 [f_1 (X_i) - \hat{f}_1 (X_i)]^2 = o_p (1)$.
    \end{itemize}
    Then, $\hat{V}_{\mathrm{ATE}} (\hat{\textbf{w}}, \hat{f}_0, \hat{f}_1) \overset{p}{\rightarrow} V_{\mathrm{ATE}} (\hat{\textbf{w}})$ where
		\begin{align*}
			\hat{V}_{\mathrm{ATE}} (\hat{\textbf{w}}, \hat{f}_0, \hat{f}_1) &= \frac{1}{n^2} \sum_{i=1}^n [\hat{f}_1 (X_i) - \hat{f}_0 (X_i)  - \hat{\tau}^{(\mathrm{ATE})}_{\mathrm{wdim}} (\hat{\textbf{w}}) ]^2  \\
			&+ \frac{1}{n_0^2}\sum_{i:D_i=0} \hat{w}_i^2 [Y_i (0) - \hat{f}_0 (X_i)]^2 \\
			&+  \frac{1}{n_1^2}\sum_{i:D_i=1} \hat{w}_i^2 [Y_i (1) - \hat{f}_1 (X_i)]^2 
		\end{align*}
	\end{proposition}
Proof for Proposition~\ref{thm:an_ate} is given in Appendix~\ref{app:thm5pf} and proof for Proposition~\ref{thm:Vhat_ate} is given in Appendix~\ref{app:thm6pf}.

We emphasize that Propositions~\ref{thm:an_ate} and~\ref{thm:Vhat_ate} are for an estimator with \textit{general} weights, not TFB specifically. However, as with the ATT, these propositions provide motivation for a closed-form variance estimator for TFB. $(100 \times \gamma) \%$ confidence intervals can then be obtained using the normal approximation.

They also motivate a closed-form variance for TFB with sample splitting. First, randomly split the full sample into two non-overlapping (sub-)samples: Samples 1 and 2. Use Sample 2 to obtain $(\hat{\beta}^{(s2)}_0, \hat{V}^{(s2)}_{\beta_0}, \hat{\beta}^{(s2)}_1, \hat{V}^{(s2)}_{\beta_1})$, and apply them to TFB in Sample 1 to form $\hat{\textbf{w}}^{(\mathrm{ATE}),(s1)}_{\mathrm{TFB}}$ and $\hat{\tau}_{\mathrm{wdim}}^{(\mathrm{ATE}),(s1)} (\hat{\textbf{w}}_{\mathrm{TFB}}^{(\mathrm{ATE}),(s1)})$. Next, switch the roles of the two sub-samples to obtain $(\hat{\beta}_0^{(s1)}, \hat{V}^{(s1)}_{\beta_0}, \hat{\beta}_1^{(s1)}, \hat{V}^{(s1)}_{\beta_1})$ from Sample 1, and $\hat{\textbf{w}}_{\mathrm{TFB}}^{(\mathrm{ATE}),(s2)}$ and $\hat{\tau}_{\mathrm{wdim}}^{(\mathrm{ATE}),(s2)} (\hat{\textbf{w}}_{\mathrm{TFB}}^{(\mathrm{ATE}),(s2)})$ from Sample 2. The final estimate is then the average of the estimates from each sub-sample:
    \begin{align}
        \hat{\tau}_{\mathrm{wdim}}^{(\mathrm{ATE}),(\mathrm{TFB})} = \frac{ \hat{\tau}_{\mathrm{wdim}}^{(\mathrm{ATE}),(s1)} (\hat{\textbf{w}}_{\mathrm{TFB}}^{(\mathrm{ATE}),(s1)}) + \hat{\tau}_{\mathrm{wdim}}^{(\mathrm{ATE}),(s2)} (\hat{\textbf{w}}_{\mathrm{TFB}}^{(\mathrm{ATE}),(s2)}) }{2}
    \end{align}
The proposed variance estimator is then:
    \begin{align}
        \widehat{\var} ( \hat{\tau}_{\mathrm{wdim}}^{(\mathrm{ATE}),(\mathrm{TFB})})  =  \frac{1}{4} \hat{V}_{\mathrm{ATE}}^{(s1)} (\hat{\textbf{w}}_{\mathrm{TFB}}^{(\mathrm{ATE}),(s1)}, \hat{f}_0^{(s2)}, \hat{f}_1^{(s2)}) + \frac{1}{4} \hat{V}_{\mathrm{ATE}}^{(s2)} (\hat{\textbf{w}}_{\mathrm{TFB}}^{(\mathrm{ATE}),(s2)}, \hat{f}_0^{(s1)}, \hat{f}_1^{(s1)}) 
    \end{align}
$(100 \times \gamma) \%$ confidence intervals can again be obtained using the normal approximation: $\hat{\tau}_{\mathrm{wdim}}^{(\mathrm{ATE}),(\mathrm{TFB})} \pm \Phi^{-1} ( \frac{1 + \gamma}{2} )  \times \sqrt{\widehat{\var} ( \hat{\tau}_{\mathrm{wdim}}^{(\mathrm{ATE}),(\mathrm{TFB})})}$.

\newpage
\clearpage

\section{Proofs}

\subsection{Proof of Proposition~\ref{prop:ewc}}\label{app:ewcprop_pf}

Decomposing the difference between $\hat{\tau}_{\mathrm{wdim}} (\mathbf{w})$ and SATT reveals Proposition~\ref{prop:ewc}:
 	\begin{align}\label{eq:taubias}
		\hat{\tau}_{\mathrm{wdim}} (\mathbf{w}) - \mathrm{SATT} = \mathrm{EWC \ Bias} \ + \ \biggr( \frac{1}{n_1} \sum_{i:D_i=1} \epsilon_i (0) - \frac{1}{n_0}\sum_{i:D_i=0} w_i \epsilon_i (0) \biggr)
	\end{align}
First, SATT is unbiased for the ATT. Then, assuming that each pair of $w_i$ and $Y_i (0) $ (and thus $\epsilon_i (0)$) are independent given $\mathbf{X}$ and $\mathbf{D}$,
taking the expectation of the both sides of (\ref{eq:taubias}) proves the result, as the terms with $\epsilon_i (0) $ are then mean-zero under Assumption~\ref{asm:ci}:
\begin{align}\label{eq:prop1_proof}
    & \E\biggr( \frac{1}{n_1} \sum_{i:D_i=1} \epsilon_i (0) - \frac{1}{n_0}\sum_{i:D_i=0} w_i \epsilon_i (0) \biggr)  \nonumber \\
    & \ \ \ \ \ \ \ \ \ =  \E \biggr[ \E\biggr( \frac{1}{n_1} \sum_{i:D_i=1} \epsilon_i (0) - \frac{1}{n_0}\sum_{i:D_i=0} w_i \epsilon_i (0)  \ \biggr| \ \mathbf{X}, \mathbf{D} \biggr) \biggr] \nonumber \\ 
    & \ \ \ \ \ \ \ \ \ = \E \biggr( \frac{1}{n_1} \sum_{i:D_i=1} \E [ \epsilon_i (0) \ | \ \mathbf{X}, \mathbf{D}] - \frac{1}{n_0}\sum_{i:D_i=0} \E[ w_i \epsilon_i (0) \ | \ \mathbf{X}, \mathbf{D}]  \biggr) \nonumber \\ 
    & \ \ \ \ \ \ \ \ \ = \E \biggr( \frac{1}{n_1} \sum_{i:D_i=1} \E [ \epsilon_i (0) \ | \ \mathbf{X}, \mathbf{D}] - \frac{1}{n_0}\sum_{i:D_i=0}  \underbrace{\E[ w_i \ | \ \mathbf{X}, \mathbf{D}] \E[ \epsilon_i (0) \ | \ \mathbf{X}, \mathbf{D}]}_{w_i \indep \epsilon_i (0) \ | \ \mathbf{X}, \mathbf{D}}  \biggr) \nonumber \\
    & \ \ \ \ \ \ \ \ \ = \E \biggr( \frac{1}{n_1} \sum_{i:D_i=1} \underbrace{\E[ \epsilon_i (0) \ | \ \mathbf{X}]}_{\text{Assumption~\ref{asm:ci}}} - \frac{1}{n_0}\sum_{i:D_i=0}  \E[ w_i \ | \ \mathbf{X}, \mathbf{D}] \underbrace{\E[ \epsilon_i (0) \ | \ \mathbf{X}]}_{\text{Assumption~\ref{asm:ci}}}  \biggr) \nonumber \\
     & \ \ \ \ \ \ \ \ \ = 0
\end{align}
where the last line of (\ref{eq:prop1_proof}) above applies that $\E[ \epsilon_i (0) \ | \ \mathbf{X}] = \E[ \epsilon_i (0) \ | \ X_i] = 0$ by definition.

\begin{flushright}
$\square$
\end{flushright}

\subsection{Proof of Proposition~\ref{prop:wc}}\label{app:wcprop_pf}

Letting the $w_i = w (X_i, D_i)$ for a weight function $w(\cdot)$, 
	\begin{align}\label{eq:wc1}
		\E[ \hat{\tau}_{\mathrm{wdim}} (\mathbf{w})  ] = \E[Y_{i} (1) \ | \ D_i = 1] - \E[w_i Y_i (0) \ | \ D_i = 0]
	\end{align}
Therefore, obtaining an unbiased estimate of the ATT using $\hat{\tau}_{\mathrm{wdim}} (\textbf{w})$ requires that 
	\begin{align}\label{eq:wc2}
		& \E[w_i Y_i (0) \ | \ D_i = 0] = \E[Y_{i} (0) \ | \ D_i = 1] 
	\end{align}
Applying the law of iterative expectations, (\ref{eq:wc2}) becomes
	\begin{align}\label{eq:wc3}
		& \E \biggr( w_i \E[  Y_i (0) \ | \ X_i, D_i=0] \ \biggr| \ D_i = 0 \biggr) = \E \biggr( \E[ Y_i (0) \ | \ X_i, D_i=1] \ \biggr| \ D_i = 1 \biggr)
	\end{align}
Assumption~\ref{asm:ci} then allows one to remove the conditioning on $D_i$ in the inner expectations of both sides of (\ref{eq:wc3}), giving
	\begin{align}\label{eq:wc4}
		& \E \biggr( w_i \E[  Y_i (0) \ | \ X_i] \ \biggr| \ D_i = 0 \biggr) = \E \biggr( \E[ Y_i (0) \ | \ X_i] \ \biggr| \ D_i = 1 \biggr)
	\end{align}
Applying the definition of $f_0 (X) = \E[ Y (0) \ | \ X]$ to (\ref{eq:wc4}) yields \ref{eq:wc5}.

\begin{flushright}
$\square$
\end{flushright}

\subsection{Proof of Proposition~\ref{thm:an_att}}\label{app:thm1pf}

$\hat{\tau}_{\mathrm{wdim}} (\hat{\mathbf{w}}) - \mathrm{ATT}$ can be written as the sum of three terms that we analyze separately:
	\begin{align}\label{eq:thm1pf_1}
		\hat{\tau}_{\mathrm{wdim}} (\hat{\mathbf{w}}) - \mathrm{ATT} = \underbrace{\biggr( \frac{1}{n_1} \sum_{i : D_i = 1}[ Y_i (1) - f_0 (X_i)  - \mathrm{ATT}] \biggr)}_{(a)} \ - \ \underbrace{\frac{1}{n_0} \sum_{i : D_i = 0} \hat{w}_i \epsilon_i (0) }_{(b)} \ + \ \text{EWC Bias} 
	\end{align}
	
For $(a)$, first notice that
	\begin{align}\label{eq:thm1pf_a_epsilon1}
		\E( \epsilon (0)  \ | \ D = 1) &= \E\biggr( \E( \epsilon (0)  \ | \ D = 1, X) \ \biggr| \ D = 1\biggr) \nonumber \\
		&= \E\biggr( \E( \epsilon (0)  \ | \ X) \ \biggr| \ D = 1\biggr) \nonumber \\
		&= 0
	\end{align}
where Assumption~\ref{asm:ci} allows the second line in (\ref{eq:thm1pf_a_epsilon1}) above. Therefore,
	\begin{align}\label{eq:thm1pf_a_epsilon2}
		\E[ Y (1) - f_0 (X)  - \mathrm{ATT} \ | \ D = 1] &= \E[ Y (1) - f_0 (X)  - \mathrm{ATT} \ | \ D = 1] - \E( \epsilon (0) \ | \ D = 1) \nonumber \\ 
		&= \E[ Y (1) - Y (0)  - \mathrm{ATT} \ | \ D = 1] \nonumber \\
		&= \E[ Y (1) - Y (0)  \ | \ D = 1] - \mathrm{ATT} \nonumber \\
		&= 0
	\end{align}
Thus, the Central Limit Theorem applies to $(a)$ by (\ref{eq:thm1pf_a_epsilon2}) and iid data, meaning that
	\begin{align}\label{eq:thm1pf_a}
		\sqrt{n_1} \times (a) \overset{d}{\rightarrow} N \biggr( 0, \E \biggr[ [ Y (1) - f_0 (X)  - \mathrm{ATT}]^2  \ \biggr| \ D = 1\biggr] \biggr)
	\end{align}
where $\E ( [ Y (1) - f_0 (X)  - \mathrm{ATT}]^2  \ | \ D = 1 )< \infty$ by condition (iv) for the proposition. Some manipulation then gives that  
	\begin{align}\label{eq:thm1pf_a_rewrite}
		\sqrt{n} \times (a) \overset{d}{\rightarrow} N \biggr( 0, \frac{ \E ( [ Y (1) - f_0 (X)  - \mathrm{ATT}]^2  \ | \ D = 1 ) }{p(D=1)}  \biggr)
	\end{align}

Moving to $(b)$, we first consider its distribution given $\mathbf{X}$ and $\mathbf{D}$. To begin, under Assumption~\ref{asm:ci}, iid data, and condition (i) for the proposition,
	\begin{align}
		\var \biggr( (b) \ \biggr| \ \mathbf{X}, \mathbf{D} \biggr) = \var(\frac{1}{n_0} \sum_{i : D_i = 0} \hat{w}_i \epsilon_i (0) \ | \ \mathbf{X}, \mathbf{D}) = \frac{1}{n_0^2}\sum_{i : D_i = 0} \hat{w}_i^2 \sigma^2 (0, X_i)
	\end{align}
Additionally, under conditions (iii) and (v) for the proposition, $\exists \delta > 0$ such that
	\begin{align}
		\frac{ \sum_{i: D_i = 0} \E (| \frac{\hat{w}_i}{n_0} \epsilon_i (0) |^{2 + \delta} \ | \ \mathbf{X}, \mathbf{D}) } {\var(\sum_{i : D_i = 0} \frac{\hat{w}_i}{n_0} \epsilon_i (0) \ | \ \mathbf{X}, \mathbf{D})^{\frac{2 + \delta}{2}}} \rightarrow 0 
	\end{align}
To see this, Assumption~\ref{asm:ci}, iid data, and conditions (i) and (iii) give
	\begin{align}\label{eq:thm1pf_2}
		\frac{ \sum_{i: D_i = 0} \E (| \frac{\hat{w}_i}{n_0} \epsilon_i (0) |^{2 + \delta} \ | \ \mathbf{X}, \mathbf{D}) } {\var(\sum_{i : D_i = 0} \frac{\hat{w}_i}{n_0} \epsilon_i (0) \ | \ \mathbf{X}, \mathbf{D})^{\frac{2 + \delta}{2}}} &= \frac{ \sum_{i: D_i = 0} \E (|\hat{w}_i \epsilon_i (0) |^{2 + \delta} \ | \ \mathbf{X}, \mathbf{D}) } {\var(\sum_{i : D_i = 0} \hat{w}_i \epsilon_i (0) \ | \ \mathbf{X}, \mathbf{D})^{\frac{2 + \delta}{2}}} \nonumber \\
		&= \frac{ \sum_{i: D_i = 0} \hat{w}_i^{2 + \delta} \E (| \epsilon_i (0) |^{2 + \delta} \ | \ \mathbf{X}) } {(\sum_{i : D_i = 0} \hat{w}_i^2 \sigma^2 (0, X_i) )^{\frac{2 + \delta}{2}}} \nonumber \\
		& \leq \biggr( \frac{1}{\underline{C}^{\frac{2 + \delta}{2}}} \biggr) \biggr( \frac{  \sum_{i: D_i = 0} \hat{w}_i^{2 + \delta} \E (| \epsilon_i (0) |^{2 + \delta} \ | \ \mathbf{X}) } {  (\sum_{i : D_i = 0} \hat{w}_i^2 )^{\frac{2 + \delta}{2}}} \biggr)
	\end{align}
Then, by condition (v) for the proposition, $\exists \delta, \overline{C}_{\delta} > 0$ that allow (\ref{eq:thm1pf_2}) to continue as
	\begin{align}
		\frac{ \sum_{i: D_i = 0} \E (| \frac{\hat{w}_i}{n_0} \epsilon_i (0) |^{2 + \delta} \ | \ \mathbf{X}, \mathbf{D}) } {\var(\sum_{i : D_i = 0} \frac{\hat{w}_i}{n_0} \epsilon_i (0) \ | \ \mathbf{X}, \mathbf{D})^{\frac{2 + \delta}{2}}} \leq \biggr( \frac{\overline{C}_{\delta}}{\underline{C}^{\frac{2 + \delta}{2}}} \biggr) \biggr( \frac{ \sum_{i: D_i = 0} \hat{w}_i^{2 + \delta} } {  (\sum_{i : D_i = 0} \hat{w}_i^2 )^{\frac{2 + \delta}{2}}} \biggr) \rightarrow 0
	\end{align}
Therefore, $(b)$ satisfies the Lyapunov condition conditional on $\mathbf{X}$ and $\mathbf{D}$, meaning that, by the Lindeberg-Feller theorem,
	\begin{align}\label{eq:thm1pf_b_cond}
		\frac{(b)}{\sqrt{ \frac{1}{n_0^2}\sum_{i : D_i = 0} \hat{w}_i^2 \sigma^2 (0, X_i) }} \ \biggr{|} \ \mathbf{X}, \mathbf{D} \overset{d}{\rightarrow} N(0, 1)
	\end{align}
which also implies here that
	\begin{align}\label{eq:thm1pf_b_uncond}
		\frac{(b)}{\sqrt{ \frac{1}{n_0^2} \sum_{i : D_i = 0} \hat{w}_i^2 \sigma^2 (0, X_i) }} \overset{d}{\rightarrow} N(0, 1)
	\end{align}
or, after some manipulation, that
	\begin{align}\label{eq:thm1pf_b_uncond_rewrite}
		\sqrt{n} \times (b) \overset{d}{\rightarrow} N \biggr( 0, \  \frac{ \underset{n \rightarrow \infty}{\mathrm{plim}} \ \frac{1}{n_0}\sum_{i : D_i = 0} \hat{w}_i^2 \sigma^2 (0, X_i)  }{p(D = 0)}  \biggr)
	\end{align}
where $\underset{n \rightarrow \infty}{\mathrm{plim}} \ \frac{1}{n_0}\sum_{i : D_i = 0} \hat{w}_i^2 \sigma^2 (0, X_i)  < \infty$ by condition (ii). 

Putting these pieces together,  $\sqrt{n} \times (a)$ and $\sqrt{n} \times (b)$ 
are asymptotically independent because they are independent given $\mathbf{X}$ and $\mathbf{D}$ and (\ref{eq:thm1pf_b_cond}) holds given $\mathbf{X}$ and $\mathbf{D}$. Therefore, by (\ref{eq:thm1pf_1}), (\ref{eq:thm1pf_a_rewrite}), and (\ref{eq:thm1pf_b_uncond_rewrite})
	\begin{align}\label{eq:thm1pf_alltogether1}
		& \sqrt{n} (\hat{\tau}_{\mathrm{wdim}} (\hat{\mathbf{w}}) - \mathrm{ATT} - \mathrm{EWC \ Bias}) \nonumber \\
		& \ \ \ \ \ = \sqrt{n} \biggr( (a) - (b) \biggr) \nonumber \\
		& \ \ \ \ \ \overset{d}{\rightarrow} N\biggr( 0, \frac{ \E ( [ Y (1) - f_0 (X)  - \mathrm{ATT}]^2  \ | \ D = 1 )  }{p(D=1)} + \frac{\underset{n \rightarrow \infty}{\mathrm{plim}} \ \frac{1}{n_0}\sum_{i : D_i = 0} \hat{w}_i^2 \sigma^2 (0, X_i) }{p(D = 0)}  \biggr)
	\end{align}
Further, $\sqrt{n} \times \mathrm{EWC \ Bias} \overset{p}{\rightarrow} 0$ by condition (vi) for the proposition. Thus, (\ref{eq:thm1pf_alltogether1}) can be rewritten as
	\begin{align}\label{eq:thm1pf_alltogether2}
		& \sqrt{n} (\hat{\tau}_{\mathrm{wdim}} (\hat{\mathbf{w}}) - \mathrm{ATT}) \nonumber \\
		& \ \ \ \ \ \overset{d}{\rightarrow} N\biggr( 0, \frac{ \E ( [ Y (1) - f_0 (X)  - \mathrm{ATT}]^2  \ | \ D = 1 )  }{p(D=1)} + \frac{\underset{n \rightarrow \infty}{\mathrm{plim}} \ \frac{1}{n_0}\sum_{i : D = 0} \hat{w}_i^2 \sigma^2 (0, X_i) }{p(D = 0)}  \biggr)
	\end{align}
Finally, examining $V_{\mathrm{ATT}} (\hat{\mathbf{w}}) $, it is then evident that
	\begin{align}\label{eq:thm1pf_V_consistent}
		n \times V_{\mathrm{ATT}} (\hat{\mathbf{w}}) &= \frac{ \E ( [ Y (1) - f_0 (X)  - \mathrm{ATT}]^2  \ | \ D = 1 )  }{\frac{n_1}{n}} + \frac{\frac{1}{n_0}\sum_{i : D_i = 0} \hat{w}_i^2 \sigma^2 (0, X_i) }{\frac{n_0}{n}} \nonumber \\
		&\overset{p}{\rightarrow} \frac{ \E ( [ Y (1) - f_0 (X)  - \mathrm{ATT}]^2  \ | \ D = 1 )  }{p(D=1)} + \frac{\underset{n \rightarrow \infty}{\mathrm{plim}} \ \frac{1}{n_0}\sum_{i : D_i = 0} \hat{w}_i^2 \sigma^2 (0, X_i) }{p(D = 0)}
	\end{align}
Thus, applying (\ref{eq:thm1pf_V_consistent}) to (\ref{eq:thm1pf_alltogether2}) completes the proof.

\begin{flushright}
$\square$
\end{flushright}

\subsection{Proof of Proposition~\ref{thm:Vhat}}\label{app:thm2pf}

$\hat{V}_{\mathrm{ATT}} (\hat{\mathbf{w}}, \hat{f}_0)$ can be rewritten as:
	\begin{align}\label{eq:thm2pf_decomp}
		\hat{V}_{\mathrm{ATT}}  (\hat{\mathbf{w}}, \hat{f}_0) &= \underbrace{ \frac{1}{n_1^2} \sum_{i: D_i = 1} \biggr( [Y_i - f_0 (X_i) - \mathrm{ATT}] + [f_0 (X_i) - \hat{f}_0 (X_i)] + [\mathrm{ATT} - \hat{\tau}_{\mathrm{wdim}} (\hat{\mathbf{w}})] \biggr)^2 }_{(a)} \nonumber \\
		&+ \underbrace{ \frac{1}{n_0^2}\sum_{i: D_i = 0} \hat{w}_i^2 \biggr( \epsilon_i (0)  + [f_0 (X_i) - \hat{f}_0 (X_i)] \biggr)^2 }_{(b)}
	\end{align}
We prove the proposition by showing that $(a) \overset{p}{\rightarrow} \frac{1}{n_1} \E ( [Y (1) - f_0 (X) - \mathrm{ATT}]^2 \ | \ D = 1 )$ and that $(b) \overset{p}{\rightarrow} \frac{1}{n_0^2} \sum_{i : D_i = 0} \hat{w}_i^2 \sigma^2 (0, X_i) $. 

Starting with $(a)$, 
	\begin{align}\label{eq:thm2pf_a}
		(a) &= \underbrace{ \frac{1}{n_1^2} \sum_{i: D_i = 1} [Y_i - f_0 (X_i) - \mathrm{ATT}]^2 }_{(a.1)} \nonumber \\ 
		&+ \underbrace{  \frac{1}{n_1^2} \sum_{i: D_i = 1}   [f_0 (X_i) - \hat{f}_0 (X_i)]^2 }_{(a.2)} \nonumber \\ 
		&+ \underbrace{  \frac{1}{n_1^2} \sum_{i: D_i = 1} [\mathrm{ATT} - \hat{\tau}_{\mathrm{wdim}} (\hat{\mathbf{w}})]^2 }_{(a.3)} \nonumber \\ 
		&+ \underbrace{  \frac{2}{n_1^2} \sum_{i: D_i = 1} [Y_i - f_0 (X_i) - \mathrm{ATT}] [f_0 (X_i) - \hat{f}_0 (X_i)] }_{(a.4)} \nonumber \\
		&+ \underbrace{  \frac{2}{n_1^2} \sum_{i: D_i = 1} [Y_i - f_0 (X_i) - \mathrm{ATT}][\mathrm{ATT} - \hat{\tau}_{\mathrm{wdim}} (\hat{\mathbf{w}})] }_{(a.5)} \nonumber \\
		&+ \underbrace{  \frac{2}{n_1^2} \sum_{i: D_i = 1} [f_0 (X_i) - \hat{f}_0 (X_i)][\mathrm{ATT} - \hat{\tau}_{\mathrm{wdim}} (\hat{\mathbf{w}})] }_{(a.6)}		
	\end{align}
Given iid data, and thus by the Law of Large Numbers,
	\begin{align}\label{eq:thm2pf_a1}
		(a.1) \overset{p}{\rightarrow} \frac{1}{n_1} \E \biggr[ [Y (1) - f_0 (X) - \mathrm{ATT}]^2 \ \biggr| \ D = 1 \biggr]
	\end{align}
where $\E ( [ Y (1) - f_0 (X)  - \mathrm{ATT}]^2  \ | \ D = 1 )< \infty$ by condition (iv) for Proposition~\ref{thm:an_att}. By condition (vii) of the proposition, 
	\begin{align}\label{eq:thm2pf_a2}
		(a.2) = o_p (n^{-1})
	\end{align}
and, by the Cauchy-Schwartz Inequality, along with (\ref{eq:thm2pf_a1}) and (\ref{eq:thm2pf_a2}),
	\begin{align}\label{eq:thm2pf_a4}
		| (a.4) | \leq 2 \sqrt{ (a.1) } \sqrt{ (a.2) } = 2 \sqrt{O_p (n^{-1})} \sqrt{o_p (n^{-1})} = o_p (n^{-1}) 
	\end{align}
Then, under the conditions for Proposition~\ref{thm:an_att}, $\hat{\tau}_{\mathrm{wdim}} (\hat{\mathbf{w}}) \overset{p}{\rightarrow} \mathrm{ATT}$, so 
	\begin{align}\label{eq:thm2pf_a3}
		(a.3) = \frac{1}{n_1} (\mathrm{ATT} - \hat{\tau}_{\mathrm{wdim}} (\hat{\mathbf{w}}))^2 = o_p (n^{-1}) 	
	\end{align}
and, by iid data, and thus by the Law of Large Numbers, 	
	\begin{align}\label{eq:thm2pf_a5}
		(a.5) \overset{p}{\rightarrow} 2 \biggr( \frac{\mathrm{ATT} - \hat{\tau}_{\mathrm{wdim}} (\hat{\mathbf{w}})}{n_1} \biggr) \E [ Y (1) - f_0 (X) - \mathrm{ATT} \ | \ D = 1] = o_p (n^{-1}) 	
	\end{align}
	\vspace{-0.3in}
	\begin{align}\label{eq:thm2pf_a6}
		(a.6) \overset{p}{\rightarrow} 2 \biggr( \frac{\mathrm{ATT} - \hat{\tau}_{\mathrm{wdim}} (\hat{\mathbf{w}})}{n_1} \biggr) \E [ f_0 (X) - \hat{f}_0 (X)  \ | \ D = 1] = o_p (n^{-1}) 
	\end{align}
Therefore, applying (\ref{eq:thm2pf_a1}), (\ref{eq:thm2pf_a2}), (\ref{eq:thm2pf_a4}), (\ref{eq:thm2pf_a3}), (\ref{eq:thm2pf_a5}), and (\ref{eq:thm2pf_a6}) to (\ref{eq:thm2pf_a}) yields
	\begin{align}\label{eq:thm2pf_alim}
		(a) \overset{p}{\rightarrow} \frac{1}{n_1} \E \biggr[ [Y (1) - f_0 (X) - \mathrm{ATT}]^2 \ \biggr| \ D = 1 \biggr] 
	\end{align}

Moving to $(b)$ in (\ref{eq:thm2pf_decomp}), 
	\begin{align}\label{eq:thm2pf_b}
		(b) &= \underbrace{ \frac{1}{n_0^2} \sum_{i: D_i = 0} \hat{w}_i^2 \epsilon_i^2 (0)  }_{(b.1)} \nonumber \\
		&+ \underbrace{ \frac{1}{n_0^2}\sum_{i: D_i = 0} \hat{w}_i^2 [f_0 (X_i) - \hat{f}_0 (X_i)]^2 }_{(b.2)} \nonumber \\
		&+  \underbrace{  \frac{2}{n_0^2} \sum_{i: D_i = 0} \hat{w}_i^2 \epsilon_i (0) [f_0 (X_i) - \hat{f}_0 (X_i)] }_{(b.3)}
	\end{align}
By condition (viii) for the proposition,
	\begin{align}\label{eq:thm2pf_b1}
		(b.1) \overset{p}{\rightarrow}  \frac{1}{n_0^2} \sum_{i : D_i =0} \hat{w}_i^2 \sigma^2 (0, X_i) 
	\end{align}
Then, by condition (ix) for the proposition, 
	\begin{align}\label{eq:thm2pf_b2}
		(b.2) = o_p(n^{-1})
	\end{align}
and using the Cauchy-Schwartz Inequality and (\ref{eq:thm2pf_b1}), 
	\begin{align}\label{eq:thm2pf_b3_pt1}
		| (b.3) | &\leq 2 \sqrt{(b.1)} \sqrt{(b.2)} \nonumber \\
		&\overset{p}{\rightarrow} 2 \sqrt{ \frac{1}{n_0^2} \sum_{i : D_i =0} \hat{w}_i^2 \sigma^2 (0, X_i) } \sqrt{(b.2)}
	\end{align}
Then, because $\underset{n \rightarrow \infty}{\mathrm{plim}} \ \frac{1}{n_0} \sum_{i : D_i =0} \hat{w}_i^2 \sigma^2 (0, X_i)   $ is finite by condition (ii) for Proposition~\ref{thm:an_att}, we find that $\frac{1}{n_0^2} \sum_{i : D_i =0} \hat{w}_i^2 \sigma^2 (0, X_i) $ in (\ref{eq:thm2pf_b3_pt1}) is $O_p (n^{-1})$. Additionally using (\ref{eq:thm2pf_b2}) allows (\ref{eq:thm2pf_b3_pt1}) to continue as
	\begin{align}\label{eq:thm2pf_b3_pt2}
		| (b.3) | \leq 2 \sqrt{O_p (n^{-1})} \sqrt{o_p (n^{-1})} = o_p (n^{-1})
	\end{align}
Therefore, applying (\ref{eq:thm2pf_b1}), (\ref{eq:thm2pf_b2}), and (\ref{eq:thm2pf_b3_pt2}) to (\ref{eq:thm2pf_b}) gives 
	\begin{align}\label{eq:thm2pf_blim}
		(b) \overset{p}{\rightarrow} \frac{1}{n_0^2} \sum_{i : D_i = 0} \hat{w}_i^2 \sigma^2 (0, X_i) 
	\end{align}
Finally, applying (\ref{eq:thm2pf_alim}) and  (\ref{eq:thm2pf_blim}) to (\ref{eq:thm2pf_decomp}) completes the proof.

\begin{flushright}
$\square$
\end{flushright}

\subsection{Proof of Proposition~\ref{thm:an_ate}}\label{app:thm5pf}

$\hat{\tau}^{(\mathrm{ATE})}_{\mathrm{wdim}} (\hat{\textbf{w}}) - \mathrm{ATE}$ can be written as the sum of four terms that we analyze separately:
	\begin{align}\label{eq:thm5pf_1}
		\hat{\tau}^{(\mathrm{ATE})}_{\mathrm{wdim}} (\hat{\mathbf{w}}) - \mathrm{ATE} &= \underbrace{\biggr( \frac{1}{n} \sum_{i=1}^n[ f_1 (X_i) - f_0 (X_i)  - \mathrm{ATE}] \biggr)}_{(a)} \nonumber \\ 
		&- \underbrace{ \frac{1}{n_0} \sum_{i: D_i = 0} \hat{w}_i \epsilon_i (0) }_{(b)} \nonumber \\
		&+ \underbrace{ \frac{1}{n_1} \sum_{i: D_i = 1} \hat{w}_i \epsilon_i (1) }_{(c)} \nonumber \\
		&+ \underbrace{ \biggr( \mathrm{imbal}^{(\mathrm{ATE, 0})} (\hat{\mathbf{w}}, f_0 (\mathbf{X}), \mathbf{D}) - \mathrm{imbal}^{(\mathrm{ATE, 1})} (\hat{\mathbf{w}}, f_1 (\mathbf{X}), \mathbf{D}) \biggr) }_{(d)}
	\end{align}

For $(a)$, first notice that
	\begin{align}\label{eq:thm5pf_a_epsilon1}
		\E( \epsilon (d)  ) &= \E [\E( \epsilon (d) \ | \  X) ] = 0 \ \ \text{for} \ \ d\in \{0, 1\}
	\end{align}
Therefore,
	\begin{align}\label{eq:thm5pf_a_epsilon2}
		\E[ f_1 (X) - f_0 (X)  - \mathrm{ATE} ] &= \E[ f_1 (X) - f_0 (X)  - \mathrm{ATE}] + \E( \epsilon (1) ) -  \E( \epsilon (0) ) \nonumber \\ 
		&= \E[ Y (1) - Y (0)  - \mathrm{ATE}] \nonumber \\
		&= \E[ Y (1) - Y (0) ] - \mathrm{ATE} \nonumber \\
		&= 0
	\end{align}
Thus, the Central Limit Theorem applies to $(a)$ by (\ref{eq:thm5pf_a_epsilon2}) and iid data, 
meaning that
	\begin{align}\label{eq:thm5pf_a}
		\sqrt{n} \times (a) \overset{d}{\rightarrow} N \biggr( 0, \E \biggr[ [ f_1 (X) - f_0 (X)  - \mathrm{ATE}]^2  \biggr] \biggr)
	\end{align}
where $\E ( [ f_1 (X) - f_0 (X)  - \mathrm{ATE}]^2  )< \infty$ by condition (iv) for the proposition.

As for $(b)$ and $(c)$, we only thoroughly derive the asymptotic distribution of $(b)$, because that of $(c)$ comes from a similar argument. We begin by considering the distribution of $(b)$ given $\mathbf{X}$ and $\mathbf{D}$. Notice that under Assumption~\ref{asm:ci}, iid data, and condition (i) for the proposition,
	\begin{align}
		\var \biggr( (b) \ \biggr| \ \mathbf{X}, \mathbf{D} \biggr) = \var(\frac{1}{n_0} \sum_{i : D_i = 0} \hat{w}_i \epsilon_i (0) \ | \ \mathbf{X}, \mathbf{D}) = \frac{1}{n_0^2}\sum_{i : D_i = 0} \hat{w}_i^2 \sigma^2 (0, X_i) 
	\end{align}
Additionally, under conditions (iii) and (v) for the proposition, $\exists \delta > 0$ such that
	\begin{align}
		\frac{ \sum_{i: D_i = 0} \E (| \frac{\hat{w}_i}{n_0} \epsilon_i (0) |^{2 + \delta} \ | \ \mathbf{X}, \mathbf{D}) } {\var(\sum_{i : D_i = 0} \frac{\hat{w}_i}{n_0} \epsilon_i (0)  \ | \ \mathbf{X}, \mathbf{D})^{\frac{2 + \delta}{2}}} \rightarrow 0 
	\end{align}
To see this, Assumption~\ref{asm:ci}, iid data, and conditions (i) and (iii) give
	\begin{align}\label{eq:thm5pf_2}
		\frac{ \sum_{i: D_i = 0} \E (| \frac{\hat{w}_i}{n_0} \epsilon_i (0) |^{2 + \delta} \ | \ \mathbf{X}, \mathbf{D}) } {\var(\sum_{i : D_i = 0} \frac{\hat{w}_i}{n_0} \epsilon_i (0)  \ | \ \mathbf{X}, \mathbf{D})^{\frac{2 + \delta}{2}}} &= \frac{ \sum_{i: D_i = 0} \E (|\hat{w}_i \epsilon_i (0) |^{2 + \delta} \ | \ \mathbf{X}, \mathbf{D}) } {\var(\sum_{i : D_i = 0} \hat{w}_i \epsilon_i (0)  \ | \ \mathbf{X}, \mathbf{D})^{\frac{2 + \delta}{2}}} \nonumber \\
		&= \frac{ \sum_{i: D_i = 0} \hat{w}_i^{2 + \delta} \E (| \epsilon_i (0) |^{2 + \delta} \ | \ \mathbf{X}) } {(\sum_{i : D_i = 0} \hat{w}_i^2 \sigma^2 (0, X_i) )^{\frac{2 + \delta}{2}}} \nonumber \\
		& \leq \biggr( \frac{1}{\underline{C}^{\frac{2 + \delta}{2}}} \biggr) \biggr( \frac{  \sum_{i: D_i = 0} \hat{w}_i^{2 + \delta} \E (| \epsilon_i (0) |^{2 + \delta} \ | \ \mathbf{X}) } {  (\sum_{i : D_i = 0} \hat{w}_i^2 )^{\frac{2 + \delta}{2}}} \biggr)
	\end{align}
Then, by condition (v) for the proposition, $\exists \delta, \overline{C}_{\delta} > 0$ that allow (\ref{eq:thm5pf_2}) to continue as
	\begin{align}
		\frac{ \sum_{i: D_i = 0} \E (| \frac{\hat{w}_i}{n_0} \epsilon_i (0) |^{2 + \delta} \ | \ \mathbf{X}, \mathbf{D}) } {\var(\sum_{i : D_i = 0} \frac{\hat{w}_i}{n_0} \epsilon_i (0)  \ | \ \mathbf{X}, \mathbf{D})^{\frac{2 + \delta}{2}}} \leq \biggr( \frac{\overline{C}_{\delta}}{\underline{C}^{\frac{2 + \delta}{2}}} \biggr) \biggr( \frac{ \sum_{i: D_i = 0} \hat{w}_i^{2 + \delta} } {  (\sum_{i : D_i = 0} \hat{w}_i^2 )^{\frac{2 + \delta}{2}}} \biggr) \rightarrow 0
	\end{align}
Therefore, $(b)$ satisfies the Lyapunov condition conditional on $\mathbf{X}$ and $\mathbf{D}$, meaning that, by the Lindeberg-Feller theorem,
	\begin{align}\label{eq:thm5pf_b_cond}
		\frac{(b)}{\sqrt{ \frac{1}{n_0^2}\sum_{i : D_i = 0} \hat{w}_i^2 \sigma^2 (0, X_i) }} \ \biggr{|} \ \mathbf{X}, \mathbf{D} \overset{d}{\rightarrow} N(0, 1)
	\end{align}
which also implies here that
	\begin{align}\label{eq:thm5pf_b_uncond}
		\frac{(b)}{\sqrt{ \frac{1}{n_0^2} \sum_{i : D_i = 0} \hat{w}_i^2 \sigma^2 (0, X_i) }} \overset{d}{\rightarrow} N(0, 1)
	\end{align}
or, after some manipulation, that
	\begin{align}\label{eq:thm5pf_b_uncond_rewrite}
		\sqrt{n} \times (b) \overset{d}{\rightarrow} N \biggr( 0, \  \frac{ \underset{n \rightarrow \infty}{\mathrm{plim}} \ \frac{1}{n_0}\sum_{i : D_i = 0} \hat{w}_i^2 \sigma^2 (0, X_i)  }{p(D = 0)}  \biggr)
	\end{align}
where $\underset{n \rightarrow \infty}{\mathrm{plim}} \ \frac{1}{n_0}\sum_{i : D_i = 0} \hat{w}_i^2 \sigma^2 (0, X_i)  < \infty$ by condition (ii). 

Moving to $(c)$, the steps to derive its asymptotic distribution are nearly identical to those for $(b)$, so we omit them here and simply state the final results. One finds that
	\begin{align}\label{eq:thm5pf_c_cond}
		\frac{(c)}{\sqrt{ \frac{1}{n_1^2}\sum_{i : D_i = 1} \hat{w}_i^2 \sigma^2 (1, X_i) }} \ \biggr{|} \ \mathbf{X}, \mathbf{D} \overset{d}{\rightarrow} N(0, 1)
	\end{align}
which also means that 
	\begin{align}\label{eq:thm5pf_c_uncond}
		\frac{(c)}{\sqrt{ \frac{1}{n_1^2} \sum_{i : D_i = 1} \hat{w}_i^2 \sigma^2 (1, X_i) }} \overset{d}{\rightarrow} N(0, 1)
	\end{align}
and, after some manipulation, that 
	\begin{align}\label{eq:thm5pf_c_uncond_rewrite}
		\sqrt{n} \times (c) \overset{d}{\rightarrow} N \biggr( 0, \  \frac{ \underset{n \rightarrow \infty}{\mathrm{plim}} \ \frac{1}{n_1}\sum_{i : D_i = 1} \hat{w}_i^2 \sigma^2 (1, X_i)  }{p(D = 1)}  \biggr)
	\end{align}
where $\underset{n \rightarrow \infty}{\mathrm{plim}} \ \frac{1}{n_1}\sum_{i : D_i = 1} \hat{w}_i^2 \sigma^2 (1, X_i)  < \infty$ by condition (ii).

Putting these pieces together, $\sqrt{n} \times (a)$, $\sqrt{n} \times (b)$, and $\sqrt{n} \times (c)$ are asymptotically independent because they are independent given $\mathbf{X}$ and $\mathbf{D}$ and (\ref{eq:thm5pf_b_cond}) and (\ref{eq:thm5pf_c_cond}) both hold given $\mathbf{X}$ and $\mathbf{D}$. Therefore, by (\ref{eq:thm5pf_1}), (\ref{eq:thm5pf_a}), (\ref{eq:thm5pf_b_uncond_rewrite}), and (\ref{eq:thm5pf_c_uncond_rewrite}),
	\begin{align}\label{eq:thm5pf_alltogether1}
		& \sqrt{n} \biggr( \hat{\tau}_{\mathrm{wdim}}^{(\mathrm{ATE})} (\hat{\mathbf{w}}) - \mathrm{ATE} - (d) \biggr) \nonumber \\
		& \ \ \ \ \ = \sqrt{n} \biggr( (a) - (b) + (c) \biggr) \nonumber \\
		& \ \ \ \ \ \overset{d}{\rightarrow} N\biggr( 0, \E ( [ f_1 (X) - f_0 (X)  - \mathrm{ATE}]^2 ) + \frac{\underset{n \rightarrow \infty}{\mathrm{plim}} \ \frac{1}{n_0}\sum_{i : D_i = 0} \hat{w}_i^2 \sigma^2 (0, X_i) }{p(D = 0)} + \frac{\underset{n \rightarrow \infty}{\mathrm{plim}} \ \frac{1}{n_1}\sum_{i : D_i = 1} \hat{w}_i^2 \sigma^2 (1, X_i) }{p(D = 1)}   \biggr)
	\end{align}
Further, $\sqrt{n} \times (d) \overset{p}{\rightarrow} 0$ by condition (vi) for the proposition. Thus, (\ref{eq:thm5pf_alltogether1}) can be rewritten as
	\begin{align}\label{eq:thm5pf_alltogether2}
		& \sqrt{n} ( \hat{\tau}_{\mathrm{wdim}}^{(\mathrm{ATE})} (\hat{\mathbf{w}}) - \mathrm{ATE}) \nonumber \\
		& \ \ \ \ \ \overset{d}{\rightarrow} N\biggr( 0, \E ( [ f_1 (X) - f_0 (X)  - \mathrm{ATE}]^2 ) + \frac{\underset{n \rightarrow \infty}{\mathrm{plim}} \ \frac{1}{n_0}\sum_{i : D_i = 0} \hat{w}_i^2 \sigma^2 (0, X_i) }{p(D = 0)} + \frac{\underset{n \rightarrow \infty}{\mathrm{plim}} \ \frac{1}{n_1}\sum_{i : D_i = 1} \hat{w}_i^2 \sigma^2 (1, X_i) }{p(D = 1)}   \biggr)
	\end{align}
Finally, examining $V_{\mathrm{ATE}} (\hat{\mathbf{w}}) $, it is then evident that
	\begin{align}\label{eq:thm5pf_V_consistent}
		& n \times V_{\mathrm{ATE}} (\hat{\mathbf{w}}) \nonumber \\
		& \ \ \ \ \ =  \E ( [ f_1 (X) - f_0 (X)  - \mathrm{ATE}]^2 )  + \frac{\frac{1}{n_0}\sum_{i : D_i = 0} \hat{w}_i^2 \sigma^2 (0, X_i) }{\frac{n_0}{n}} + \frac{\frac{1}{n_1}\sum_{i : D_i = 1} \hat{w}_i^2 \sigma^2 (1, X_i) }{\frac{n_1}{n}}\nonumber \\
		& \ \ \ \ \ \overset{p}{\rightarrow} \E ( [ f_1 (X) - f_0 (X)  - \mathrm{ATE}]^2 ) + \frac{\underset{n \rightarrow \infty}{\mathrm{plim}} \ \frac{1}{n_0}\sum_{i : D_i = 0} \hat{w}_i^2 \sigma^2 (0, X_i) }{p(D = 0)} + \frac{\underset{n \rightarrow \infty}{\mathrm{plim}} \ \frac{1}{n_1}\sum_{i : D_i = 1} \hat{w}_i^2 \sigma^2 (1, X_i) }{p(D = 1)}  
	\end{align}
Thus, applying (\ref{eq:thm5pf_V_consistent}) to (\ref{eq:thm5pf_alltogether2}) completes the proof.

\begin{flushright}
$\square$
\end{flushright}

\subsection{Proof of Proposition~\ref{thm:Vhat_ate}}\label{app:thm6pf}

$\hat{V}_{\mathrm{ATE}}  (\hat{\mathbf{w}}, \hat{f}_0)$ can be rewritten as:
	\begin{align}\label{eq:thm6pf_decomp}
		\hat{V}_{\mathrm{ATE}}  (\hat{\mathbf{w}}, \hat{f}_0) &= \underbrace{ \frac{1}{n^2} \sum_{i=1}^n \biggr( [f_1 (X_i) - f_0 (X_i) - \mathrm{ATE}] + [f_0 (X_i) - \hat{f}_0 (X_i)] + [\hat{f}_1 (X_i) - f_1 (X_i)] + [\mathrm{ATE} - \hat{\tau}^{(\mathrm{ATE})}_{\mathrm{wdim}} (\hat{\mathbf{w}})] \biggr)^2 }_{(a)} \nonumber \\
		&+ \underbrace{ \frac{1}{n_0^2} \sum_{i: D_i = 0} \hat{w}_i^2 \biggr( \epsilon_i (0)  + [f_0 (X_i) - \hat{f}_0 (X_i)] \biggr)^2 }_{(b)} \nonumber \\
		&+ \underbrace{ \frac{1}{n_1^2} \sum_{i: D_i = 1} \hat{w}_i^2 \biggr( \epsilon_i (1)  + [f_1 (X_i) - \hat{f}_1 (X_i)] \biggr)^2 }_{(c)}
	\end{align}
We prove the proposition by showing that $(a) \overset{p}{\rightarrow} \frac{1}{n} \E ( [f_1 (X) - f_0 (X) - \mathrm{ATE}]^2 )$,  $(b) \overset{p}{\rightarrow} \frac{1}{n_0^2} \sum_{i : D_i = 0} \hat{w}_i^2 \sigma^2 (0, X_i) $, and $(c) \overset{p}{\rightarrow} \frac{1}{n_1^2} \sum_{i : D_i = 1} \hat{w}_i^2 \sigma^2 (1, X_i) $.

Starting with $(a)$, 
	\begin{align}\label{eq:thm6pf_a}
		(a) &= \underbrace{ \frac{1}{n^2} \sum_{i=1}^n [f_1 (X_i) - f_0 (X_i) - \mathrm{ATE}]^2 }_{(a.1)} \nonumber \\ 
		&+ \underbrace{  \frac{1}{n^2} \sum_{i=1}^n  [f_0 (X_i) - \hat{f}_0 (X_i)]^2 }_{(a.2)} \nonumber \\ 
		&+ \underbrace{  \frac{1}{n^2} \sum_{i=1}^n  [f_1 (X_i) - \hat{f}_1 (X_i)]^2 }_{(a.3)} \nonumber \\ 
		&+ \underbrace{  \frac{1}{n^2} \sum_{i=1}^n [\mathrm{ATE} - \hat{\tau}^{(\mathrm{ATE})}_{\mathrm{wdim}} (\hat{\mathbf{w}})]^2 }_{(a.4)} \nonumber \\ 
		&+ \underbrace{  \frac{2}{n^2} \sum_{i=1}^n [f_1 (X_i) - f_0 (X_i) - \mathrm{ATE}] \biggr( [f_0 (X_i) - \hat{f}_0 (X_i)] + [\hat{f}_1 (X_i) - f_1 (X_i)] \biggr) }_{(a.5)} \nonumber \\
		&+ \underbrace{  \frac{2}{n^2} \sum_{i=1}^n [f_1 (X_i) - f_0 (X_i) - \mathrm{ATE}][\mathrm{ATE} - \hat{\tau}^{(\mathrm{ATE})}_{\mathrm{wdim}} (\hat{\mathbf{w}})] }_{(a.6)} \nonumber \\
		&+ \underbrace{ \frac{2}{n^2} \sum_{i=1}^n [f_0 (X_i) - \hat{f}_0 (X_i)][\hat{f}_1 (X_i) - f_1 (X_i)] }_{(a.7)} \nonumber \\
		&+ \underbrace{  \frac{2}{n^2} \sum_{i=1}^n \biggr( [f_0 (X_i) - \hat{f}_0 (X_i)] + [\hat{f}_1 (X_i) - f_1 (X_i)] \biggr)[\mathrm{ATE} - \hat{\tau}^{(\mathrm{ATE})}_{\mathrm{wdim}} (\hat{\mathbf{w}})] }_{(a.8)}		
	\end{align}
Given iid data, and thus by the Law of Large Numbers,
	\begin{align}\label{eq:thm6pf_a1}
		(a.1) \overset{p}{\rightarrow} \frac{1}{n} \E \biggr[ [f_1 (X) - f_0 (X) - \mathrm{ATE}]^2 \biggr]
	\end{align}
where $\E ( [f_1 (X) - f_0 (X) - \mathrm{ATE}]^2 )$ is finite by condition (iv) for Proposition~\ref{thm:an_ate}. 
By condition (vii) of the proposition, 
	\begin{align}\label{eq:thm6pf_a2}
		(a.2) = o_p (n^{-1})
	\end{align}
	\begin{align}\label{eq:thm6pf_a3}
		(a.3) = o_p (n^{-1})
	\end{align}
and, by the Cauchy-Schwartz Inequality, along with (\ref{eq:thm6pf_a1}), (\ref{eq:thm6pf_a2}), and (\ref{eq:thm6pf_a3})
	\begin{align}\label{eq:thm6pf_a5}
		| (a.5) | &\leq 2 \sqrt{ (a.1) } \sqrt{ (a.2) } + 2 \sqrt{ (a.1) } \sqrt{ (a.3) } \nonumber \\
		&= 2 \sqrt{O_p (n^{-1})} \sqrt{o_p (n^{-1})} + 2 \sqrt{O_p (n^{-1})} \sqrt{o_p (n^{-1})} \nonumber \\
		& = o_p (n^{-1}) 
	\end{align}
	\vspace{-0.45in}
	\begin{align}\label{eq:thm6pf_a7}
		| (a.7) | &\leq 2 \sqrt{ (a.2) } \sqrt{ (a.3) } = 2 \sqrt{o_p (n^{-1})} \sqrt{o_p (n^{-1})} = o_p (n^{-1}) 
	\end{align}
Then, under the conditions for Proposition~\ref{thm:an_ate}, $\hat{\tau}^{(\mathrm{ATE})}_{\mathrm{wdim}} (\hat{\mathbf{w}}) \overset{p}{\rightarrow} \mathrm{ATE}$, so 
	\begin{align}\label{eq:thm6pf_a4}
		(a.4) = \frac{1}{n} (\mathrm{ATE} - \hat{\tau}^{(\mathrm{ATE})}_{\mathrm{wdim}} (\hat{\mathbf{w}}))^2 = o_p (n^{-1}) 	
	\end{align}
and, by iid data, and thus by the Law of Large Numbers, 	
	\begin{align}\label{eq:thm6pf_a6}
		(a.6) \overset{p}{\rightarrow} 2 \biggr( \frac{\mathrm{ATE} - \hat{\tau}^{(\mathrm{ATE})}_{\mathrm{wdim}} (\hat{\mathbf{w}})}{n} \biggr) \E [  f_1 (X) - f_0 (X) - \mathrm{ATE} ] = o_p (n^{-1}) 	
	\end{align}
	\vspace{-0.3in}
	\begin{align}\label{eq:thm6pf_a8}
		(a.8) \overset{p}{\rightarrow} 2 \biggr( \frac{\mathrm{ATE} - \hat{\tau}^{(\mathrm{ATE})}_{\mathrm{wdim}} (\hat{\mathbf{w}})}{n} \biggr) \E \biggr( [f_0 (X) - \hat{f}_0 (X)] + [\hat{f}_1 (X) - f_1 (X)] \biggr) = o_p (n^{-1}) 
	\end{align}
Therefore, applying (\ref{eq:thm6pf_a1}), (\ref{eq:thm6pf_a2}), (\ref{eq:thm6pf_a3}), (\ref{eq:thm6pf_a5}), (\ref{eq:thm6pf_a7}), (\ref{eq:thm6pf_a4}), (\ref{eq:thm6pf_a6}), and (\ref{eq:thm6pf_a8}) to (\ref{eq:thm6pf_a}) yields
	\begin{align}\label{eq:thm6pf_alim}
		(a) \overset{p}{\rightarrow} \frac{1}{n} \E \biggr[ [f_1 (X) - f_0 (X) - \mathrm{ATE}]^2  \biggr] 
	\end{align}

Moving to $(b)$ in (\ref{eq:thm6pf_decomp}), 
	\begin{align}\label{eq:thm6pf_b}
		(b) &= \underbrace{ \frac{1}{n_0^2} \sum_{i: D_i = 0} \hat{w}_i^2 \epsilon_i^2 (0)  }_{(b.1)} \nonumber \\
		&+ \underbrace{ \frac{1}{n_0^2}\sum_{i: D_i = 0} \hat{w}_i^2 [f_0 (X_i) - \hat{f}_0 (X_i)]^2 }_{(b.2)} \nonumber \\
		&+  \underbrace{  \frac{2}{n_0^2} \sum_{i: D_i = 0} \hat{w}_i^2 \epsilon_i (0) [f_0 (X_i) - \hat{f}_0 (X_i)] }_{(b.3)}
	\end{align}
By condition (viii) for the proposition,
	\begin{align}\label{eq:thm6pf_b1}
		(b.1) \overset{p}{\rightarrow}  \frac{1}{n_0^2} \sum_{i : D_i =0} \hat{w}_i^2 \sigma^2 (0, X_i) 
	\end{align}
Then, by condition (ix) for the proposition, 
	\begin{align}\label{eq:thm6pf_b2}
		(b.2) = o_p(n^{-1})
	\end{align}
and using the Cauchy-Schwartz Inequality and (\ref{eq:thm6pf_b1}), 
	\begin{align}\label{eq:thm6pf_b3_pt1}
		| (b.3) | &\leq 2 \sqrt{(b.1)} \sqrt{(b.2)} \nonumber \\
		&\overset{p}{\rightarrow} 2 \sqrt{ \frac{1}{n_0^2} \sum_{i : D_i =0} \hat{w}_i^2 \sigma^2 (0, X_i)  } \sqrt{(b.2)}
	\end{align}
Then, because $\underset{n \rightarrow \infty}{\mathrm{plim}} \ \frac{1}{n_0} \sum_{i : D_i =0} \hat{w}_i^2 \sigma^2 (0, X_i)   $ is finite by condition (ii) for Proposition~\ref{thm:an_ate}, we find that $\frac{1}{n_0^2} \sum_{i : D_i =0} \hat{w}_i^2 \sigma^2 (0, X_i) $ in (\ref{eq:thm6pf_b3_pt1}) is $O_p (n^{-1})$. Additionally using (\ref{eq:thm6pf_b2}) allows (\ref{eq:thm6pf_b3_pt1}) to continue as
	\begin{align}\label{eq:thm6pf_b3_pt2}
		| (b.3) | \leq 2 \sqrt{O_p (n^{-1})} \sqrt{o_p (n^{-1})} = o_p (n^{-1})
	\end{align}
Therefore, applying (\ref{eq:thm6pf_b1}), (\ref{eq:thm6pf_b2}), and (\ref{eq:thm6pf_b3_pt2}) to (\ref{eq:thm6pf_b}) gives 
	\begin{align}\label{eq:thm6pf_blim}
		(b) \overset{p}{\rightarrow} \frac{1}{n_0^2} \sum_{i : D_i = 0} \hat{w}_i^2 \sigma^2 (0, X_i) 
	\end{align}
	
As for $(c)$, deriving its probability limit follows nearly identical steps as does deriving the probability limit of $(b)$. Therefore, we omit the derivation for $(c)$ and simply state the result. One finds that
	\begin{align}\label{eq:thm6pf_clim}
		(c) \overset{p}{\rightarrow} \frac{1}{n_1^2} \sum_{i : D_i = 1} \hat{w}_i^2 \sigma^2 (1, X_i) 
	\end{align}

Finally, applying (\ref{eq:thm6pf_alim}), (\ref{eq:thm6pf_blim}), and (\ref{eq:thm6pf_clim}) to (\ref{eq:thm6pf_decomp}) completes the proof.

\begin{flushright}
$\square$
\end{flushright}

\end{document}